Featured in Physics

# Hybrid Oscillator-Qubit Quantum Processors: Simulating Fermions, Bosons, and Gauge Fields

Eleanor Crane,[1,2,3,4,*] Kevin C. Smith,[5,6,7] Teague Tomesh,[8,9] Alec Eickbusch,[6,7,†] John M. Martyn,[10,11] Stefan Kühn,[12] Lena Funcke,[13,1,10,11] Michael Austin DeMarco,[5,1] Isaac L. Chuang,[1,2] Nathan Wiebe,[14,15,16] Alexander Schuckert,[3,17,‡] and Steven M. Girvin[6,7,§]

[1]*Department of Physics, Co-Design Center for Quantum Advantage, Massachusetts Institute of Technology, Cambridge, Massachusetts 02139, USA*
[2]*Department of Electrical Engineering and Computer Science, Massachusetts Institute of Technology, Cambridge, Massachusetts 02139, USA*
[3]*Joint Quantum Institute and Joint Center for Quantum Information and Computer Science, University of Maryland and NIST, College Park, Maryland 20742, USA*
[4]*Department of Physics, King's College London, Strand, London, WC2R 2LS, United Kingdom*
[5]*Brookhaven National Laboratory, Upton, New York 11973, USA*
[6]*Yale Quantum Institute, PO Box 208 334, 17 Hillhouse Avenue, New Haven, Connecticut 06520-8263, USA*
[7]*Departments of Applied Physics and Physics, Yale University, New Haven, Connecticut 06511, USA*
[8]*Department of Computer Science, Princeton University, Princeton, New Jersey 08544, USA*
[9]*Infleqtion, Chicago, Illinois 60604, USA*
[10]*Center for Theoretical Physics, Massachusetts Institute of Technology, Cambridge, Massachusetts 02139, USA*
[11]*The NSF AI Institute for Artificial Intelligence and Fundamental Interactions*
[12]*CQTA, Deutsches Elektronen-Synchrotron DESY, Platanenallee 6, 15738 Zeuthen, Germany*
[13]*Transdisciplinary Research Area "Building Blocks of Matter and Fundamental Interactions" (TRA Matter) and Helmholtz Institute for Radiation and Nuclear Physics (HISKP), University of Bonn, Nussallee 14-16, 53115 Bonn, Germany*
[14]*Department of Computer Science, University of Toronto, Canada*
[15]*Pacific Northwest National Laboratory, Richland, Washington, USA*
[16]*Canadian Institute for Advanced Studies, Toronto, Canada*
[17]*DIENS, École Normale Supérieure, PSL University, CNRS, INRIA, 45 rue d'Ulm, Paris 75005, France*

Key to solving many societally relevant problems in materials science, quantum chemistry, and nuclear physics is understanding their fundamental microscopic quantum behavior. All of these problems can be phrased as fermionic (e.g., electrons, quarks) and bosonic particles (e.g., lattice vibrations, photons) interacting with each other. However, solving these systems of strongly interacting fermions and bosons is out of reach for classical computers and also non-error-corrected qubit-only quantum computers. In this paper, we introduce a hybrid qubit and bosonic-oscillator quantum-simulation framework, which is so efficient that it could simulate some of these strongly correlated problems even without error correction. Specifically, we leverage the available universal bosonic gate set with $\mathcal{O}(1)$ operations per gate as a function of the bosonic cutoff. This enables our framework to provide exact decompositions of higher-order fermion-boson interactions, including notoriously challenging terms such as $(2+1)$D gauge theories. We show how to compile dynamics, ancilla-free partial error detection, nonlocal observable measurements, and ground-state energy estimation using both qubit-boson variational quantum eigensolvers and quantum signal processing. We introduce a "post-qubit quantum advantage" experiment which is out of reach for both near-term qubit-only quantum computers and classical computers. We justify this proposal by estimating bosonic gate fidelities through high–boson-occupation Lindblad simulation and by comparing to qubit-only algorithms. For the latter, we develop a new qubit-based algorithm for simulating a beam splitter with a boson number cutoff $N_{\max}$, which scales as $\tilde{\mathcal{O}}(N_{\max}^{1+o(1)})$, asymptotically outperforming previously known qubit methods. However, to reach the same circuit fidelity as qubit-boson hardware with 99.9% per-gate fidelity, a qubit quantum computer using the best-known method would still need fidelities of at least 99.9999%, likely requiring substantial error correction. Our framework can be used in superconducting hardware, which we focus on, as well as trapped-ion and neutral-atom hardware. This work establishes digital hybrid qubit-boson quantum simulation as a promising and advantageous method for solving the many challenging, strongly correlated, fermion-boson models of relevance to science and technology.

*Contact author: eleanor.crane@kcl.ac.uk
†Present address: Google Quantum AI, Santa Barbara, California, USA.
‡Contact author: alexander.schuckert@ens.fr
§Contact author: steven.girvin@yale.edu

## I. INTRODUCTION

Underlying many mechanisms in the natural world is the interaction between fermions and bosons, such as phonons mediating electron attraction, leading to superconductivity or the downconversion of light during photosynthesis. Examples of fermionic particles include electrons, holes, fermionic atoms, baryons, protons, neutrons, quarks, and leptons; examples of bosonic particles include photons, phonons, excitons, plasmons, Cooper pairs, bosonic atoms, mesons, W bosons, Z bosons, Higgs bosons, and gluons. When the interactions among these particles are strong, analytical and numerical methods fail, leading to a large range of open problems in high-energy physics, chemistry, and condensed-matter physics. In particular, quantum Monte Carlo simulations are effective only for static equilibrium properties (or imaginary-time dynamics) and only in special cases where there is no fermion sign problem. Tensor network methods work well only for systems with limited entanglement. Some of the hardest examples to simulate can be found in the lattice gauge theory (LGT) formulation of the standard model of particle physics [1]. LGTs constitute an especially difficult problem because, in addition to containing fermionic or bosonic matter, they also contain gauge fields that require the implementation of $N$-body interactions with $N > 2$, such as gauge-invariant kinetic energy ($N = 3$) and the plaquette magnetic-field term ($N = 4$).

Quantum simulators [2–4] offer an attractive alternative to study these problems as they can naturally host entangled states, which can be strongly correlated [5–7]; in addition, they can be used to study real-time equilibrium [8–10] and nonequilibrium dynamics [11–14]. Analog simulations of LGTs have been proposed [15–26] and carried out using, for example, cold atom [27–31], trapped-ion [32,33], and superconducting qubit platforms [34,35], but analog quantum simulators are limited in terms of the Hamiltonians that can be implemented and observables that can be measured. In particular, the multibody interactions appearing in LGTs make their analog simulation challenging, but they have recently been made accessible through inclusion of digital components [36]. Digital simulators can, in principle, realize multiqubit gates either by decomposition into two-qubit gates or directly using multiqubit gates (see, e.g., Refs. [37,38]), and proof-of-principle digital simulations of LGTs have been performed. However, the overheads encountered in mapping bosons and fermions to qubits make these problems challenging to implement on qubit-based discrete-variable (DV) hardware, where ion-trap and superconducting hardware, in particular, have led the way [35,39–51]. While digital qudit-based quantum processors offer a promising approach toward efficiently encoding the large Hilbert space of bosons and gauge fields [42,52–55], the qudit dimensions explored thus far offer only a slight increase in Hilbert space dimensions over qubits. More importantly, neither qubit nor qudit systems natively implement the bosonic operations necessary for creating bosonic Hamiltonian terms—in particular, synthesis of the occupation-dependent square-root factors needed for bosonic raising and lowering operators in qubit-based encodings leads to a large overhead for simulating bosonic models [56,57]. Simulation of purely bosonic $(1+1)$D theories using continuous-variable hardware has been proposed [58]—for example, using microwave cavities [59,60] or photonic devices [61]—but such an approach does not easily facilitate the study of models involving interactions between bosons and fermionic matter.

The $\mathbb{Z}_2$ lattice gauge theory with bosonic matter is reminiscent of the Higgs sector coupled to the non-Abelian $SU(2)$ electroweak gauge sector (W and Z bosons) of the $(3+1)$D standard model of particle physics with the two simplifications of a lower dimensionality and a $\mathbb{Z}_2$ gauge field instead of the non-Abelian $SU(2)$ electroweak gauge fields.

In this work, we investigate the promise of hybrid qubit-boson quantum processors [57] for solving the above challenges. In particular, we leverage this hardware to realize a straightforward mapping between model and computational degrees of freedom, as illustrated in Fig. 1. By encoding both the bosonic matter and gauge fields in native oscillator modes, we avoid the costly overheads incurred by boson-to-qubit encodings. More specifically, the primary goal of our approach is to explore the advantage given by the *native* availability of bosonic gate sets, a crucial component for simulation of Hamiltonians containing bosons yet costly to implement in qubit-based hardware. While the universal control of oscillator modes is challenging due to the equal level spacing of linear oscillators, here this issue is resolved by leveraging a hybrid qubit-boson architecture [62] with nonlinearity provided by the qubits, leading to non-Gaussian quantum states [63,64]. To that end, we build on the universal qubit-boson gate sets recently developed in a companion paper [57]. In particular, we focus on a circuit quantum electrodynamics (cQED) architecture with high-Q 3D cavities, although many of the hybrid qubit-boson operations and techniques we discuss can be implemented in other platforms such as ion traps [57,65–70], where a similar qubit-boson approach for gauge theories has been proposed [71], or neutral atoms in tweezer arrays [57,72,73] in which a coupling to digital error-corrected fermionic modes has also been proposed [74].

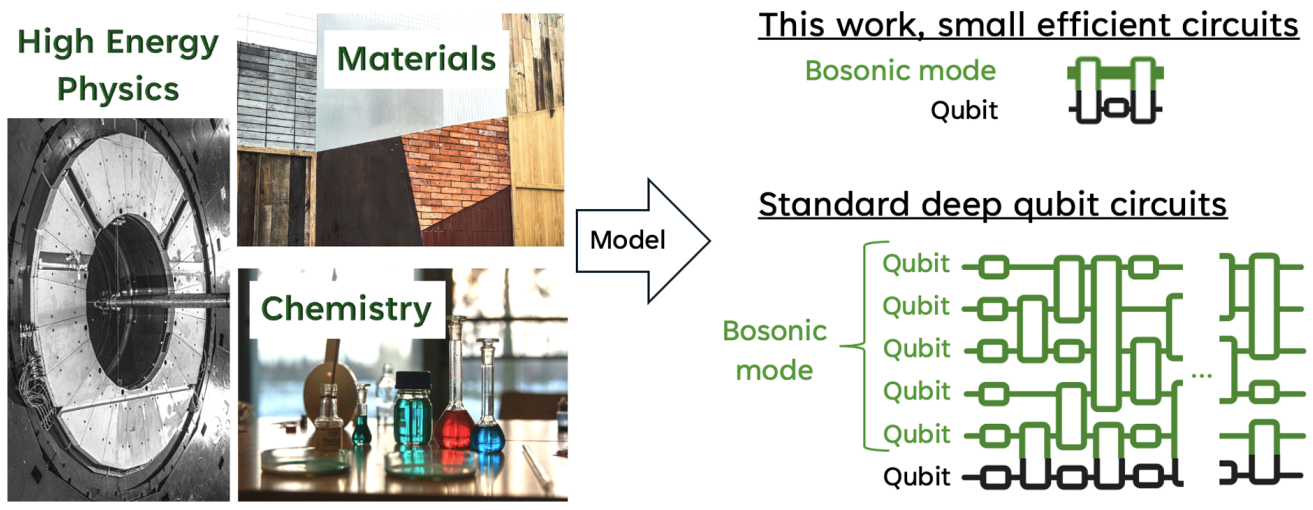

FIG. 1. Digital hybrid qubit-boson circuits. Mapping a physical fermion-boson problem to an all-qubit (discrete variable, DV) circuit leads to large qubit and gate counts. Instead, we propose to directly implement hybrid oscillator (continuous variable, CV)-qubit circuits.

We focus on the task of simulating dynamics, i.e., evolving an initial state under a time-dependent Hamiltonian $\hat{H}(t)$, using a digital approach in which we compile evolution under that Hamiltonian into elementary, native qubit-boson operations. Many quantum algorithms can be formulated with (controlled) time evolution under the Hamiltonian [75] as a fundamental subroutine, and we adapt some of them to qubit-boson processors to determine ground states and observables. The goal of our work is to find these compilations for the most important models in quantum many-body physics involving fermions, bosons, and gauge fields and to show that this decomposition leads to advantages compared to compilations using qubit-only hardware.

### A. Structure of the article

This work is structured as follows. We start by reviewing elementary concepts (Sec. I). We then present our main results on compilation strategies, which is divided into fundamental primitives (Sec. II), methods for matter fields (Sec. III), and gauge fields and their interactions with matter (Sec. IV). We then perform an end-to-end resource estimation of our methods and compare to all-qubit hardware (Sec. V). Finally, we introduce methods to measure observables, algorithms to perform dynamics simulation and ground-state preparation, and numerical benchmarks (Sec. VI). We conclude in Sec. VII.

As a guide to the reader, we provide more detail about each of the seven sections of this paper, in sequence. In Sec. I, i.e., the remainder of this introductory section, we give context by broadly introducing LGTs (Sec. I E) and the challenges associated with their simulation. We choose paradigmatic models [27,76] as a case study, the $\mathbb{Z}_2$-Higgs model and the $U(1)$ quantum link model, which we later use in Sec. VI to benchmark qubit-boson algorithms. We then introduce the superconducting hybrid qubit-bosonic oscillator hardware comprising transmon qubits coupled to high-Q microwave cavity resonators hosting bosonic oscillator modes on which we base our study (Sec. I F). Finally, we introduce Trotter algorithms (Sec. I G), which are the basic subroutines used in our qubit-boson algorithms.

In Sec. II, we introduce a toolset of hybrid qubit-boson compilation strategies used throughout this work. We review previously introduced methods but also develop exact methods for synthesizing bosonic parity-dependent and density-density operations, as well as oscillator-mediated qubit-qubit entangling gates that we extend to multi-qubit gates.

In Sec. III, we leverage this toolset to develop compilation strategies for implementing the time-evolution operator for common bosonic (Sec. III A) and fermionic (Sec. III B) Hamiltonian terms, including single-site potentials, nearest-neighbor and on-site interactions, and hopping terms in hybrid superconducting hardware. We consider both $(1+1)$D and $(2+1)$D. For bosonic matter sites, we utilize the direct mapping to a single bosonic oscillator mode of a microwave cavity resonator. For fermionic matter, we leverage a Jordan-Wigner mapping using either transmon qubits or dual-rail (DR) qubits in microwave cavities.

In Sec. IV, we present compilation strategies for implementing pure gauge and gauge-matter Hamiltonian terms, specializing to the case of $\mathbb{Z}_2$ and $U(1)$ gauge fields in $(2+1)$D.

In Sec. V, we introduce explicit Trotter circuits for the $\mathbb{Z}_2$-Higgs model and $U(1)$ quantum link model and perform an analysis of the Trotter errors, allowing us to derive the end-to-end asymptotic scaling of the number of gates required for a dynamics simulation, including the errors encountered in the compilation. We then perform an explicit gate count comparison between our qubit-boson and an all-qubit simulation of the bosonic hopping Hamiltonian. For this purpose, we develop a qubit algorithm in the Fock-binary encoding to simulate bosonic hopping.

In Sec. VI, we adapt algorithms for ground-state preparation and dynamics to our qubit-boson approach, utilizing the compilation methods from the previous sections. We use the $(1+1)$D $\mathbb{Z}_2$-Higgs model and $U(1)$ quantum link model as example models for numerical benchmarks using Bosonic Qiskit [77]. In particular, we develop a qubit-boson variational quantum eigensolver (VQE) (Secs. VI C and VI D) ansatz. Within the VQE approach and in the context of the $\mathbb{Z}_2$-Higgs model, we investigate the effects of hardware and shot noise on VQE ground-state preparation (Sec. VI D) and demonstrate the utility of postselection upon Gauss's law in improving ground-state preparation. Next, we show how to measure observables in these ground states, in particular, nonlocal observables such as string-order correlators and the superfluid stiffness (Sec. VI E). Finally, we discuss ground-state preparation with quantum signal processing (QSP) and compare the error in ground-state preparation as a function of the time to solution between QSP and the VQE (Sec. VI F).

In Sec. VII, we conclude with a summary of our main results, a discussion of the prospects for a quantum advantage using our approach, and an outlook on future research directions.

Before continuing, we briefly comment on the vocabulary and notation used throughout this work. We use the term bosonic mode (or just mode) to refer to bosonic-oscillator modes (qumodes); more broadly, we use the term qubit-boson as a synonym for qubit-oscillator. We use the term cavity to refer to experimental cavity resonators containing bosonic oscillator modes. In general, we refer to certain hybrid qubit-boson gates as being "controlled," meaning that the generator of the gate contains either a qubit or a mode number operator; i.e., the phase is proportional to the number of excitations in the state (the gate does not affect $|0\rangle$). We generally call gates "conditioned" when they obtain a state-dependent phase, i.e., when the generator contains a qubit $\hat{Z}$ or mode Fock-state-dependent phase, which could, for example be implemented using a projector.

### B. Key results

In this section, we provide the reader with key results that can be found in this work (see Fig. 2). These results relate to novel compilation tricks and the resulting quantum advantage of our new method over classical and qubit methods, new and improved qubit-only circuits for simulating bosons, and preliminary results on dynamics, ground-state generation, and observable measurement in noisy hardware with the new method.

First, we show that $(2+1)$D quantum link models can be efficiently mapped onto the bosonic cavity-qubit architecture that is currently being built in the context of quantum error correction (encoding qubits into bosonic oscillator modes) (Fig. 4 and Sec. I F).

The results on qubit-boson compilation are as follows.

(i) Arbitrary bosonic operations acting on one mode, controlled on the Fock space occupations of another mode, can be compiled into $\mathcal{O}(\log(N_{\max}))$ qubit-boson gates, where $N_{\max}$ is a maximum level on which the operator is chosen to act (Secs. II C and II D). This feature, in particular, overcomes the previous barrier to implementing mode-mode density-density interactions digitally in qubit-boson hardware (Sec. III A 4).

(ii) A multiqubit gate can be implemented by mediation via modes in a digital version of the Mølmer-Sørensen construction (Sec. II F).

(iii) A dual-rail $ZZ$ gate can be implemented in a reduced time of $2(\pi+\theta)/\chi$; here, $\chi$ is the dispersive interaction, which is shorter than the previous $4\pi/\chi$ even for the maximally entangling case $\theta = \pi/2$ (Sec. II H).

Next, we look at compiling simulation subroutines in qubit-boson architectures. We note that these decompositions could also be useful for qubit-only architectures by combining our results with qubit-only compilations of the bosonic operations in Table II. Our findings are as follows.

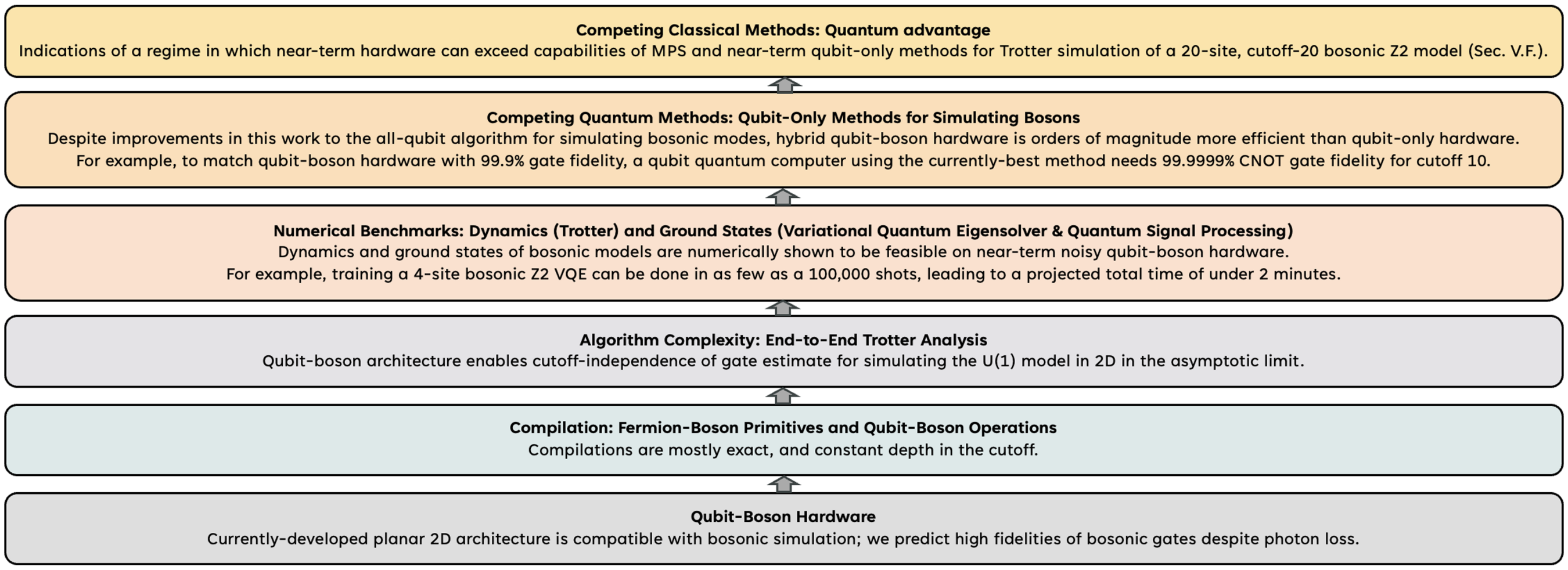

FIG. 2. Overview of the structure. See Sec. I B for an overview of the key results.

(i) The bosonic on-site interaction can be implemented via a single selective number-dependent arbitrary phase (SNAP) gate at the cost of a single ancilla (Sec. III A 3), and the ancilla can be removed when simulating the $\mathbb{Z}_2$ lattice gauge theory (Appendix B).

(ii) A Jordan-Wigner fermion-to-qubit mapping can be implemented in planar dual-rail architectures by employing the $ZZ$ gate introduced in Sec. II F. We compare dual-rail qubit and transmon qubit implementations for simulating fermions using current fidelity estimates, finding that dual-rail qubits achieve an order-of-magnitude-per-gate higher fidelity at the cost of a shot overhead of a factor of around 10 for 100 gates (Sec. III B).

(iii) We show how to implement phonon-matter as well as light-matter interactions using qubit-boson hardware, with fermions encoded in either transmons or dual-rail qubits (Secs. III C 1 and III C 2).

(iv) We find an explicit construction for how to implement the $\mathbb{Z}_2$ magnetic plaquette field term in dual-rail qubits in constant depth (Sec. IV A 2).

(v) We recast the $U(1)$ electric-field term into a form that can be implemented with two SNAP gates (Sec. IV B 1).

(vi) One of our major results is a constant-in-cutoff construction for the $U(1)$ magnetic-field term (Sec. IV B 2) with scaling $\mathcal{O}(t^{5/4}g^{-5/2})$, where $t$ is the Trotter step and $g$ is the coupling strength. This finding is opposed to the naive expectation that because of its eight bosonic raising or lowering mode operators, it should have a complexity of the fourth power of the cutoff and is the first compilation in any platform to achieve constant depth.

(vii) We show that all gauge-invariant hopping terms in $\mathbb{Z}_2$ and $U(1)$ LGT can be implemented in constant depth for bosonic and fermionic matter (Sec. IV D) using square-lattice cavity-qubit arrays (Fig. 15), utilizing both transmon qubits and dual-rail qubits to encode fermions. In this context, we also introduce a novel scheme using three-wave mixing (Appendix C), which reduces the number of total gates needed compared to two-wave mixing.

Our results for Trotter simulation end-to-end complexity analysis are listed below.

(i) We introduce a general framework for analyzing qubit-boson Trotter simulation errors based on introducing a virtual cutoff (Sec. V A). To our knowledge, the gate estimates below are the first end-to-end estimates for simulating lattice gauge theories in qubit-boson hardware.

(ii) Using this framework, we show that the number of gates to simulate bosonic $\mathbb{Z}_2$ LGT up to time $T$, a 2D $L\times L$ lattice, and error $\epsilon$ scale as $\mathcal{O}(L^{11}T^{1.5}/\epsilon^{0.5})$ for a second-order Trotter formula and constant filling fraction (Sec. V B).

(iii) The number of second-order Trotter steps for simulating 2D $U(1)$ LGT in the large cutoff limit scales as $\mathcal{O}(S^3L^3T^{1.5}/\epsilon^{0.5})$ (Sec. V C). However, it is important to note that the cutoff dependence of the number of gates drops out in the asymptotic limit of large $L$, $T$, and small $\epsilon$ because of the interplay between Trotter and compilation errors [cf. Eq. (145) in Sec. V D], even for a large polynomial scaling of $S\propto L^\alpha$ with $\alpha\leq 7$. Instead, we find that the number of gates scales as $\mathcal{O}(T^5L^{10}/\epsilon^4)$ with respect to these parameters. To our knowledge, this finding is the first gate estimate for simulating this model that is cutoff independent in the asymptotic limit.

Next, we list the prospects for a quantum advantage, as well as the advantage of qubit-boson hardware over qubit-only hardware.

(i) In order to compare to an all-qubit architecture, we compute the gate counts necessary for the simulation of a beam splitter in all-qubit hardware (Sec. V E). Previously known results required $\mathcal{O}(N_{\max}^2\log_2(N_{\max}))$ CNOT gates to simulate a beam splitter with a bosonic cutoff in Hilbert space of $N_{\max}$ [56]. We improve on these results in Appendix D by developing two new algorithms employing arithmetic circuits (for calculating square roots on the fly) and look-up tables (for avoiding on-the-fly square-root calculation at the cost of asymptotically higher depth and gate count), finding an algorithm for simulating a bosonic beam splitter in all-qubit hardware that scales as $N_{\text{gates}}^{(\text{qubit})}=\mathcal{O}(N_{\max}^{1+\frac{1}{2p}}\log_2(N_{\max}))$, where $p$ is the Trotter order.

(ii) We perform an end-to-end gate complexity analysis and show that, using the currently best-available qubit-only methods for simulating bosons at small cutoffs, the number of gates for a single beam splitter with cutoff 10 is more than 3 orders of magnitude higher using all-qubit hardware than for oscillator qubit hardware (Fig. 17 in Sec. V E). In other words, to reach the same circuit fidelity as qubit-boson hardware with 99.9% per-gate fidelity, a qubit quantum computer needs 99.9999% fidelity, which likely requires substantial error correction.

(iii) A key result is the demonstration of a crossover to quantum advantage (Fig. 19 in Sec. V F). We show that qubit-boson hardware can likely exceed the reach of matrix-product-state TEBD simulations for simulating time evolution of a 20-site, cutoff 20 $\mathbb{Z}_2$ model, starting in a high-occupation initial state. Near-term conditional-parity fidelities of 99% could yield quantum advantage, and longer-term fidelities of 99.9% would do so at extremely short times. For this purpose, we predict the cutoff dependence of qubit-boson gates using a Lindblad simulation of conditional-parity and beam-splitter gates in the presence of photon loss and qubit decay and dephasing, and we apply those to predict the circuit fidelity of these high-occupation $\mathbb{Z}_2$ model Trotter circuits (Appendix H). As this is the most favorable regime for matrix-product simulations (2D would be much harder), these results show the prospect for quantum advantage of this hardware even without error correction and for a small number of physical elements.

Qubit-boson quantum algorithms and numerical benchmarking results are as follows.

(i) We show that current noisy hardware can simulate Trotterized dynamics of the soft-core bosonic $\mathbb{Z}_2$ model on five sites up to 85 Trotter steps by numerically simulating the circuit in the presence of currently realistic hardware noise (Fig. 21 in Sec. VI B). This result should be compared with current state-of-the art analog simulators [94,95], which only realize two sites and qubit digital simulations [96], only simulating the hard-core-boson version.

(ii) We design a qubit-boson variational quantum eigensolver ansatz for the $\mathbb{Z}_2$ model, which is able to reach less than $10^{-3}$ relative error for two sites and four bosons in four layers (Sec. VI C 1), and for the $U(1)$ model, which is able to reach a relative error of $10^{-7}$ for three sites and two gauge fields and two layers and reproduce its first-order phase transition (Fig. 23 in Sec. VI C 2).

(iii) We propose a method for measuring the expectation value of the beam splitter [Eq. (161) in Sec. VI C 1], which we use to propose an experimental method for measuring the energy of the $\mathbb{Z}_2$ and $U(1)$ lattice gauge theory for VQE training.

(iv) We simulate the VQE training of the $\mathbb{Z}_2$ model under the influence of both shot noise and realistic hardware noise in the most challenging regime of competing Hamiltonian terms (Fig. 24 in Sec. VI D). To do so, we design an iterative procedure, increasing shot count as gradients flatten and using a carefully tuned averaging procedure. Without hardware noise, we reach a fidelity greater than 95% in 2500 iterations in as few as 115,000 shots. In the presence of hardware noise, an energy well below the first excited state is reached in a similar number of iterations.

(v) We also find an ancilla-free circuit using midcircuit measurement, which enables us to partially post-select on Gauss's law during the energy measurement (Sec. VI D 5). We numerically show that this halves the energy error in our noise model (Fig. 25).

(vi) We introduce compilation methods for measuring many relevant correlation functions, including the superfluid density that can be extracted from ground-state energies under twisted boundary conditions (Sec. VI E). Most importantly, we show how to measure string-order correlators using a perturbative scheme that has a trade-off between measurement precision and number of shots, showing that a value of 10,000 shots enables good resolution on the string-order correlator on the numerically prepared noisy VQE ground state (Fig. 27).

(vii) We use these methods to numerically show that the VQE state obtained above reproduces the exact correlation functions to high precision (Fig. 26).

(viii) We introduce qubit-boson quantum signal processing a method with provable performance guarantees to prepare ground states (Appendixes E and F). We numerically show that the time-to-solution to prepare the ground state of the $\mathbb{Z}_2$ model to a specified precision is comparable to the VQE (Fig. 28), which is surprising given that the VQE is heuristic and is therefore naively expected to be faster. However, because of the deep QSP circuits, it only becomes competitive once qubit-boson gate infidelities are reduced well below $10^{-4}$, favoring the shot trade-off of the VQE in the near term, which can tolerate lower fidelity gates.

(ix) We finally compare the influence of noise on qubit-boson hardware and all-qubit hardware for simulating bosonic models (Sec. VII and Appendix G), pointing out that in all-qubit hardware with binary encoding, single-qubit bit-flip noise can lead to an effective loss of $2^n$ bosons, which could even pose challenges for error-corrected all-qubit hardware.

Taken together, these results establish hybrid oscillator-qubit processors as a scalable and experimentally viable route to post-qubit quantum advantage for strongly correlated lattice gauge theories and related fermion-boson models.

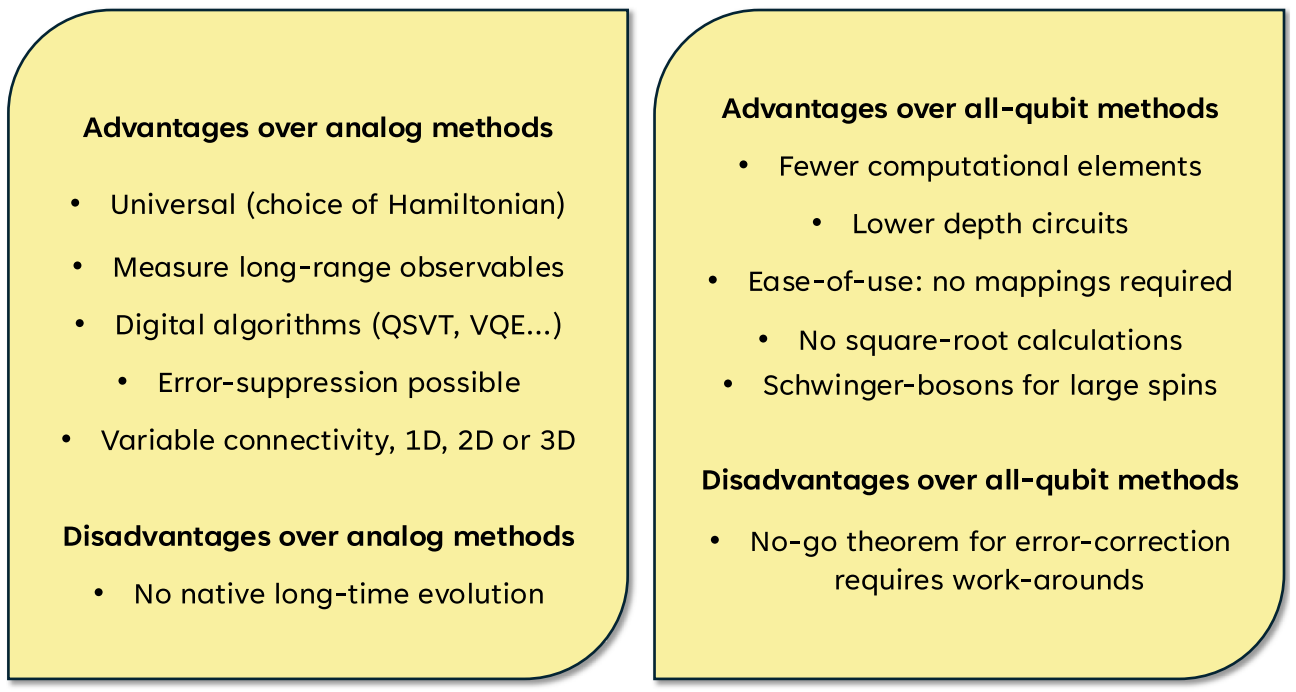

FIG. 3. Advantages of digital qubit-boson quantum simulation over other approaches for the simulation of bosonic quantum simulation problems.

### C. Advantages over classical, analog, and qubit approaches

We summarize the advantages of qubit-boson processors in Fig. 3 and Table III, and we explain them below.

The basic classical difficulty of bosonic problems with large local cutoff $N_{\max}$ is that the number of amplitudes that can fluctuate grows very quickly. Exact diagonalization must store and evolve the amplitudes for every allowed boson-occupation pattern, giving a Hilbert space that scales as $N_{\max}^L$ for $L$ unrestricted bosonic sites. Sparse and fixed-number methods reduce this basis, but when bosons bunch or fluctuate strongly across sites, the remaining sector can still be combinatorially large; for example, the fixed-number sector used in Sec. V F already has dimension $\binom{32+20-1}{19} \approx 2^{46}$. These sizes are useful for benchmarking, but they quickly exceed the reach of brute-force dynamics brute-force dynamics, with the largest comparable sparse simulations reaching Hilbert spaces of order $2^{40}$ [98].

Tensor-network methods reduce this cost by keeping only the entanglement structure that is actually generated. This approach is powerful for low-entanglement one-dimensional states, but for bosons, the local physical index itself is large: Optimized time-evolving block decimation (TEBD) for short-range one-dimensional bosonic dynamics scales as $LN_{\max}^2\chi^3$ [99], where the $N_{\max}^2$ comes from applying two-site gates on two local Hilbert spaces and $\chi$ is the bond dimension. Standard singular-value-decomposition (SVD)-based implementations can scale more poorly in $N_{\max}$, and $\chi$ must grow when real-time dynamics, quenches, or scattering generate entanglement [100,101]. In higher dimensions, the bottleneck is even more brutal because tensor contractions must sum over networks of virtual bonds not just local physical indices; projected entangled pair state (PEPS) algorithms therefore have steep bond-dimension costs, up to $\mathcal{O}(\chi^{10})$ in common contractions [102–105].

Other classical methods are complementary but rely on approximations that are stressed by the same physics. Quantum Monte Carlo (QMC) methods are efficient for some equilibrium and sign-problem-free systems; however, real-time amplitudes are oscillatory rather than positive probabilities, and fermions or frustration lead to cancellations. Density-functional theory (DFT), density-functional perturbation theory (DFPT), Migdal-Eliashberg theory, and configuration-coordinate approaches are practical because they reduce the many-body problem to effective single-particle, perturbative, or reaction-coordinate descriptions; however, strong anharmonicity, polarons, defects, and correlated electrons can undermine these reductions [82–84,106]. Keldysh, nonequilibrium mean-field, and variational methods can treat driven systems, but they require truncating correlations or choosing restricted ansatz classes; nonlinear phonons and long time evolution can generate correlations outside those closures [85,86]. Likewise, hierarchical equations of motion (HEOM), multi-configuration time-dependent Hartree (MCTDH), surface-hopping, and related nonadiabatic molecular methods are effective for selected modes or bath structures, but many active modes, long bath memory, low temperature, and branching nuclear wave packets rapidly multiply the required hierarchy elements, configurations, or trajectories [79,81,107,108]. Qubit-boson hardware is therefore targeted at regimes where several of these bottlenecks coincide: large $N_{\max}$, strong anharmonicity, real-time dynamics, fermions, and high entanglement.

There are many advantages of a digital qubit-boson approach over analog bosonic quantum simulation. First and foremost, it is possible to rely on the extensive work that enables error correction for qubits, and there have been proposals toward error suppression for the bosonic modes [109–119]. We note that Gaussian resources alone cannot correct Gaussian errors, as established by the no-go theorem of Ref. [120]; the proposals above circumvent this issue by employing non-Gaussian elements, as in the oscillator-into-oscillators encoding of Ref. [114]. A digital approach enables a flexible implementation and study of different Hamiltonians, as well as measurement of long-range observables such as the string-order correlator. We can also adapt existing quantum simulation algorithms to the qubit-boson platform (Trotter, phase estimation, variational quantum eigensolver, quantum signal processing, etc.). A straightforward generalization to higher-dimensional systems is also possible. A particular advance of our study is to show how to simulate the plaquette term in 2D with a gate complexity independent of the boson cutoff $N_{\max}$, which had been an open challenge in the field.

We note that it would also be possible to implement these methods in different physical platforms. While we have focussed on superconducting hardware, our approach can be directly implemented in trapped ions and neutral atoms as well, which have seen rapid progress in the coherent manipulation of bosonic modes (see, e.g., Ref. [73]).

Finally, to simulate bosonic problems on a qubit device, the bosonic modes need to be mapped to qubits. We make comparisons and discuss this issue in Sec. V. The mappings that are often used are the unary or one-hot encodings, which require $O(N_{\max})$ qubits; the binary or Gray encodings, which only require $\log_2(N_{\max})$ qubits; or the Jordan-Lee-Preskill encoding for which a discretization must be chosen in phase space that renders it approximate.

More important than the overhead in terms of computational elements is the problem of implementing natively bosonic operations. In particular, this case requires the calculation of square roots from the action of the creation or annihilation operators, e.g., $a|n\rangle = \sqrt{n}|n-1\rangle$. In all-qubit hardware, they are best calculated using Newton iterations [121] (see Sec. V), which are costly and approximate. As we discuss in Sec. V E and Appendix D, the algorithm we introduce for simulating a single bosonic unitary has the best-known asymptotic gate cost, scaling at least linearly with respect to $N_{\max}$. For a bosonic quantum computer, only a single natively available operation is required. Therefore, it is also easier and more straightforward to use qubit-boson processors for the quantum simulation of qubit-boson quantum simulation problems as the question of mapping the bosons to qubits does not arise. Furthermore, it is possible to use a native implementation of the Schwinger-boson mapping for the simulation of large spins. In the same way as for qubit devices, qubit-boson processors can benefit from digital algorithms such as Trotter, VQE, QSP, midcircuit-measurement, and feed-forward, etc. It would be possible to use error correction to improve the qubit lifetimes, which are the limiting factor in qubit-boson processors. There have also been proposals for boson-to-boson error suppression [109–118]; however, more work needs to be done in this direction.

### D. Open fermion-boson problems

Qubit-boson devices of the sort proposed in this work would be particularly suited for the simulation of systems that involve large boson fluctuations. This is because for both classical numerical simulations and quantum simulations using qubit quantum computers, simulations of systems with small fluctuations are easy: They require only a small number of discrete states per site to be represented and, correspondingly, only a small number of operations. While we make these statements more precise in Appendixes D and G, here we discuss several important challenges in quantum simulation, which involve large boson fluctuations and would benefit from qubit-boson devices. The general difficulty of simulating large bosonic cutoffs and anharmonicity with classical, analog, and qubit-based methods is discussed in Sec. I C. In this section, we sometimes mention which classical method has been used to tackle aspects of the respective problem.

*Chemistry and biology.* A crucial area is the class of quantum chemistry problems in which the interactions of the vibrations of the ion cores with the electronic wave function play an important role, i.e., situations in which the Born-Oppenheimer approximation breaks down [122]. For example, understanding how energy transfer can occur with such high efficiency (around 99%) in photosynthetic systems—given they are disordered, liquid, open systems—could inform future energy-transfer technology design for biological medium [123]. Aspects of toy models have been solved using HEOM and related vibronic models [79,124,125]; however, challenging large cutoffs appear because many vibronic modes in the protein environment can jointly participate in ultrafast energy transport, and effective anharmonicity is generated by qubitlike excitonic degrees of freedom strongly coupled to bosonic modes.

Similarly, an open chemistry problem that could raise the efficiency of solar cells above the Shockley-Queisser limit (around 33%) is to quantitatively predict which materials undergo efficient singlet fission and separate the resulting triplet excitons [126–128], which would enable one high-energy photon to generate two electronic excitations instead of one. Aspects of this problem have been studied with MCTDH and quantum-chemistry screening methods [80,107,129–133]; however, many phase-space points appear because triplet-pair formation, separation, and charge transfer depend on strongly coupled molecular vibrations, and anharmonicity appears through electronic-state-dependent vibrational potentials.

In photodynamic therapy, it is not yet possible to predict which photosensitizers simultaneously optimize absorption, intersystem crossing, photostability, and reactive oxygen generation; solving this issue would accelerate the discovery of targeted light-activated cancer treatments [134]. Aspects of this problem have been approached with medical-physics models, quantum chemistry, and nonadiabatic dynamics [81,135–137], but many phase-space points are needed because triplet-yield dynamics depend on coupled vibrational and spin-vibronic coordinates, with anharmonicity arising from excited-state molecular distortions.

More broadly, vibronic effects are central to nonradiative relaxation and ultrafast energy-transfer processes [138,139], nonadiabatic charge dynamics in photovoltaic materials and organic conjugated systems [140,141], conical intersections [142,143] (see, e.g., small-scale qubit-boson demonstrations in trapped ions [69] and superconducting qubits [144]), Jahn-Teller and pseudo-Jahn-Teller interactions [145], molecular photomagnets [146–150], thermally activated delayed fluorescence materials for OLEDs [151–154], sunscreen molecules [155–158], and DNA photostability and repair [159–161].

*Materials science.* It is not yet possible to microscopically predict nonradiative recombination and defect-assisted carrier capture in perovskite halides; solving this issue would guide the design of cheaper, more stable, and more efficient solar cells. Aspects of this problem have been treated with DFT, DFPT, spectroscopy-guided polaron models, and nonadiabatic carrier-capture calculations [82], but large phonon cutoffs appear because soft polar and anharmonic lattice modes can be strongly displaced during carrier relaxation and defect-assisted recombination.

Similarly, it is not yet possible to quantitatively predict defect-mediated carrier capture and persistent photoconductivity in semiconductors; solving this issue would improve photodetectors, optoelectronic devices, and defect-engineered materials. Aspects of this problem have been studied using configuration-coordinate and first-principles defect calculations [83,84], but strong anharmonicity and large cutoffs arise from large lattice relaxations around defects.

In cavity-coupled materials, it is not yet possible to predict nonequilibrium phase diagrams and fluctuation-engineered phases; solving this issue would turn structured electromagnetic environments into a design tool for superconducting, magnetic, moiré, and topological matter. Aspects of this problem have been treated with nonequilibrium mean-field, Keldysh, and material-specific nonlinear-response calculations [85], but bosonic occupation can range from vacuum fluctuations to intense coherent drives, while the effective anharmonicity depends on the driven phonon or polariton mode.

TABLE I. Overview of areas of application for qubit-boson quantum simulation. The boson-cutoff column indicates representative bosonic Hilbert-space truncation cutoffs or phase-space requirements derived from theoretical or numerical treatments of the corresponding systems.

| Area | Boson cutoff $n$ or phase-space points $K$ | Anharmonicity | Open computational question (examples) |
|---|---|---|---|
| Photosynthetic energy transfer | $n > 100$ [79] | Weak | Nonequilibrium vibronic transport with multiple phonon modes, disorder, and multi-excitation effects |
| Singlet fission solar cells | $K \sim 16$ [80] | Strong | Triplet-pair formation, separation, and harvesting |
| Photodynamic therapy | $K \sim 10$–$100$ [81] | Weak–moderate | Prediction of absorption, intersystem crossing, spin-vibronic effects, and triplet yield |
| Perovskite halides | $n > 300$ [82] | Strong | Microscopic prediction of nonradiative recombination and defect-assisted carrier capture |
| Photoconductivity in semiconductors | $n > 50$ [83,84] | Strong | Quantitative defect-mediated carrier capture and persistent photoconductivity |
| Cavity quantum materials | $n \sim 1$–$10^{11}$ [85] | Weak–strong | Prediction of nonequilibrium phase diagrams and fluctuation-engineered phases |
| Light-induced superconductivity | $n < 1$ [86] | Strong due to drive | Microscopic prediction of transient superconductivity from driven nonlinear phonons, including electron heating, phonon squeezing, and material-specific nonlinear couplings |
| Lattice gauge theories | $n \sim 10$–$100$ [87,88] | Strong | Real-time dynamics, e.g., scattering |

More broadly, it is not yet clear how much phonons control conventional, high-temperature, and light-induced superconductivity at the microscopic level; solving this issue would clarify how to design better superconductors and transient superconducting phases [162]. Aspects of equilibrium electron-phonon models such as the Holstein-Hubbard model have been studied with determinant quantum Monte Carlo and tensor-network methods [163–165], while driven superconductivity has been modeled with nonequilibrium variational and Keldysh methods [86]; however, strong effective anharmonicities appear because phonon fluctuations, squeezing, and nonlinear driven phonon coordinates can couple strongly to fermionic electrons even when the mean occupation is small.

*High-energy and nuclear physics.* It is not yet possible to compute generic real-time dynamics of lattice gauge theories in regimes relevant to scattering, early-universe dynamics, heavy-ion collisions, and dense nuclear matter; solving this issue would connect quantum simulation directly to otherwise inaccessible questions in nuclear and high-energy physics [43]. In this setting, the field theory of the strong force is discretized into lattice gauge theories in order to predict, for example, the properties of nuclei [166]. Aspects of these theories have been studied using Euclidean lattice gauge theory, quantum Monte Carlo, tensor-network methods, and quantum-simulation resource estimates; however, the bosonic cutoff appears when continuous gauge fields are truncated to finite-dimensional link Hilbert spaces, and relevant gauge-field occupations can grow in real-time scattering and strong-field regimes [87,88]. As lattice gauge theories are strongly interacting models of fermions and bosons, they represent one of the most natural applications of qubit-boson quantum computers.

### E. Lattice gauge theories

In the following, we briefly introduce the LGTs we will use as paradigmatic examples, a $\mathbb{Z}_2$-Higgs model and a $U(1)$ quantum link model; we show how to compile them onto qubit-boson hardware in the rest of the study. Here, we explain why we choose these models and introduce the Hamiltonians in order to prepare our discussion on how to implement them. In Sec. VI A, we introduce the physics in the context of our numerical experiments, especially emphasizing past work. Note that these models only serve as examples; because of the digital nature of our methods, our study enables the simulation of a wide range of models involving bosons, fermions, and gauge fields.

In lattice gauge theories, matter fields reside on sites and gauge fields reside between sites, usually (for the case of $2+1$D) in a square lattice geometry. We choose a model with bosonic matter similar to the Higgs model, called the $\mathbb{Z}_2$-Higgs model, and a model with fermionic matter, the $U(1)$ quantum link model.

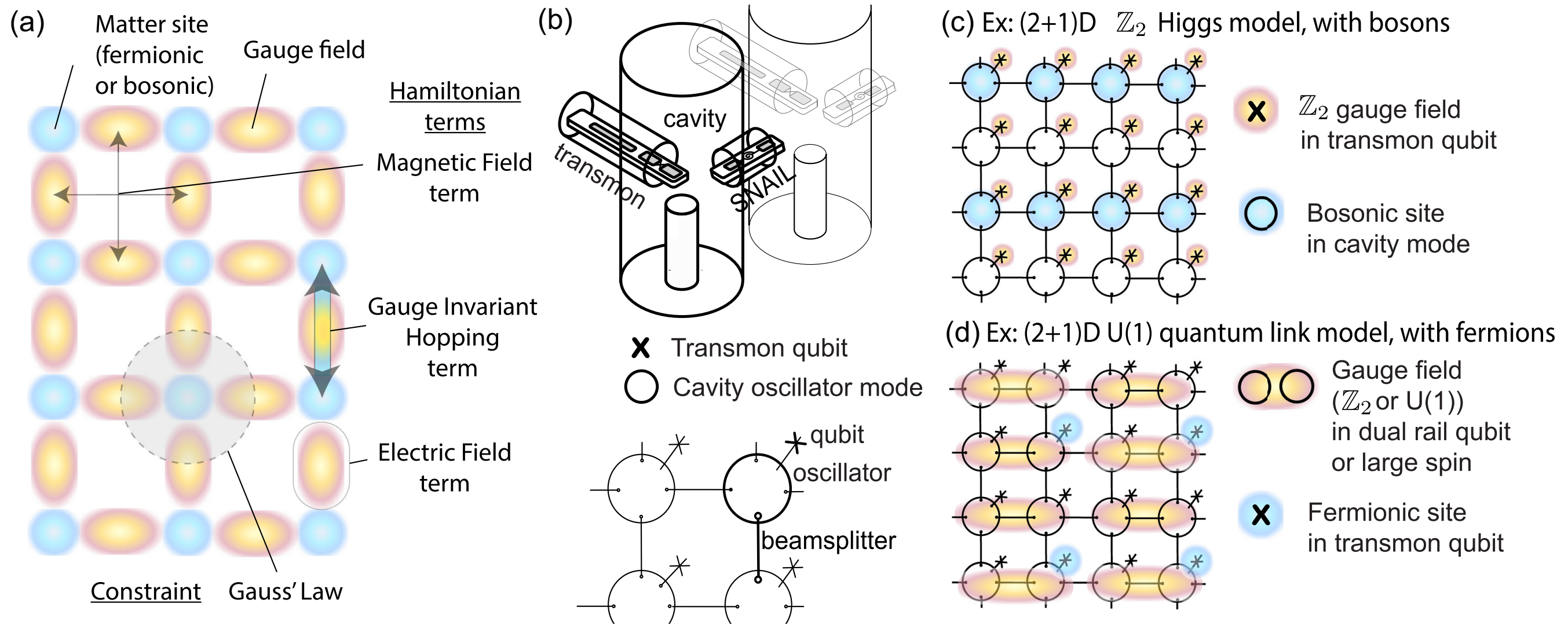

FIG. 4. Mapping lattice gauge theory models to qubit and oscillator degrees of freedom in circuit QED. (a) Lattice gauge theories. Components (matter sites and gauge field), Hamiltonian terms (gauge-invariant hopping, electric field, magnetic field), and constraints (Gauss's law) of lattice gauge theory models. (b) Top diagram: 3D cavity architecture. Transmon qubits couple to the lowest-energy mode of the cavity by overlapping with the mode. SNAILs couple two cavities to each other. The cavities are approximately 50 mm long [78], with a diameter of approximately 10 mm. Bottom diagram: corresponding schematic of an array of 3D cavities. Concentric circles represent cavities, implementing oscillators; crosses represent transmon qubits, implementing qubits; lines denote SNAILs, which are used to implement beam-splitter operations between cavities. Examples of models mapped to this architecture are shown in panels (c) and (d). (c) $\mathbb{Z}_2$-Higgs model. The gauge fields are mapped to the transmon qubit degrees of freedom, and there is one bosonic matter site per cavity, represented by one of the bosonic microwave modes in that cavity. (d) $U(1)$ quantum link model. Using the Schwinger-boson encoding, the gauge fields are mapped to a pair of modes in two adjacent cavities, and the fermionic matter is mapped to transmon qubits. In $(2+1)$D, there are twice as many gauge-field links as there are matter sites, which can be seen on the illustration. High-fidelity SWAP operations between cavity modes enable the requested connectivity for this model.

The $\mathbb{Z}_2$-Higgs model lattice gauge theory was introduced by Wegner [167] as a simple model showing signatures of confinement and deconfinement at the critical point [168], making it an ideal test bed for early testing of novel methods for LGT simulations. In two spatial dimensions in the absence of an electric field, its ground state encodes a spin-$1/2$ degree of freedom and can therefore be used for quantum error correction [169,170]. Note that $\mathbb{Z}_2$ lattice gauge theory with bosonic matter is reminiscent of the Higgs sector coupled to the non-Abelian $SU(2)$ electroweak gauge sector (W and Z bosons) of the $(3+1)$D standard model of particle physics with the two simplifications of a lower dimensionality and a $\mathbb{Z}_2$ gauge field instead of the non-Abelian $SU(2)$ electroweak gauge fields.

The $U(1)$ quantum link model can be connected to a continuum field theory by introducing a lattice spacing $a$ by $J \to J/a$ and $g \to g\sqrt{a}$ and taking the limit $a \to 0$. In $(1+1)$D, this model shares many properties with quantum chromodynamics in $(3+1)$D: chiral symmetry breaking, confinement, a $U(1)_A$ quantum anomaly, and a topological $\tau$-term, which gives rise to a nontrivial topological vacuum structure [171,172].

*The $\mathbb{Z}_2$-Higgs model.* In this model, the gauge fields are spin-$1/2$ degrees of freedom. The Hamiltonian is written as follows [167,173]:

$$\hat{H}_{\mathbb{Z}_2} = -g\sum_{\langle i,j\rangle}\hat{X}_{i,j} - B\sum_{i,j,k,l\in\square}\hat{Z}_{i,j}\hat{Z}_{j,k}\hat{Z}_{k,l}\hat{Z}_{l,i} - J\sum_{\langle i,j\rangle}(\hat{m}_i^\dagger\hat{Z}_{i,j}\hat{m}_j + \text{H.c.}) + U\sum_i\hat{n}_i^2, \quad (1)$$

where $i$, $j$ indicates the link between lattice site $i$ and $j$ on a square lattice, $\sum_{\langle i,j\rangle}$ indicates a sum over all combinations of $i$, $j$ such that $i$ and $j$ are nearest neighbors, and $\square$ is a plaquette of four sites [see Fig. 4(a)]. The annihilation operator at matter site $i$ is given by $\hat{m}_i$ and the density or occupation by $\hat{n}_i = \hat{m}_i^\dagger\hat{m}_i$. The matter can be either bosonic ($\hat{m} = \hat{a}$) or fermionic ($\hat{m} = \hat{c}$), with corresponding commutation and anticommutation relations $[\hat{a}_i, \hat{a}_j^\dagger] = \delta_{ij}$ and $\{\hat{c}_i, \hat{c}_j^\dagger\} = \delta_{ij}$. Note that $\hat{X}_{i,j}$ and $\hat{Z}_{i,j}$ are Pauli operators that act on the gauge field at the link between sites $i$ and $j$. The first, second, and third terms in Eq. (1) represent the electric-field (coupling $g$), magnetic-field (coupling $B$), and gauge-invariant hopping (coupling $J$), respectively. The last term is an on-site interaction of strength $U$, which is the simplest possible interaction term for bosonic matter; however, it does not contribute as an interaction term for (spinless) fermions since $\hat{n}^2 = \hat{n}$, due to the Pauli exclusion principle.

The defining feature of gauge theories is invariance under local gauge transformations. This feature leads to the presence of Gauss's law dividing the Hilbert space of the theory into sectors that are not coupled by the Hamiltonian. This constraint is in analogy to the relation between the integral of the electric-field flux through a surface and the charge enclosed by the surface in classical electrodynamics, with a product of field operators replacing the integral. Note that $(-1)^{n_i}$ takes the role of the charge. In quantum mechanics, Gauss's law is formulated as a constraint on the eigenstates $|\Psi\rangle$; for all $i$, $\hat{G}_i|\psi\rangle = q_i|\psi\rangle$, where $q_i \in \mathbb{Z}$ and

$$\hat{G}_i = \prod_{\langle i,j\rangle} \hat{X}_{i,j}(-1)^{\hat{n}_i}, \tag{2}$$

in $(2+1)$D, where the product is over the four links $\langle i,j\rangle$ that connect to site $i$ [see Figs. 4(a) and 4(c)]. We choose $q_i = 1$ throughout this work. For open boundary conditions, terminated by matter sites, the Gauss law constraints on the boundary are fixed by imagining virtual qubit sites around the system and fixing them in a $\hat{X}$ product state. We set $\hat{X} = 1$ for these qubits when necessary.

*The $U(1)$ quantum link model.* The $U(1)$ LGT, which is equivalent to quantum electrodynamics in the limit of vanishing lattice spacing, hosts continuous gauge fields. The quantum link formalism [174,175] replaces the gauge fields with discrete spin-$S$ degrees of freedom with commutation relations $[\hat{S}_i^z, \hat{S}_j^+] = \delta_{ij}\hat{S}_i^+$ and $(\hat{S}_i^z)^2 + \frac{1}{2}(\hat{S}_i^+\hat{S}_i^- + \hat{S}_i^-\hat{S}_i^+) = S(S+1)$ [176–179], where $S$ is the spin length leading to a Hilbert space cutoff of $2S+1$. The physics of continuous $U(1)$ gauge fields are recovered for $S \to \infty$ [180]. Choosing (large) finite $S$, this encoding approximately preserves the bosonic algebra. The Hamiltonian in this formalism reads [175]

$$\begin{aligned}\hat{H}_{U(1)} = {}& \frac{g^2}{2}\sum_{\langle i,j\rangle}\left(\hat{S}_{i,j}^z + \frac{\tau}{2\pi}\right)^2 \\ & - \frac{1}{4g^2(S(S+1))^2}\sum_{i,j,k,l\in\square}(\hat{S}_{i,j}^+\hat{S}_{j,k}^+\hat{S}_{k,l}^-\hat{S}_{l,i}^- + \text{H.c.}) \\ & + \frac{J}{2\sqrt{S(S+1)}}\sum_{\langle i,j\rangle}(\hat{m}_i^\dagger\hat{S}_{i,j}^+\hat{m}_j + \text{H.c.}) \\ & + M\sum_i(-1)^i\hat{n}_i,\end{aligned} \tag{3}$$

where the first term is the electric-field term, the second term is the magnetic-field term, and the third is the gauge-invariant hopping. Note that $g$ is the gauge-matter coupling strength, and $J$ is the hopping strength. In one spatial dimension, the constant background electric field $\tau$ corresponds to a topological term [172,181–183]. The last term is the staggered mass term with mass $M$, which comes from the mapping of the continuous Dirac fermion fields onto the lattice and is therefore only necessary for fermions [184]. We set the lattice spacing to unity here. Also note that, in order to take the above Hamtiltonian to the continuum limit of QED, the phases in the hopping and staggered mass terms need to be slightly altered, cf. [180]. This process does not introduce any additional challenges in our implementation, so we choose the above convention for simplicity.

In addition, the physical states of the Hamiltonian, i.e., the gauge-invariant ones, have to fulfill Gauss's law: For all $\mathbf{i}$, $\hat{G}_\mathbf{i}|\psi\rangle = q_\mathbf{i}|\psi\rangle$, where $q_\mathbf{i} \in \mathbb{Z}$ and

$$\hat{G}_\mathbf{i} = \hat{S}^z_{\mathbf{i}-\mathbf{e}_x} + \hat{S}^z_{\mathbf{i}-\mathbf{e}_y} - \hat{S}^z_{\mathbf{i}+\mathbf{e}_x} - \hat{S}^z_{\mathbf{i}+\mathbf{e}_y} - \hat{Q}_\mathbf{i}. \tag{4}$$

To clarify the spatial structure of Gauss's law, we specify the lattice site as a two-dimensional vector $\mathbf{i}$ in this paragraph only. Here, $\mathbf{e}_{x/y}$ is a unit vector in the $x/y$ direction of the lattice, respectively. In the above expression, $\hat{Q}_\mathbf{i} = \hat{m}_\mathbf{i}^\dagger\hat{m}_\mathbf{i} - (1-(-1)^{r_\mathbf{i}})/2$ is the staggered charge operator, where $r_\mathbf{i}$ is the Manhattan distance of the site $\mathbf{i}$ from the arbitrarily chosen origin. The fixed values $q_\mathbf{i}$ correspond to a choice of static background charge configuration, and for the rest of the paper, we restrict ourselves to the sector of vanishing static charges, i.e., $q_\mathbf{i} = 0$ for all $\mathbf{i}$.

In order to encode the spin $S$ degrees of freedom representing $U(1)$ gauge fields in the $U(1)$ quantum link model into bosonic modes, we use the Schwinger boson mapping [15,76]. It consists of two bosonic modes, labeled $a$ and $b$, which fulfill the constraint $\hat{n}^a_{i,j} + \hat{n}^b_{i,j} = 2S$. The electric-field operator in this encoding is represented by

$$\hat{S}^z_{i,j} = \frac{\hat{n}^a_{i,j} - \hat{n}^b_{i,j}}{2}, \tag{5}$$

and the electric-field raising operator $S^+_{i,j}$ becomes

$$\hat{S}^+_{i,j} = \hat{a}^\dagger_{i,j}\hat{b}_{i,j}, \tag{6}$$

$$\hat{S}^-_{i,j} = \hat{a}_{i,j}\hat{b}^\dagger_{i,j}. \tag{7}$$

Compared to an encoding in just a single oscillator [185], our encoding allows for a smaller occupation of the modes to represent electric field zero, reducing the impact of mode decay.

We show a possible encoding of this model in our architecture in Fig. 4(d).

### F. Circuit QED platform

We now outline the main features of the platform we consider in the rest of the study—a lattice of microwave cavities coupled to superconducting transmon qubits [see Fig. 4(b)]. In particular, we discuss the native Hamiltonian of this platform and the gates that can be implemented via microwave pulses.

Within the circuit QED framework, long-lived bosonic modes can be realized in three-dimensional superconducting microwave cavities with typical lifetimes on the order of one to ten milliseconds (ms) in aluminum [78,186] [Fig. 4(b)]. Recently, niobium cavities have been used to improve coherence times, with observed single-photon lifetimes of 34 ms [187] to over 2 s [188]. The cavity-mode decay rate increases linearly with photon number; i.e., if $|1\rangle$ has decay time of 1 ms, $|20\rangle$ has a decay time of 50 μs. Separately, planar microwave cavity resonators have been attracting attention as they may be more scalable due to their smaller size and fabrication advantages [189], and recent work has demonstrated the ability to reach 1-ms single-photon lifetimes [190]. In this platform, it is common to utilize one mode per cavity, although recent work has demonstrated that multimode cavities are also feasible [191].

Superconducting transmon qubits are routinely coupled to on-chip microwave cavity resonators for control and readout [192]. The relaxation $T_1$ and dephasing times $T_2$ of transmons when coupled to microwave cavities are typically on the order of 250 μs [193–195], with some recent coherence times in tantalum-based transmons reaching 500 μs [196]. When simulating the operation in hardware in Sec. VI, we take into account the dominant source of hardware error, qubit decay, and dephasing, adopting a conservative value of $T_1 = T_2 = 200$ μs if not stated otherwise.

The Superconducting Nonlinear Asymmetric Inductive eLement (SNAIL) [91] is an attractive choice for realizing a programmable coupling between cavities. The SNAIL consists of three large junctions shunted by one small junction. When biased with a nonzero dc magnetic flux, the SNAIL (or SNAIL array) generates a three-wave mixing Hamiltonian that enables a parametrized beam-splitter interaction of strength $g(t)$ between the cavities while minimizing unwanted nonlinear static interactions, such as the cross-Kerr interaction [91,197–199].

Our proposed architecture is an array of "qubit-bosonic oscillator pairs," each comprising a high-Q mode of a cavity dispersively coupled to a transmon qubit [see Fig. 4(b), top panel]. These units are arranged into a 2D square lattice, where adjacent cavities are connected via a SNAIL, which can be used to implement a beam splitter [see Fig. 4(b), bottom panel]. In the remainder of this work, we represent cavities (qubits) by circles (crosses). This symbolic representation will later prove useful for denoting the various possibilities for mapping systems of fermions, bosons, and gauge fields onto this 2D hardware layout.

While this architecture is physically large in size (each cavity is about 50 mm long [78]) and might therefore exhibit scaling limitations, planar cavity resonators, while at present exhibiting shorter coherence times compared to 3D cavity resonators, are smaller and therefore more scalable. Thus, the proposed architecture in Fig. 4(b) can also be implemented with planar cavity resonators, and all results presented herein extend to this case (with the caveat that Hamiltonian parameters, loss rates, and gate speeds are informed by recent experiments in 3D cavities and would need to be adjusted to appropriately reflect the planar case).

In this architecture, each transmon-cavity pair in the array is in the strong-dispersive coupling regime [192]. In the rotating-wave approximation, the effective Hamiltonian of the 2D array can be written as

$$\hat{H} = \hat{H}_0 + \hat{H}_1, \tag{8}$$

$$\hat{H}_0 = \sum_{i=1}^{N} \frac{\chi_i}{2} \hat{Z}_i \hat{n}_i + \sum_{i=1}^{N} (\varepsilon_i(t)\hat{a}_i^\dagger + \Omega_i(t)\hat{\sigma}_i^+ + \mathrm{H.c.}) \tag{9}$$

$$+ \sum_{\langle i,j\rangle} (g_{i,j}(t)\hat{a}_i^\dagger \hat{a}_j + \mathrm{H.c.}), \tag{10}$$

$$\hat{H}_1 = \sum_{i=1}^{N} K_i \hat{n}_i^2 + \sum_{\langle i,j\rangle} \chi_{i,j} \hat{n}_i \hat{n}_j. \tag{11}$$

Here, $\hat{a}_i$ ($\hat{a}_i^\dagger$) is the annihilation (creation) operator for the $i$th cavity mode, $\hat{n}_i = \hat{a}_i^\dagger \hat{a}_i$ is the corresponding number operator, and $\hat{\sigma}_i^-$ ($\hat{\sigma}_i^+$) is the lowering (raising) operator for the transmon coupled to the $i$th mode. Note that $\hat{H}_0$ includes the dispersive shift ($\chi_i$), cavity drives [$\varepsilon_i(t)$], transmon Rabi drives [$\Omega_i(t)$], and tunable beam-splitter couplings [$g_{i,j}(t)$] implemented via SNAILs. The time dependence in all parameters can be used to engineer different couplings. Here, $\hat{H}_1$ includes additional unwanted static couplings such as self-Kerr ($K_i$) and cross-Kerr ($\chi_{i,j}$) couplings. These nonlinear couplings incur contributions from the hybridization of each cavity mode with the transmon and SNAIL couplers. With careful engineering, it is possible to design the components such that the Kerr contributions from transmon and SNAIL cancel or nearly cancel, and $\chi_{i,i+1} \approx 0$, $K_i \approx 0$ could be achieved at a single flux point, with small residual couplings not likely to contribute to the principal results of this work [91]. When the beam-splitter interaction is not activated ($g_{i,j} = 0$), each qubit-bosonic oscillator pair can be independently controlled. We calculate the fidelity of the conditional-parity and the beam-splitter gates as a function of Fock state in Appendix H. The gates available in the proposed superconducting hybrid qubit-boson platform are also discussed in Ref. [57], where the requirements for universal instruction set architectures are explained. The native gates arising directly from the Hamiltonian of the system are collected in Table II. Note that, in some parameter regimes, certain transmon-conditioned bosonic operations, such as the conditional displacement (sometimes referred to as "controlled displacement") [89], can be implemented natively.

TABLE II. Native transmon-mode gates. Note that transmon-mode gates are only native if $i = j$, i.e., if both indices correspond to a directly coupled cavity-transmon pair. The case $i \neq j$ can be synthesized by combining native gates in sequence. Similarly, the beam splitter is native if $i$ and $j$ are nearest neighbors, i.e., if the modes are directly coupled by a SNAIL, otherwise additional SWAP operations must be used. Note that $CR_{i,i}(\chi_i t)$ corresponds to the "always on" dispersive interaction between the $i$th cavity and transmon (though it can be effectively "turned off" by echoing away its effects via dynamical decoupling—see discussion below). For the SNAP and SQR gates, $|n\rangle$ denotes the Fock state with occupation number $n$, fulfilling $\hat{n}|n\rangle = n|n\rangle$, while $\theta_n$ and $\varphi_n$ are the entries of the vectors $\vec{\theta} = \{\theta_0, \theta_1, \ldots\}$, $\vec{\varphi} = \{\varphi_0, \varphi_1, \ldots\}$. CD($\alpha$) is a conditional displacement gate [89]. The parameter $\alpha$ is complex, while $\theta$ is real for all gates. Duration refers to demonstrated gate times, where each gate was demonstrated in an architecture with parameters optimized for that gate. We calculate the fidelity of the conditional-parity and the beam-splitter gates as a function of Fock state in Appendix H.

| Gate name | Operation | Duration |
|---|---|---|
| | Mode gates | |
| $\mathrm{R}_i(\theta)$ | $\exp(-i\theta\hat{n}_i)$ | "in software" |
| $\mathrm{D}_i(\alpha)$ | $\exp(\alpha\hat{a}_i^\dagger - \alpha^*\hat{a}_i)$ | ~10 ns [89] |
| $\mathrm{BS}_{i,j}(\varphi,\theta)$ | $\exp(-i\theta(e^{i\varphi}\hat{a}_i^\dagger\hat{a}_j + e^{-i\varphi}\hat{a}_i\hat{a}_j^\dagger))$ | 9 ns [90] –100 ns [91] |
| | Transmon gates | |
| $\mathrm{R}_i^z(\theta)$ | $\exp(-i\frac{\theta}{2}\hat{Z}_i)$ | ~10 ns [89,92] |
| $\mathrm{R}_i^\varphi(\theta)$ | $\exp(-i\frac{\theta}{2}\hat{\sigma}_i^\varphi)$ $\hat{\sigma}_i^\varphi = \hat{X}_i\cos\varphi + \hat{Y}_i\sin\varphi$ | ~10 ns [89,92] |
| | Transmon-mode gates | |
| $\mathrm{CR}_{i,j}(\theta)$ | $\exp(-i\frac{\theta}{2}\hat{Z}_i\hat{n}_j)$ | ≤ 200 ns [90] –15 μs [89] |
| $\mathrm{C\Pi}_{i,j}$ | $\exp(-i\frac{\pi}{2}\hat{Z}_i\hat{n}_j)$ | 200 ns [90] 15 μs [89] |
| $\mathrm{CD}_{i,j}(\alpha)$ | $\exp(\hat{Z}_i(\alpha\hat{a}_j^\dagger - \alpha^*\hat{a}_j))$ | 1 μs [89] |
| $\mathrm{SNAP}_{i,j}(\vec{\theta})$ | $\exp(-i\hat{Z}_i\sum_n\theta_n\|n\rangle\langle n\|_j)$ | 500 ns [93] –50 μs [89] |
| $\mathrm{SQR}_{i,j}(\vec{\theta},\vec{\varphi})$ | $\sum_n \mathrm{R}_i^{\varphi_n}(\theta_n)\otimes\|n\rangle\langle n\|_j$ | 500 ns [93] –50 μs [89] |

All phase-space rotations $\mathrm{R}_i(\theta)$ can be realized "in software" by absorbing them into the phases of the subsequent microwave pulses because all operations are carried out in the rotating frame of the free evolution of the cavity oscillator mode $\omega\hat{a}^\dagger\hat{a}$ (each transmon is only ever connected to a single cavity). Therefore, a phase-space rotation can be applied by changing the phase of the subsequent microwave drive carrying out a gate (such as a beam splitter or displacement) on the corresponding mode. Arbitrary qubit $Z$ rotation gates can similarly be carried out in software.

Arbitrary initial states for each cavity can be prepared using optimal control [89,200]: Both the qubit and the cavity are initialized to the ground state; the qubit prepares the cavity in the required Fock state and is then measured to validate this preparation [201]. In particular, Fock states $|n\rangle$ with low photon number can be prepared with typical state-of-the-art Fock preparation infidelities of $10^{-2}$ [201–203].

The native Hamiltonian parameter $\chi$ representing the dispersive coupling of the transmon qubit and the cavity mode can vary over orders of magnitude, from kHz to tens of MHz [204]. The magnitude of $\chi$ is primarily determined by the qubit-cavity coupling $g$ and detuning $\Delta$ via $\chi \approx g^2/\Delta$ (neglecting higher-level corrections) [205]. In the weak dispersive regime, $\chi/2\pi$ is typically on the order of 10–100 kHz [89,195]. For transmon qubits in the strong dispersive regime, $\chi/2\pi$ has been experimentally demonstrated in this platform to reach 1–4 MHz [90,204,206,207]. Dynamical control of qubit-boson interactions with squeezing have also been demonstrated [208]. Recent experiments with "dynamical-$\chi$ (LINC)" couplers use a three-wave mixing element to couple the cavity to the qubit on demand. They can modulate the dispersive shift by 4 orders of magnitude (0.3 KHz to 2.5 MHz) in about 20 ns [90]. However, $\chi$ cannot be made arbitrarily large without violating the dispersive approximation ($g \ll \Delta$) or introducing unwanted nonlinearities. Leaving this native interaction on for a duration of $t = \pi/\chi$ leads to the implementation of a conditional-parity gate, which leads to conditional-parity gate durations ranging from 1 μs at $\chi = 2\pi \times 500$ kHz [209] (which is taken in the rest of the text as a lower bound for $\chi$) down to 200 ns at the demonstrated $\chi = 2\pi \times 2.5$ MHz [90].

Because mode decay increases linearly with Fock state, the fidelity of operations also varies with mode occupation. We implemented a Lindblad calculation of the fidelity of a conditional parity, which includes photon jump operators such as decay and dephasing for the transmon qubit, and decay (dephasing has been shown to be negligible) for the cavity modes. The results of these calculations are given in Appendix H.

As mentioned above, it is possible to effectively turn off the "always on" dispersive interaction using dynamical decoupling. Specifically, the transmons can be driven with Rabi rate $\Omega_i \gg \chi_i\langle\hat{a}_i^\dagger\hat{a}_i\rangle$ during a time $T = 2\pi k/\Omega_i$, where $k$ is an integer, to decouple the cavity mode from the transmon and leave the transmons in their initial state. Thus, the beam-splitter gate $\mathrm{BS}_{i,j}(\varphi,\theta)$ is performed by simultaneously driving transmons $i$ and $j$ and driving the SNAIL coupler linking modes $i$ and $j$ such that $g_{i,j} > 0$. Alternative schemes based on transmon echoing can also be used to decouple and perform transmon-state-independent beam splitting [210].

In addition to unitary control, the transmons can be leveraged for measurements of the cavity modes. In particular, it is possible to measure the boson occupation in the mode in a total duration that is logarithmic in the maximum Fock state. This process is completed by reading out the photon number bit by bit in its binary representation. Each bit is mapped onto the transmon qubit dispersively coupled to the cavity, which is measured and then reset [144,211].

### G. Trotter simulations

In this work, we focus on a Trotter decomposition of dynamics as our basic simulation algorithm. In such simulations, the time-evolution operator $\exp(-i\hat{H}T)$ until time $T$ with Hamiltonian $\hat{H} = \sum_{\gamma=1}^{\Gamma} \hat{H}_\gamma$ is divided into $r$ time steps of duration $t = T/r$, i.e., $\exp(-i\hat{H}T) = (\exp(-i\hat{H}t))^r$. If $t$ is chosen sufficiently small, a product formula approximation to $\exp(-i\hat{H}t)$ can then be employed to decompose the Hamiltonian into unitaries that only act on a few qubits and oscillator modes. For example, a $(2p)$th-order Trotter formula, in general, can be written as a sequence of $N_{\rm exp}$ exponentials of Hamiltonian terms $\hat{H}_{\gamma_i}$ drawn from $\{\hat{H}_1, \ldots, \hat{H}_\Gamma\}$ and a sequence of times $t_i$ with $|t_i| \leq t$, which satisfies

$$\exp(-i\hat{H}t) = e^{-i\hat{H}_{\gamma_1} t_1} \cdots e^{-i\hat{H}_{\gamma_{N_{\rm exp}}} t_{N_{\rm exp}}} + \mathcal{O}(t^{2p+1}). \quad (12)$$

There is great freedom in the choice of this decomposition; however, certain orderings can be cheaper, more accurate, or more natural. We choose our decomposition to be given by the local terms inside the sums over lattice sites in many-body Hamiltonians, cf. Eqs. (1) and (3). For example, the first-order Trotter decomposition of the $\mathbb{Z}_2$-Higgs model for $U = 0$ in one spatial dimension, where $B = 0$, is given by

$$e^{-i\hat{H}_{\mathbb{Z}_2} t} = \prod_{\langle i,j\rangle} e^{igt\hat{X}_{i,j}} \prod_{\langle i,j\rangle} e^{iJt(\hat{m}_i^\dagger \hat{Z}_{i,j} \hat{m}_j + \text{H.c.})} + \mathcal{O}(t^2). \quad (13)$$

One of the main goals of this paper is to show how to compile unitaries acting on two or more sites, such as the gauge-invariant hopping $\exp[iJt(\hat{m}_i^\dagger \hat{Z}_{i,j} \hat{m}_j + \text{H.c.})]$, into the native operations available in circuit QED hardware or other hardware with equivalent instruction set architectures.

In Sec. V, we discuss the errors incurred from the Trotter approximation and their interplay with compilation errors.

## II. IMPLEMENTATION PRIMITIVES

In this section, we introduce the key compilation strategies we use to synthesize bosonic oscillator–bosonic oscillator, qubit-qubit, and qubit-bosonic oscillator gates from native operations. Throughout, we use the word “qubit” to refer either to a transmon qubit or a dual-rail qubit encoded in two cavities [201,210,212–216], and we only distinguish between the two possibilities when clarification is necessary. A summary of the gates we obtain through these compilation strategies is provided in Table III. These primitives will then be used in Secs. III and IV to compile Hamiltonian terms comprising fermions, bosons, and gauge fields to native operations.

Specifically, in Sec. II A, we discuss the bosonic SWAP gate, a particularly useful primitive that enables the synthesis of entangling gates between non-neighboring modes. In Sec. II B, we demonstrate how one can synthesize qubit-conditioned bosonic operations using conditional-parity gates, and, in Secs. II C and II D, we develop strategies to realize bosonic operations conditioned or controlled on the Fock-space occupancy of another mode. Building upon these “Fock-projector-conditioned” and “Fock-projector-controlled” operations, we then discuss how one can apply them in sequence to implement more complex density-controlled gates in Sec. II E. Next, we discuss a technique for implementing oscillator-mediated, multiqubit gates in Sec. II F, which we follow with a discussion on approximate methods for synthesizing multimode gates in Sec. II G. Finally, we discuss the dual-rail encoding in Sec. II H as an alternative scheme for realizing qubits in the proposed hybrid qubit-boson architecture.

### A. Bosonic SWAP

Our architecture in Fig. 4 is not all-to-all connecting, and hence it is useful to be able to implement SWAP operations between the bosonic sites. In particular, bosonic SWAP gates enable quantum communication and entangling operations between remote bosonic modes and additionally allow for entangling operations between qubits using the bosonic modes as a quantum bus [57].

To that end, a bosonic SWAP gate can be realized using a beam splitter and a pair of phase-space rotations,

$$\text{SWAP}_{i,j} = R_i\left(-\frac{\pi}{2}\right) \text{R}_j\left(-\frac{\pi}{2}\right) \text{BS}_{i,j}\left(0, \frac{\pi}{2}\right), \quad (14)$$

with $\text{BS}_{i,j}(\varphi, \theta)$ and $\text{R}_j(\theta)$ defined in Table II. Here, the role of the phase-space rotations is to cancel the spurious phase obtained from the action of the beam splitter, $\text{BS}_{i,j}[0, (\pi/2)]|\Psi_i, \Psi_j\rangle = e^{-i\frac{\pi}{2}[\hat{n}_i + \hat{n}_j]}|\Psi_j, \Psi_i\rangle$. As mentioned previously, all such phase-space rotations $\text{R}_i(\theta)$ can be realized in software by absorbing them into the phases of the subsequent microwave drives. In the proposed architecture with SNAILs linking adjacent cavities, SWAP operations have been demonstrated with a duration of around 100 ns and a fidelity of around 0.999 [91,217], and recent work has demonstrated SWAP operations with a duration of 9 ns [90].

As discussed in Sec. I F, our proposed architecture assumes native entangling gates between each qubit-boson pair, and beam splitters are only available between adjacent sites. However, as discussed in Ref. [57], it is possible to use a beam-splitter network (i.e., a sequence of bosonic SWAPs) to synthesize non-native bosonic oscillator–bosonic oscillator and qubit-bosonic oscillator gates, e.g., $\mathrm{BS}_{i,j}(\varphi, \theta)$ for any modes $i \neq j$, and $\mathrm{CR}_{i,j}(\theta)$ for any transmon $i$ and mode $j$. Thus, we will henceforth assume access to such non-native gates, abstracting away the underlying bosonic SWAPs.

TABLE III. Non-native primitives used in this study and section references where their synthesis is explained. Row $\mathrm{CU}^{Z}_{i,j}$ is a qubit-conditional variant of the unitary $\hat{U} = \exp(\hat{\Theta}\hat{a}_j^\dagger - \hat{\Theta}^\dagger \hat{a}_j)$, where $\hat{\Theta}$ is any operator that commutes with $\hat{a}_j$ and $\hat{Z}_j$. For example, $\hat{\Theta} \to \alpha$ corresponds to a conditional displacement [see Eq. (18)], while $\hat{\Theta} \to \theta e^{i\varphi}\hat{a}_k$ yields a conditional beam splitter [see Eq. (19)]. In the third row, we give an ancilla-mediated variant of the unitary $\hat{U} = \exp(-i\hat{O}_j)$ on mode $j$ that is conditioned by an operator $\hat{P}_i$ on mode $i$. Here, $\hat{P}_i$ can be any operator that is diagonal in Fock space with eigenvalues $\pm 1$. The angles $\vec{\theta}_P$ are chosen according to Eq. (32). The fourth row shows an ancilla-mediated variant of $\hat{U} = \exp(-i\hat{O}_j)$ conditioned on the photon number $\hat{n}_i$ of mode $i$. This example leverages an iterative sequence of the primitive $\mathrm{CU}^{P}_{i,j}$ with the $k$th iteration conditioning on the $k$th bit of $\hat{n}_i$, requiring $K = \lceil \log_2(n_{\max} + 1) \rceil$ total iterations, where $n_{\max}$ is the maximum photon-number cutoff for mode $i$. The fifth row shows a useful instance of $\mathrm{CU}^{\hat{n}}_{i,j}$, where $\hat{O}_j \to \theta\hat{n}_j$. Here, $\vec{\pi}_k$ is defined in Eq. (41). The sixth row is an example of an oscillator-mediated entangling gate between two transmons. In the seventh and eighth rows, we give the approximate methods (see relevant sections for discussions and Ref. [97] for the BCH method). The superscripts $X$ and $Y$ in CU denote the qubit axis on which the operator $\hat{U}$ is conditioned. This axis can be controlled by conjugating $\mathrm{CU}^Z$ by single-qubit rotations. Here, $\vec{I}$ and $\vec{J}$ indicate support on a vector of mode indices.

| Gate | Definition | Gate decomposition | Reference |
|---|---|---|---|
| $\mathrm{SWAP}_{i,j}$ | $\sum_{n,m} \lvert m\rangle\langle n\rvert_i \otimes \lvert n\rangle\langle m\rvert_j$ | $R_i(-\frac{\pi}{2})R_j(-\frac{\pi}{2})\mathrm{BS}_{i,j}(0,\frac{\pi}{2})$ | Sec. II A |
| $\mathrm{CU}^{Z}_{i,j}$ | $\exp(\hat{Z}_i(\hat{\Theta}\hat{a}_j^\dagger - \hat{\Theta}^\dagger\hat{a}_j))$ | $\mathrm{C\Pi}_{i,j} e^{i(\hat{\Theta}\hat{a}_j^\dagger + \hat{\Theta}^\dagger\hat{a}_j)} \mathrm{C\Pi}^\dagger_{i,j}$ | Fig. 5, Sec. II B |
| $\mathrm{CU}^{P}_{i,j}$ | $\exp(-i\hat{Z}_{\mathrm{anc}}\hat{P}_i\hat{O}_j)$ | $SQR_{\mathrm{anc},i}(\vec{\theta}_P, \vec{0}) e^{-i\hat{Z}_{\mathrm{anc}}\hat{O}_j} SQR_{\mathrm{anc},i}(-\vec{\theta}_P, \vec{0})$ | Sec. II C |
| $\mathrm{CU}^{\hat{n}}_{i,j}$ | $\exp(-i\hat{Z}_{\mathrm{anc}}\hat{n}_i\hat{O}_j)$ | $\prod_{k=0}^{K-1} SQR_{\mathrm{anc},i}(\vec{\pi}_k, \vec{0}) e^{-i2^{k-1}\hat{O}_j} e^{i2^{k-1}\hat{Z}_{\mathrm{anc}}\hat{O}_j} SQR_{\mathrm{anc},i}(-\vec{\pi}_k, \vec{0})$ | Fig. 6, Sec. II E |
| $\mathrm{RR}_{i,j}(\theta)$ | $\exp(-i\theta\hat{Z}_{\mathrm{anc}}\hat{n}_i\hat{n}_j)$ | $\prod_{k=0}^{K-1} SQR_{\mathrm{anc},i}(\vec{\pi}_k, \vec{0}) R_j(\frac{2^k\theta}{2}) CR^\dagger_{i,j}(\frac{2^k\theta}{2}) SQR_{\mathrm{anc},i}(-\vec{\pi}_k, \vec{0})$ | Sec. II E |
| $ZZ_{i,j}(\theta)$ | $\exp(-i\frac{\theta}{2}\hat{Z}_i\hat{Z}_j)$ | $\mathrm{CD}_{i,j}(-i\sqrt{\theta}/2)\mathrm{CD}_{i,i}(\sqrt{\theta}/2)\mathrm{CD}_{i,j}(+i\sqrt{\theta}/2)\mathrm{CD}_{i,i}(-\sqrt{\theta}/2)$ | Fig. 7, Sec. II F |
| BCH | $\exp(-2i\hat{Z}\hat{O}_{\vec{I}}\hat{O}_{\vec{J}}\theta^2)$ | $\mathrm{CU}^{X}_{k,\vec{I}}(\theta)\mathrm{CU}^{Y}_{k,\vec{J}}(\theta)\mathrm{CU}^{X}_{k,\vec{I}}(-\theta)\mathrm{CU}^{Y}_{k,\vec{J}}(-\theta) + \mathcal{O}(\theta^3)$ | Fig. 8, Sec. II G |
| Trotter | $\exp(-i\hat{Z}(\hat{O}_{\vec{I}} + \hat{O}_{\vec{J}})\theta)$ | $\mathrm{CU}^{Z}_{i,\vec{I}}(\theta)\mathrm{CU}^{Z}_{j,\vec{J}}(\theta) + \mathcal{O}(\theta^2)$ | Fig. 8, Sec. II G |

### B. Qubit-conditioned bosonic operations

Certain bosonic operations, such as beam splitters or displacements, can be conditioned on qubits by conjugating them with conditional-parity gates, $\mathrm{C\Pi}_{i,j}$ [57]. The intuition for this approach is provided by the relation

$$\mathrm{C\Pi}_{i,j}\hat{a}_j\mathrm{C\Pi}^\dagger_{i,j} = i\hat{Z}_i\hat{a}_j, \tag{15}$$

where we used the Baker-Campbell-Hausdorff formula

$$e^{\hat{A}}\hat{B}e^{-\hat{A}} = \hat{B} + [\hat{A}, \hat{B}] + \frac{1}{2!}[\hat{A}, [\hat{A}, \hat{B}]] + \ldots, \tag{16}$$

and $\mathrm{C\Pi}_{i,j}$ is the conditional-parity gate defined in Table II, experimentally realized by evolving under the dispersive interaction for a time $T = \pi/\chi$ [see Eq. (9)], where $\chi$ is expressed as an angular frequency such as $\chi = 2\pi \times 1.24$ MHz [209].

It is possible to implement the Hermitian adjoint $\mathrm{C\Pi}^\dagger_{i,j}$ by simply appending a bosonic phase-space rotation,

$$\mathrm{C\Pi}^\dagger_{i,j} = \mathrm{R}_j(-\pi)\mathrm{C\Pi}_{i,j} = \mathrm{C\Pi}_{i,j}\mathrm{R}_j(-\pi), \tag{17}$$

because $\mathrm{C\Pi}^\dagger_{i,j} = e^{i\frac{\pi}{2}\hat{Z}_i\hat{n}_j} = e^{i\pi\hat{Z}_i\hat{n}_j}e^{-i\frac{\pi}{2}\hat{Z}_i\hat{n}_j}$ and $e^{i\pi\hat{Z}_i\hat{n}_j} = \cos(\pi\hat{n})\mathbb{1} + i\sin(\pi\hat{n})\hat{Z} = \cos(\pi\hat{n}) = e^{i\pi\hat{n}_j}$. In effect, applying a $\mathrm{C\Pi}^\dagger_{i,j}$ gate requires the application of $\mathrm{C\Pi}_{i,j}$ and a phase-space rotation $\mathrm{R}_j(-\pi) = \mathrm{R}_j(\pi)$, which, as explained in Sec. I F, can be implemented in software via modification of the phases of the subsequent microwave drives.

The relation in Eq. (15) allows us to "convert" the generator of certain bosonic operations into one that depends upon the state of a qubit. For example, the conditional displacement gate can be synthesized from an unconditional displacement and a pair of conditional-parity gates via

$$\begin{aligned}\mathrm{CD}_{i,j}(\alpha) &= e^{\hat{Z}_i(\alpha\hat{a}_j^\dagger - \alpha^*\hat{a}_j)} \\ &= e^{-i\frac{\pi}{2}\hat{Z}_i\hat{n}_j}e^{i(\alpha\hat{a}_j^\dagger + \alpha^*\hat{a}_j)}e^{i\frac{\pi}{2}\hat{Z}_i\hat{n}_j} \\ &= \mathrm{C\Pi}_{i,j}D_i(i\alpha)\mathrm{C\Pi}^\dagger_{i,j},\end{aligned} \tag{18}$$

where $\alpha \in \mathbb{C}$. Note the modified phase in the unconditional displacement, which is due to the factor of $i$ inherited from the transformation in Eq. (15). Furthermore, we emphasize that while this sequence produces exactly the conditional displacement gate as defined in Table II, this gate can be more efficiently realized natively using the techniques developed in Ref. [89]. In the native implementation, this gate has been demonstrated to take 1 μs [89]. This compiled approach would require a duration corresponding to two conditional-parity operations that can each be as fast as 200 ns at $\chi = 2\pi \times 2.5$ MHz [90], and one displacement that takes 10 ns [89], leading to a total implementation time of 410 ns.

As another example, we can exactly compile the conditional beam-splitter gate as follows:

$$\begin{aligned}\mathrm{CBS}_{i,j,k}(\varphi,\theta) &= e^{-i\theta\hat{Z}_i(e^{i\varphi}\hat{a}_j^\dagger\hat{a}_k + e^{-i\varphi}\hat{a}_j\hat{a}_k^\dagger)}\\ &= e^{-i\frac{\pi}{2}\hat{Z}_i\hat{n}_j}e^{\theta(e^{i\varphi}\hat{a}_j^\dagger\hat{a}_k - e^{-i\varphi}\hat{a}_j\hat{a}_k^\dagger)}e^{i\frac{\pi}{2}\hat{Z}_i\hat{n}_j}\\ &= \mathrm{C}\Pi_{i,j}\mathrm{BS}_{j,k}(\varphi+\pi/2,\theta)\mathrm{C}\Pi_{i,j}^\dagger, \end{aligned} \tag{19}$$

where $\{\varphi,\theta\} \in \mathbb{R}$. This gate will play an instrumental role in Sec. IV for the compilation of interaction terms between matter and gauge fields. Its duration is experimentally limited by the speed of each conditional-parity gate, requiring roughly 1 μs each. As such, the conditional beam-splitter gate requires about 2.1 μs, including the worst-case scenario, around 100 ns, required for a full SWAP beam splitter [217].

More generally, as shown in Fig. 5(a), we can employ this strategy to realize a conditional unitary $\mathrm{CU}_{i,j}^Z$ from its unconditional counterpart:

$$\begin{aligned}\mathrm{CU}_{i,j}^Z &= e^{\hat{Z}_i(\hat{\Theta}\hat{a}_j^\dagger - \hat{\Theta}^\dagger\hat{a}_j)}\\ &= \mathrm{C}\Pi_{i,j}e^{i(\hat{\Theta}\hat{a}_j^\dagger + \hat{\Theta}^\dagger\hat{a}_j)}\mathrm{C}\Pi_{i,j}^\dagger. \end{aligned} \tag{20}$$

This decomposition holds for any choice of operator $\hat{\Theta}$ that satisfies $[\hat{\Theta},\hat{a}_j] = [\hat{\Theta},\hat{Z}_i] = 0$. For instance, $\hat{\Theta} \to \alpha$ corresponds to the conditional displacement in Eq. (18) while $\hat{\Theta} \to \theta e^{i\varphi}\hat{a}_k$ yields the condition beam splitter in Eq. (19).

This strategy can be iterated to realize bosonic gates conditioned on multiple qubits. For example, one can realize a doubly conditioned gate of the form

$$\mathrm{CU}_{i,j,k}^{ZZ} = e^{\hat{Z}_i\hat{Z}_j(\hat{\Theta}\hat{a}_k^\dagger - \hat{\Theta}^\dagger\hat{a}_k)} \tag{21}$$

for any operator $\hat{\Theta}$ that commutes with $\hat{Z}_i$, $\hat{Z}_j$, and $\hat{a}_k$. In this case, conditional-parity gates are implemented on mode $k$ for qubits $j$ and $i$ in sequence with the aid of SWAP gates. Alternatively, one can apply conditional-parity gates acting on different qubit-boson pairs in parallel as shown in Fig. 5(b).

We note that, in the above, we use "qubit-conditional" and "qubit-controlled" interchangeably to refer to a gate that enacts a bosonic operation that depends upon the $Z$ basis state of a qubit. In particular, distinct nontrivial operations are enacted for both qubit states $|0\rangle$ and $|1\rangle$. However, in certain contexts, it is useful to synthesize a "1-controlled" bosonic operation, i.e., one that acts only if the qubit is in $|1\rangle$. In this work, we will refer to this as an $n$-controlled gate, referring to the operator that appears as a multiplicative factor in the generator. For example, for an $n$-controlled mode rotation, we have

$$e^{-i\theta\hat{n}_{\mathrm{qubit}}\hat{n}_{\mathrm{mode}}} = \mathrm{R}_i\left(-\frac{\theta}{2}\right)\mathrm{CR}_{i,i}(\theta), \tag{22}$$

where $\hat{n}_{\mathrm{qubit}} = (\mathbb{1} - \hat{Z})/2$. More generally, we have

$$e^{i\theta\hat{n}_{\mathrm{qubit}}\hat{O}_i} = e^{i\frac{\theta}{2}\hat{O}_i}e^{-i\frac{\theta}{2}\hat{Z}_i\hat{O}_i}, \tag{23}$$

for any $\hat{O}_i$ that commutes with operations on the qubit.

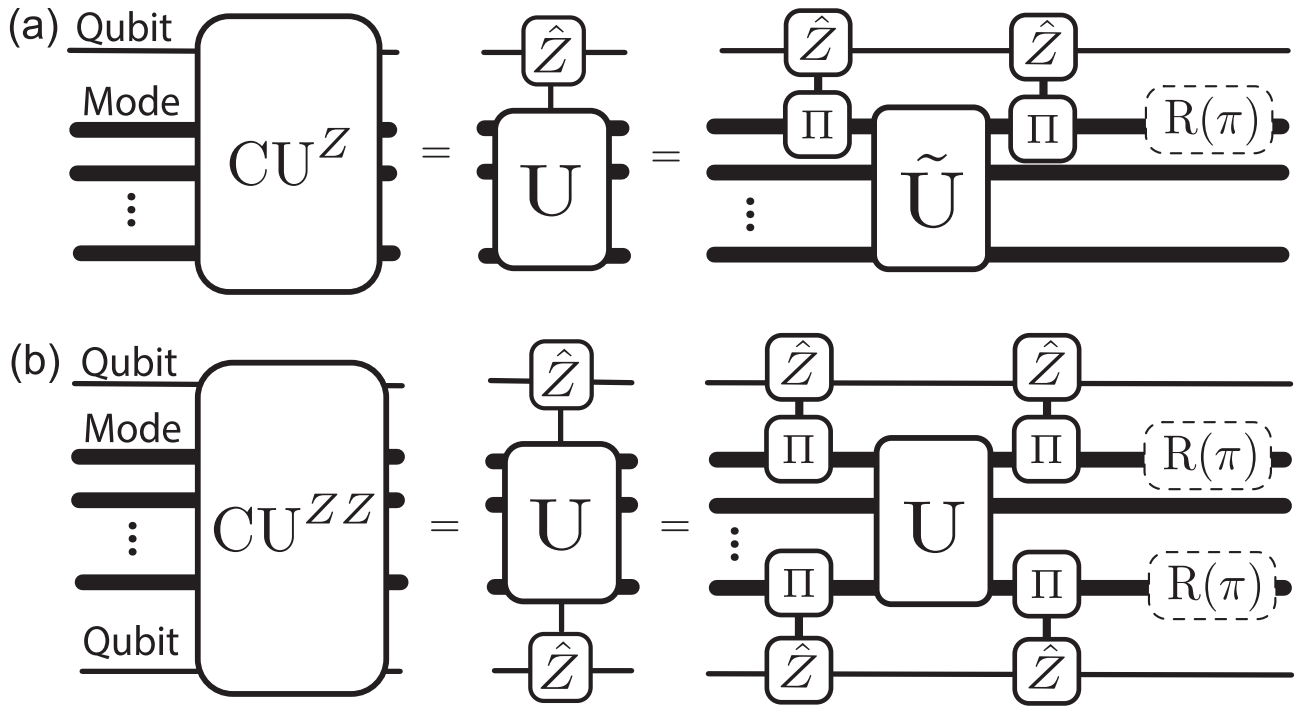

FIG. 5. Qubit-conditional bosonic gates. Circuit diagrams containing qubits (thin line) and modes (thick lines), with time going from left to right. All conditional gates represent unitaries conditional on $\hat{Z}$. (a) Qubit-conditional gate of the form $\mathrm{CU}_{i,j}^Z$ defined in Eq. (20). Here, $\tilde{U}$ denotes the unconditional variant of $\mathrm{CU}^Z$. The tilde indicates that the phase is modified by $\pi/2$—see, e.g., Eq. (18). Gates in a dashed box can be implemented in software as explained in Sec. I F. (b) Multicontrolled bosonic operations, implemented by repeating the primitive in panel (a) [see Eq. (21)].

### C. Fock-projector-conditioned bosonic operations

An important primitive in the hybrid qubit-boson architecture of Fig. 4 is the synthesis of nontrivial two-mode entangling gates. In this section, we develop a technique to realize bosonic operators that are conditioned on the Fock-space information of a second mode:

$$\mathrm{CU}_{i,j}^{\bar{P}} = e^{-i\hat{Z}_{\mathrm{anc}}\bar{P}_i\hat{O}_j}, \tag{24}$$

where $\bar{P}_i$ is an operator with eigenvalues $\pm 1$ defined below (hence the name "conditioned"), "anc" refers to an ancillary transmon qubit, and $\hat{O}_j$ is an arbitrary (Hermitian) operation acting on mode $j$. For an arbitrary subspace $\mathcal{P}$ of the oscillator, $\bar{P}_i$ gives states inside $\mathcal{P}$ a phase of $-1$ and states outside $\mathcal{P}$ a phase of 1. Mathematically,

$$\bar{P}_i = \sum_{n \notin \mathcal{P}} |n\rangle\langle n|_i - \sum_{n \in \mathcal{P}} |n\rangle\langle n|_i, \tag{25}$$

where each projector acts on mode $i$. Note that we define the operator $\bar{P}_i$ without a hat for simplicity of notation.

Because $\bar{P}_i$ has eigenvalues $\pm 1$, the core idea is to map this information to an ancillary qubit, which is then used to mediate this information to a second mode via a hybrid qubit-boson gate. This approach effectively realizes the Fock-projector-conditioned bosonic gate in Eq. (24) and requires the ancillary qubit to be initialized to an eigenstate of $\hat{Z}_{\rm anc}$. This ancillary qubit can correspond to the transmon qubit dispersively coupled to mode $i$ or $j$, or any other qubit via the appropriate use of bosonic SWAP gates.

To further explain, we begin with the particular example of a parity-conditioned gate, defined as

$$\mathrm{CU}^{\bar{\Pi}}_{i,j} = e^{-i\hat{Z}_{\rm anc}\bar{\Pi}_i\hat{O}_j}, \tag{26}$$

where

$$\bar{\Pi}_i = e^{i\pi\hat{n}_i} \tag{27}$$

is the parity operator acting on mode $i$ with eigenvalues $\pm 1$. Therefore, for an ancilla initialized to $|0\rangle_{\rm anc}$, $\mathrm{CU}^{\bar{\Pi}}_{i,j}$ applies the unitary $e^{-i\hat{O}_j}$ ($e^{i\hat{O}_j}$) to mode $j$ if mode $i$ has even (odd) parity. To realize this gate, it is helpful to first note that

$$\mathrm{SQR}_{{\rm anc},i}(\vec{\pi}_0, \vec{0}) = e^{-i\frac{\pi}{2}\hat{X}_{\rm anc}\sum_{n\ \rm odd}|n\rangle\langle n|_i}, \tag{28}$$

where $\vec{\pi}_0 = \{0, \pi, 0, \pi, \ldots\}$ and $\vec{0} = \{0, 0, 0, \ldots\}$, and $\hat{\Pi}_i$ (as opposed to $\bar{\Pi}_i$) is the projector over odd Fock states. In other words, an SQR gate is simply a Fock-projector-controlled qubit rotation. By leveraging the Baker-Campbell-Hausdorff formula in Eq. (16), it can be shown that

$$\mathrm{SQR}_{{\rm anc},i}(\vec{\pi}_0, \vec{0})\hat{Z}_{\rm anc}\mathrm{SQR}^{\dagger}_{{\rm anc},i}(\vec{\pi}_0, \vec{0}) = \hat{Z}_{\rm anc}\bar{\Pi}_i. \tag{29}$$

Thus, in direct analogy to the strategy for synthesizing qubit-conditional bosonic operations in Sec. II B, we can leverage a pair of SQR gates to condition a bosonic operation on the parity of a second mode:

$$\mathrm{CU}^{\bar{\Pi}}_{i,j} = \mathrm{SQR}_{{\rm anc},i}(\vec{\pi}_0, \vec{0})e^{-i\hat{Z}_{\rm anc}\hat{O}_j}\mathrm{SQR}_{{\rm anc},i}(-\vec{\pi}_0, \vec{0}), \tag{30}$$

where we have used the fact that $\mathrm{SQR}^{\dagger}_{{\rm anc},i}(\vec{\theta}, \vec{\varphi}) = \mathrm{SQR}_{{\rm anc},i}(-\vec{\theta}, \vec{\varphi})$. This method requires the realization of the intermediary qubit-boson gate $\exp(-i\hat{Z}_{\rm anc}\hat{O}_j)$, either natively or, e.g., using the technique in Sec. II B.

Returning to the general form in Eq. (24), it is possible to generalize this technique by substituting $\bar{\Pi}_i$ with $\bar{P}_i$, which allows for any arbitrary set of Fock states $\mathcal{P}$ by choosing appropriate angles for the SQR gate in Eq. (30). In particular, Eq. (29) generalizes as

$$\mathrm{SQR}_{{\rm anc},i}(\vec{\theta}_P, \vec{0})\hat{Z}_{\rm anc}\mathrm{SQR}^{\dagger}_{{\rm anc},i}(\vec{\theta}_P, \vec{0}) = \hat{Z}_{\rm anc}\bar{P}_i, \tag{31}$$

where $\vec{\theta}_P = \{\theta_0, \theta_1, \ldots, \theta_n \ldots\}$ and

$$\theta_n = \begin{cases} \pi & \text{if } |n\rangle \in \mathcal{P} \\ 0 & \text{otherwise.} \end{cases} \tag{32}$$

Akin to Eq. (28), this approach leverages the fact that $\mathrm{SQR}_{{\rm anc},i}(\vec{\theta}_P, \vec{0})$ realizes a projector-controlled qubit rotation:

$$\mathrm{SQR}_{{\rm anc},i}(\vec{\theta}_P, \vec{0}) = e^{-i\frac{\pi}{2}\hat{X}_{\rm anc}\sum_{n\in\mathcal{P}}|n\rangle\langle n|_i}. \tag{33}$$

Thus, one can realize the generalized Fock-projector-conditional gate,

$$\begin{aligned} \mathrm{CU}^{\bar{P}}_{i,j} &= e^{-i\hat{Z}_{\rm anc}\bar{P}_i\hat{O}_j} \\ &= \mathrm{SQR}_{{\rm anc},i}(\vec{\theta}_P, \vec{0})e^{-i\hat{Z}_{\rm anc}\hat{O}_j}\mathrm{SQR}_{{\rm anc},i}(-\vec{\theta}_P, \vec{0}). \end{aligned} \tag{34}$$

This sequence is illustrated in Fig. 6(a).

### D. Fock-projector-controlled bosonic operations

In this section, we build on the technique developed in the previous section in order to realize bosonic operators that are *controlled* on the Fock-space information of another mode,

$$\mathrm{CU}^{\hat{P}}_{i,j} = e^{-i\hat{Z}_{\rm anc}\hat{P}_i\hat{O}_j}, \tag{35}$$

where $\hat{P}_i$ is a Fock-space projector onto the subspace $\mathcal{P}$,

$$\hat{P}_i = \sum_{n\in\mathcal{P}}|n\rangle\langle n|_i \tag{36}$$

$$= (\mathbb{1} - \bar{P}_i)/2. \tag{37}$$

The second equality reminds us that the projector and its reflection operator about the orthogonal complement $\bar{P}_i$ are reminiscent of $\hat{n}_{\rm qubit}$ and $\hat{Z}$ for qubits, where $\hat{n}_{\rm qubit} = (\mathbb{1} - \hat{Z})/2$. It follows that while $\mathrm{CU}^{\bar{P}}_{i,j}$ applies a distinct, nontrivial operation for each value of the mode $i \in \mathcal{P}$, the gate $\mathrm{CU}^{\hat{P}}_{i,j}$ acts with $\hat{O}_j$ *only* if mode $i \in \mathcal{P}$ (hence the "controlled" nomenclature).

In fact, using Eq. (37), we can easily construct the target gate (35), similar to the construction in Eq. (23):

$$\begin{aligned} \mathrm{CU}^{\hat{P}}_{i,j} &= e^{-i\hat{Z}_{\rm anc}((\mathbb{1}_i - \bar{P}_i)/2)\hat{O}_j} \\ &= e^{-i\hat{Z}_{\rm anc}\hat{O}_j/2}e^{i\hat{Z}_{\rm anc}\bar{P}_i\hat{O}_j/2}. \end{aligned} \tag{38}$$

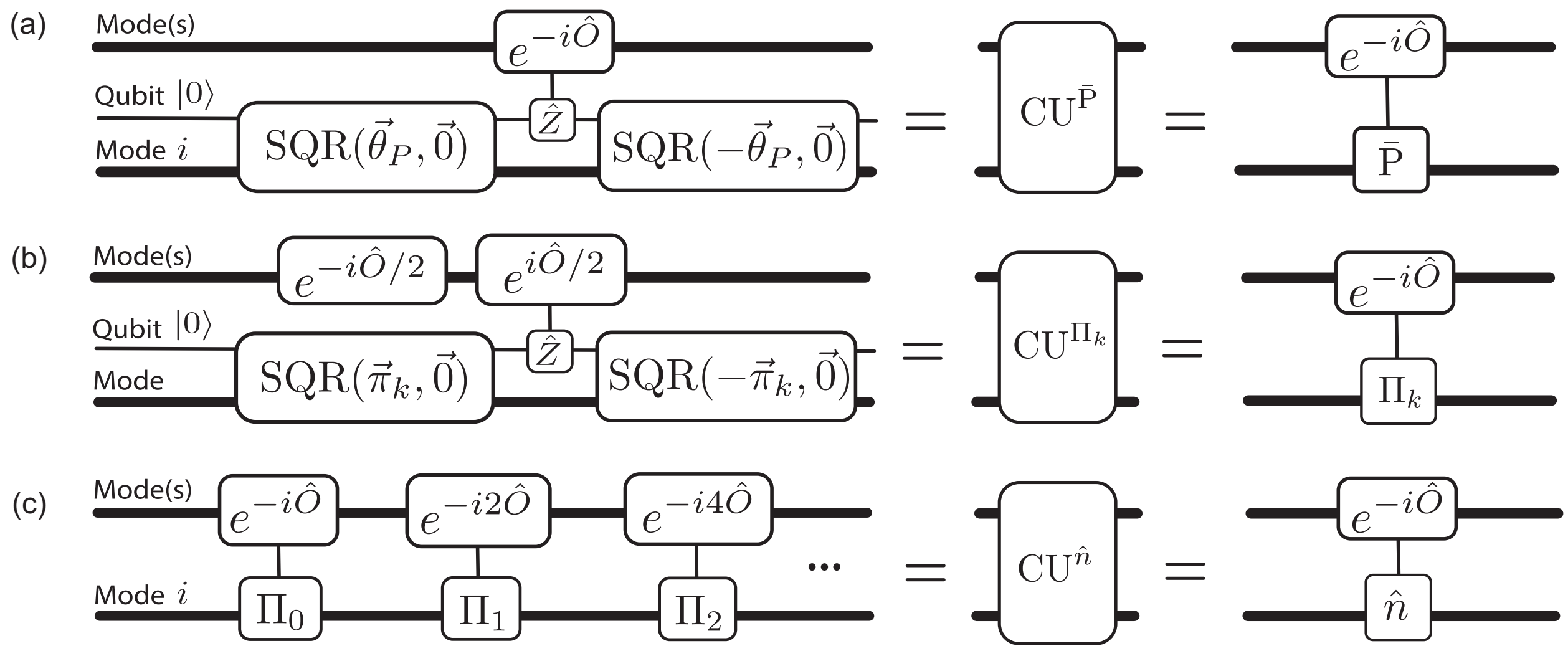

FIG. 6. Mode-conditioned and controlled unitaries. (a) Bosonic Fock-projector-conditioned unitary, see Eq. (34). (b) Bosonic parity-controlled unitary. (c) Compilation of a bosonic controlled or occupation-controlled unitary by binary phase acquisition [cf. Eq. (39)]. The upper mode denoted by "mode(s)" may contain any set of modes or qubits for which $e^{-i\hat{O}}$ commutes with the lower qubit and modes on which SQR operates.

### E. Boson density-controlled operations

A powerful use case of the generalized controlled gate in Eq. (38) is to apply many of them in sequence to synthesize complex entangling gates between two modes, mediated by an ancillary qubit in a known eigenstate of $\hat{Z}_{\mathrm{anc}}$. In this section, we leverage this technique to exactly compile a bosonic density-controlled operation:

$$\mathrm{CU}^{\hat{n}}_{i,j} = e^{-i\hat{Z}_{\mathrm{anc}}\hat{n}_i\hat{O}_j}. \tag{39}$$

To obtain this gate, we follow the iterative procedure illustrated in Fig. 6(c). The core idea is to sequentially condition the qubit-boson gate $U = \exp(-i\hat{Z}_{\mathrm{anc}}\hat{O}_j)$ on each bit of the binary representation of $\hat{n}_i$. With each iteration, an appropriate rotation angle is chosen such that the sequence altogether produces the desired operation $\mathrm{CU}^{\hat{n}}_{i,j}$. This strategy is reminiscent of binary photon number readout [144,211], and it requires a circuit depth that is logarithmic in the boson number cutoff $n_{\max}$.

Specifically, as a particular instance of Eq. (34), we define the operator generated by $\hat{O}_j$ and conditioned on the $k$th superparity of mode $i$,

$$\begin{aligned}\mathrm{CU}^{\hat{\Pi}_k}_{i,j}(\theta) &= e^{-i\theta\hat{Z}_{\mathrm{anc}}\hat{\Pi}_{ki}\hat{O}_j}\\ &= \mathrm{SQR}_{\mathrm{anc},i}(\vec{\pi}_k,\vec{0})e^{-i\theta\hat{Z}_{\mathrm{anc}}\hat{O}_j}\mathrm{SQR}_{\mathrm{anc},i}(-\vec{\pi}_k,\vec{0}),\end{aligned} \tag{40}$$

where we have made the variable angle $\theta$ explicit for reasons that will become clear shortly. Similarly, we define $\mathrm{CU}^{\bar{\Pi}_k}_{i,j}(\theta)$ in analogy to Eq. (38) [see Fig. 6(b)]. Here, $\hat{\Pi}_k$ is the projection operator that returns the $k$th bit of $\hat{n}_i$ (and $\bar{\Pi}_k$ its corresponding reflection operator, returning $\pm 1$), and $\vec{\pi}_k = \{\theta_0, \theta_1, \theta_2, \ldots\}$, with

$$\theta_n = \begin{cases}\pi & \text{if } \lfloor n/2^k \rfloor \bmod 2 = 1\\ 0 & \text{otherwise.}\end{cases} \tag{41}$$

For example, $\vec{\pi}_0 = \{0, \pi, 0, \pi, \ldots\}$ is equivalent to the parity vector defined below Eq. (28), converting the SQR gate into a qubit rotation controlled on the least-significant (zeroth) bit of $\hat{n}_i$. We note that $\bar{\Pi}_0 \equiv \bar{\Pi}$, where the latter is defined in Eq. (27). Likewise, the choice $\vec{\pi}_1 = \{0, 0, \pi, \pi, \ldots\}$ corresponds to qubit rotation controlled on the first bit of $\hat{n}_i$. In general,

$$\mathrm{SQR}_{\mathrm{anc},i}(\vec{\pi}_k, \vec{0}) = e^{-i\frac{\pi}{2}\hat{X}_{\mathrm{anc}}\hat{\Pi}_{ki}}, \tag{42}$$

and the choice $\vec{\pi}_k$ enables a qubit rotation controlled on the $k$th bit of $\hat{n}_i$.

Applying successive operations controlled on each bit of $\hat{n}_i$ in sequence then yields the general form

$$\begin{aligned}\mathrm{CU}^{f(\hat{\Pi}_0,\hat{\Pi}_1,\hat{\Pi}_2,\ldots)}_{i,j} &= \prod_{k=0}^{K-1}\mathrm{CU}^{\hat{\Pi}_k}_{i,j}(\theta_k)\\ &= \exp\left(-iZ_{\mathrm{anc}}\sum_{k=0}^{K-1}\theta_k\hat{\Pi}_{ki}\hat{O}_j\right),\end{aligned} \tag{43}$$

requiring $K = \lceil \log_2(n_{\max}+1)\rceil$ iterations of (super)-parity controlled gates and where $f(\hat{\Pi}_0, \hat{\Pi}_1, \hat{\Pi}_2, \ldots)$ is any linear function in the bit operators $\hat{\Pi}_k$. Returning to the initial goal of synthesizing $\mathrm{CU}^{\hat{n}}_{i,j}$ in Eq. (39), we note that

$$\hat{n}_i = \sum_{k=0}^{K-1} 2^k \hat{\Pi}_{ki}. \tag{44}$$

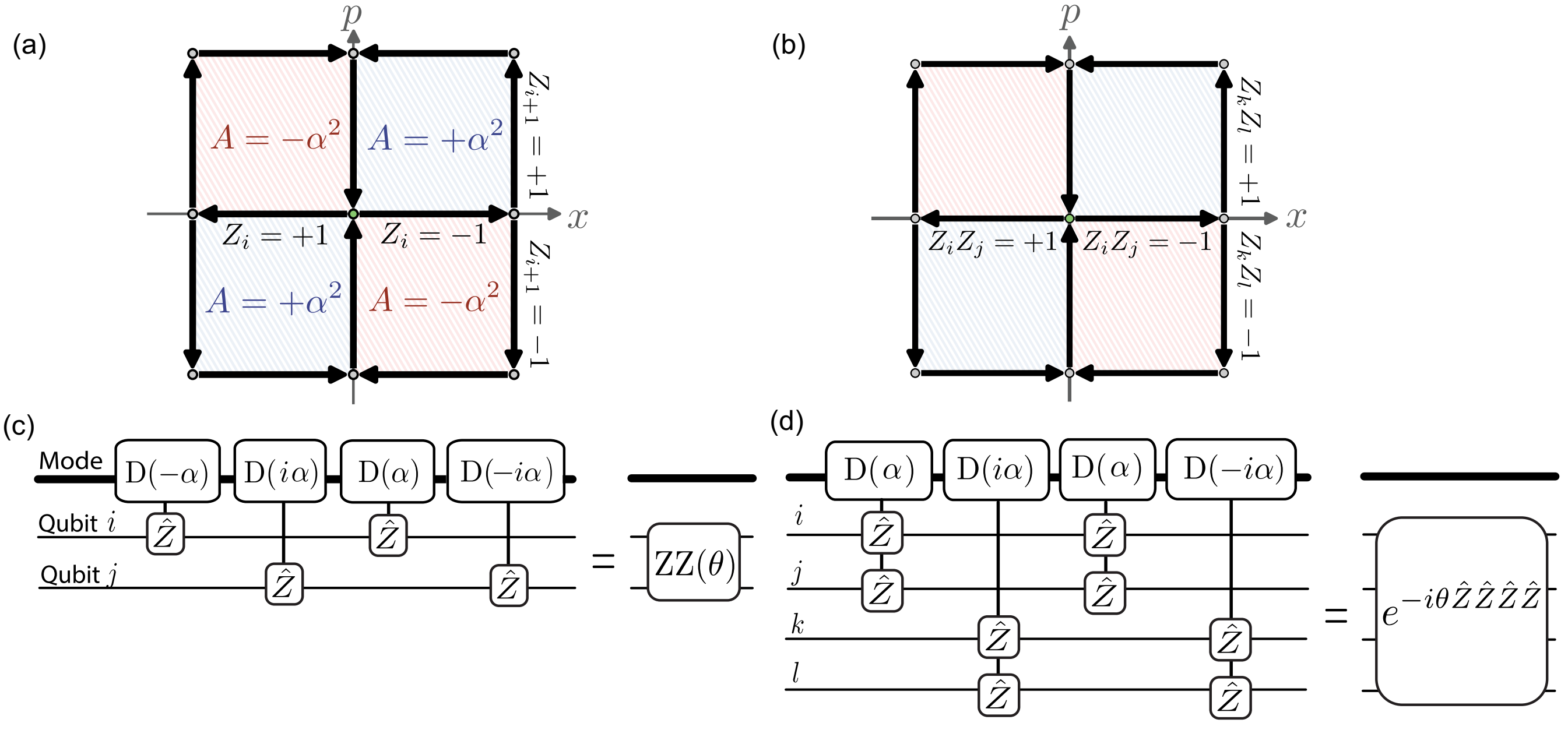

FIG. 7. Oscillator-mediated multiqubit gates. Multiqubit gates can be implemented in the proposed architecture via a sequence of conditional displacement gates on an oscillator, such that the total enclosed area $A$ corresponds to a geometric phase $e^{2iA}$ imparted on the oscillator. (a) Phase-space trajectories arising from the sequence of conditional displacements in Eq. (47), enclosing an area $A = \pm\alpha^2 Z_i Z_j$ and enacting the gate $ZZ_{i,j} = \exp[-i(\theta/2)\hat{Z}_i\hat{Z}_j]$, where $\alpha = \sqrt{\theta}/2$. The green circle indicates the starting point. Crucially, the implementation of the gate does not depend upon or alter the initial state of the mode, as all points in phase space undergo translations of the trajectory shown. (b) Circuit diagram corresponding to panel (a). Importantly, the implementation of the gate does not depend on or alter the initial state of the modes. (c) Phase-space trajectories that implement the four-qubit gate $Z_\square$ via the sequence in Eq. (50). In contrast to panel (a), all displacements are conditioned on pairs of qubits by leveraging the primitives discussed in Sec. II B. (d) Circuit diagram to realize $e^{-i\frac{\theta}{2}\hat{Z}_i\hat{Z}_j\hat{Z}_k\hat{Z}_l}$ requiring four doubly conditional displacements in sequence.

Consequently, choosing $\theta_k = 2^k$ realizes the desired gate. This sequence is shown in Fig. 6(c).

As a particular example of this gate that will prove useful in Sec. III A 4 for implementing density-density interactions, one can choose $\hat{O}_j \to \theta\hat{n}_j$ to realize the phase-space rotation of one mode controlled on the density of a second,

$$\mathrm{RR}_{i,j}(\theta) = e^{-i\theta\hat{Z}_{\mathrm{anc}}\hat{n}_i\hat{n}_j}. \tag{45}$$

For a SQR gate duration of 500 ns [93] and a conditional rotation duration less than or equal to 200 ns with $\chi = 2\pi \times 2.5$ MHz [90], one iteration towards the $\mathrm{RR}(\theta)$ gate would take 1.2 μs.

Finally, we emphasize that while our focus has been on synthesizing the density-controlled operator $\mathrm{CU}_{i,j}^{\hat{n}}$, the general form $\mathrm{CU}_{i,j}^{f(\hat{\Pi}_0,\hat{\Pi}_1,\hat{\Pi}_2,\ldots)}$ enables further possibilities beyond this particular choice. Specifically, it allows for operations conditioned on any function $f(\hat{\Pi}_0, \hat{\Pi}_1, \hat{\Pi}_2, \ldots)$ that is linear in the bit operators $\hat{\Pi}_{ki}$. Moreover, we note that an arbitrary function $f(\hat{n}_i)$ can be realized by iterating the primitive $\mathrm{CU}_{ij}^{\hat{P}}$ a number of times linear in $n_{\mathrm{max}}$. This possibility is discussed in our companion work, Ref. [57].

### F. Oscillator-mediated multiqubit gates

As the architecture illustrated in Fig. 4 does not include direct couplings between transmons, their utility for encoding fermionic and gauge-field degrees of freedom relies on the ability to synthesize transmon-transmon entangling gates using native gates. Here, we adopt an oscillator-mediated approach proposed in our companion work, Ref. [57], to realize such entangling gates that bear close similarity to Mølmer-Sørensen gates in trapped-ion platforms [218,219], and also some parallels to mediated gates in silicon [220–222]. Crucially, this approach allows for the exact analytical compilation of multiqubit gates, independent of (and without altering) the initial state of the mediating oscillators in the absence of noise. The core idea is to use a sequence of conditional oscillator displacements that form a closed phase-space path that depends upon the state of the qubits [see, for example, Fig. 7(a)]. Upon completion, this closed path leaves the oscillator unchanged yet imparts a geometric phase on the system that depends upon the state of the qubits, thus enacting the target gate.

To realize these oscillator-mediated multiqubit gates, we require the ability to enact displacements of the $j$th oscillator conditioned on the $i$th transmon,

$$\mathrm{CD}_{i,j}(\alpha) = e^{\hat{Z}_i(\alpha\hat{a}_j^\dagger - \alpha^*\hat{a}_j)}. \tag{46}$$

For $i = j$, such conditional displacements can be realized either natively [89] or compiled using conditional-parity gates and unconditional displacements, as shown in Eq. (18); furthermore, as discussed in Ref. [57] and summarized in Sec. II A, it is possible to synthesize conditional displacements for arbitrary $i$ and $j$ by additionally leveraging a beam-splitter network that links the two sites.

As a particular example, the two-qubit entangling gate $ZZ_{i,j}(\theta) = \exp[-i(\theta/2)\hat{Z}_i\hat{Z}_j]$ can be realized via a sequence of four conditional displacements,

$$ZZ_{i,j}(\theta) = \mathrm{CD}_{i,j}(-i\alpha)\mathrm{CD}_{i,i}(\alpha)\mathrm{CD}_{i,j}(+i\alpha)\mathrm{CD}_{i,i}(-\alpha). \tag{47}$$

The corresponding phase-space trajectories and circuit diagram are shown in Figs. 7(a) and 7(b), respectively. Four conditional displacements each taking 1 μs [89] would amount to a total sequence duration of 4 μs. As explained in Ref. [57], this sequence can be further optimized by leveraging conditional displacements on both modes $i$ and $j$ such that each accumulates a geometric phase in parallel, leading to a reduction in gate duration by a factor of $\sqrt{2}$ compared to the above single-mode approach, while also reducing the potential impact of nonidealities such as self-Kerr interactions (anharmonicities). See Ref. [57] for more details.

This strategy can be extended to realize oscillator-mediated $N$-qubit gates that are useful for studying lattice gauge theories. For example, as will be discussed in Sec. IV A 2, the four-qubit gate

$$e^{-i\frac{\theta}{2}\hat{Z}_i\hat{Z}_j\hat{Z}_k\hat{Z}_l} \tag{48}$$

is useful for implementing a magnetic-field term in the $\mathbb{Z}_2$ LGT introduced in Eq. (1). To realize this gate, we modify the sequence in Eq. (47) such that each displacement is conditioned on the joint state of *pairs* of qubits, e.g., $\hat{Z}_i\hat{Z}_j$ or $\hat{Z}_k\hat{Z}_l$, as shown in Fig. 7(c). To that end, we define the doubly conditional displacement as

$$\mathrm{CD}^{ZZ}_{i,j,r}(\alpha) = e^{\hat{Z}_i\hat{Z}_j(\alpha\hat{a}_r^\dagger - \alpha^*\hat{a}_r)}. \tag{49}$$

This gate can be implemented via the technique described in Sec. II B. As illustrated in Fig. 7(d), the four-qubit term can be exactly realized by sequencing four such doubly conditional displacements:

$$e^{-i\frac{\theta}{2}\hat{Z}_i\hat{Z}_j\hat{Z}_k\hat{Z}_l} = \mathrm{CCD}_{i,j,r}(-i\alpha)\mathrm{CCD}_{k,l,r}(\alpha) \times \mathrm{CCD}_{i,j,r}(+i\alpha)\mathrm{CCD}_{k,l,r}(-\alpha), \tag{50}$$

with $\alpha = \sqrt{\theta}/2$.

Similar to the implementation of $ZZ_{i,j}(\theta)$, it is beneficial to use multiple modes in parallel to accumulate the necessary geometric phase and reduce the displacement parameter $\alpha$. Moreover, this strategy can be extended to realize arbitrary operations of the form $e^{-i\theta\hat{W}}$, where $\hat{W}$ is an arbitrary-weight Pauli operator. See Ref. [57] for a more complete discussion.

Finally, we note that a similar strategy for oscillator-mediated $N$-body entangling gates has been proposed [223] and demonstrated [37] in trapped-ion platforms. There, spin-dependent squeezing operations are used in place of conditional displacement operators, but the overall framework bears similarity to the strategy described here for the superconducting architecture described in Fig. 4.

### G. Approximate synthesis of multimode gates

To approximately implement arbitrary multimode operations that fall outside of the form in Secs. II C and II E, we use the method introduced in Ref. [97]. It relies on a variant of the Baker-Campbell-Hausdorff (BCH) formula,

$$e^{\hat{A}t}e^{\hat{B}t}e^{-\hat{A}t}e^{-\hat{B}t} = e^{[\hat{A},\hat{B}]t^2 + O(t^3)}. \tag{51}$$

The key insight is to choose the operators $\hat{A}$ and $\hat{B}$ such that they enact commuting bosonic operations $\hat{O}_1$ and $\hat{O}_2$ (possibly acting on distinct sets of different modes) but conditioned on the *same* qubit with respect to noncommuting Pauli operators, e.g., using the method described in Sec. II B. For example, we can leverage this idea as follows:

$$\begin{aligned}\Gamma_{1,2} &= e^{i\hat{X}\hat{O}_1 t}e^{i\hat{Y}\hat{O}_2 t}e^{-i\hat{X}\hat{O}_1 t}e^{-i\hat{Y}\hat{O}_2 t}\\ &= e^{[\hat{X}\hat{O}_1,\hat{Y}\hat{O}_2](-it)^2 + O(t^3)}\\ &= e^{-2i\hat{Z}\hat{O}_1\hat{O}_2 t^2 + O(t^3)}.\end{aligned} \tag{52}$$

See also Fig. 8(a) for the circuit diagram. In this way, one can implement operator multiplication between $O_1$ and $O_2$. Combining this case with the Trotter formula to implement addition, following Ref. [97] and Fig. 8(b), we can implement exponentials of nonlinear functions in the creation and annihilation operators.

When the commutator obtained through the above BCH formula contains both desired and undesired high-order mode terms, we can cancel the latter by applying Eq. (52) a second time with a different choice of operators. Via the Trotter formula, we then obtain

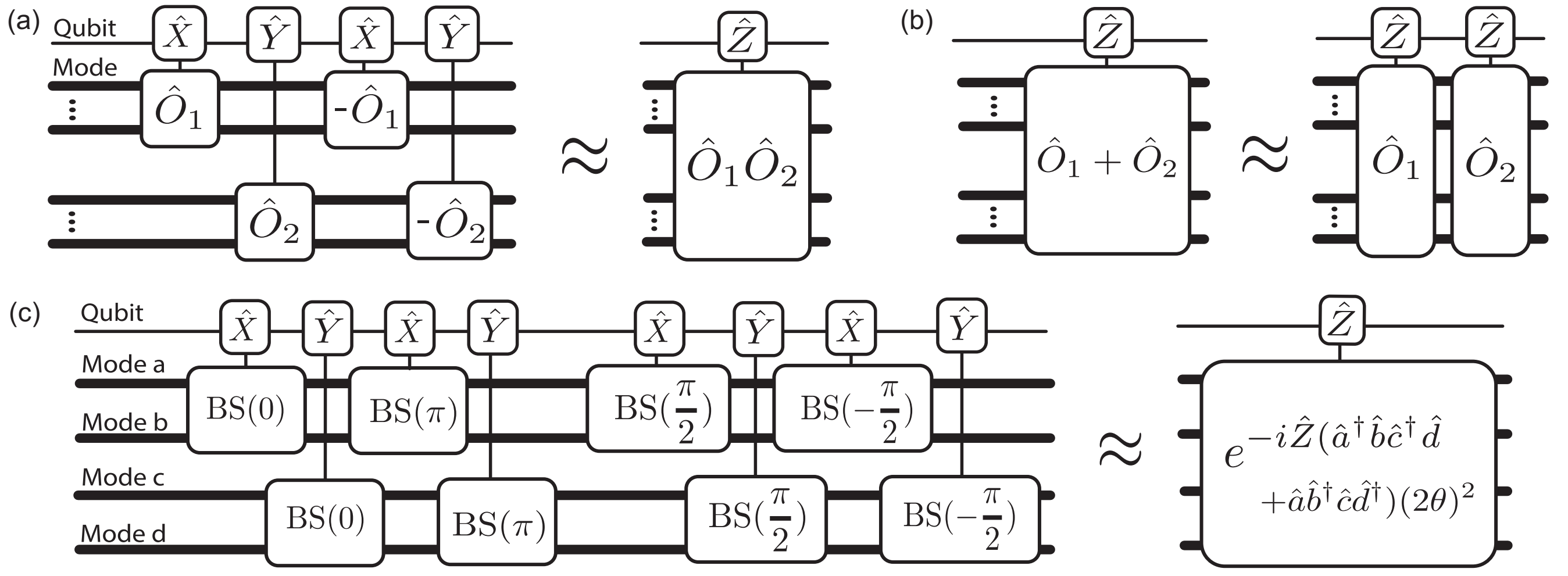

FIG. 8. Arbitrary higher-order operator compilation. (a) Operator multiplication using the Baker-Campbell-Hausdorff formula. (b) Operator addition using the Trotter formula. (c) Example: compilation of qubit-conditional quartic term. Note that these methods are only exact if $[\hat{O}_1, \hat{O}_2] = 0$. The variable parameter $\theta$ is omitted from the beam splitters BS for ease of notation as it is the same parameter for all of them. Ancilla-conditional operations can be synthesized using the method described in Sec. II B.

$$\Delta_{1,2} = e^{-2i\hat{Z}\hat{O}_1\hat{O}_2t^2} e^{-2i\hat{Z}\,\hat{O}_1'\,\hat{O}_2't^2} + O(t^3)$$
$$= e^{-2i\hat{Z}(\hat{O}_1\hat{O}_2+\hat{O}_1'\hat{O}_2')t^2} + O(t^3), \quad (53)$$

where the error term will be different between the two lines and we rewrote the error as an additive error by Taylor expanding the exponential. Here, each term is conditioned on the same qubit, either natively or via SWAP operations between the qubits or the modes.

In Fig. 8(c), we use this technique to approximately synthesize the four-mode operator,

$$e^{-i\hat{Z}(\hat{a}^\dagger\hat{b}\hat{c}^\dagger\hat{d}+\mathrm{H.c.})(2\theta)^2}, \quad (54)$$

which will later be used in Sec. IV B 2 to realize the plaquette term in the $U(1)$ quantum link model.

### H. Dual-rail qubits

While so far we have discussed primitives for mapping the model to the hardware with transmons being used as data or ancillary qubits and the modes being used to encode bosonic degrees of freedom, another possibility is to use the modes to encode qubits. In this section, we discuss such an encoding: the dual-rail qubit. The core idea is to encode a single qubit using a pair of cavity modes that share a single photon. The location of the photon then represents the qubit state [201,214,216]

$$|0\rangle_{\mathrm{DR}} = |0, 1\rangle, \quad (55)$$

$$|1\rangle_{\mathrm{DR}} = |1, 0\rangle. \quad (56)$$

Denoting the cavity photon numbers as $|n^a, n^b\rangle$, we see that the Pauli $Z$ operator has various equivalent representations,

$$\hat{Z}^{\mathrm{DR}} = \hat{n}^b - \hat{n}^a \quad (57)$$

$$= \mathbb{1} - 2\hat{n}^a = 2\hat{n}^b - \mathbb{1} \quad (58)$$

$$= e^{i\pi\hat{n}^a} = -e^{i\pi\hat{n}^b}, \quad (59)$$

where we have used the fact that $\hat{n}^a + \hat{n}^b = \mathbb{1}$. The remaining Pauli operators are given by

$$X = a^\dagger b + ab^\dagger, \quad (60)$$

$$Y = i[a^\dagger b - ab^\dagger]. \quad (61)$$

Note that this is simply the Schwinger boson representation of a spin $1/2$.

The primary advantage of the dual-rail encoding is that photon loss in either the left or right cavity results in an erasure error that is detectable via joint-parity measurements [201,214,216]. Furthermore, single-qubit operations are straightforward: rotations about any axis in the azimuthal plane can be carried out with a beam splitter, and rotations about the $Z$ axis correspond to a phase-space rotation of one of the modes (up to a global phase). Furthermore, it has been shown that one can construct a universal gate set that enables detection of ancillary transmon dephasing and relaxation errors, the latter made possible by leveraging three levels of the transmon qubit [210]. Consequently, for models that require many qubit-qubit gates (such as 2D fermionic problems that necessitate fermionic SWAP networks), the dual-rail encoding is potentially advantageous for mitigating errors with post-selection in near-term experiments.

*The* $*ZZ^{\mathrm{DR}}_{i,j}(\theta)$ *case.* As an example of an entangling gate between dual-rail qubits, we consider the compilation of $ZZ^{\mathrm{DR}}_{i,j}(\theta) = e^{-i(\theta/2)\hat{Z}^{\mathrm{DR}}_i\hat{Z}^{\mathrm{DR}}_j}$. Using the fact that $\hat{Z}^{\mathrm{DR}}_i = \mathbb{1} - 2\hat{n}^a_i$, this gate can be expressed as

$$\begin{aligned} ZZ^{\mathrm{DR}}_{i,j}(\theta) &= e^{-i\frac{\theta}{2}}e^{-i2\theta\hat{n}^a_i\hat{n}^a_j}e^{i\theta\hat{n}^a_i}e^{i\theta\hat{n}^a_j} \\ &= e^{-i\frac{\theta}{2}}(\langle 0|_{\mathrm{anc}}\otimes\mathbb{1})\mathrm{RR}_{i,j}(2\theta)(|0\rangle_{\mathrm{anc}}\otimes\mathbb{1}) \\ &\quad\times R_i(-\theta)\mathrm{R}_j(-\theta), \end{aligned} \tag{62}$$

where, for simplicity of notation, we have dropped the superscript $a$. Therefore, this gate can be realized by combining a pair of phase-space rotations with the technique introduced in Sec. II E to realize $\mathrm{RR}(2\theta)$—see Eq. (45).

However, note that in the dual-rail encoding, the boson number of mode $i$ is restricted to either 0 or 1, which means that the SQR gate used to implement Eq. (45) of Sec. II E can be reduced to a conditional phase-space rotation up to single-qubit operations. For the choice of $\vec{\theta} = \vec{\pi}_0$, this conditional rotation in the $\{|0\rangle, |1\rangle\}$ subspace corresponds to a conditional-parity gate:

$$\mathrm{SQR}_{i,j}(\vec{\pi}_0, \vec{0}) \Rightarrow H_i\mathrm{C\Pi}_{i,j}H_i = \mathrm{C\Pi}^X_{i,j}, \tag{63}$$

where $H_i$ is a Hadamard gate on qubit $i$ and we have defined the shorthand $\mathrm{C\Pi}^X_{i,j}$ to denote the $X$-conditional-parity gate. Therefore, we can rewrite the decomposition of $ZZ^{\mathrm{DR}}_{i,j}(\theta)$ as

$$\begin{aligned} ZZ^{\mathrm{DR}}_{i,j}(\theta) = e^{-i\frac{\theta}{2}}(\langle 0|_{\mathrm{anc}}\otimes\mathbb{1})\mathrm{C\Pi}^X_{\mathrm{anc},i}CR_{\mathrm{anc},j}(-2\theta)\mathrm{C\Pi}^X_{\mathrm{anc},i} \\ \times(|0\rangle_{\mathrm{anc}}\otimes\mathbb{1})R_i(\pi)R_j(\theta)R_i(-\theta)R_j(-\theta), \end{aligned} \tag{64}$$

where the top line, up to the global phase, corresponds to $\mathrm{RR}(2\theta)$ in the truncated dual-rail subspace. The complete circuit (with single-qubit rotations simplified) is presented in Fig. 9(a). We note that sequential ancilla-conditional gates using the same transmon (but distinct modes) require implicit SWAP operations not shown.

While exact, the circuit in Fig. 9(a) does not enable the detection of transmon dephasing errors. To that end, a scheme to synthesize error-detectable entangling gates was presented in Ref. [210]. It leverages an "exponentiation circuit" that interleaves ancilla-controlled unitaries (controlled-$\hat{P}$) with ancilla rotations to construct any unitary $\hat{P}(\theta) = \exp[-i(\theta/2)\hat{P}]$, provided $\hat{P}^2 = \hat{\mathbb{1}}$. Returning to the case of $ZZ^{\mathrm{DR}}_{i,j}(\theta)$, this gate can be constructed following the prescription in the previous paragraph, choosing $\hat{P} = \hat{Z}\otimes\hat{Z}$ and implementing controlled-$\hat{P}$ using a pair of conditional-parity gates. We illustrate the full implementation in terms of our discrete gate set in Fig. 9(b). Notably, this latter circuit can be understood as a further decomposition of its counterpart in Fig. 9(a), where the middle $\mathrm{CR}(-2\theta)$ gate is broken into conditional-parity gates and single-qubit operations using the identities in Sec. II C.

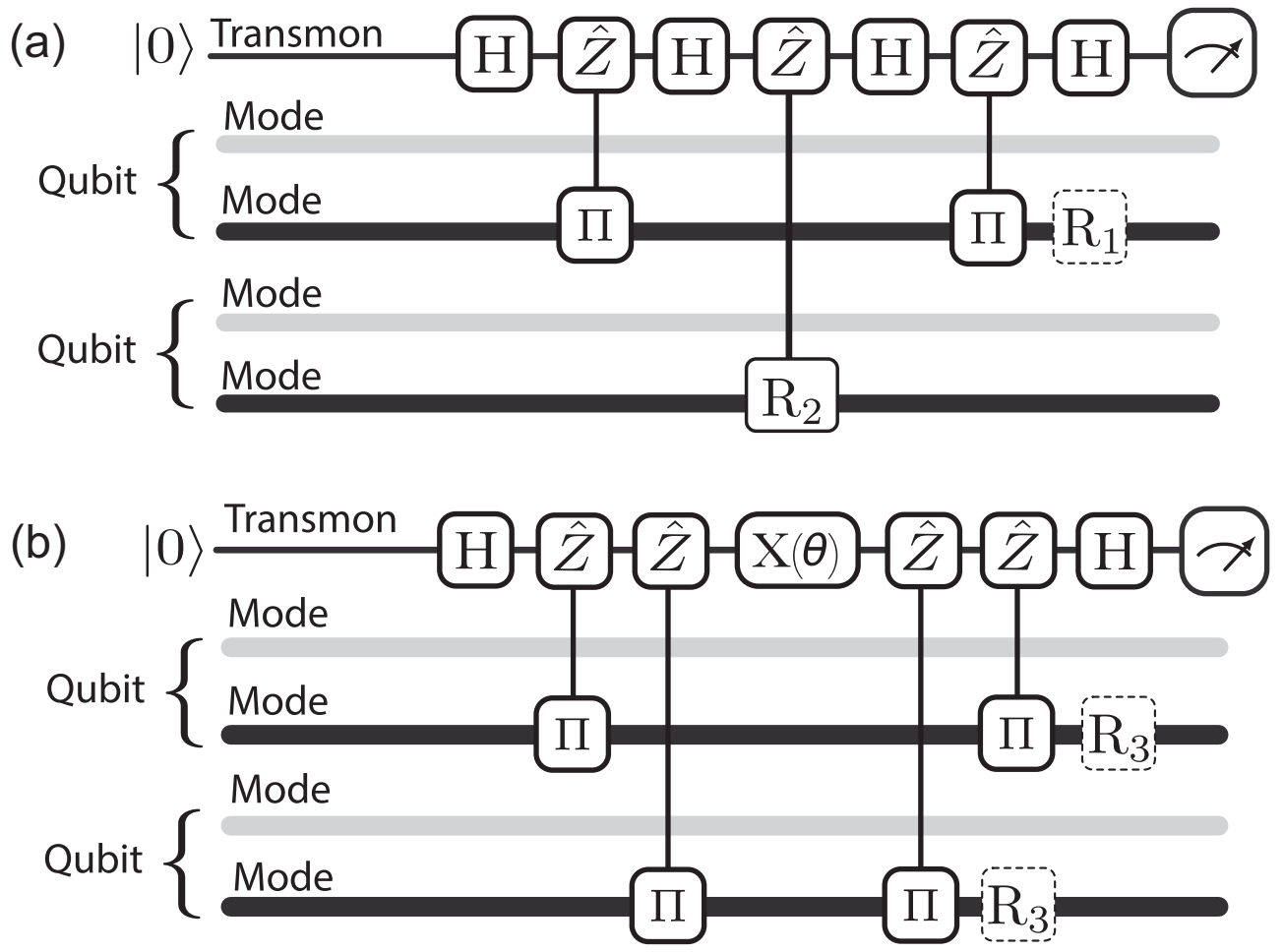

FIG. 9. Dual-rail entangling gate. Both panels show compilation methods for $ZZ(\theta) = \exp[-i(\theta/2)\hat{Z}_i\hat{Z}_j]$. The gray andcwires correspond to the location of the boson when the dual-rail qubit is in state $|0\rangle$ or $|1\rangle$, respectively. In both methods, the ancilla transmon must start out in $|0\rangle$. (a) Circuit as compiled using the methods shown in Sec. II C. Note that $\mathrm{R}_1 = R(\pi - \theta)$ and $\mathrm{R}_2 = R(-2\theta)$. The measurement is optional—it can help detect amplitude damping errors. (b) Method presented in Ref. [210] for implementing $ZZ(\theta)$. Note that $\mathrm{R}_3 = \mathrm{R}(\pi)$. The $Z$-conditional rotation gates refer to $\mathrm{CR}(\theta)$. The measurement will help detect amplitude damping errors as well as phase errors [210].

While both circuits are mathematically equivalent, the structure of Fig. 9(b) enables partial error detection of some of the dominant errors, including transmon dephasing. However, it comes at the cost of requiring four conditional-parity gates, each with a duration $T_{\mathrm{C\Pi}} = \pi/\chi$. Thus, ignoring the cost of SWAPs and single-qubit and mode operations, the minimum duration of $ZZ^{\mathrm{DR}}_{i,j}(\theta)$ using the method in Fig. 9(b) is $T_{b,\min} = 4\pi/\chi$, independent of $\theta$. In contrast, the method in Fig. 9(a) requires $T_{a,\min} = 2(\pi + \theta)/\chi$, a reduction for even the maximal entangling case $\theta = \pi/2$. Thus, there is a trade-off in error detectability and gate duration, and the approach in Fig. 9(a) may therefore be beneficial in certain contexts, particularly for repeated applications of $ZZ^{\mathrm{DR}}_{i,j}(\theta)$ for small $\theta$ (e.g., for Trotterized circuits).

## III. IMPLEMENTATION OF MATTER FIELDS

In this section, we leverage the compilation strategies of the previous section to realize the unitary time-evolution operator $e^{-i\hat{H}t}$ for a range of common Hamiltonian terms involving bosonic and fermionic matter. For the purely bosonic matter terms, in Sec. III A, we show that some of the common dynamics terms correspond to native gates and others to terms that can be realized using the compilation techniques in the previous section. For the purely fermionic matter terms, in Sec. III B, we demonstrate how to encode fermionic SWAP networks into various qubit encodings possible in hybrid qubit-boson hardware. For the fermion-boson interaction terms, in Sec. III C, we present novel compilation strategies specific to hybrid qubit-boson hardware. We summarize these results in Tables IV and V. All terms involving gauge fields will be treated in Sec. IV.

TABLE IV. Bosonic Hamiltonian terms and the implementation of $e^{-i\hat{H}t}$ in the proposed architecture. Note that for the on-site interaction, an ancillary qubit is required—see Appendix B for an ancilla-free implementation that leverages the Gauss law constraint in a $\mathbb{Z}_2$ lattice gauge theory.

| $\hat{H}$ | $e^{-i\hat{H}t}$ gates | Details |
|---|---|---|
| $\sum_{\langle i,j\rangle} J_{i,j}(e^{i\varphi}\hat{a}_i^\dagger \hat{a}_j + \text{H.c.})$ | $\text{BS}_{i,j}(\varphi,\theta)$ | Table I F, Sec. III A 1 |
| $\sum_i \mu_i\ \hat{n}_i$ | $\text{R}_i(\theta)$ | Table I F, Sec. III A 2 |
| $\sum_i U_i \hat{n}_i^2$ | $\text{SNAP}_i(\vec{\theta})$ | Table I F, Sec. III A 3 |
| $\sum_{\langle i,j\rangle} V_{i,j}\hat{n}_i\hat{n}_j$ | $\text{RR}_{i,j}(\theta)$ | Fig. 6, Sec. III A 4 |

### A. Bosonic matter

In this section, we discuss the most common Hamiltonian terms for bosonic matter fields (which typically involve small powers of the creation and annihilation operators). The results are summarized in Table IV. We first discuss hopping as this is the most elementary term coupling bosonic modes. We then discuss terms related to powers and products of number operators $\hat{n}$: the on-site potential (linear in $\hat{n}$), the on-site interaction (quadratic in $\hat{n}$), and the intersite interaction (bilinear, $\hat{n}_i\hat{n}_j$). We map the bosonic matter fields to the native bosonic modes supported in the hybrid qubit-boson hardware in Fig. 4.

#### *1. Hopping*

The time-evolution operator $e^{-i\hat{H}t}$ for the hopping interaction $\hat{H} = \sum_{\langle i,j\rangle} J_{i,j}(\hat{a}_i^\dagger \hat{a}_j + \text{H.c.})$ between bosonic matter sites can be implemented via Trotterization using a sequence of beam-splitter gates $\text{BS}_{i,j}(\varphi,\theta)$ (defined in Table II) by choosing $\theta = Jt$. Furthermore, a static gauge field (vector potential) appears as a complex phase in the hopping $J \to Je^{i\varphi}$ [224], and it is important to simulate models of the fractional quantum Hall effect [225]. To implement this term, one can simply adjust the phase $\varphi$ of the beam splitter on each link such that a net phase is accumulated in moving around each plaquette.

#### *2. On-site potential*

The on-site potential is $\hat{H} = \sum_i \mu_i \hat{n}_i$, with $\hat{n}_i = \hat{a}_i^\dagger \hat{a}_i \mu_i \in \mathbb{R}$. The time evolution of this term can be implemented on each site separately using a phase-space rotation gate (defined in Table II) as is shown in Table IV.

As mentioned previously in Sec. I F, it is possible to absorb phase-space rotations into operators that include creation or annihilation operators such as beam splitters. In particular, a constant site energy shift can be created with a fixed frequency detuning of the microwave tones that activate the beam splitters.

#### *3. On-site interaction*

The bosonic on-site interaction is $\sum_i U_i \hat{n}_i^2$, with $U_i \in \mathbb{R}$. To implement the time-evolution operator for this term, one can use the SNAP gate [200]. As defined in Table II, $\text{SNAP}_i(\vec{\theta})$ operates in the strong-dispersive coupling regime where the qubit frequencies depend on the mode occupation such that the SNAP gate effectively imparts an independently chosen phase to the system comprising the qubit and mode, for each value of the mode occupation. Choosing $\theta_k = U_i k^2 t$, where $k$ is the photon number corresponding to a particular Fock state, implements the on-site interaction $\hat{n}_i^2$ on the $i$th mode. This procedure additionally requires an ancillary qubit initialized to the state $|0\rangle$.

TABLE V. Fermionic Hamiltonian terms. We list native gates required for the implementation of essential Hamiltonian terms in the proposed architecture. Here, $\lambda$ is proportional to time and the Hamiltonian parameter. The fourth column only lists the necessary gates; it does not provide the complete circuits—these are described in the relevant sections listed in the last column of the table. The on-site term $\sum_i U_i \hat{n}_{i,\gamma}\hat{n}_{i,\delta}$ contains terms that depend on $\gamma$ and $\delta$, which refer to the fermion spin. Mapping these two spin species to two different fermionic sites $i$ and $j$ casts this term into the same form as the nearest-neighbor term in the bottom row of the table.

| $\hat{H}$ | Intermediary step | Implementation | $e^{-i\hat{H}t}$ gates | Details |
|---|---|---|---|---|
| $\sum_{\langle i,j\rangle} J_{i,j}(\hat{c}_i^\dagger \hat{c}_j + \text{H.c.})$ | (1) FSWAP<br>(2) iSWAP($\lambda$) | Dual rail | (1) $\text{SWAP}_{i,j}$, $\text{C}\Pi_{i,j}$, $\text{R}^{\varphi}{}_i(\theta)$, $\text{R}_i(\theta)$<br>(2) $\text{BS}_{i,j}(\varphi,\theta)$, $\text{C}\Pi_{i,j}$, $\text{R}^{\varphi}{}_i(\theta)$ | Fig. 10, Appendix A 1 |
| | | Transmon | (1) $\text{SWAP}_{i,j}, \text{CD}_{i,j}(\theta), \text{R}_i^Z(\theta)$<br>(2) $\text{CD}_{i,j}(\theta)$, $\text{R}_i^{\varphi}(\theta)$ | Fig. 7, Sec. II F |
| $\sum_i \mu_i \hat{n}_i$ | $\exp(-i\lambda \hat{Z}_i)$ | Dual rail | $\text{R}_i(\theta)$ | Table I F, Appendix A 3 |
| | | Transmon | $\text{R}_i^z(\theta)$ | Table I F |
| $\sum_{\langle i,j\rangle} V_{i,j}\hat{n}_i\hat{n}_j$ | $\exp(-i\frac{\lambda}{2}\hat{Z}_i\hat{Z}_j)$ | Dual rail | $\text{C}\Pi_{i,j}$, $\text{R}_i^{\varphi}(\theta)$ | Appendix A 3 |
| | | Transmon | $\text{CD}_{i,j}(\theta), \text{C}\Pi_{i,j}$ | Fig. 7, Sec. II F |

#### 4. Intersite interaction

The time evolution of a nearest-neighbor density-density interaction $\hat{H} = \sum_{\langle i,j \rangle} J_{i,j} \hat{n}_i \hat{n}_j$ between two modes in separate cavities, which is useful, for example, in simulation of the extended Hubbard model or dipolar interacting systems, can be implemented with the method proposed in Sec. II E via the RR($\theta$) gate defined in Eq. (45):

$$e^{-iJ_{i,j}\hat{n}_i\hat{n}_j t} = (\langle 0|_{\text{anc}} \otimes \mathbb{1}) \text{RR}_{i,j}(J_{i,j}t)(|0\rangle_{\text{anc}} \otimes \mathbb{1}). \quad (65)$$

Here, we have used an ancillary qubit that begins and (deterministically) ends in the state $|0\rangle_{\text{anc}}$. As discussed in Sec. II E and shown in Table III, this gate requires a native gate count that is logarithmic in the bosonic cutoff $n_{\text{max}}$. Long-range interactions between sites $i$ and $j$ can be implemented at the cost of a bosonic SWAP gate count that is linear in the Manhattan-metric distance between the sites.

### B. Fermionic matter

In this section, we show that it is possible to treat purely fermionic terms within our proposed hybrid qubit-boson architecture. Although we do not expect this area to have an explicit advantage over qubit-only hardware, we show that no significant cost is added compared to existing qubit platforms. Note that many of the compilation strategies in this work are hardware agnostic and could be used in concatenation with digital fermionic modes [74]. In the following, however, we focus on using qubit degrees of freedom to represent the fermionic modes.

We treat fermionic matter by mapping the fermions to qubits and rely on FSWAP networks [226] of Jordan-Wigner (JW) strings for a low-depth implementation (a logarithmic depth implementation was proposed here [74]). For an overview of the set of operations required to implement $e^{-i\hat{H}t}$ for a 2D fermionic Hamiltonian $\hat{H}$ in qubit-based hardware, see Appendix A. In summary, the two-dimensional lattice is virtually mapped to a one-dimensional chain. Nearest-neighbor hoppings in the 2D lattice become either nearest-neighbor hoppings in the 1D JW chain (which we call "JW adjacent") or long-range hoppings (which we call "non-JW adjacent"). The former hoppings only require iSWAP operations, whereas the latter also require FSWAP operations. Gates that involve the fermionic number density are mapped to gates involving the $\hat{Z}$ operator and do not add any overhead associated with the Jordan-Wigner encoding.

We first summarize the possibilities for encoding fermions (i.e., implementing iSWAP and FSWAP operations) either in transmons (Sec. III B 1) or in cavity dual-rail qubits (Sec. III B 2). While the former is more hardware efficient, the latter enables error detection and, as a result, is a particularly promising avenue for near-term simulations on noisy hardware. The high-level results for both encodings are summarized in Table V. In Sec. III B 3, we discuss the fidelity of gates that underly the simulation of fermions in the dual-rail encoding.

#### 1. Transmon qubits

In the proposed 3D post-cavity circuit QED setup, encoding the fermion into the transmon enables the use of the cavities for a separate purpose, such as for representing the phonon modes in the Hubbard-Holstein model or for encoding bosonic gauge fields, as will be demonstrated in the context of a $\mathbb{Z}_2$ lattice gauge theory later in this work.

As the architecture illustrated in Fig. 4 does not include direct couplings between transmons, we rely on the oscillator-mediated compilation strategy discussed in Sec. II F to realize entangling gates, $ZZ_{i,j}(\theta)$, in particular. We reemphasize that this approach does not require that the mediating oscillators are in a known, particular state. Thus, they can simultaneously be used to encode matter or gauge degrees of freedom.

#### 2. Dual-rail qubits

We now discuss the separate possibility to map fermions to dual-rail qubits. As discussed in Sec. II H, dual-rail qubits present a number of advantages, including the possibility to detect photon loss and ancilla relaxation and dephasing errors [210].

*FSWAP operation.* For the non-JW-adjacent hopping terms, a full $\text{SWAP}_{i,j}$ operation is required between the dual-rail qubits as part of the FSWAP operation. This operation can be implemented by separately applying a $\text{SWAP}_{i,j}$ operation between the cavities representing the $|1\rangle$ and $|0\rangle$ states, as shown in Fig. 10. As described in Appendix A 1, the FSWAP also requires a $\text{CZ}_{i,j}(\theta)$ gate that is equivalent to the $ZZ_{i,j}(\theta)$ gate up to single-qubit phase gates. As discussed in Sec. II H (cf. Appendix A 1), these single-qubit phase gates correspond to phase-space rotations on the mode that hosts the boson when the dual-rail qubit is in state $|1\rangle$. The corresponding circuit is shown in Fig. 10(a).

*The* iSWAP($\theta$) *gate.* For the JW-adjacent hopping terms, following Eq. (A3), we only need to implement a variable-angle iSWAP gate defined in Fig. 10(b), which requires only $ZZ_{i,j}(\theta)$ gates and $H$ and $S$ single-qubit rotations. An error-detectable implementation of $ZZ_{i,j}(\theta)$ between dual-rail qubits was presented in Ref. [210] and is summarized here in Sec. II H. We show the circuit for implementing $e^{i\theta(\hat{X}_i\hat{X}_{i+1}+\hat{Y}_i\hat{Y}_{i+1})}$ at the level of the modes supporting the dual-rail qubits in Fig. 10(b).

#### 3. Comparing advantages of different platforms for encoding fermions

We note that different platforms have different advantages for implementing non-JW-adjacent hopping.

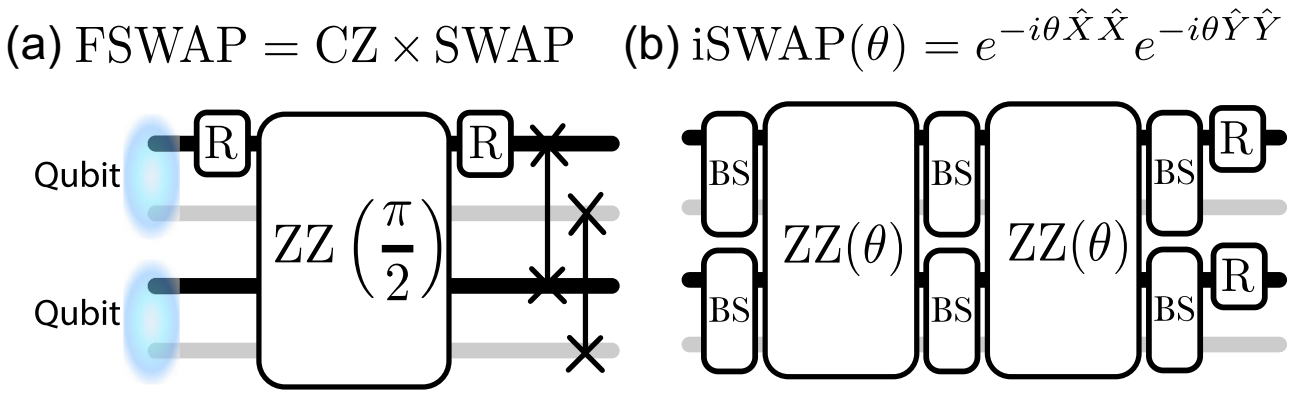

FIG. 10. Fermionic SWAP network operations in dual-rail qubits. (a) FSWAP gate. (b) Arbitrary-angle iSWAP gate for implementing the hopping on Jordan-Wigner adjacent sites. The vertical lines connecting the crosses are SWAP gates between the bosonic modes (with the pair together realizing a SWAP gate between dual-rail qubits), $R$ is shorthand for a phase-space rotation gate $\mathrm{R}(\pi/4)$, BS denotes a beam splitter $\mathrm{BS}(\varphi,\tilde{\theta})$ defined in Table II, and the $ZZ(\theta)$ gate acting on the dual-rail qubits is defined in Sec. II H. The gray and black mode wires correspond to the location of the boson when the dual-rail qubit is in state $|0\rangle$ or $|1\rangle$, respectively.

In a planar transmon qubit array with tunable couplers, FSWAP operations can be implemented using the FSIM gate using $\mathrm{FSIM}[\theta=(\pi/2),\phi=0]$ with high fidelity (approximately 0.9930) [227]. The $e^{i\theta(\hat{X}\hat{X}+\hat{Y}\hat{Y})}$ operations can be implemented using $\mathrm{FSIM}(\theta,\phi=0)$, also with high fidelity (at present, approximately 0.9962) [227].

In the ion-trap QCCD architecture [228,229], FSWAP operations can be implemented using a SWAP operation and a $ZZ(\pi/2)$. Because the ions can be physically moved around the circuit to achieve all-to-all connectivity, the fidelity of the FSWAP operation reduces to the fidelity of $ZZ(\pi/2)$, which is currently 0.9980 [228,229], yielding a fidelity advantage compared to the planar transmon implementation. Furthermore, $e^{i\theta(\hat{X}\hat{X}+\hat{Y}\hat{Y})}$ can be implemented with two $ZZ(\theta)$, which currently has a fidelity of approximately 0.9960 [228].

In the circuit QED setup using dual-rail qubits, as discussed, we can use the $ZZ(\theta)$ gates to implement fermionic SWAP networks similar to ion-trap QCCD architectures. Beam-splitter operations in the high-Q cavity setup required for a SWAP operation between dual-rail qubits have a very high fidelity of 0.9992 [91,217] (comparable to single-qubit gates on transmons). Importantly, we can use the error-detection capability to increase the fidelity of the operations with postselection, as discussed in Refs. [210,214]. This process results in $ZZ(\pi/2)$ gates with a theoretically calculated fidelity in the presence of noise and postselection of approximately 0.9999 [210], for a pure dephasing channel with time $T_\phi = 200\,\mu\mathrm{s}$. The error detection comes at the price of a postselection requirement, leading to a shot overhead. For the above value of $T_\phi$, it was estimated that an error would be detected in 2%[210] of shots. While this percentage increases exponentially with circuit depth, even for 100 $ZZ_{i,i+1}(\theta)$ gates, only $1-(1-0.02)^{100}\approx 86.7\%$ of shots would have to be discarded due to a detected error, which leads to a shot overhead of a factor of $1/(1-86.7\%)\approx 7.5$.

### C. Fermion-boson interactions

Here, we compile the unitary time-evolution operator for two paradigmatic fermion-boson models: phonon-electron interactions in solids and the coupling of strong light fields with electrons. Employing the techniques of the previous sections, we discuss the use of either transmons or dual-rail qubits to encode the fermionic matter, and we employ modes to represent the bosonic matter.

#### *1. Phonon-matter interactions*

The Holstein model [230] describes (dispersionless) optical phonons in the solid state coupled to the density of electrons. It consists of a kinetic energy term for the electrons (cf. first row in Table V), an on-site potential term for the phonons (cf. Sec. III A 2), and an electron-phonon interaction term

$$\hat{H} = \sum_i g_i \hat{c}_i^\dagger \hat{c}_i (\hat{a}_i^\dagger + \hat{a}_i). \tag{66}$$

To implement the latter, note that in the Jordan-Wigner encoding, independent of the dimension, this Hamiltonian term becomes equivalent to the spin-Holstein coupling [231,232],

$$\hat{H} = \sum_i \frac{g_i}{2} (\hat{Z}_i + 1)(\hat{a}_i^\dagger + \hat{a}_i). \tag{67}$$

Time evolution under this Hamiltonian can be easily implemented since all terms in the sum commute and each has a simple implementation; the term proportional to $\hat{Z}_i$ generates a conditional displacement gate, and the remaining term can be eliminated by a simple coordinate frame shift for each oscillator (resulting in a constant chemical potential shift for the fermions).

*Transmon qubits.* If the qubits are mapped to transmon qubits, conditional displacements can be compiled using conditional parities $\mathrm{C}\Pi_{i,j}$ (Table II) and a displacement $D_i(g_i t/2)$. This control was discussed in Sec. II B and is illustrated in Fig. 5(a).

*Dual-rail qubits.* In the case of dual-rail qubits, $\hat{Z}_i^{\mathrm{DR}} = e^{i\pi\hat{n}_{i,1}}$, where the subscripts $i,1$ indicates the mode that hosts the boson when the dual-rail qubit is in state $|1\rangle$. Thus, the required dual-rail qubit-conditional displacement will take the form $e^{-i\theta e^{i\pi\hat{n}_{i,1}}(\hat{a}_j^\dagger+\hat{a}_j)}$, where $\theta = g_i t/2$. This displacement is conditioned on the parity of another mode $\{i,1\}$ and can be realized by $\mathrm{CU}_{i,j}^{\Pi}$, whose compilation is presented in Sec. II C. The full operation illustrated in Fig. 11 is as follows:

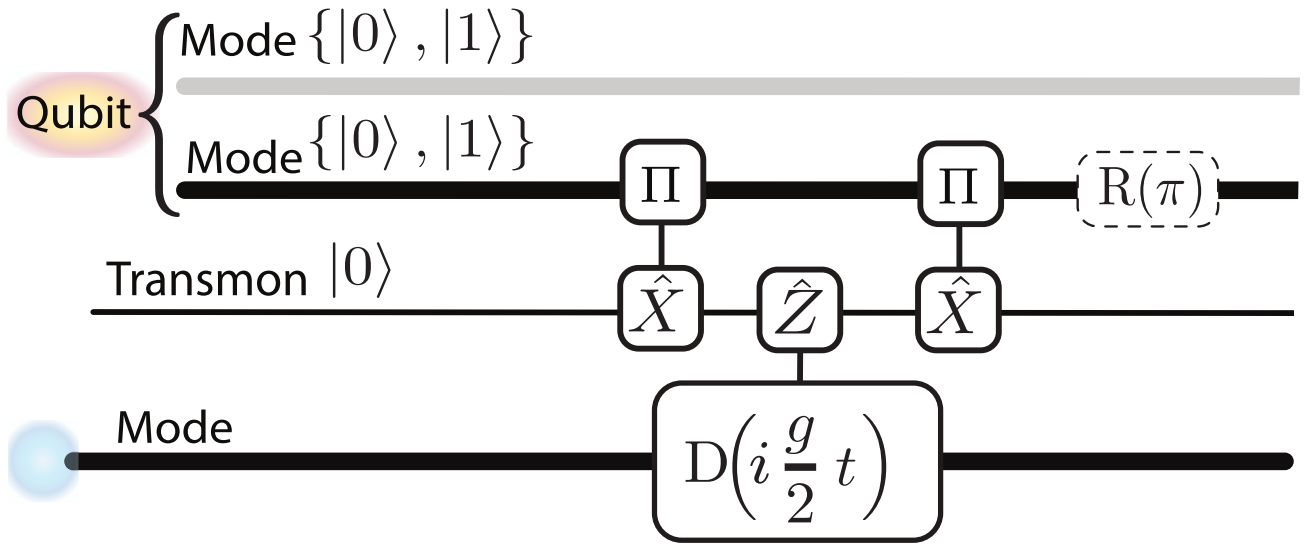

FIG. 11. Phonon-matter interactions. We show a circuit that implements a displacement on a mode conditioned on $\hat{Z}^{\mathrm{DR}}$ of a dual-rail qubit. This circuit can be used to realize time evolution for a duration $t$ under the Hamiltonian in Eq. (67) (up to an unconditional displacement).

$$e^{-i\theta\hat{Z}_i^{\mathrm{DR}}(\hat{a}_j^\dagger+\hat{a}_j)} = \langle 0|_{\mathrm{anc}}\mathrm{C}\Pi^X_{\mathrm{anc},i}\mathrm{CD}_{\mathrm{anc},j}(i\theta)\mathrm{C}\Pi^X_{\mathrm{anc},i}\mathrm{R}_i(\pi)|0\rangle_{\mathrm{anc}}, \tag{68}$$

where $\mathrm{CD}(\alpha)$ corresponds to a conditional displacement gate as defined in Table II and $\mathrm{C}\Pi^X$ denotes the Hadamard-conjugated conditional-parity gate introduced in Eq. (63). The subscript "anc" refers to the ancilla transmon qubit that is initialized to $|0\rangle_{\mathrm{anc}}$ and guaranteed to end in $|0\rangle_{\mathrm{anc}}$ at the end of the sequence unless an error occurs—a fact that could be used for error-detection purposes as previously mentioned.

### 2. *Light-matter interactions*

When light interacts with charged matter on a lattice, for instance, when a material is placed in an optical cavity, the Hamiltonian is given by [233]

$$\hat{H} = \omega\hat{a}^\dagger\hat{a} - J\sum_{\langle i,j\rangle}\hat{c}_i^\dagger\hat{c}_j e^{i\frac{g}{\sqrt{L}}(\hat{a}^\dagger+\hat{a})} + \mathrm{H.c.} \tag{69}$$

for a single light mode $\hat{a}$, where $\omega$ corresponds to the free evolution of the photons, $J$ is the strength of the hopping, $g$ is a coupling constant that depends on the specifics of the system (such as the geometry and material composition of the cavity), and $L$ is the number of lattice sites. Within the Peierls substitution approximation, the exponential term, $\tilde{A} = (g/\sqrt{L})(\hat{a}^\dagger + \hat{a})$, is given by the particle charge multiplied by the line integral of the electromagnetic vector potential along the link from site $j$ to site $i$.

For small $g/\sqrt{L}$, the exponent in the second term can be expanded:

$$\hat{H} = \omega\hat{a}^\dagger\hat{a} - \frac{J}{2}\sum_{\langle i,j\rangle}(\hat{c}_i^\dagger\hat{c}_j + \mathrm{H.c.}) + \frac{gJ}{\sqrt{L}}(\hat{a}^\dagger + \hat{a})\sum_{\langle i,j\rangle}\hat{c}_i^\dagger\hat{c}_j + O\left(J\frac{g^2}{L}\right). \tag{70}$$

The first term can be implemented using a bosonic phase-space rotation as discussed in Sec. III A 2. The second term corresponds to fermionic hopping and can be implemented using the strategy of Appendix A 1 for fermions encoded in either transmons or dual-rail qubits. The second term can be reexpressed as $\exp[-it(gJ/\sqrt{L})(\hat{a}^\dagger + \hat{a})\hat{X}_i\hat{X}_{i+1}]\exp[-it(gJ/\sqrt{L})(\hat{a}^\dagger + \hat{a})\hat{Y}_i\hat{Y}_{i+1}]$ for Jordan-Wigner adjacent sites. This expression corresponds to two doubly conditional displacements, i.e., conditioned on two qubits, with single-qubit rotations.

*Transmon qubits.* If the qubits are mapped to transmon qubits, doubly conditional displacements can be compiled using the technique described in Sec. II B and illustrated in Fig. 5(b). It requires a pair of conditional-parity gates $\mathrm{C}\Pi$ (Table II) and a conditional displacement $\mathrm{CD}(\alpha)$ (or, alternatively, two pairs of conditional-parity gates and an *unconditional* displacement).

*Dual-rail qubits.* If the qubits are mapped to dual-rail qubits, doubly conditional displacements can be synthesized by iterating the compilation strategy for the singly conditional displacement in Eq. (68), developed in the context of phonon-matter interactions. The full sequence is as follows:

$$e^{-i\theta\hat{Z}_i^{\mathrm{DR}}\hat{Z}_j^{\mathrm{DR}}(\hat{a}_k^\dagger+\hat{a}_k)} = (\langle 0|_{\mathrm{anc}}\otimes\mathbb{1})\mathrm{C}\Pi^X_{\mathrm{anc},i}\mathrm{C}\Pi^X_{\mathrm{anc},j}\mathrm{CD}_{\mathrm{anc},k}(i\theta) \times\mathrm{C}\Pi^X_{\mathrm{anc},j}\mathrm{C}\Pi^X_{\mathrm{anc},i}R_i(\pi)R_j(\pi)(|0\rangle_{\mathrm{anc}}\otimes\mathbb{1}), \tag{71}$$

where, similarly to previous gates, we have projected out the ancillary qubit, which is initialized to $|0\rangle_{\mathrm{anc}}$ and guaranteed (in the absence of noise) to be disentangled at the end of the sequence. With the above gate realized, single dual-rail qubit gates (i.e., beam splitters) can be used to rotate the operator $\hat{Z}^{\mathrm{DR}}\hat{Z}^{\mathrm{DR}}$ to $\hat{X}^{\mathrm{DR}}\hat{X}^{\mathrm{DR}}$ or $\hat{Y}^{\mathrm{DR}}\hat{Y}^{\mathrm{DR}}$, enabling the implementation of all desired gates.

## IV. IMPLEMENTATION OF GAUGE FIELDS AND GAUGE-MATTER COUPLING

In this section, we present implementations of $e^{-i\hat{H}t}$ for pure gauge Hamiltonian terms—i.e., for the electric and magnetic fields—and for the gauge-invariant hopping that couples the gauge fields to matter. In particular, we discuss the implementation of these terms for the two illustrative examples in $(2+1)$D: a $\mathbb{Z}_2$ gauge field and a $U(1)$ gauge field, each playing a vital role in the paradigmatic LGTs discussed in Sec. I E. However, we emphasize that due to the digital nature of our implementations, more complex models can be implemented beyond those in Sec. I E using techniques in this section, for instance, models that contain both $\mathbb{Z}_2$ and $U(1)$ fields coupled to the same matter fields.

We present the implementation of pure $\mathbb{Z}_2$ and $U(1)$ gauge fields in Secs. IV A and IV B, respectively. Next, we briefly discuss the possibility to realize static gauge fields (which do not possess any of their own dynamics) in Sec. IV C. We then return to our two examples and describe the implementation of gauge-invariant hopping terms in Secs. IV D and IV E for the $\mathbb{Z}_2$ and $U(1)$ fields, respectively. Finally, we present a summary of the implemented terms in Tables VI and VII.

TABLE VI. Pure gauge Hamiltonian terms and their implementation using native gates. We use $\lambda$ as the gate angle in the intermediate step to emphasize that it is different from the angle $\theta$.

| $\hat{H}$ | $e^{-i\hat{H}t}$ intermediary step | $e^{-i\hat{H}t}$ gates | Details |
|---|---|---|---|
| Electric field | $\mathbb{Z}_2$: $\exp\left(-i\lambda X_{i,j}\right)$ | $\mathrm{R}^{\varphi}{}_i(\theta)$ | Table II |
| | $U(1)$: $\exp\left(-i\lambda\left(\frac{\hat{n}^a_{i,j}+\hat{n}^b_{i,j}}{2}-\frac{\tau}{2\pi}\right)^2\right)$ | $\mathrm{SNAP}_i(\vec{\theta})$ | Sec. IV B 1 |
| Magnetic field | $\mathbb{Z}_2$: $\exp\left(-i\lambda\hat{Z}_{i,j}\hat{Z}_{j,k}\hat{Z}_{k,l}\hat{Z}_{l,i}\right)$ | $\mathrm{CD}_{i,j}(\theta)$, $\mathrm{SWAP}_{i,j}$ | Fig. 12<br>Sec. IV A 2 |
| | $U(1)$: $\exp\left(-i\lambda(\hat{a}^\dagger\hat{b}\hat{c}^\dagger\hat{d}\hat{f}\hat{e}^\dagger\hat{h}\hat{g}^\dagger+\mathrm{H.c.})\right)$ | $\mathrm{BS}_{i,j}(\varphi,\theta)$, $\mathrm{C\Pi}_{i,j}$, $\mathrm{R}^{\varphi}_i(\theta)$ | Fig. 13<br>Sec. IV B 2 |

### A. Pure gauge dynamical $\mathbb{Z}_2$ fields

In this section, we present two strategies for representing $\mathbb{Z}_2$ gauge fields in our architecture—one that uses the transmon qubits and another that relies on the dual-rail encoding discussed in Sec. II H. Also see Sec. I E for the introduction of the $\mathbb{Z}_2$ Hamiltonian and the notation—in particular, how we label the links on which the gauge fields reside.

#### *1. Electric field*

The time-evolution operator of the electric field defined in Table VI is $e^{-ig\hat{X}_{i,j}t}$, where the coupling strength is $g$ as defined in Eq. (1). This result corresponds to a single-qubit rotation.

*Transmon qubits.* This term is directly realized through on-resonant driving of each transmon using the $R_i^{\phi=0}(\theta)$ gate defined in Table II.

*Dual-rail qubits.* This term is a single-qubit rotation. As discussed in Sec. II H, single-qubit rotations (defined in Table II) around the $\hat{X}^{\mathrm{DR}}$ and $\hat{Y}^{\mathrm{DR}}$ axes are carried out using beam splitters $\mathrm{BS}_{i,j}(\varphi,\theta)$ between the two cavities composing the dual-rail qubit.

#### *2. Magnetic field*

The time-evolution operator associated with the magnetic-field term defined in Table VI is $e^{(-it(-B)\hat{Z}_{i,j}\hat{Z}_{j,k}\hat{Z}_{k,l}\hat{Z}_{l,i})}$, corresponding to a four-qubit entangling gate.

*Transmon qubits.* The exact gate sequence that implements this term is discussed in Sec. II F and illustrated in Fig. 7. Multibody gates in the transmon qubits in our proposed circuit QED architecture have a compact form: The required displacements can all be carried out using a single ancilla mode. This method removes the requirement for a costly Pauli gadget [235], which implements this four-qubit gate using six two-qubit gates. However, it instead requires bosonic SWAP gates to mediate the multiqubit interaction. However, as previously mentioned, the bosonic SWAP operation has a very high fidelity in the proposed architecture (see Sec. II A).

*Dual-rail qubits.* In Sec. III B 2, we discussed how to obtain $ZZ_{i,j}(\theta)$ gates using the exponentiation gadget presented in Ref. [210], which implements $e^{i\theta\hat{P}}$ for any operator $\hat{P}$ if $\hat{P}^2 = \mathbb{1}$. Therefore, using the same method, we can extend this result to include higher-weight Pauli strings in $\hat{P}$. This circuit is illustrated in Fig. 12. As mentioned previously, ancilla relaxation and dephasing errors, in addition to photon loss, can be detected using this general gate scheme [210].

TABLE VII. Compilation strategy for gauge-invariant hopping Hamiltonian terms into the native gates in the proposed architecture (see Fig. 4). We use $\hat{U}_{i,j}$ as a placeholder for a parallel transporter.

| $\hat{H}$ | $e^{-i\hat{H}t}$ intermediary step | $e^{-i\hat{H}t}$ gates | Definition |
|---|---|---|---|
| Bosons<br>$\sum_{\langle i,j\rangle} J_{i,j}(\hat{a}_i^\dagger\hat{U}_{i,j}\hat{a}_j+\mathrm{H.c.})$ | $\mathbb{Z}_2$ $\exp\left(-iJ_{i,j}t(\hat{a}_i^\dagger Z_{i,j}\hat{a}_j+\mathrm{H.c.})\right)$ | Transmon $\mathbb{Z}_2$ $\mathrm{C\Pi}_{i,j}$, $\mathrm{BS}_{i,j}(\varphi,\theta)$<br>Dual-rail $\mathbb{Z}_2\mathrm{C\Pi}_{i,j}$, $\mathrm{BS}_{i,j}(\varphi,\theta)$, $\mathrm{SQR}_{i,j}(\theta)$ | Sec. IV D 1 |
| | $U(1)$ $\exp\left(-iJ_{i,j}t(\hat{a}_i^\dagger\hat{a}_{i,j}^\dagger\hat{b}_{i,j}\hat{a}_j+\mathrm{H.c.})\right)$ | Dual-rail $U(1)$: $\mathrm{BS}_{i,j}(\varphi,\theta)$, $\mathrm{C\Pi}$, $R_i^{\varphi}(\theta)$ | Sec. IV E 1 |
| Fermions<br>$\sum_{\langle i,j\rangle} J_{i,j}(\hat{c}_i^\dagger\hat{U}_{i,j}\hat{c}_j+\mathrm{H.c.})$ | $\mathbb{Z}_2$ (1) SWAP, FSWAP<br>(2) $\exp\left(-iJ_{i,j}t(\hat{\sigma}_i^+\hat{Z}_{i,j}\hat{\sigma}_j^-+\mathrm{H.c.})\right)$ | Transmon $\mathbb{Z}_2$ and dual-rail fermions<br>(1) $\mathrm{SWAP}_{i,j}$, $\mathrm{C\Pi}_{i,j}$, $\mathrm{R}^{\varphi}{}_i(\theta)$, $\mathrm{R}_i(\theta)$<br>(2) $\mathrm{BS}_{i,j}(\varphi,\theta)$, $\mathrm{C\Pi}_{i,j}$, $\mathrm{R}^{\varphi}{}_i(\theta)$ | Sec. IV D 2 |
| | | Dual-rail gauge and transmon fermions<br>(1) $\mathrm{SWAP}_{i,j}, \mathrm{CD}_{i,j}(\theta), R_i^Z(\theta)$<br>(2) $\mathrm{C\Pi}_{i,j}$, $e^{i\theta\hat{E}}$, $e^{i\theta\hat{F}}$, $\mathrm{R}_i^{\varphi}(\theta)$ | Sec. IV E 2 |
| | $U(1)$ (1) SWAP, FSWAP<br>(2) $\exp\left(-iJ_{i,j}t(\hat{\sigma}_i^+\hat{a}_{i,j}^\dagger\hat{b}_{i,j}\hat{\sigma}_j^-+\mathrm{H.c.})\right)$ | Dual-rail gauge and dual-rail fermions<br>$\mathrm{SWAP}_{i,j}$, $\mathrm{CR}_{i,j}(\theta)$, $\mathrm{R}_i^{\varphi}(\theta)$<br>(2) $\mathrm{BS}_{i,j}(\varphi,\theta)$, $\mathrm{C\Pi}$, $\mathrm{R}_i^{\varphi}(\theta)$ | Sec. IV E 2 |

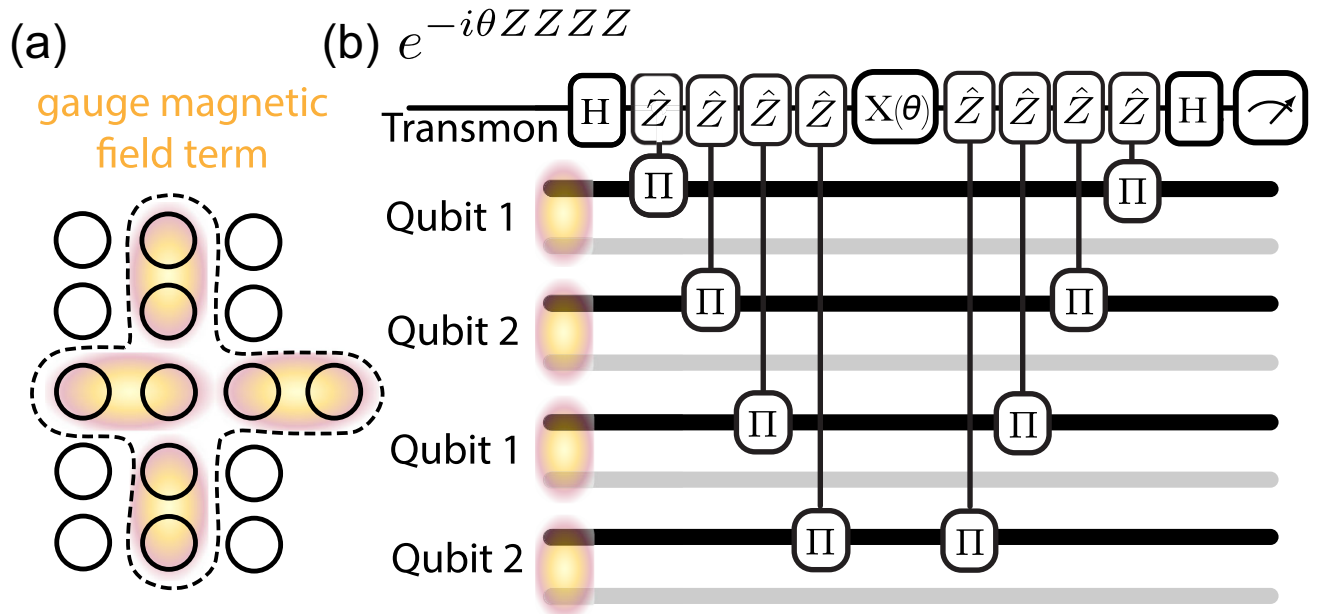

FIG. 12. *Dual-rail magnetic-field term* $e^{(-i\theta \hat{Z}_{i,j}\hat{Z}_{j,k}\hat{Z}_{k,l}\hat{Z}_{l,i})}$. (a) Representation of the cavities taking part in the magnetic-field term. (b) Corresponding circuit. The gray and black mode wires correspond to the location of the boson when the dual-rail qubit is in state $|0\rangle$ or $|1\rangle$, respectively.

### B. Pure gauge dynamical $U(1)$ fields

Next, we turn to the $U(1)$ quantum link model described by the Hamiltonian in Eq. (3). As described in Sec. I E, we utilize the Schwinger-boson representation to encode the link degrees of freedom, requiring two bosonic modes per link. We note that the Schwinger-boson representation may be understood as an extension of the dual-rail qubit encoding to spin $S > 1/2$ [57].

#### 1. Electric field

In the Schwinger-boson representation, the electric-field energy can be written as

$$\hat{H}_{\rm E} = \frac{g^2}{2}\sum_{\langle i,j\rangle}\left(\frac{\hat{n}^a_{i,j} - \hat{n}^b_{i,j}}{2} - \frac{\tau}{2\pi}\right)^2, \qquad (72)$$

where $g$ is the strength of the electric field and $\tau$ is the topological parameter. When multiplying out the square, the cross-Kerr terms $-(g^2/2)\hat{n}^a_{i,j}\hat{n}^b_{i,j}$ appear. Although we can, in principle, implement these terms, cf. Sec. III A 4, they are harder to implement than self-Kerr terms on the same site. Luckily, we can remove the cross-Kerr terms by using the Schwinger-boson constraint $\hat{n}^a_{i,j} + \hat{n}^b_{i,j} = 2S$. Because of this identity, we may add a term $(g^2/2)\sum_{i,j}[(\hat{n}^a_{i,j} + \hat{n}^b_{i,j})/2]^2$—which, due to the constraint, is just a constant—to the Hamiltonian in Eq. (72) to obtain

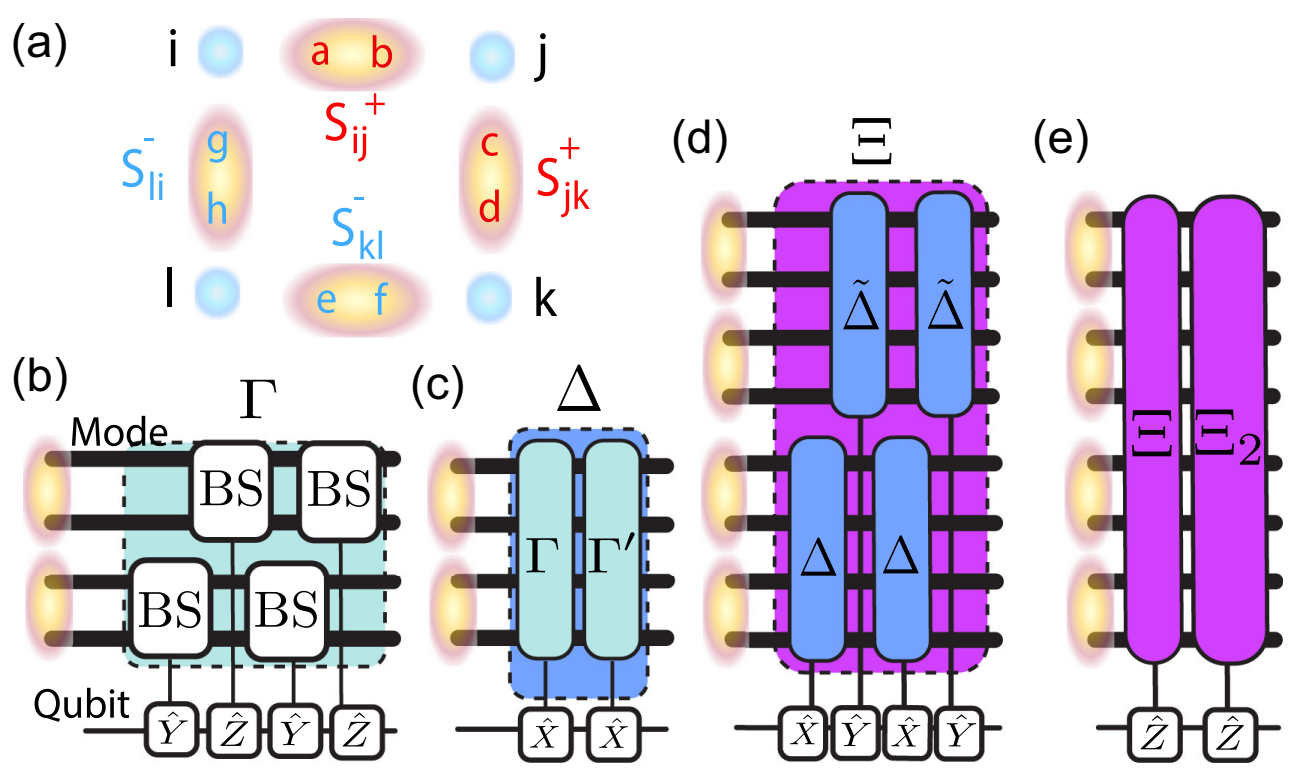

FIG. 13. Baker-Campbell-Hausdorff synthesis of the plaquette term for the quantum link model in $(2+1)$D. (a) BCH circuit corresponding to the creation of a qubit-conditional quartic term. (b) Use of the Trotter formula to remove additional unwanted quartic terms. (c) BCH circuit enabling the compilation of a term containing a multiplication of eight terms (octic). (d) Final step in the circuit, which uses the Trotter formula to remove unwanted octic terms. (e) Representation of participating modes. The modes composing the gauge fields are labeled with letters $a$ through $g$, and the matter fields are labeled $i$ through $l$. Depending on the geometry, not all CΠ will be able to be directly conditioned on the same qubit, leading to the requirement for additional SWAP gates.

$$\hat{H}_{\rm E} = \frac{g^2}{2}\sum_{\langle i,j\rangle}\left(\frac{1}{2}((\hat{n}^a_{i,j})^2 + (\hat{n}^b_{i,j})^2) - \frac{\tau}{2\pi}(\hat{n}^a_{i,j} - \hat{n}^b_{i,j})\right) \qquad (73)$$

$$\equiv \sum_{\langle i,j\rangle}\hat{h}_{E,i,j} \qquad (74)$$

up to a constant, thereby removing the cross-Kerr term.

The time evolution under $\hat{H}_E$ fulfills $\exp(-i\hat{H}_E t) = \prod_{\langle i,j\rangle}\exp(-i\hat{h}_{E,i,j}t)$. Because the $\hat{a}$ and $\hat{b}$ modes commute, $\exp(-i\hat{h}_{E,i,j}t) = e^{-i\frac{g^2}{2}(\frac{1}{2}\hat{n}^{a2}_{i,j} - \frac{\tau}{2\pi}\hat{n}^a_{i,j})t}e^{-i\frac{g^2}{2}(\frac{1}{2}\hat{n}^{b2}_{i,j} + \frac{\tau}{2\pi}\hat{n}^b_{i,j})t}$. This result can be implemented exactly using two SNAP$(\vec{\theta})$ gates defined in Table II on modes $a$ and $b$:

$$\exp(-i\hat{h}_{E,i,j}t) = {\rm SNAP}_a(\vec{\theta}^a){\rm SNAP}_b(\vec{\theta}^b), \qquad (75)$$

where $\theta^a_n = (g^2/2)\{\frac{1}{2}n^2 - [\tau/(2\pi)]n\}t$ and $\theta^b_n = (g^2/2)\{\frac{1}{2}n^2 + [\tau/(2\pi)]n\}t$.

#### 2. Magnetic field

In $(2+1)$D, the magnetic-field Hamiltonian, illustrated in Fig. 13(a), acts on the four gauge-field raising operators linking a square (also called a plaquette) of fermionic sites labeled $i$, $j$, $k$, $l$. We discuss the compilation of a single plaquette exponential term $\exp(-i\hat{H}_\square t)$, where

$$\hat{H}_\square = -\frac{1}{4g^2(S(S+1))^2}(\hat{S}^+_{i,j}\hat{S}^+_{j,k}\hat{S}^-_{k,l}\hat{S}^-_{l,i} + \text{H.c.}). \quad (76)$$

Using our Schwinger-boson mapping of the gauge fields in Eq. (6), we obtain a product of creation and annihilation operators, which we relabel $a, b$ for the modes of link $\langle i,j\rangle$, $c, d$ for link $\langle j,k\rangle$, $e, f$ for link $\langle k,l\rangle$, and $g, h$ for link $\langle l,i\rangle$, for simplicity of notation, such that

$$\hat{H}_\square = -\frac{1}{4g^2(S(S+1))^2}(\hat{a}^\dagger\hat{b}\hat{c}^\dagger\hat{d}\hat{f}\hat{e}^\dagger\hat{h}\hat{g}^\dagger + \text{H.c.}). \quad (77)$$

We show here how to realize the plaquette term using the method introduced in Ref. [97] and summarized in Sec. II G, which makes use of the group commutator relations given by the BCH formula, specifically, the relation

$$e^{i\hat{A}\theta}e^{i\hat{B}\theta}e^{-i\hat{A}\theta}e^{-i\hat{B}\theta} = e^{-[\hat{A},\hat{B}]\theta^2 + O(\theta^3)}. \quad (78)$$

Through an appropriate choice of hybrid qubit-boson operators $\hat{A}$ and $\hat{B}$, we obtain the means to implement multiplication between commuting bosonic operators. We also use Trotter formulas to implement addition.

We rely on the above method and present a possible implementation that leverages the conditional beam-splitter gate introduced in Eq. (19) as the fundamental building block. This approach has the advantage of relying on simple hardware primitives, at the cost of Trotter error. Next, we briefly comment on the possibility to instead leverage a three-wave mixing term realized on the hardware level, cf. Appendix C. This latter approach would enable the compilation of the desired term without Trotter error, but it uses a gate that is more complicated to engineer on the hardware level.

The key idea for this synthesis is to apply the BCH formula to conditional beam splitters acting on the four pairs of modes around the plaquette, conditioned on the same ancillary qubit. Intuitively, this process leads to multiplication of creation and annihilation operators for the four pairs of modes, yielding the eight-mode term in Eq. (77).

We first show how to synthesise a product of four-mode operators. Choosing $e^{-i\hat{A}\theta} = \text{CBS}^Y_{a,b}[-(\pi/2),\theta] = e^{-i\theta(i\hat{a}^\dagger\hat{b} - i\hat{a}\hat{b}^\dagger)\hat{Y}}$ and $e^{-i\hat{C}\theta} = \text{CBS}^Z_{c,d}(0,\theta) = e^{-i\theta(\hat{c}^\dagger\hat{d} + \hat{c}\hat{d}^\dagger)\hat{Z}}$, where $\hat{H}$ is a Hadamard gate on the qubit and $\text{CBS}(\varphi,\theta)$ is defined in Eq. (19), we define the primitive

$$V(A\theta, C\theta) = e^{i\hat{A}\theta}e^{i\hat{C}\theta}e^{-i\hat{A}\theta}e^{-i\hat{C}\theta}. \quad (79)$$

This set of operations is illustrated in Fig. 13(b). In order to reduce the error sufficiently, we use this primitive in a higher-order product formula [236],

$$\Gamma_{A,C}(\theta) = V(A\gamma\theta, C\gamma\theta)V(-A\gamma\theta, -C\gamma\theta)V(C\gamma^2\theta, A\gamma^2\theta) \times V(-C\gamma^2\theta, -A\gamma^2\theta)V(A\gamma\theta, C\gamma\theta)V(-A\gamma\theta, -C\gamma\theta) \quad (80)$$

$$= e^{-2\theta^2(\hat{a}^\dagger\hat{b}\hat{c}^\dagger d + \hat{a}^\dagger\hat{b}\,\hat{c}\,\hat{d}^\dagger - \hat{a}\hat{b}^\dagger\hat{c}^\dagger\hat{d} - \hat{a}\hat{b}^\dagger\hat{c}\hat{d}^\dagger)\otimes\hat{X} + \mathcal{O}(\theta^5)}, \quad (81)$$

where $\gamma = 1/\sqrt{4-2\sqrt{2}}$.

This expression contains two terms that are not part of the plaquette operator. In order to remove the unwanted terms, we apply the same four-mode synthesis again with different phases. This process yields a new term $\Gamma_{A',C'}$ obtained using the primitive $V(A'\theta, C'\theta)$ with $e^{-i\hat{A}'\theta} = \text{CBS}^Y_{a,b}(0,\theta)$ and $e^{-i\hat{C}'\theta} = \text{CBS}^Z_{c,d}[-(\pi/2),\theta]$. Combining $\Gamma_{A,C}$ and $\Gamma_{A',C'}$ in a second-order Trotter formula, we remove the unwanted terms,

$$\Delta_{A,C} = \Gamma_{A,C}(\theta/2)\Gamma_{A',C'}(\theta)\Gamma_{A,C}(\theta/2) = e^{((-2i\hat{a}^\dagger\hat{b}\hat{c}^\dagger\hat{d} + 2i\hat{a}\hat{b}^\dagger\hat{c}\hat{d}^\dagger)\otimes 2iX)(-i\theta)^2} + \mathcal{O}(\theta^5). \quad (82)$$

This set of operations is illustrated in Fig. 13(c), and we dropped the $\mathcal{O}(\theta^6)$ Trotter error as the $\mathcal{O}(\theta^5)$ BCH error dominates. Using a single-qubit gate on the qubit, one can also obtain $(\Delta_{A,C})^\dagger$. The same expression can be obtained on the other four modes that are also involved in the plaquette term:

$$\tilde{\Delta}_{E,G} = e^{((-2i\hat{e}^\dagger\hat{f}\hat{g}^\dagger\hat{h} + 2i\hat{e}\hat{f}^\dagger\hat{g}\hat{h}^\dagger)\otimes 2i\hat{Y})(-i\theta)^2 + \mathcal{O}(\theta^5)}. \quad (83)$$

Note that $\tilde{\Delta}_{E,G}$ is chosen such that $\hat{Y}$ is carried out on the qubit rather than $\hat{X}$ to enable the further use of the BCH formula, which we explain next.

In order to obtain the eight-wave mixing required to realize the plaquette term, we concatenate four four-mode terms according to

$$\Xi_{\{A,C\},\{E,G\}} = \Delta_{A,C}\tilde{\Delta}_{E,G}\Delta^\dagger_{A,C}\tilde{\Delta}^\dagger_{E,G}. \quad (84)$$

This set of operations is illustrated in Fig. 13(d). Similarly to the four-mode term, Eq. (84) yields terms that are not part of the plaquette term. In order to remove them, we multiply $\Xi_{\{A,C\},\{E,G\}}$ with a term that is obtained in the same way but with a different choice of phases in the beam splitters:

$$\Xi_{2,\{A_2,C_2\},\{E_2,G_2\}} = \Delta_{A_2,C_2}\tilde{\Delta}_{E_2,G_2}\Delta^\dagger_{A_2,C_2}\tilde{\Delta}^\dagger_{E_2,G_2}. \quad (85)$$

The constituent operation parts of this term are the following, where we use the same notation with primes and tildes as in the discussion of the synthesis of $\Xi_{\{A,C\},\{E,G\}}$. On modes $a$, $b$, $c$, $d$, we use

$$A_2 = \hat{H}\mathrm{CBS}_{a,b}(\pi,1)\hat{H}, \quad (86)$$

$$C_2 = \mathrm{CBS}_{c,d}(0,1), \quad (87)$$

$$A'_2 = \hat{H}\mathrm{CBS}_{a,b}\left(\frac{\pi}{2},1\right)\hat{H}, \quad (88)$$

$$C'_2 = \mathrm{CBS}_{c,d}\left(\frac{\pi}{2},1\right). \quad (89)$$

On modes $e$, $f$, $g$, $h$, we use

$$\tilde{E}_2 = \hat{H}\,\mathrm{CBS}_{e,f}(0,1)\,\hat{H}, \quad (90)$$

$$\tilde{G}_2 = \hat{H}\,\hat{S}\,\mathrm{CBS}_{g,h}(0,1)\hat{H}\,\hat{S}^\dagger, \quad (91)$$

$$\tilde{E}'_2 = \hat{H}\,\mathrm{CBS}_{e,f}\left(\frac{\pi}{2},1\right)\hat{H}, \quad (92)$$

$$\tilde{G}'_2 = \hat{H}\,\hat{S}\,\mathrm{CBS}_{g,h}\left(-\frac{\pi}{2},1\right)\hat{H}\,\hat{S}^\dagger. \quad (93)$$

In total, we obtain

$$\Xi_{\{A,C\},\{E,G\}}\Xi_{2,\{A_2,C_2\},\{E_2,G_2\}} = \quad (94)$$

$$e^{-i(16(\hat{a}^\dagger\hat{b}\hat{c}^\dagger\hat{d}\hat{f}\hat{e}^\dagger\hat{h}\hat{g}^\dagger+\hat{a}\hat{b}^\dagger\hat{c}\hat{d}^\dagger\hat{f}^\dagger\hat{e}\hat{h}^\dagger\hat{g})\otimes\hat{Z})\theta^4+O(\theta^5)}, \quad (95)$$

illustrated in Fig. 13(e). Choosing $\theta = (1/[64g^2(S(S+1))^2]t)^{1/4}$, we obtain time evolution under a single plaquette Hamiltonian for a time step $t$. Each CBS gate can be implemented using two conditional-parity operations and one beam splitter, cf. Eq. (19). We count conditional parities in the following. Each $\Gamma$ can be implemented using six $V$ operations [see Eq. (79)], which require eight conditional parities to implement, and hence each $\Gamma$ requires 48 conditional parities. Each $\Delta$ operation requires three $\Gamma$ operations using the symmetric Trotter formula, implying that it requires 144 conditional parities. Each $\Xi$ is built from four $\Delta$ terms, and hence each requires 576 conditional parities. Then, the final approximation $V_\square$ requires 1152 conditional parities as it is composed of two $\Xi$ operations.

The conditional parity would take 200 ns using a LINC coupler at $\chi = 2\pi \times 2.5$ MHz [90]; therefore, the full 1152 conditional parities would take 230.4 μs, an order of magnitude smaller than recently experimentally demonstrated tantalum lifetimes of $T_1 \approx T_2/2 \approx 2$ ms [237]. In Appendix H, for these parameters, we find a conditional-parity infidelity for up to Fock state 10 of $6.8 \times 10^{-5}$, which, taken 1152 times, leads to a fidelity of 93% (for $\chi = 2\pi \times 1$ MHz [209] we obtain 82%). This result is important considering that qubit circuits for implementing a similar term require orders of magnitude higher gate counts (see Sec. VI and Ref. [238]). Therefore, the radically smaller number of gates required in our qubit-boson hardware brings the simulation of these multibody qubit-boson interactions into the nearer term (especially as high-fidelity local observables do not always require high-fidelity states). This finding shows the promise of this hardware, which deserves more attention and development. In addition, we believe that further improvements of our algorithm are possible, e.g., through pulse engineering. Furthermore, an advantage of the digital approach is the potential for including error suppression from oscillators to oscillators, which has already been proposed [109–119].

We close with a discussion of the error term. Because each pair of mode operators has norm $\|a^\dagger b\| = \sqrt{S(S+1)}$ in the Schwinger-boson encoding of a large spin and the error term contains five such pairs, our result implies that the distance between the thus-implemented unitary $V_\square$ and the exact plaquette exponential is

$$\|V_\square - e^{-i(16(\hat{a}^\dagger\hat{b}\hat{c}^\dagger\hat{d}\hat{f}\hat{e}^\dagger\hat{h}\hat{g}^\dagger+\hat{a}\hat{b}^\dagger\hat{c}\hat{d}^\dagger\hat{f}^\dagger\hat{e}\hat{h}^\dagger\hat{g})\otimes\hat{Z})t}\| = \mathcal{O}\left(\frac{t^{5/4}}{g^{5/2}}\right). \quad (96)$$

In particular, this error is independent of $S$.

Another way of synthesising this term would be to use the three-wave mixing term introduced in Appendix C. Because the additional terms appearing in the beam-splitter approach mentioned above would not appear using three-wave mixing, we could save on Trotter error and reduce the number of native operations needed by a prefactor. However, because the Trotter error is subleading, this approach will not change the asymptotic error scaling shown above.

### C. Static gauge fields coupled to matter

Static gauge fields appear when a magnetic field [224] is coupled to matter and are crucial when, for example, considering the fractional quantum Hall effect [225]. In this case, the hopping term of the Hamiltonian has a fixed complex amplitude on each link, i.e.,

$$-J\sum_{\langle i,j\rangle}(e^{i\varphi}\hat{a}_i^\dagger\hat{a}_j + \mathrm{H.c.}), \quad (97)$$

where we have specialized to the case of bosonic matter. This term can be directly implemented by a beam splitter $\mathrm{BS}_{i,j}(\varphi,\theta)$ defined in Table II via the appropriate choice for the phase $\varphi$ for each link.

For fermionic matter, the hopping term generalizes as

$$-J\sum_{\langle i,j\rangle}(e^{i\varphi}\hat{c}_i^\dagger\hat{c}_j + \mathrm{H.c.}). \quad (98)$$

The phase can be implemented as a part of the gates necessary for the Jordan-Wigner encoding described in Appendix A 1, for either choice of qubit type. The $F$SWAP operation would remain unchanged, but the phase can be incorporated by conjugating the fermionic hopping by $e^{i\varphi\hat{Z}}$ on one of the qubits.

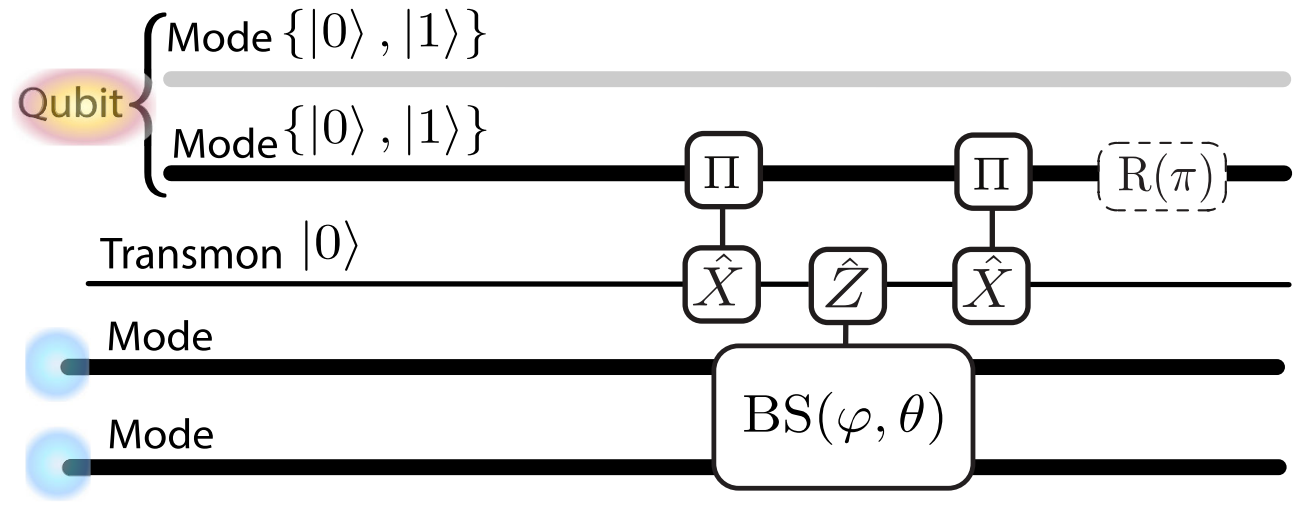

FIG. 14. The $\mathbb{Z}_2$ gauge-invariant hopping circuit. Implementation of the gauge-invariant hopping with dual-rail qubits as $\mathbb{Z}_2$ gauge fields and modes as the matter sites: $e^{-i\theta e^{i\pi\hat{n}_{i,j}}(e^{i\varphi}a_i^\dagger a_j + \mathrm{H.c.})}$. The gray and black mode wires participating in the dual-rail qubit (two upper modes) correspond to the location of the boson when the dual-rail qubit is in state $|0\rangle$ or $|1\rangle$, respectively. The beam splitter is conditioned on $\hat{Z}$.

## D. Dynamical $\mathbb{Z}_2$ fields coupled to matter

Here, we discuss $\mathbb{Z}_2$-Higgs model gauge fields coupled to bosonic and fermionic matter. We first discuss the bosonic gauge-invariant hopping in Sec. IV D 1, mapping the gauge fields to transmon and then dual-rail qubits. Next, we discuss the fermionic gauge-invariant hopping in Sec. IV D 2, mapping the fermions and the gauge fields to transmon and dual-rail qubits.

### 1. Bosonic gauge-invariant hopping

The time-evolution operator of the gauge-invariant bosonic hopping mediated by a $\mathbb{Z}_2$ gauge field takes the form

$$e^{-it\hat{Z}_{i,j}(\hat{a}_i^\dagger \hat{a}_j + \hat{a}_i \hat{a}_j^\dagger)}. \tag{99}$$

In the following, we discuss the implementation of this operator using either transmons or DR qubits. For both cases, we use modes to directly encode the bosonic matter.

*Field—transmons; matter—modes*. For the situation where we map the gauge fields onto transmon qubits, Eq. (99) corresponds to a conditional beam splitter, discussed in Sec. II B.

*Field—dual rail; matter—modes*. For the case where we map the gauge fields onto dual-rail qubits, we leverage the fact that the parity of the photon number $\hat{n}$ in one of the rails is equivalent to the Pauli operator $\hat{Z}^{\mathrm{DR}}$ [see Eq. (59)]. Therefore, the time evolution of the hopping between sites $i$ and $j$ takes the form

$$e^{-ite^{i\pi\hat{n}_{i,j}}(a_i^\dagger a_j + \hat{a}_i \hat{a}_j^\dagger)} \tag{100}$$

and can be compiled using the methods described in Sec. II C and shown in Fig. 14.

### 2. Fermionic gauge-invariant hopping

The fermionic SWAP networks used to implement fermionic gauge-invariant hopping require the exact same FSWAP operations as discussed in Appendix A 1, where the gauge field also needs to be swapped. To make the hopping gauge invariant, we replace the iSWAP operation with

$$\begin{aligned} &e^{-it(-J)\hat{Z}_{i,j}(\hat{X}_i\hat{X}_j + \hat{Y}_i\hat{Y}_j)} \\ &\quad = e^{-it(-J)\hat{Z}_{i,j}(\hat{X}_i\hat{X}_j)} e^{-it(-J)\hat{Z}_{i,j}(\hat{Y}_i\hat{Y}_j)} \\ &\quad = R_i^{\frac{\pi}{2}}\left(\frac{\pi}{2}\right) R_j^{\frac{\pi}{2}}\left(\frac{\pi}{2}\right) e^{-it(-J)\hat{Z}_{i,j}(\hat{Z}_i\hat{Z}_j)} R_i^{\frac{\pi}{2}}\left(-\frac{\pi}{2}\right) R_j^{\frac{\pi}{2}}\left(-\frac{\pi}{2}\right) \\ &\qquad \times R_i^0(\tfrac{\pi}{2}) R_j^0(\tfrac{\pi}{2}) e^{-it(-J)\hat{Z}_{i,j}(\hat{Z}_i\hat{Z}_j)} R_i^0(-\tfrac{\pi}{2}) R_j^0(-\tfrac{\pi}{2}), \end{aligned} \tag{101}$$

using the fact that $\hat{X} = R^{\frac{\pi}{2}}(\pi/2)\hat{Z}R^{\frac{\pi}{2}}(-\pi/2)$ and $\hat{Y} = R^{\pi}(\pi/2)\hat{Z}R^{\pi}(-\pi/2)$, where $R^{\varphi}(\theta)$ is defined in Table II. We note that this gate sequence can be further compressed by combining successive single-qubit gates into a single rotation.

We do not discuss the all-transmon qubit implementation of this model as this case does not leverage the advantages of our platform.

*Field—transmon; matter—dual rail.* Following Eq. (101), single-qubit rotations can be performed on the dual-rail qubits using beam-splitter gates, while $e^{-it(-J)\hat{Z}_{i,j}(\hat{Z}_i^{\mathrm{DR}}\hat{Z}_j^{\mathrm{DR}})}$ is a transmon $Z$-conditional dual-rail $ZZ^{\mathrm{DR}}$ operation. This latter gate is compiled in Eq. (64) using an ancillary transmon initialized to $|0\rangle_{\mathrm{anc}}$, which is ultimately traced out. However, we can adapt this same sequence to realize the desired gate by replacing the ancilla with the transmon encoding the gauge field in an unknown state [and removing the matrix element projection in Eq. (64)].

*Field—dual rail; matter—transmon.* This term requires the implementation of $e^{-it(-J)e^{i\pi\hat{n}_{i,j}}(\hat{X}_i\hat{X}_j + \hat{Y}_i\hat{Y}_j)} = e^{-it(-J)e^{i\pi\hat{n}_{i,j}}\hat{X}_i\hat{X}_j} e^{it(-J)e^{-i\pi\hat{n}_{i,j}}\hat{Y}_i\hat{Y}_j}$, which can be implemented using additional single-qubit rotations from $e^{-i\theta e^{i\pi\hat{n}_{i,j}}\hat{Z}_i\hat{Z}_j}$. The latter is an instance of the parity-controlled gate $\mathrm{CU}_{i,j}^{\Pi}$ introduced in Sec. II C. Specifically, the intermediate gate in Eq. (30) would be replaced by the two-transmon-qubit gate $ZZ(\theta)$ implemented in our architecture via the technique in Sec. II F.

*Field—dual rail; matter—dual rail.* Following Eq. (101), aside from single-qubit rotations, we require implementation of $e^{-it(-J)\hat{Z}_{i,j}^{\mathrm{DR}}(\hat{Z}_i^{\mathrm{DR}}\hat{Z}_j^{\mathrm{DR}})}$. This type of multi-dual-rail gate can be realized using the Pauli-exponentiation gadget as discussed in Sec. IVA 2 and shown in Fig. 12 for the slightly more complex case of four dual-rail qubits. Alternatively, one can iterate the parity-control synthesis technique of Sec. II C, similarly to the doubly conditional displacement realized in Eq. (71).

### E. Dynamical $U(1)$ fields coupled to matter

Next, we discuss the implementation of dynamical $U(1)$ fields coupled to bosonic (Sec. IV E 1) and fermionic matter (Sec. IV E 2). We show how to implement the dynamics of the hopping described in Eq. (3), which has the form

$$\exp\left(-i\frac{Jt}{2\sqrt{S(S+1)}}\sum_{\langle i,j\rangle}(\hat{m}_i^\dagger\hat{S}_{i,j}^+\hat{m}_j+\mathrm{H.c.})\right). \tag{102}$$

In the remainder of the section, we will assume large $S$ and write $\sqrt{S(S+1)}\approx S$. As in Sec. IV B, we leverage the Schwinger-boson encodings that represent the $U(1)$ gauge fields. Furthermore, for the case of fermionic matter, we separately consider transmons and dual-rail qubits.

#### 1. Bosonic gauge-invariant hopping

For a bosonic matter site $i$ denoted by operator $\hat{d}$, the gauge-invariant hopping between site $i$ and site $j$ is given by

$$e^{-it\frac{J}{2S}(\hat{d}_i^\dagger\hat{a}_{i,j}^\dagger\hat{b}_{i,j}\hat{d}_j+\mathrm{H.c.})}, \tag{103}$$

where $\hat{a}_{i,j}^\dagger\hat{b}_{i,j}$ is the Schwinger boson representation of the link operator $\hat{S}_{i,j}^+$. This unitary is generated by a four-mode mixing term and can be synthesized using the BCH formula [see Eq. (54)].

#### 2. Fermionic gauge-invariant hopping

For fermions in $(2+1)$D, we first use FSWAP networks to move the sites to neighboring sites within the JW encoding. Then, to perform the gauge-invariant hopping, we also SWAP the corresponding gauge fields as depicted in Fig. 15. The remaining operation is then given by the nearest-neighbor hopping

$$e^{-it\frac{J}{2S}(\hat{\sigma}_i^+\hat{a}_{i,j}^\dagger\hat{b}_{i,j}\hat{\sigma}_j^-+\mathrm{H.c.})}, \tag{104}$$

where $j$ is a neighboring site to $i$. In $(1+1)$D, SWAP and FSWAP operations are not necessary. We now discuss how this term is synthesized for the cases where the fermions are encoded in transmons or dual-rail qubits.

*Fermionic matter—transmons.* We present two schemes, a naive implementation that contains Trotter error and a scheme that avoids Trotter error (for this term alone) but requires a particular experimental adaptation of the hardware.

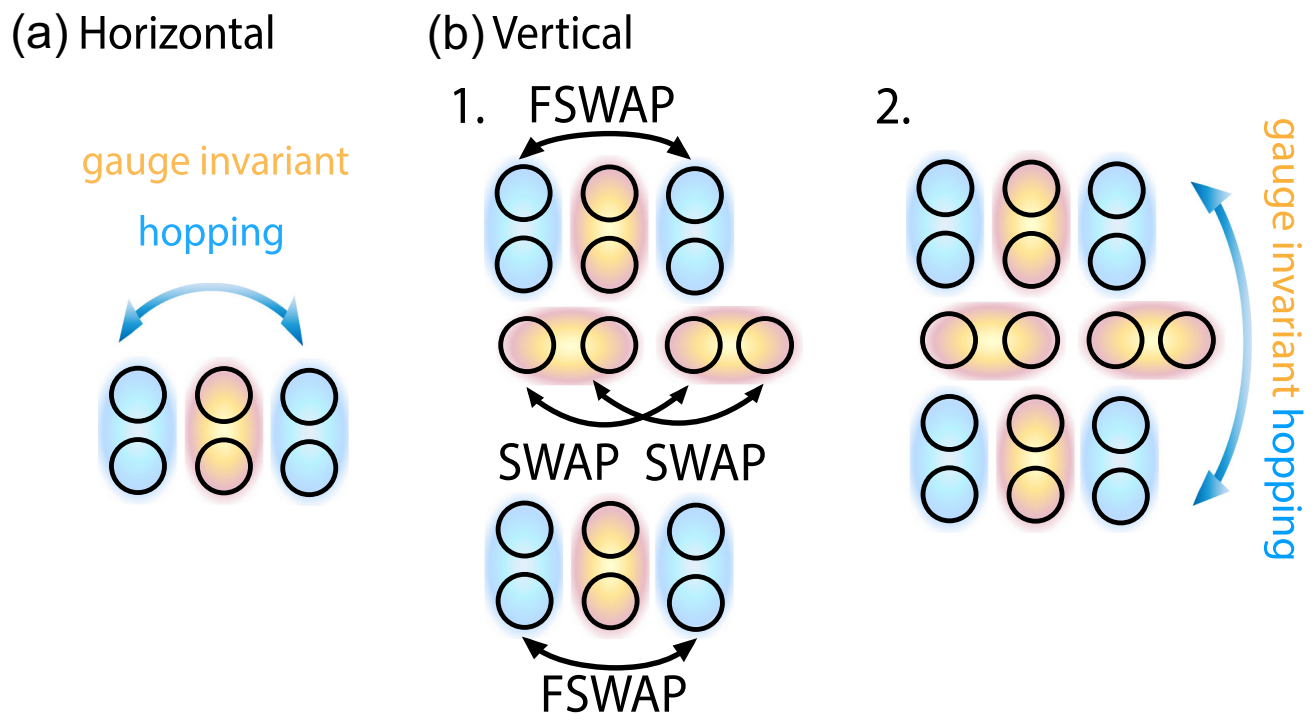

FIG. 15. Gauge-invariant hopping of fermions across $U(1)$ gauge fields. (a) Horizontal hopping. This hopping only requires a nearest-neighbor gauge-invariant hopping operation where fermion operators are replaced with qubit operators. (b) Vertical hopping. FSWAP networks are generalized by first SWAPing the gauge-field modes at the same time as performing FSWAPs and then applying the same operation as for the horizontal hopping on the vertical bond. The gauge-field modes involved in the horizontal hoppings are left unchanged.

For scheme 1, we rewrite the gauge-invariant hopping term as

$$e^{-i\frac{Jt}{2S}(\hat{H}_1+\hat{H}_2)}\approx e^{-i\frac{Jt}{2S}\hat{H}_1}e^{-i\frac{Jt}{2S}\hat{H}_2}+\mathcal{O}\left(\left(\frac{Jt}{2S}\right)^2\right), \tag{105}$$

where we used Trotterization and defined

$$\hat{H}_1=(\hat{X}_i\hat{X}_j+\hat{Y}_i\hat{Y}_j)(\hat{a}_{i,j}^\dagger\hat{b}_{i,j}+\mathrm{H.c.}), \tag{106}$$

$$\hat{H}_2=(\hat{Y}_i\hat{X}_j-\hat{X}_i\hat{Y}_j)(e^{i\pi/2}\hat{a}_{i,j}^\dagger\hat{b}_{i,j}+\mathrm{H.c.}). \tag{107}$$

Noting that $\hat{X}_i\hat{X}_j$ and $\hat{Y}_i\hat{Y}_j$ commute, as do $\hat{Y}_i\hat{X}_j$ and $\hat{X}_i\hat{Y}_j$, we are then left with implementing doubly conditional beam splitters, which is accomplished by sandwiching a beam splitter with conditional-parity operations conditioned on the qubits $i$ and $j$, as shown in Eq. (19). However, $[\hat{H}_1,\hat{H}_2]\neq 0$, and this term will therefore acquire Trotter error.

Scheme 2 does not introduce Trotter error but requires additional engineering at the hardware level. We define the unitary as $e^{-i\theta\hat{H}}$ with $\hat{H}=\sigma_i^+a^\dagger b\sigma_j^-+\sigma_i^-ab^\dagger\sigma_j^+$, where, for simplicity, we have dropped the indices $i$, $j$ on the modes $a$ and $b$ and we defined $\theta=[(Jt)/(2S)]$. We start by breaking up $\hat{H}$ into $\hat{H}=\hat{C}+\hat{D}$, where

$$\hat{C}=\hat{X}_i(\hat{X}_j(a^\dagger b+ab^\dagger)+i\hat{Y}_j(a^\dagger b-ab^\dagger))/4\equiv\hat{X}_i\hat{E}_j, \tag{108}$$

$$\hat{D}=\hat{Y}_i(-i\hat{X}_j(a^\dagger b-ab^\dagger)+\hat{Y}_j(a^\dagger b+ab^\dagger))/4\equiv\hat{Y}_i\hat{F}_j. \tag{109}$$

Indeed, $[\hat{C},\hat{D}]=0$ because $\{\hat{E}_j,\hat{F}_j\}=0$.

It is helpful to rewrite

$$\hat{E}_j = (ab^\dagger\sigma_j^+ + a^\dagger b\sigma_j^-)/2, \tag{110}$$

$$\hat{F}_j = i(ab^\dagger\sigma_j^+ - a^\dagger b\sigma_j^-)/2. \tag{111}$$

Thus, we can now express the desired time-evolution operator as

$$\begin{aligned} e^{-i\theta\hat{H}} &= e^{-i\theta\hat{D}}e^{-i\theta\hat{C}} \\ &= e^{-i\theta\hat{Y}_i\hat{F}_j}e^{-i\theta\hat{X}_i\hat{E}_j} \\ &= R_i^0\left(-\frac{\pi}{2}\right)\mathrm{C\Pi}_{i,j}e^{-i\theta\hat{F}}\mathrm{C\Pi}_{i,j}R_i^0\left(\frac{\pi}{2}\right) \\ &\quad\times R_i^{\frac{\pi}{2}}\left(-\frac{\pi}{2}\right)\mathrm{C\Pi}_{i,j}e^{-i\theta\hat{E}}\mathrm{C\Pi}_{i,j}R_i^{\frac{\pi}{2}}\left(\frac{\pi}{2}\right), \end{aligned} \tag{112}$$

where $R^\varphi(\theta)$ (a single transmon qubit rotation gate) and $\mathrm{C\Pi}$ gates are defined in Table II. This reduces the problem of realizing four-body interactions to the more manageable problem of realizing the three-body interaction Hamiltonians—the latter being more natural to realize in the circuit QED architecture given the primary source of nonlinearity is the four-wave mixing property of Josephson junctions. We derive one possible implementation of these interactions in Appendix C.

*Fermionic matter—dual rail.* In this case, we must implement a many-body gate that involves the two Schwinger-boson modes representing the gauge fields $a_{i,j}$ and $b_{i,j}$, and the four modes composing the two dual-rail qubits $c_i$, $d_i$ and $c_j$, $d_j$:

$$e^{-i\theta(\hat{c}_i^\dagger\hat{d}_i\hat{a}_{i,j}^\dagger\hat{b}_{i,j}\hat{c}_j\hat{d}_j^\dagger + \mathrm{H.c.})}. \tag{113}$$

To implement such a gate, we use the BCH formula synthesis methods presented in Sec. IV B 2 for pure $U(1)$ gauge fields. In particular, we apply Eq. (52), where we set $e^{i\hat{X}\hat{O}_1\theta} = e^{((-2ic_i^\dagger d_i a_{i,j}^\dagger b_{i,j} + 2ic_i d_i^\dagger a_{i,j} b_{i,j}^\dagger)\otimes 2iX)(-i\theta)^2 + O(\theta^3)}$ for which the compilation was shown in Eq. (82), and we set $e^{i\hat{Y}\hat{O}_2\theta} = e^{-i\hat{Y}_k(c_j^\dagger d_j + \mathrm{H.c.})\theta}$. This process yields unwanted terms, which we cancel using the Trotter formula in Eq. (53) through use of the BCH formula with appropriate phases in the beam splitters.

## V. QUANTUM ADVANTAGE PROPOSAL: GATE COMPLEXITY AND COMPARISON TO ALL-QUBIT ALGORITHMS

In this section, we analyze the gate complexity of our qubit-boson Trotter approach. In particular, we discuss the interplay of Trotter errors with errors introduced by some of our compilation schemes. We first review the general principles of Trotter error analysis. We then discuss a particular Trotter implementation of the $\mathbb{Z}_2$-Higgs model and $U(1)$ quantum link model and its error. While all compilations are exact for the $\mathbb{Z}_2$-Higgs model, the compilation of the magnetic-field term in the $U(1)$ quantum link model (cf. Sec. IV B 2) introduces additional errors. We therefore discuss the interplay of Trotter error with this compilation error. While the following sections focus on the asymptotic scaling of the whole Trotter algorithm, we close with a comparison of the explicit gate counts of a single Trotter step in our qubit-boson approach to an all-qubit approach. For this purpose, we introduce a qubit algorithm for simulating beam splitters between bosonic modes.

### A. General Trotter error analysis

We write the Hamiltonian as a sum of terms $\hat{H} = \sum_{\gamma=1}^{\Gamma}\hat{H}_\gamma$. We first decompose the time evolution $\exp(-i\hat{H}T)$ until time $T$ into $r$ time steps of duration $t = T/r$, i.e., $\exp(-i\hat{H}T) = (\exp(-i\hat{H}t))^r$. An estimate of the error

$$\epsilon_{\mathrm{Trotter}} \equiv \|(\mathcal{S}(t))^r - e^{-i\hat{H}T}\| \tag{114}$$

of a $2p$th order Trotter formula $\mathcal{S}(t)$ as introduced in Sec. I G is [239]

$$\epsilon_{\mathrm{Trotter}} = \mathcal{O}\left(\left(\sum\nolimits_\gamma\|\hat{H}_\gamma\|\right)^{2p+1}\frac{T^{2p+1}}{r^{2p}}\right). \tag{115}$$

The number of Trotter steps for a target Trotter error is hence

$$r = \mathcal{O}\left(\left(\sum\nolimits_\gamma\|\hat{H}_\gamma\|\right)^{\frac{2p+1}{2p}}\frac{T^{\frac{2p+1}{2p}}}{\epsilon_{\mathrm{Trotter}}^{\frac{1}{2p}}}\right) \tag{116}$$

$$= \mathcal{O}\left(\left(\sum\nolimits_\gamma\|\hat{H}_\gamma\|T\right)^{1+\frac{1}{2p}}\epsilon_{\mathrm{Trotter}}^{-\frac{1}{2p}}\right). \tag{117}$$

From this expression, we obtain the total number of gates as

$$N_{\mathrm{gates},2p} = r5^{p-1}N_{\mathrm{gates},2p=2} \tag{118}$$

$$= r5^{p-1}2\sum_\gamma N_{\mathrm{gates},\gamma}, \tag{119}$$

where $N_{\mathrm{gates},\gamma}$ is the number of gates required for the individual exponentials $\exp(-i\hat{H}_\gamma t)$. We used the fact that a Trotter formula of order $2p$ requires 5 times the number gates of a Trotter formula of order $2p-2$ when defined recursively [239]. The second-order Trotter formula requires twice as many gates as the first-order Trotter formula. Finally, the number of gates required for the first-order Trotter formula is the sum of the number of gates, $N_\gamma$, required for each exponential. In cases where the individual exponentials can only be synthesized with errors (dependent on $t$), $N_{\mathrm{gates},\gamma}$ contributes to the overall complexity. Note that, in general, the complexity of Trotter formulas is evaluated at fixed $p$. Hence, factors such as $5^{p-1}$ above are dropped for estimating the asymptotic complexity in $\mathcal{O}$ notation.

Approximating the Trotter error in terms of the norms of the Hamiltonian terms overestimates the error because most terms in many-body Hamiltonians commute with each other. We therefore expect the Trotter error to be much smaller in practice; see, e.g., Ref. [234] for a tighter estimate for the $(1+1)$D fermionic $U(1)$ quantum link model.

Bosonic Hamiltonians have, in principle, an unbounded norm and therefore naively exhibit an unbounded product-formula error. However, in practice, infinitely high Fock states are not occupied during the dynamics such that the error stays bounded. In particular, in the $U(1)$ quantum link model, the spectral norm of bosonic operators is bounded by $\sqrt{S}$. Similarly, in the $\mathbb{Z}_2$-Higgs model, using an initial state with a total number of $N$ bosons in the system, bosonic operators are bounded by $\sqrt{N}$.

We detail specific Trotter decompositions of the $U(1)$ quantum link model and $\mathbb{Z}_2$-Higgs model and their errors next. We specify open boundary conditions on an $L\times L$ square lattice and use the triangle inequality $\|\hat{H}_0+\hat{H}_1\|\leq\|\hat{H}_0\|+\|\hat{H}_1\|$ as well as $\|a\hat{H}_0\|=|a|\|\hat{H}_0\|$. For bosonic matter, we assume evolution within a sector of total boson number $N$ to estimate the Trotter error.

### B. Trotter decomposition and gate complexity of the $\mathbb{Z}_2$-Higgs model

The electric field and on-site interaction term commute with each other such that we can choose

$$\hat{H}_0=-g\sum_{\langle i,j\rangle}\hat{X}_{i,j}+U\sum_{i=1}^{L^2}\hat{n}_i^2, \tag{120}$$

where $\langle i,j\rangle$ indicates summation over the $2L(L-1)$ bonds on a square lattice with $L^2$ sites. It follows from the triangle inequality and $\|\hat{X}_i\|=1$, $\|\hat{n}_i\|=N$ in a fixed particle sector with $N$ particles that

$$\|\hat{H}_0\|\leq 2gL(L-1)+|U|L^2N^2. \tag{121}$$

Next, we split the gauge-invariant hopping into four noncommuting terms $\xi=1, 2, 3, 4$, where 1 (2) denotes the even (odd) hoppings in the horizontal direction and 3 (4) the even (odd) hoppings in the vertical direction, cf. Fig. 16(a). We write this case as

$$\hat{H}_1^\xi=-J\sum_{\langle i,j\rangle_\xi}(\hat{m}_i^\dagger\hat{Z}_{i,j}\hat{m}_j+\mathrm{H.c.}). \tag{122}$$

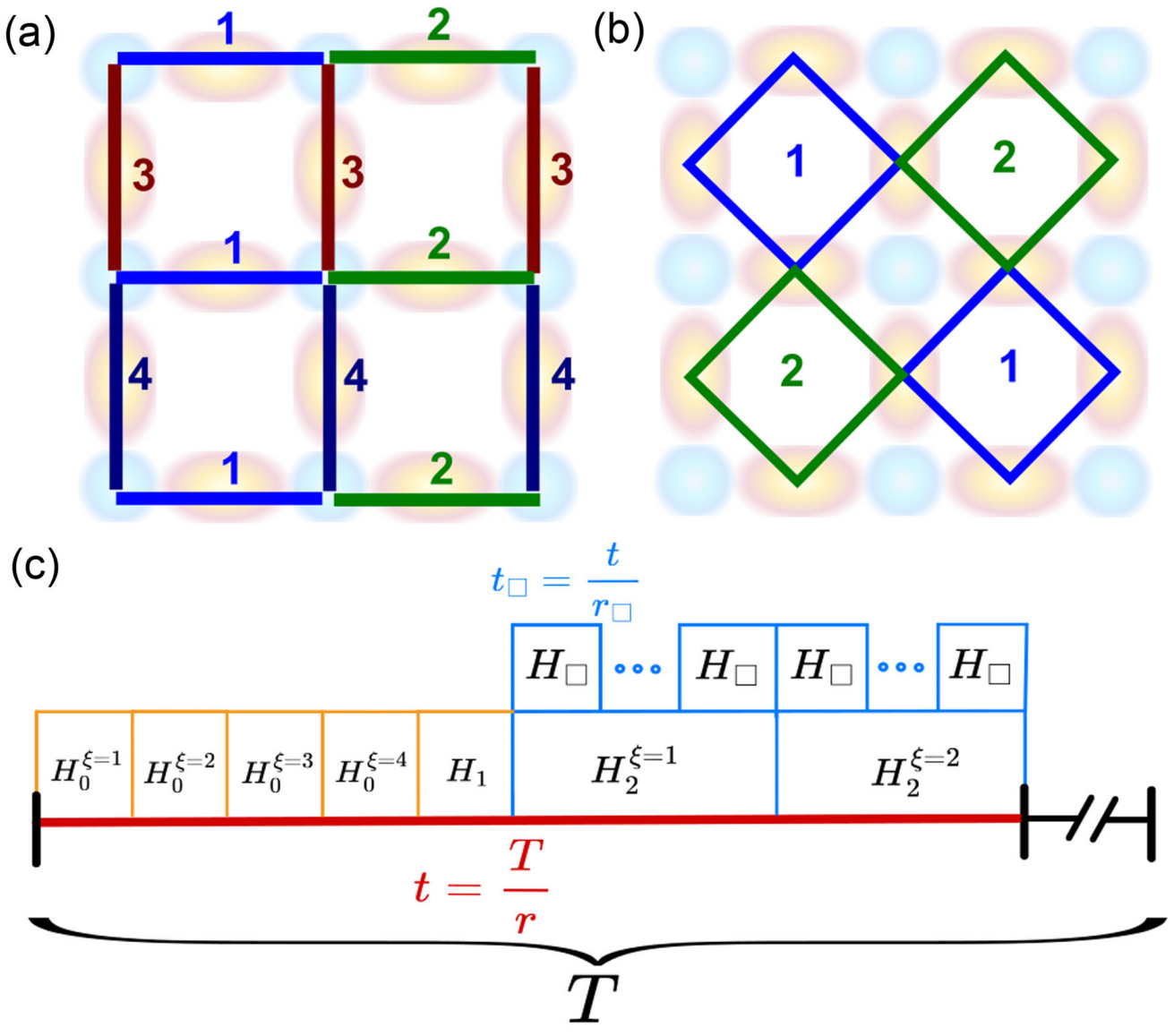

FIG. 16. Illustration of the Trotter decomposition. (a) Gauge-invariant hopping, split into four components $\xi=1, 2, 3, 4$ such that all terms within the component commute (b) Plaquette term. We split the square lattice into two disjoint checkerboard patterns. In both panels, the numbers correspond to the $\xi$ indices. (c) Illustration of the partition of time evolution until time $T$ into different sub-Hamiltonians. For the $\mathbb{Z}_2$-Higgs model [Eqs. (120)–(124)], a further decomposition of Eq. (124) into $r_\square$ evolutions under $H_\square$ is not necessary, but it is necessary for Eq. (132) of the $U(1)$ quantum link model [Eqs. (128)–(132)].

Next, we find $\|\hat{m}_i^\dagger\hat{Z}_{i,j}\hat{m}_j+\mathrm{H.c.}\|\leq 2\|\hat{m}_i^\dagger\|\|\hat{Z}_{i,j}\|\|\hat{m}_j\|=2N$, where we used the fact that $\|\hat{m}_i^\dagger\|=\sqrt{N}=\|\hat{m}_i\|$. From here, we obtain

$$\sum_{\xi=1}^{4}\|\hat{H}_1^\xi\|\leq 4JNL(L-1), \tag{123}$$

where $N=1$ for fermions. Finally, we divide up the plaquette terms into the two nonoverlapping checkerboard patterns of the square lattice [cf. Fig. 16(b)], which we label with $\xi=1, 2$, such that

$$\hat{H}_2^\xi=-B\sum_{i,j,k,l\in\square_\xi}(\hat{Z}_{i,j}\hat{Z}_{j,k}\hat{Z}_{k,l}\hat{Z}_{l,i}), \tag{124}$$

and from the triangle inequality,

$$\sum_\xi\|\hat{H}_2^\xi\|\leq B(L-1)^2, \tag{125}$$

where we used the fact that there are $(L-1)^2$ plaquettes.

Inserting Eqs. (121), (123), and (125) into Eq. (117), we find a number of Trotter steps given by

$$r = \mathcal{O}\left( ((2g + |U|N^2 + 4JN + B)L^2T)^{1+\frac{1}{2p}} \frac{1}{(\epsilon_{\text{Trotter}})^{\frac{1}{2p}}} \right), \tag{126}$$

where $N = 1$ and $U = 0$ for fermionic matter.

In particular, we find that, for bosonic matter, the interaction term dominates the error. If $U = 0$, instead, the hopping term is the leading source of error. For fermionic matter, both hopping and field terms contribute equally to the Trotter error.

All our compilations of the terms in the $\mathbb{Z}_2$-Higgs model are exact and therefore $N_{\text{gates},\gamma} = \mathcal{O}(1)$. There are $(L-1)^2$ plaquette terms on a square lattice. Therefore, we find that the total number of gates is given by

$$\begin{aligned} N_{\text{gates},2p} &= r(L-1)^2 5^{p-1} 2 \sum_\gamma N_{\text{gates},\gamma} \\ &= \mathcal{O}\bigg( ((2g + |U|N^2 + 4JN \\ &\quad + B)L^2T)^{1+\frac{1}{2p}} \frac{L^2}{(\epsilon_{\text{Trotter}})^{\frac{1}{2p}}} \bigg) \end{aligned} \tag{127}$$

for fixed $p$. Thus, for a large value of $p$ and $N = \mathcal{O}(1)$ in $L$, the number of gates approximately scales as $L^4$, i.e., the square of the total number of sites. This result makes sense because of the number of terms in the Hamiltonian, which is linear in the number of sites, and in our crude estimate, the number of Trotter steps $r$ for a certain target error is linear in the number of terms in the Hamiltonian. We note that, in most interesting situations, $N \propto L^2$, such that the scaling in $L$ will be higher; in particular, the on-site term dominates the scaling with $L$ in that case, i.e., approximately $L^8$ for large $p$. However, we expect this scaling to be much improved when considering tighter bounds on the Trotter error [239].

### C. Trotter decomposition of the $U(1)$ quantum link model

The electric-field and mass terms commute with each other and between lattice sites. Hence, we choose

$$\hat{H}_0 = \frac{g^2}{2} \sum_{\langle i,j \rangle} \left( \hat{S}^z_{i,j} + \frac{\tau}{2\pi} \right)^2 + M \sum_i (-1)^i \hat{n}_i. \tag{128}$$

Using the fact that spin $S$ operators fulfill $\|\hat{S}^z\| = S$ and $\|\hat{n}_i\| = N$, from the triangle inequality, we obtain

$$\|\hat{H}_0\| \leq \frac{g^2}{2} 2L(L-1) \left( S + \frac{\tau}{2\pi} \right)^2 + MNL^2. \tag{129}$$

Furthermore, we split the gauge-invariant hopping in the same way as for the $\mathbb{Z}_2$-Higgs model, i.e., into the two checkerboard patterns of the square lattice, writing

$$\hat{H}^\xi_1 = \frac{J}{2\sqrt{S(S+1)}} \sum_{\langle i,j \rangle_\xi} (\hat{m}^\dagger_i \hat{S}^+_{i,j} \hat{m}_j + \text{H.c.}). \tag{130}$$

Because $\|\hat{S}^+\| = \|\hat{S}^-\| = \sqrt{S(S+1)}$, the $S$ dependence of the norms is canceled by the prefactors, and we find that

$$\sum_\xi \|\hat{H}^\xi_1\| \leq 2NJL(L-1). \tag{131}$$

Finally, akin to the $\mathbb{Z}_2$-Higgs model, we use the checkerboard decomposition of the plaquette term, which we write as

$$\hat{H}^\xi_2 = \frac{-1}{4g^2(S(S+1))^2} \sum_{i,j,k,l \in \square_\xi} (\hat{S}^+_{i,j} \hat{S}^+_{j,k} \hat{S}^-_{k,l} \hat{S}^-_{l,i} + \text{H.c.}). \tag{132}$$

Again, there is no $S$ dependence of the norm. Because there are $(L-1)^2$ plaquettes, the norm is

$$\sum_\xi \|\hat{H}^\xi_2\| \leq \frac{1}{2g^2} (L-1)^2. \tag{133}$$

Inserting Eqs. (129), (131), and (133) into Eq. (117), we find

$$\begin{aligned} r = \mathcal{O}\Bigg( &\left( \left( \frac{1}{2} g^2 S^2 + MN + NJ + \frac{1}{2g^2} \right) L^2 T \right)^{1+\frac{1}{2p}} \\ &\times \frac{1}{(\epsilon_{\text{Trotter}})^{\frac{1}{2p}}} \Bigg), \end{aligned} \tag{134}$$

where we dropped some subleading contributions in $S$ and $L$. Again, $N = 1$ for fermionic matter. For large $S$, we find

$$r = \mathcal{O}\left( ((gSL)^2 T)^{1+\frac{1}{2p}} \frac{1}{(\epsilon_{\text{Trotter}})^{\frac{1}{2p}}} \right). \tag{135}$$

Note that a further decomposition of the $\hat{H}^{(\xi)}_n$ into terms that only act locally yields no further Trotter error as the constituent Hamiltonians commute with each other. For example, $\exp(-i\hat{H}^\xi_2 t) = \prod_{i,j,k,l \in \square_\xi} \exp\{-i[t/4g^2(S(S+1))^2](\hat{S}^+_{i,j} \hat{S}^+_{j,k} \hat{S}^-_{k,l} \hat{S}^-_{l,i} + \text{H.c.})\}$ with no error.

Contrary to the $\mathbb{Z}_2$-Higgs model, not all our compilations are exact for the $U(1)$ quantum link model. Specifically, the magnetic-field term is compiled using the BCH formula; hence leading to error dependent on the time step $T/r$. Because $N_{\text{gates},\gamma}$ for the magnetic-field term contributes to the overall gate count, we need to include this error in the analysis, which is the topic of the following subsection.

### D. Gate complexity of the $U(1)$ quantum link model

We focus on the simulation of the pure gauge theory, i.e., only including electric and magnetic-field terms to make the equations simpler to read. The gauge-invariant hopping can be synthesized without error, and the Trotter error is, in any case, dominated by the electric-field term for large $S$; therefore, the analysis below generalizes trivially to the $U(1)$ quantum link model with matter. We discuss the BCH synthesis error encountered in the plaquette term and then combine this with the Trotter error.

First, consider the error of the synthesis of a single plaquette term, which we showed in Sec. IV B 2 to be given by $\mathcal{O}(g^{-5/2}t^{5/4})$. This error can be further reduced by splitting the evolution in $r_\square$ steps of time $t/r_\square$ [cf. Fig. 16(c)], yielding

$$\epsilon_\square = r_\square \mathcal{O}(g^{-5/2}t^{5/4}r_\square^{-5/4}) \tag{136}$$

$$= \mathcal{O}(g^{-5/2}t^{5/4}r_\square^{-1/4}) \tag{137}$$

for the error $\epsilon_\square$ to evolve under the Hamiltonian of a single plaquette until time $t$. Solving for $r_\square$, we obtain

$$r_\square = \mathcal{O}((g^{-5/2}t^{5/4}\epsilon_\square^{-1})^4) \tag{138}$$

$$= \mathcal{O}(g^{-10}t^5\epsilon_\square^{-4}). \tag{139}$$

Because the BCH error depends on the time step $t$ chosen in the Trotterization scheme, $r_\square$ depends on the relation chosen between the target Trotter error and the BCH error. We make the choice that the total BCH error incurred to synthesize all plaquette terms for all Trotter steps scales the same as the Trotter error $\epsilon_{\rm Trotter}$. Because there are $(L-1)^2$ plaquette terms in the magnetic-field Hamiltonian, and a $2p$ Trotter formula calls the magnetic-field term $2\times 5^{p-1}$ times in each of the $r$ Trotter steps, this total error is given by

$$\epsilon_{\rm BCH} = 25^{p-1}r(L-1)^2\epsilon_\square. \tag{140}$$

Solving this equation for $\epsilon_\square$ and demanding $\epsilon_{\rm BCH} = \epsilon_{\rm Trotter}$, we find

$$\epsilon_\square = \mathcal{O}\left(\frac{\epsilon_{\rm Trotter}}{rL^2}\right). \tag{141}$$

Inserting this equation into Eq. (139), we find

$$r_\square = \mathcal{O}\left(\frac{t^5r^4L^8}{\epsilon_{\rm Trotter}^4 g^{10}}\right) \tag{142}$$

$$= \mathcal{O}\left(\frac{T^5L^8}{r\epsilon_{\rm Trotter}^4 g^{10}}\right), \tag{143}$$

where in the second step we inserted $t = T/r$.

There are $\mathcal{O}(L^2)$ plaquette and electric-field terms in each Trotter step. The electric-field term can be synthesized with $\mathcal{O}(1)$ gates, while the plaquette term requires $\mathcal{O}(r_\square)$ gates. Hence, the total gate complexity for implementing the $U(1)$ gauge theory is given by

$$N_{{\rm gates},2p} = \mathcal{O}(rL^2(1+r_\square)). \tag{144}$$

Inserting Eq. (135) for $r$ and Eq. (143) for $r_\square$, we find

$$N_{{\rm gates},2p} = \mathcal{O}\left(L^2\frac{(gSLT)^{1+\frac{1}{2p}}}{(\epsilon_{\rm Trotter})^{\frac{1}{2p}}} + \frac{T^5L^{10}}{\epsilon_{\rm Trotter}^4 g^{10}}\right), \tag{145}$$

where the first (second) term inside the brackets is the electric-field (magnetic-field) contribution.

This expression shows that the contribution to the total number of gates coming from the magnetic-field term (the second term) is independent of the cutoff $S$. This result is in contrast to qubit methods in the Fock-binary encoding, where the contribution scales as $\mathcal{O}(\log(S))$ [238]. This improvement comes at the cost of a high polynomial scaling with $\epsilon_{\rm Trotter}$, $T$, and $L$. However, this scaling can be systematically improved by using higher-order BCH formulas in the synthesis of the magnetic-field term [236].

### E. Gate-complexity advantage over all-qubit hardware

Here, we compare our hybrid qubit-boson compilation to an all-qubit compilation by calculating explicit gate counts in the qubit architecture for implementing single Hamiltonian terms used in a Trotter decomposition.

For this purpose, we must first choose a mapping of bosons to qubits. A number of different boson-to-qubit mappings exist: the unary encoding [240], which is wasteful in number of qubits but somewhat straightforward for gate application; the Jordan-Lee-Preskill encoding for phase-space-formulated bosonic Hamiltonians [241], which is suitable for initial states that can be expressed as a superposition of a small number of coherent states; the Gray encoding, which requires minimal numbers of bit flips when transferring between Fock states [240]; and the binary encoding [242] of the Fock-state basis, which is efficient in qubit count. See also Ref. [88] for a discussion of the complexity of simulating bosons in gauge theories.

Here, we focus on the Fock-binary encoding, which is the most efficient in terms of qubit number. For example, Fock state $|3\rangle$ is encoded with cutoff 7 into three qubits, such that $|3\rangle \to |011\rangle$. The encoding of the state in the binary requires $n = \lceil \log_2(N_{\rm max}+1)\rceil$ qubits, where $N_{\rm max}$ is the maximum Fock state that can be accessed during the simulation. The idea of our protocol is to entangle this register with an ancilla qubit. The ancilla qubit stores the relative amplitude of Fock states $2j$ and $2j+1$ such that a single-qubit gate on the ancilla implements an operation on this subspace. The simpler case of a displacement is discussed in Sec. VIII of Ref. [57].

One severe limitation of simulating bosonic operations with qubits is the prefactor coming from the action of the creation and annihilation operators: $\hat{a}^\dagger|k\rangle = \sqrt{k+1}|k+1\rangle$ and $\hat{a}|k\rangle = \sqrt{k}|k-1\rangle$, where $|k\rangle$ is the Fock state. In the following, we discuss three methods to apply them: by brute-force decomposition of the whole $2^n \times 2^n$ gate unitary into qubit gates using any of the above encodings, by including them as phases in the one-hot representation of the Fock basis, or by direct application of the square roots by either calculating them on the fly using arithmetic or by explicitly applying a Fock-state-dependent phase.

First, we discuss brute-force decomposition of bosonic unitaries into two- and single-qubit gates. This case would naturally include the square-root values, but brute-force decomposition of arbitrary-sized unitaries has an exponential scaling in $n$ making exact decomposition intractable for $n \gtrsim 5$ [243]. For small systems, the best-known upper bound for a brute-force decomposition of a six qubit gate, e.g., a beam-splitter matrix between two modes with cutoff Fock state 7 encoded in the binary, yields 1783 two-qubit gates [243]. At a two-qubit gate error of $10^{-6}$ (believed to be only possible with error correction), 1783 gates results in approximately 99.8% fidelity for one beam splitter. By contrast, a circuit QED beam splitter can be implemented for Fock state 7 with fidelity 99.9% according to the Lindblad calculation in Appendix H, which uses currently demonstrated mode and qubit lifetimes. Hence, non-error-corrected circuit-QED beam splitters are on par with error-corrected qubit-only beam splitters. While this finding relies on the worst-case upper bound for brute-force gate decomposition of the beam splitter, the currently best-known schemes still use hundreds of CNOT gates (cf. Fig. 17) [56] and even thousands for cutoffs greater than 7.

Second, the one-hot encoding enables precalculation of the square roots because it pre-attributes a value to each individual qubit and can therefore assign gates with specific phases to each of them. The number of qubits scales linearly with the Hilbert space cutoff, in contrast to the binary encoding, which scales logarithmically. Thus, because a bosonic gate such as the beam splitter changes the state of all qubits encoding a mode, the number of gates also scales at least linearly. As a result, the best-known algorithm for a beam splitter between two modes of cutoff $|7\rangle$ [56] takes around the same number of gates as the brute-force decomposition, even when using further gate optimizations through compilation, which are not scalable to large qubit numbers.

In the following, we introduce two new methods for time evolving under the beam-splitter Hamiltonian using the Fock-binary encoding in qubits. They are fully scalable in the cutoff and therefore also allow asymptotic scaling analysis. Our two methods are discussed in detail in Appendix D; we provide an overview here. Both methods rely on a splitting of the beam-splitter matrix into one-sparse components, which are then implemented using a helper qubit. Because these components do not commute, in general, a Trotter scheme is used to guarantee precision $\epsilon$.

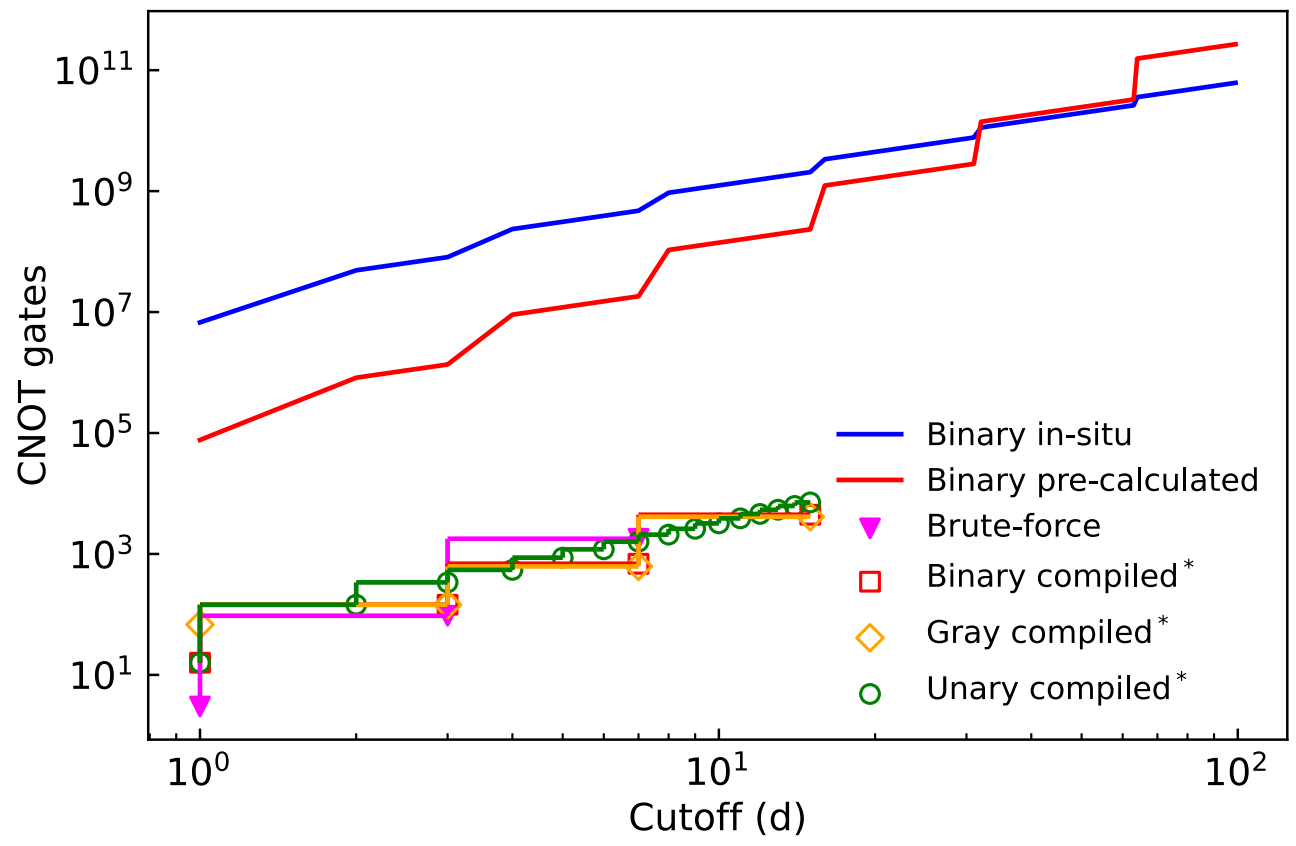

FIG. 17. CNOT gate counts for implementing a beam splitter. Comparison between mappings of bosonic modes to qubit-only hardware. Binary precalculated ($\alpha = 1, \epsilon = 10^{-3}, p = 2$) and binary *in situ* ($\alpha = 1, \epsilon = 10^{-3}, p = 2$) are methods for mapping a beam splitter which are developed in this work (see Appendix D). Brute-force decomposition results are known for up to six-qubit matrices [243], and all methods indicated with a * on the label are from Ref. [56].

One method calculates the square roots appearing from the action of the creation and annihilation operators *in situ*, using the quantum computer to calculate them on the fly, and the other method precalculates them and applies them as gate angles. It may sound more efficient to avoid the complex calculation of a square root and simply apply it as part of the gate angle. However, if the square root is to be applied as the angle of a gate, the state must undergo much testing to check which square root is appropriate. This process requires a circuit depth proportional to the number of Fock states in order to check each one and apply the correct square root, which is $K^2$ for a beam splitter acting on two modes. This approach leads to a number of CNOT gates for a single application of the beam-splitter exponential (without Trotter), which scales as $\mathcal{O}(N_{\max}^2 \log_2(N_{\max}))$. Now, taking into account the application of a (2p) Trotter formula [244] to reduce the error arising from the noncommutativity of the terms into which we break down the beam-splitter exponential, we find that, asymptotically, this method scales as $\mathcal{O}(N_{\max}^{3+\frac{1}{2p}} \log_2(N_{\max}))$.

By contrast, to calculate the square roots *in situ* with Newton iterations [121], one first calculates the inverse square root using Newton iterations and then multiplies the result with the initial value to obtain the square root. This method results in $N_{\mathrm{CNOT/SQRT}} = (270m + 126)n^2 + (228m + 96)n - 12m$ gates per square-root calculation [121], where $n = \log_2(N_{\max})$ is the number of qubits required to represent the maximum Fock state in binary and $m$ is the number of Newton iterations required for a set precision. As Newton's method converges quadratically, $m = 2$ or $m = 3$ is usually sufficient to obtain errors less than $10^{-4}$ [121]. For a single exponential (without Trotter), this method scales as $\mathcal{O}(\log_2(N_{\max})^2)$. Taking into account the (2p) Trotter formula to reduce the error from the noncommutativity of the terms into which we break down the beam-splitter operation, we find that, asymptotically, our method scales as

$$\mathcal{O}(N_{\max}^{1+\frac{1}{2p}} \log_2(N_{\max})^2), \tag{146}$$

assuming that, for large $N_{\max}$, the scaling of the number of Newton iterations, $m$, is independent of $N_{\max}$. Surprisingly, this scaling is far more optimal than the precalculation of the square roots. However, despite being among the best asymptotic methods for simulating bosonic modes with qubits, to our knowledge—prior methods for which we can find asymptotic CNOT gate counts are $\mathcal{O}(N_{\max}^2 \log_2(N_{\max}))$ [56]—it remains far from the naive assumption of logarithmic scaling for the binary encoding due to the Trotter formula used to reduce the error of the noncommuting terms into which the large Hilbert space operator is broken down into.

The two methods are compared in Fig. 17. Note that, for small cutoffs, it is advantageous to precalculate the square roots and check which one to apply because the prefactors for calculating them *in situ* are so large. We also note a more recent work, which may provide more optimal gate counts (no CNOT gate counts provided) [245]. Therefore, at least when using a naive arithmetic approach, the square-root requirement leads to a large overhead of thousands of CNOT gates.

We now calculate the number of CNOT gates required to perform a single Trotter step of the $\mathbb{Z}_2$-Higgs model in $(1+1)$D on all-qubit hardware using the Fock-binary encoding, keeping the gauge fields. We do not consider the magnetic-field term that appears in higher dimensions. However, for $\mathbb{Z}_2$ gauge fields that can be mapped to qubits, the qubit-only and the qubit-boson circuits to implement this term will be the same. Therefore, no additional Fock-binary gate count calculation or comparison would need to be made in higher dimensions for this model. The CNOT gate count originates from expressing the gauge-invariant hopping term of the bosonic matter in the qubit representation because the electric-field term for $\mathbb{Z}_2$ gauge fields corresponds only to single qubit gates (see Sec. IVA 1). We do not treat the bosonic Hubbard on-site interaction; however, it could be implemented by imprinting the phase in the binary with a small (polylog in $N_{\max}$) overhead. The gauge-invariant hopping corresponds to a qubit-conditional beam-splitter term. We convert this term to an all-qubit circuit using the Fock-binary encoding for which the gate counts are explained in Appendix D.

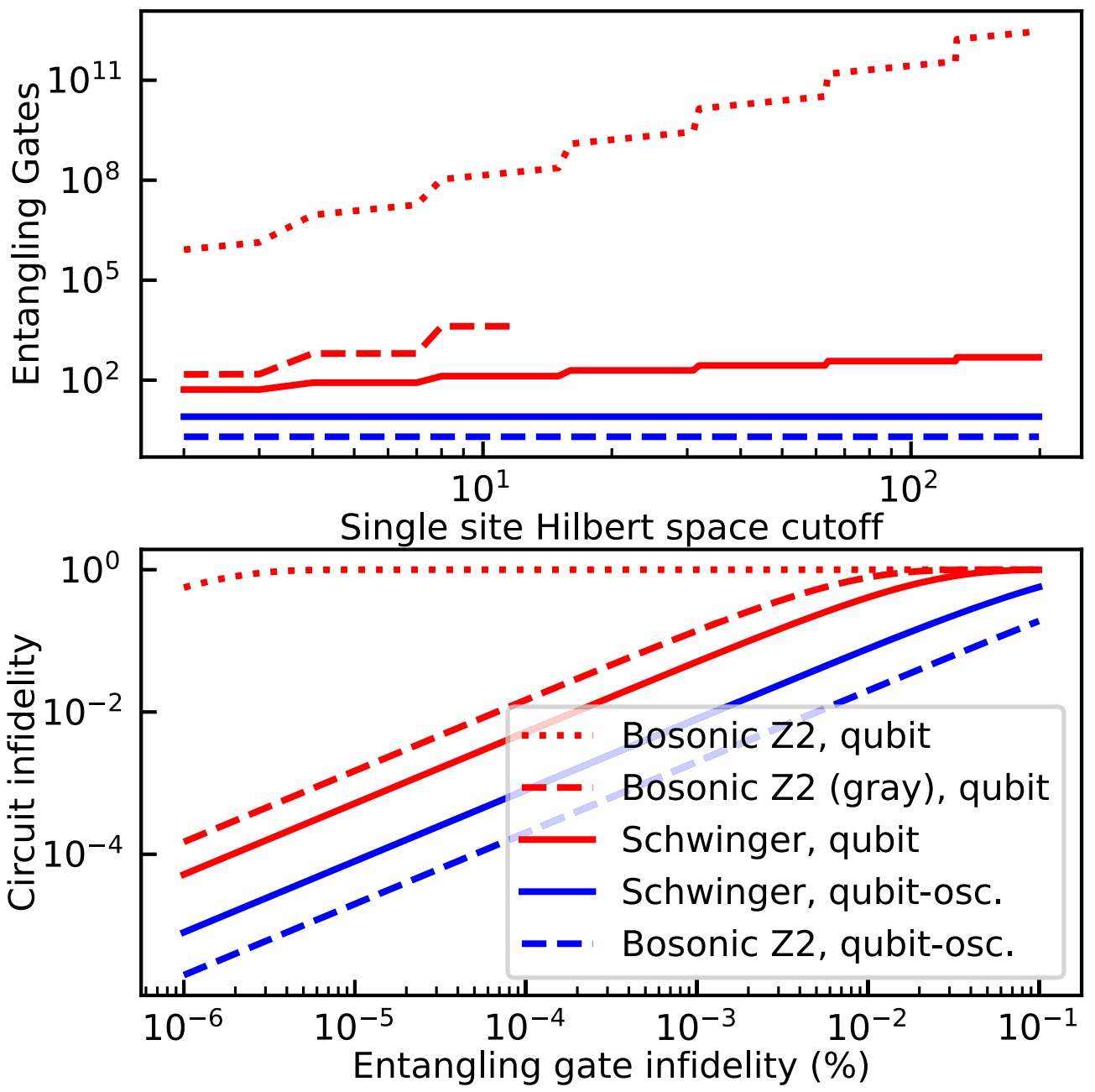

FIG. 18. Advantage of hybrid qubit-boson hardware over qubit-only hardware for the simulation of gauge theories, assuming simplistic qubit circuit construction. Entangling gate count as a function of single-site Hilbert space cutoff (upper plot) and circuit fidelity as a function of entangling gate fidelity (lower plot) for the $(1+1)$D $\mathbb{Z}_2$-Higgs model (dotted lines) and $U(1)$ quantum link model (referred to as the Schwinger model—solid lines) including gauge fields for qubit-only hardware (red) using the Fock binary encoding compared to hybrid qubit-boson hardware (blue) for one Trotter step for two matter sites. The qubit $\mathbb{Z}_2$-Higgs model requires square roots that are calculated with two Newton iterations. Gate counts are reported in the text; we count CNOT gates for all-qubit hardware and CΠ for hybrid qubit-boson hardware (single qubit gates and beam splitters are assumed to be ideal). We set the Hilbert space cutoff to be equal to the number of sites in the system. Trotter error is not taken into account in the circuit fidelity as it will be the same for circuits considered here.

In Fig. 18, we compare gate counts for the $U(1)$ quantum link model and $\mathbb{Z}_2$-Higgs model in both types of hardware; the asymptotic scalings are summarized in Table VIII. The jumps in the qubit-only lines are due to the logarithmic scaling of the qubit register required to represent increasing Hilbert space cutoffs. The $U(1)$ quantum link model in $(1+1)$D corresponds to the Schwinger model, for which we report the number of CNOT gates required for one Trotter step to be $N_{\mathrm{CNOT/U(1)}} = 9n^2 - 7n + 34$ [234]. In comparison, the number of entangling gates required to implement these models in hybrid qubit-boson hardware is

$$\mathcal{O}(1). \tag{147}$$

TABLE VIII. Asymptotic scaling of the number of entangling gates with Hilbert space cutoff for one Trotter step in qubit-only and hybrid qubit-boson hardware in $(1+1)$D for the $\mathbb{Z}_2$-Higgs model with bosonic matter and the $U(1)$ quantum link model (referred to as the Schwinger model), keeping the gauge fields. Note that $N_{\max}$ is the maximum Fock state at each bosonic matter lattice site or the spin length leading to a Hilbert space cutoff of $2N_{\max}+1$. The decay time of the bosonic modes increases linearly with Fock state in hardware.

| | Qubit (Fock-binary) | Qubit-boson |
|---|---|---|
| $\mathbb{Z}_2$-Higgs model | $\mathcal{O}(N_{\max}^{\frac{3}{2}}\log_2(N_{\max})^2)$ $(p=1)$ | $\mathcal{O}(1)$ |
| Schwinger model | $\mathcal{O}(\log_2(N_{\max})^2)$ [234] | $\mathcal{O}(1)$ |

The number of conditional parities CΠ required for the $\mathbb{Z}_2$-Higgs model and $U(1)$ quantum link model is

$$N_{\mathrm{C\Pi}/\mathbb{Z}_2} = 2, \tag{148}$$

$$N_{\mathrm{C\Pi/U(1)}} = 8. \tag{149}$$

Finally, we consider the corresponding circuit fidelities for the execution of one Trotter step for a system size of two sites. This consideration is of utmost importance in the NISQ era where entangling gate fidelities rarely reach above 99.99% [228]. In the fidelity calculation, we assume single qubit gates and beam splitters have ideal fidelity. We plot the circuit fidelity with regards to the entangling gate fidelity, which, in the all-qubit case, corresponds to the CNOT gate fidelity and, in the hybrid qubit-boson case, to the CΠ fidelity. Both types of hardware are based on the same qubit-boson circuits and will therefore encounter the same Trotter error (the bosonic $\mathbb{Z}_2$-Higgs model mapping is exact). Therefore, Trotter error is not taken into account in the circuit fidelity.

In the fault-tolerant era, ideal fidelities can only be approached with large distance codes, which directly increase the space-time volume of the computation. Therefore, even in the fault-tolerant era for all-qubit hardware, entangling gate fidelities will be of concern, and non-error-corrected hybrid qubit-boson hardware may provide advantages. This case is especially true considering the extremely large circuit depths required when mapping gauge fields or bosonic modes to qubits. In the case of the $\mathbb{Z}_2$-Higgs model that contains bosonic matter, the square-root factors lead to the requirement for extremely high-fidelity operation of a qubit-only device, unless we see substantial algorithmic improvements in arithmetic. In this case, the experimental improvements in hybrid qubit-boson hardware may enable progress to be made on bosonic models even in relatively large system sizes such as the ones presented here. Conversely, however, we must note that the harmonic approximations used for native bosonic hardware may break down if large occupation cutoffs are required. For these reasons, this work suggests the existence of a regime of advantage, but experimental and algorithmic realities need to be considered in greater detail to verify this advantage.

These results suggest that, for Trotter-based simulations, an advantage may be seen for hybrid hardware; however, a full comparison of qubit to qubit-boson devices is challenging to make. First, a full comparison between the two requires that we compare the best implementations of the best algorithms for both hardware in an agnostic fashion. This comparison is especially significant because alternative methods of simulation, such as qubitization or LCU [246,247], do not require square roots at the price of an increased number of ancillary qubits. A full comparison would need to consider optimized versions of all available algorithms for both hardware platforms. Second, a more detailed discussion of the approximation errors incurred in the approaches (such as the large $S$ limit or the use of BCH approximations and the different errors in the Trotter formula based on the form of the Hamiltonian) is needed to properly compare different approaches. For these reasons, we cannot claim that this work suggests a universal advantage for simulating $\mathbb{Z}_2$ or $U(1)$ LGTs, but it does strongly suggest advantages are likely to exist in certain regimes, and subsequent work is needed to understand the shape and size of these regions of advantage.

### F. Prospect for quantum advantage

In this section, we present a small-scale bosonic problem that would be extremely challenging to simulate using classical numerical methods as well as qubit methods. It contains large boson number fluctuations, which require a large on-site Hilbert space cutoff $D$. Examples of similar problems in socially relevant areas of physics and chemistry are discussed in Sec. I D.

We propose simulating a large Trotter step quantum dynamics with large boson fluctuations as an example problem. The total Hilbert space dimension is given by $D^L$ such that ten bosonic sites with a $D=20$ cutoff already exceeds the Hilbert space of the largest qubit simulations of around 40 qubits possible using exact diagonalization. The nonexact TEBD algorithm, the most efficient tensor-network time-evolution algorithm for short-range 1D systems, scales like $LD^2\chi^3$ [99], where $\chi$ is the bond dimension. Thus, far fewer sites are necessary to hit the boundaries of such classical simulations for large $D$ than for qubits: Assuming a constant bond-dimension simulation, a single site for $D=20$ incurs the same cost as 100 qubit sites.

To go beyond asymptotic scaling, i.e., find the system size at which classical TEBD numerics are practically challenging, we study a one-dimensional Z2 lattice gauge theory with initial state $|0+0+\ldots0+16+16+0+0+0\rangle$, i.e., 16 bosons on two neighboring sites in the middle of the system and qubit sites initialized in $|+\rangle=(|0\rangle+|1\rangle)/\sqrt{2}$ in between the boson sites. This initial state is motivated by a bunching phase recently discovered in the lattice gauge theory [248] when the electric-field strength dominates the hopping and the hopping dominates the on-site interaction. We then evolve by alternating between layers of $\exp[-i(\pi/4)\hat{Z}_{i,i+1}(\hat{a}_i^\dagger \hat{a}_{i+1} + \mathrm{H.c.})]$ and $\exp[-i(\pi/4)X]$ gates, where $\hat{a}$ is the bosonic destruction operator, simulating a Floquet-type quantum simulation, i.e., a large Trotter step time evolution. We choose a Hilbert space cutoff of $D = 20$ and $L = 20$ boson sites and $L = 20$ qubit sites. Because of the Gauss law constraint present in the gauge theory, the qubit state is completely fixed by the bosonic state, and therefore the total Hilbert space dimension is equivalent to the 32 boson number sector on 20 sites, which is $\binom{32+20-1}{19} \approx 2^{46}$. Thus, exact sparse matrix simulation is not possible.

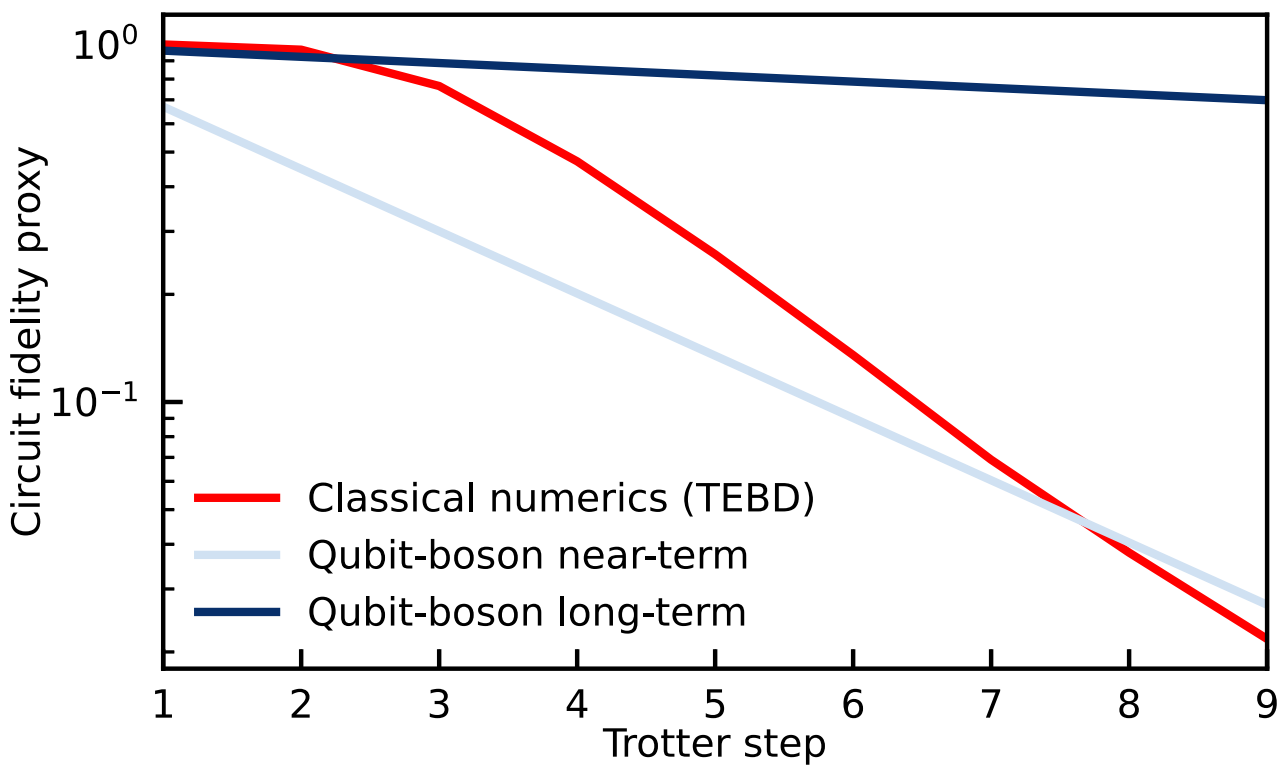

FIG. 19. Crossover to quantum advantage for long time evolution of the $Z_2$ lattice gauge theory. For multiple Trotter steps, each with a long time step, the entanglement growth is challenging for the classical MPS (red, bond dimension 400, one-day simulation time, fidelity proxy is TEBD loss of norm) while noisy hybrid qubit-boson hardware (blue) maintains higher fidelity (fidelity proxy is circuit infidelity, gate fidelity predictions include photon loss and dephasing, single-qubit gates are assumed ideal; near term: $f_{\rm BS} = 0.999, f_{\rm C\Pi} = 0.99$, long term: $f_{\rm BS} = 0.9999, f_{\rm C\Pi} = 0.999$). For all-qubit architectures, one Trotter step of this simulation with the best current method would require more than $10^5$ entangling gates.

We show the loss of norm $1 - \langle\psi|\psi\rangle$ as a function of the number of Trotter steps in Fig. 19 to determine the number of Trotter layers at which TEBD fails. The norm is an upper bound for the global fidelity $|\langle\psi|\psi_{\rm exact}\rangle|^2$ of the simulation by the Cauchy-Schwarz inequality $|\langle\psi|\psi_{\rm exact}\rangle|^2 \leq \langle\psi|\psi\rangle\langle\psi_{\rm exact}|\psi_{\rm exact}\rangle = \langle\psi|\psi\rangle$, where we use the fact that the exact state is normalized, $\langle\psi_{\rm exact}|\psi_{\rm exact}\rangle = 1$. We find that three Trotter steps suffice to push the fidelity below 90% for a bond dimension of 400. This finding is an extremely small number of viable steps, furthermore requiring nearly a day of simulation time on a single CPU.

In order to provide a fidelity of the time-evolved state simulated on an existing, near-term or state-of-the-art qubit-boson processor, we need to have an idea of the fidelities of a conditional-parity gate and a beam-splitter gate for the higher Fock states, such as $|20\rangle$, which will occur in the bunching phase. The fidelity estimates of these gates at high Fock states varies because mode decay increases linearly with Fock state; i.e., if $|1\rangle$ has a decay time of 1000 μs, $|20\rangle$ has a decay time of 50 μs. Therefore, we implement a Lindblad calculation that includes photon jump operators, such as decay and dephasing for the qubit, and decay (dephasing has been shown to be negligible) for the cavity modes. The results of these calculations are given in Appendix H. We find that, using currently achievable mode and qubit decay times, high fidelities are possible even for those high Fock-state cutoffs.

In passing, we note that the transmon qubits are the leading cause of noise in the simulations due to their much shorter decay time than the cavities. Therefore, including error correction for the transmons could help solve this challenge. On the other hand, developing error-suppression methods for digital bosonic modes is an active area of research [109–118]. Further methods to explore include advanced error mitigation strategies, such as probabilistic error cancellation, zero-noise extrapolation, readout error mitigation, learning-based mitigation, purity constraints, or approaches to mitigate parasitic gates.

In Fig. 19, we show the expected global fidelity of simulating dynamics of the bosonic $Z_2$ lattice gauge theory in the bunching phase using these calculated non-error-mitigated qubit-boson gate fidelities.

Remarkably, we find that, already in a window as small as 5–10 Trotter steps, the qubit-boson quantum computer will outperform the classical TEBD simulations using only moderate improvements in gate and coherence times over currently demonstrated ones. This finding is surprising considering that such shallow 1D circuits are usually easily classically simulable for *qubit* NISQ gate fidelities, and it shows strong promise for noisy qubit-boson hardware to reach quantum advantage very soon for relevant problems in physics and chemistry—which may even be challenging on qubit quantum computers. Indeed, for qubit-only quantum computers to simulate this problem, we would require a factor of around 1000 larger two-qubit gate count (cf. Fig. 17), meaning that only highly error-corrected qubit quantum computers could reach the fidelities achievable by uncorrected qubit-boson quantum computers.

It may be possible to create a better tensor-network algorithm. For example, we have not used the QR-based $D^2$ algorithm from Ref. [99] but instead the SVD algorithm from iTensor, which scales like $D^3$; in addition, our CPU-based implementation is far inferior to current GPU approaches. However, the bond dimension needed to simulate this problem increases so rapidly with the number of Trotter steps that, from extrapolations of our results shown here, we expect that, to reach 90% fidelity for five Trotter steps, bond dimensions in excess of current GPU records of about $\chi = 10000$ [99] are necessary, meaning that a classical simulation of this problem seems out of reach, at least for these methods. Very recently, a functional MPS method has been proposed that simulates bosonic modes with certain input states more natively [249], but its performance for qubit-boson circuits is unclear.

We compare our results with TEBD, which, to our knowledge, is the most efficient known state-vector-based algorithm for 1D quantum systems. It is an open research direction to explore whether Heisenberg picture methods (e.g., displacement propagation [250]) might yield better results for qubit-boson systems. Note, though, that a method based on non-Gaussian state simulation would generally incur exponential cost in the number of non-Gaussian gates. Nevertheless, it has been shown that, in certain cases, non-Gaussianity can actually make noisy simulations easier. Whether that remains true for the simulation under study is a potential avenue for future research.

We do not include other sources of noise, most importantly, Kerr and cross-Kerr terms (i.e., nonlinearity of the oscillator) in our simulations, which could alter the advantage claims here. However, when models including nonlinearities are the simulation target (e.g., the $\mathbb{Z}_2$-Higgs model studied in Ref. [248] or the Bose-Hubbard model), the presence of these terms would even be beneficial as they would not have to be synthesized. Note that it is challenging to reach the low-nonlinearity regime in pure transmon implementations; see, e.g., Ref. [251].

Finally, 1D is probably the "easiest" choice for classical simulations. We generalized digital qubit-boson circuits to 2D terms in Secs. III and IV, and we expect 2D circuits to be even harder for classical simulation.

## VI. MEASUREMENTS, ALGORITHMS, AND NUMERICAL BENCHMARKS

In the previous sections, we have introduced all of the necessary compilation methods to simulate both the $\mathbb{Z}_2$-Higgs model and $U(1)$ quantum link model (with either bosonic or fermionic matter) in $(2+1)$D. As an explicit demonstration of these compilation methods in practice, in this section, we apply them to the $(1+1)$D variants of these two models. In particular, we present schemes to measure short- and long-range observables, and near-term (Trotter and variational approaches) and longer-term (quantum signal processing) algorithms for dynamics and ground-state preparation, which we benchmark numerically. In particular, to enable approximate ground-state preparation using the VQE approach, we show how to measure the expectation value of the Hamiltonian. We also develop an ancilla-free method for using postselection on the Gauss law gauge constraint for partial error detection.

We use Bosonic Qiskit for our simulations [77]. Each mode is represented with $\lceil \log_2(N_b+1) \rceil + 1$ qubits, with $N_b$ the total number of bosons in the system (which is conserved for our Hamiltonians), in order to make sure that the commutation relation for the truncated mode operators obeys $[a, a^\dagger] = 1$ within the accessible Hilbert space. We also include single-qubit amplitude and phase damping error channels using Qiskit Aer [252] and neglect photon loss and dephasing of the cavities. For transmon qubits, $T_2 \sim T_1$ [196]. In tantalum transmons, it was shown that $T_1 > 500$ μs is possible [196]; however, we choose a conservative value of $T_1 = T_2 = 200$ μs [195], which implies a pure dephasing time of $T_\phi = ((1/T_2) - [1/(2T_1)])^{-1} = 400\ \mu s$. Cavities have very little dephasing and photon loss times exceeding 1 ms [187], and most of our numerical examples do not access high Fock states, which justifies us neglecting these processes.

This section is structured as follows. In Sec. VI A, we discuss the $(1+1)$D variants of the $\mathbb{Z}_2$-Higgs and $U(1)$ quantum link models, their physics, and challenges in previous approaches to solve them. In Sec. VI B, we investigate time dynamics, numerically benchmarking a simulation on the $\mathbb{Z}_2$-Higgs model while including the dominant source of noise. In Sec. VI C, we show that ground states of these models can be prepared using a VQE, for which we introduce an optimization technique better adapted to noise. In Sec. VI D, we analyze in detail how to mitigate the influence of shot noise and decoherence. In Sec. VI E, we show how to measure observables relevant to identifying phase transitions, such as the pair gap, the superfluid density, and the string-order correlator with respect to the VQE ground states. In Sec. VI F, we present a perspective on how to implement a ground-state preparation technique with provable success probabilities, known as quantum signal processing.

### A. Models and physics

To begin, we present the two models that we use for explicit demonstration throughout this section: the $\mathbb{Z}_2$-Higgs model containing bosonic matter sites and the $U(1)$ quantum link model containing fermionic matter sites, both in $(1+1)$D (see Fig. 20). These models are special cases of the more general LGT Hamiltonians presented in Sec. I E; we discuss the details and implementation of each below. In particular, we can use the architectures presented in Figs. 4(c) and 4(d) for one spatial dimension by winding a "snake" pattern through the square lattice. Thus, every cavity and qubit represents either a gauge field or matter site.

#### 1. $\mathbb{Z}_2$-Higgs model

The $\mathbb{Z}_2$ lattice gauge theory was introduced by Wegner [167] as a simple model showing signatures of confinement and deconfinement at the critical point [168], making it an ideal test bed for early testing of novel methods for LGT simulations. In two spatial dimensions in the absence of an electric field, its ground state encodes a spin-$1/2$ degree of freedom and can therefore be used for quantum error correction [169,170]. The $\mathbb{Z}_2$ lattice gauge theory with bosonic matter is reminiscent of the Higgs sector coupled to the non-Abelian $SU(2)$ electroweak gauge sector (W and Z bosons) of the $(3+1)$D standard model of particle physics with the two simplifications of a lower dimensionality and a $\mathbb{Z}_2$ gauge field instead of the non-Abelian $SU(2)$ electroweak gauge fields. The Hamiltonian in $(1+1)$D is written as follows:

$$\hat{H}_{\mathbb{Z}_2} = -g\sum_{i=1}^{L-1}\hat{X}_{i,i+1} + U\sum_{i=1}^{L}\hat{n}_i^2 - J\sum_{i=1}^{L-1}(\hat{a}_i^\dagger \hat{Z}_{i,i+1}\hat{a}_{i+1} + \mathrm{H.c.}), \qquad (150)$$

where the symbols are the same as described in the Introduction (see Sec. I E) and $L$ is the number of sites in the chain.

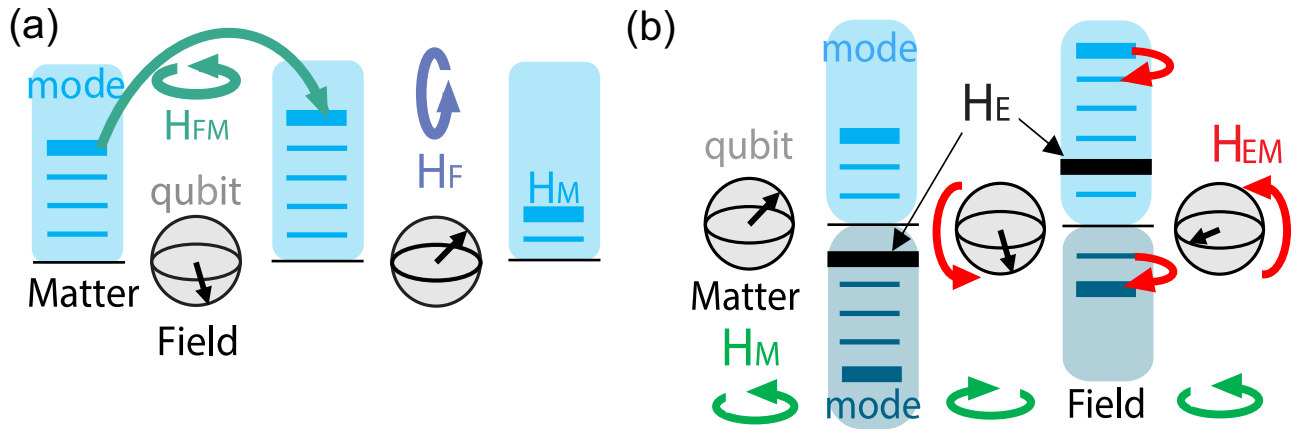

FIG. 20. Illustration of the lattice gauge theories considered in Sec. VI. (a) $\mathbb{Z}_2$-Higgs model. Bosonic matter (blue) is represented by bosonic modes on the sites, whereas a qubit represents the gauge field on the links (gray). The gauge-invariant coupling between the two (green) hops a boson from one site to a neighboring site and phase flips the qubit on the link (by acting with $\hat{Z}$). The electric-field term acts on the qubits with $\hat{X}$. Here, $\hat{H}_{\mathrm{F}}$, $\hat{H}_{\mathrm{M}}$, and $\hat{H}_{\mathrm{FM}}$ are the first, second, and third terms in Eq. (150), respectively. (b) $U(1)$ quantum link model. Matter sites are represented by qubits, and the gauge fields on the links by a pair of bosonic modes realizing the Schwinger-boson encoding. An example of their occupation number is illustrated with a thick line; the occupation number increases towards the top or bottom for the top or bottom mode. The value of the electric field is the difference in occupation between the constituting field modes (thick black line). The gauge-invariant hopping (red) exchanges an excitation between two neighboring qubits while incrementing or decrementing the gauge field (equivalent to transferring a boson from one of the composing modes to the other). A $\hat{Z}$ term on each matter qubit with opposite sign represents a staggered mass. Here, $\hat{H}_{\mathrm{E}}$, $\hat{H}_{\mathrm{M}}$, and $\hat{H}_{\mathrm{EM}}$ are the first, second, and third terms in Eq. (151), respectively.

The $\mathbb{Z}_2$ model in $(1+1)$D in the hardcore boson limit $U\to\infty$ has previously been investigated in depth [253], revealing intriguing phase transitions between insulating and Luttinger liquid phases [168,254]. A digital simulation of this model for hardcore bosons has been performed in Ref. [49]. A basic building block of two matter and one gauge-field site has been implemented in an experiment with ultracold atoms in a Floquet approach [27]. However, this approach is challenging to generalize away from the single-particle limit in the presence of on-site interactions $U$ due to Floquet heating, which is important when considering parameter regimes away from the hardcore-boson limit. Our digital implementation enables a fully flexible study of this model, including regimes away from the hardcore limit. While here we focus on the implementation aspects of the model, enabling a study away from the hardcore boson limit, we study its physics and phase diagram in a separate publication [248].

### 2. *U*(1) *quantum link model*

The $U(1)$ quantum link model constrained to $(1+1)$D and fermionic matter [175–179] reads

$$\hat{H}_{U(1)} = \frac{g^2}{2}\sum_{i=1}^{L-1}\left(\hat{S}^z_{i,i+1} + \frac{\tau}{2\pi}\right)^2 + M\sum_{i=1}^{L}(-1)^i\hat{c}_i^\dagger\hat{c}_i + \frac{\tilde{J}}{2S}\sum_{i=1}^{L-1}(\hat{c}_i^\dagger\hat{S}^+_{i,i+1}\hat{c}_{i+1} + \mathrm{H.c.}), \qquad (151)$$

where, compared to Eq. (3), the definition of $J$ differs by $\tilde{J} = J/2$ and we set $\sqrt{S(S+1)}\to S$ in the denominator of the hopping term; i.e., we consider the limit $S\to\infty$.

This model can be connected to a continuum field theory by introducing a lattice spacing $a$ by $J\to J/a$ and $g\to g\sqrt{a}$ and taking the limit $a\to 0$.

This model shares many properties with quantum chromodynamics in $(3+1)$D—chiral symmetry breaking, confinement, a $U(1)_A$ quantum anomaly, and a topological $\tau$ term—which gives rise to a nontrivial topological vacuum structure [171,172]. Because of the sign problem, this $\tau$ dependence of the model cannot be studied with quantum Monte Carlo methods. One possible way to circumvent the sign problem is using the tensor network (TN) approach, in particular, matrix product states (MPS) in $(1+1)$D. Using MPS, the spectrum of the $U(1)$ quantum link model has been computed, and the model has been studied at nonzero temperature and nonzero chemical potential, and with a nonzero $\tau$ term [181–183,255,256] (see Ref. [257] for a review). In particular, Refs. [181,182] studied the first-order phase transition at $\tau=\pi$, in which the $CP$ symmetry is spontaneously broken. The first-order phase transition line terminates at a second-order critical point [181,182]. This model has been extensively simulated in superconducting qubits [44–48,51,258].

Despite these successes in studying sign-problem afflicted regimes with TN, this approach has several limitations. First, highly entangled states such as those with volume-law entanglement can occur in the evolution after a quench or in thermal states, and they cannot be efficiently described using matrix product states [101], the most widely used TN. This limitation is particularly apparent in multistage thermalization dynamics such as after instabilities [259]. Second, the computational cost of TN algorithms increases with increasing dimensionality of the lattice field theory under consideration. For MPS, a particular kind of one-dimensional TN, the leading-order computational cost of the variational ground-state optimization in the case of open boundary conditions scales as $\mathcal{O}(D_b^3)$ [102–104], where $D_b$, called the bond dimension, indicates the tensor size in the MPS. For the generalization to $(2+1)$D, called projected entangled pair states (PEPS) [260], computational costs can scale up to $\mathcal{O}(D_b^{10})$ [105]. These costs are still polynomial in the bond dimension but substantially limit the values of $D_b$ that can be reached in practice. There have been TN studies of gauge theories in $(2+1)$D and $(3+1)$D [261–263], but designing efficient TN algorithms in higher dimensions remains a fundamental challenge.

For this Hamiltonian [Eq. (151)], Refs. [182,264] explored whether the gauge degrees of freedom can be efficiently represented with finite-dimensional systems. For intermediate values of the coupling, $g\sqrt{a} \sim 0.1$, Ref. [264] demonstrated that the results converge rapidly, implying that a reasonable accuracy can be obtained even with small cutoffs $N_{\max}$ of the gauge-field truncation. To be more precise, the truncated $U(1)$ quantum link model yielded a fast convergence to the exact ground state for $N_{\max}$ ranging from 3 to 9, reaching subpermille precision in the energy density for $N_{\max} = 9$ [264]. However, when approaching the continuum limit, $g\sqrt{a} \to 0$, the increasing electric-field fluctuations require an increasingly large number of states and therefore an increasingly large (up to exponentially large) cutoff, which strongly motivates applying hybrid qubit-boson quantum devices to $U(1)$ lattice gauge theory.

### B. Dynamics

To illustrate the power of our digital approach to qubit-boson quantum simulation of lattice gauge theories, we show that some gauge-constraint-induced dynamics can be observed in near-term noisy hardware. Here, we focus on the $\mathbb{Z}_2$-Higgs model for $U = 0$.

We first need to prepare an initial state of interest. As discussed in the exposition of the cQED platform in Sec. I F, it is straightforward to prepare the modes in a product state of Fock states. We consider a $L = 5$-site chain prepared in the states $|\psi_0\rangle = |00100\rangle$ and $|\psi_1\rangle = |00200\rangle$. The limitation to five sites is not restrictive for short timescales; because the interactions are short-range, excitations spread with a finite velocity, and as long as the "light cone" does not reach the boundary, the results obtained here are the same as in a larger system. The qubits representing the $\mathbb{Z}_2$-Higgs model gauge fields must then be initialized according to Gauss's law, which requires consistency between the qubits and the parity in each mode: Choosing the leftmost gauge-field qubit to be in $|+\rangle$ and the sites to be in Fock state $|00100\rangle$, the gauge-field qubits situated on the links between the bosonic sites need to be initialized in $|++--\rangle$. Likewise, for bosonic sites in $|00200\rangle$, the corresponding gauge-field state is $|++++\rangle$. The fact that the state of the gauge fields is fully determined by the state of the sites shows us that the gauge fields could be integrated out, which is usually the case for lattice gauge theories in $(1+1)$D. The reverse is not true, however: The state of the gauge fields does not fully determine the state of the bosonic sites, only their parity. This finding is in contrast to hard-core bosons, where the parity fully determines the occupation number [168]. See also Ref. [265] for a more in-depth consideration of these dynamics in the context of hybrid simulation in trapped ions.

We evolve the initial states $|\psi_0\rangle$ and $|\psi_1\rangle$ under the $\mathbb{Z}_2$-Higgs model with $U = 0$ Using even a first-order Trotterization $e^{-i\hat{H}_{\mathbb{Z}(2)}t} \approx (\hat{U}_1\hat{U}_2)^r$ and time step $\Delta t = t/r$ with

$$\hat{U}_1 = \prod_i^L e^{+ig\hat{X}_{i,i+1}\Delta t}, \tag{152}$$

$$\hat{U}_2 = \prod_{i=1}^{L-1} e^{+iJ\hat{Z}_i(\hat{a}_i^\dagger \hat{a}_{i+1} + \hat{a}_i \hat{a}_{i+1}^\dagger)\Delta t}. \tag{153}$$

In Sec. IV, we have discussed how the time-evolution operator for the individual Hamiltonian terms in Eq. (150) can be implemented using native gates in our proposed architecture.

At the end of the simulation, we measure the state of the qubits and modes. The measurement of the mode occupations are carried out bit by bit in a binary representation, by first mapping the boson number parity onto the qubit using an SQR gate, making a midcircuit measurement, resetting the qubit, measuring the super-parity, and so on, as has been demonstrated experimentally [144,211]. This process requires a circuit with a depth that is logarithmic in the maximum boson occupation cutoff. If the qubits representing the gauge fields are read out first, they can be reused to iteratively read out the Fock state of the cavity mode to which they are dispersively coupled.

To show that these dynamics can be probed in near-term hardware, we implement the $(1+1)$D gate sequences in Bosonic Qiskit [77,266], assuming an implementation in which the $\mathbb{Z}_2$ gauge fields are encoded in transmons. In order to simulate the dynamics in the presence of the noise, we need to specify the gate durations that we assume for the potential experimental implementation using the hardware in Sec. I F. We assume a value of $\chi \approx 2\pi \times 3$ MHz for the dispersive interaction strength. The duration of each Trotter step is then dominated by the conditional beam-splitter operation, which is on the order of 0.43 μs, consisting of two conditional-parity gates, which take $T = \pi/\chi \approx 167$ ns each, and one beam splitter, which has a duration on the order of 100 ns [91,217]. Single qubit gates take on the order of 10 ns. The Trotter step size in our numerical simulation, $\Delta t = 0.1/J$, is chosen to be small enough to recover the expected physical behavior in comparison to exact diagonalization. We choose 1000 shots per circuit, which leads to both low shot-noise errors and a realistic experimental runtime: Because each Trotter step has depth 2, a depth-1 circuit takes less than 0.45 μs, and we need 85 Trotter steps to evolve to $Jt = 8.5$. Each circuit therefore takes at most $2 \times 85 \times 0.45$ μs $= 76.5$ μs. We measure 85 circuits; therefore, using 1000 shots per circuit leads to a total projected time to take the experimental data of Fig. 21(b) for $g = 5J$ of $1000 \times 85 \times 76.5$ μs $\approx 6.5$ s.

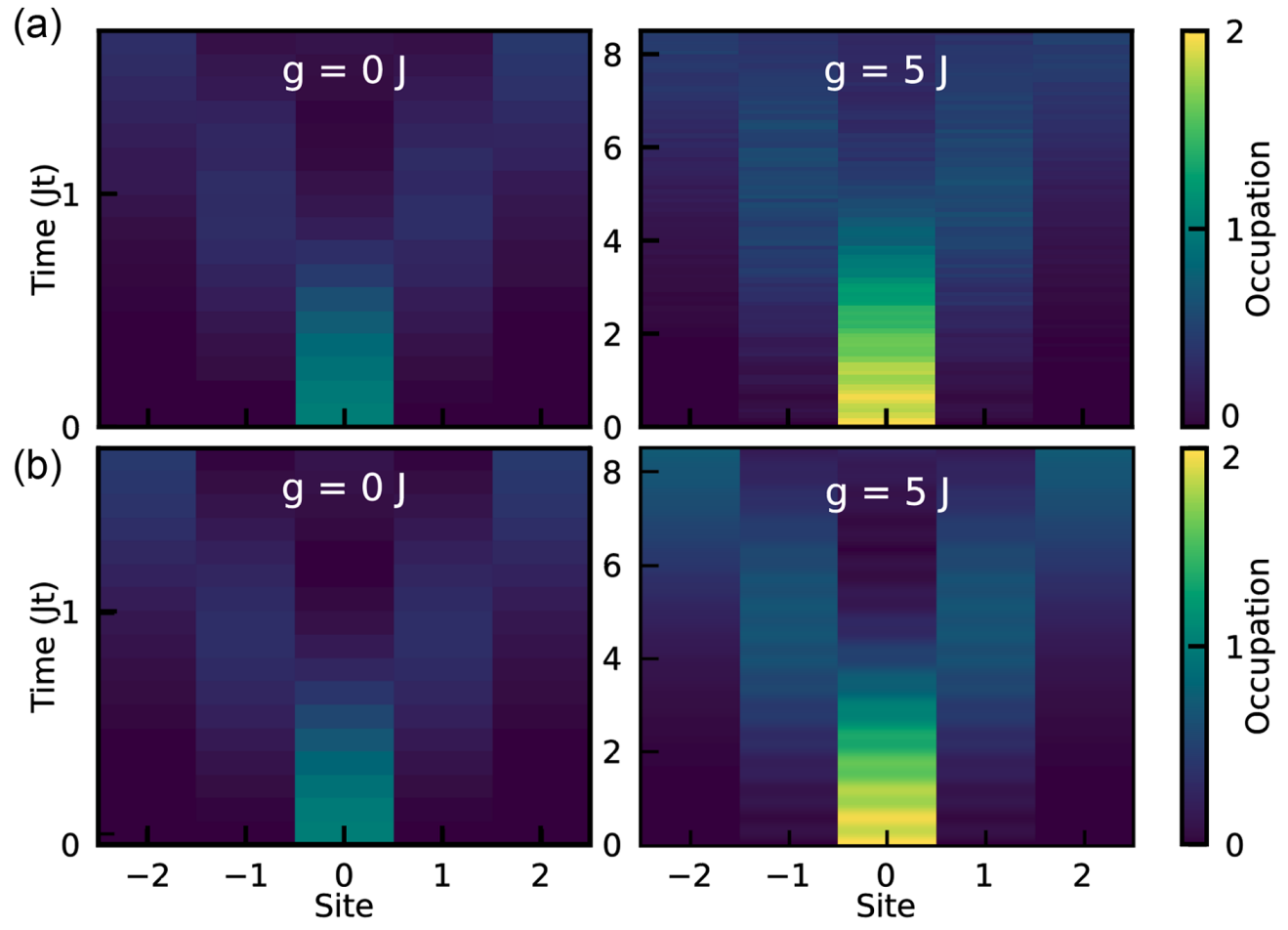

FIG. 21. Trotterized dynamics in the $\mathbb{Z}_2$ model in the presence of transmon decoherence. Left panels: free hopping regime $U/J = g/J = 0$, with initial bosonic state $|00100\rangle$. Right panels: large field regime $U = 0$, $g/J = 5$, with initial bosonic state $|00200\rangle$. (a) Numerically simulated results using Bosonic Qiskit, 1000 shots chosen to reduce shot-noise errors to a few percent, first-order Trotterization, and Trotter step $0.1/J$. (b) Exact diagonalization results.

On the one hand, in the regime of infinite hopping strength (i.e., $g = 0$), starting from the Fock state $|00100\rangle$ with open boundary conditions, we find a linear light-cone-like spreading of the boson, cf. Fig. 21. On the other hand, large field strengths prohibit the gauge fields from flipping during hopping events, which results in the boson remaining confined to the central site, as predicted and observed in Refs. [27,49,253] in cold atoms. Strikingly, for large field strengths, when starting from the Fock state with an even initial number of bosons such as $|00200\rangle$, we again find a linear light cone even in the large field strength regime. We find that the hopping takes place on a much slower timescale. This hopping is enabled by a perturbative energy-conserving pair-hopping process of strength $J^2/2g$ in which a qubit is flipped by a first boson hopping and then flipped back to its original value by the second boson hopping [248]. Despite the transmon decoherence and the much longer timescale (the circuit to evolve to the latest time point for $g = 5J$ takes around $2 \times 85 \times 2.2$ μs $\approx$ 374 μs experimental time), these coherent dynamics are observable and match exact diagonalization.

### C. Qubit-boson ground-state preparation with VQE

In this section, we show how the VQE [267,268] (see Refs. [269–272] for important follow-up works) can be implemented not only on oscillators or modes [273–275] but also on our proposed hybrid qubit-boson architecture to prepare ground states of the $\mathbb{Z}_2$-Higgs and U(1) quantum link model. The VQE is a variational ansatz for the ground-state wave function in which layers of unitaries $\hat{U}_k$ act on some initial state. The unitaries in our approach, sometimes called the quantum approximate optimization algorithm, are related to the Hamiltonian

$$\hat{H} = \sum_{k,i} \hat{h}_k(i), \tag{154}$$

where $\hat{h}_k(i)$ are terms of the Hamiltonian, with $i$ the site index on which they are applied. We choose

$$U_k(\theta, i) = e^{i\theta \hat{h}_k(i)}, \tag{155}$$

where $\theta$ is a variational parameter that is optimized. Each gate in the ansatz has an independent angle; we denote the vector of angles with $\boldsymbol{\theta}$.

In the VQE, the classical and quantum processors work together to reach the ground state by executing an optimization algorithm that updates the vector of ansatz angles $\boldsymbol{\theta} \to \boldsymbol{\theta}'$ and uses a quantum computer to measure the expected energy of the ansatz characterized by those angles $\langle \psi(\boldsymbol{\theta}')|\hat{H}|\psi(\boldsymbol{\theta}')\rangle$. The intuition behind this VQE procedure comes from the fact that, for our choice of ansatz, a sufficiently large number of layers, and a particular choice of $\boldsymbol{\theta}$, the VQE realizes adiabatic state preparation (note the parallel between choosing the gate variable angles to be proportional to time for dynamics and choosing them to be the optimization parameter for the ground-state search). Therefore, we know that, in principle, the VQE is able to prepare the ground state if the Hamiltonian has a large enough gap such that the necessary evolution time is within the coherence time of the quantum computer. In the absence of first-order phase transitions, the gap is always polynomial in the system size, and hence an efficient state preparation path can be found. In the gauge theory setting, constructing a variational ansatz this way also has the advantage of fulfilling Gauss's law, which is exactly the case if every term in the Hamiltonian conserves Gauss's law individually (which will be the case for us below). If this is not true, Gauss's law can be fulfilled approximately by choosing angles small enough to essentially realize the Trotter limit of adiabatic time evolution.

The hope behind the VQE is that, by minimizing the energy with respect to the $\boldsymbol{\theta}$ angles, a circuit can be found that prepares the ground state in a shallower circuit than would be required for adiabatic state preparation, which, in this digital setting, would require Trotterization. In the following, we discuss the VQE in the cQED architecture for the $\mathbb{Z}_2$-Higgs model and the $U(1)$ quantum link model in $(1+1)$D.

#### 1. $\mathbb{Z}_2$-Higgs model

In the case of the $\mathbb{Z}_2$-Higgs model, the $\hat{U}_k$ required for the ansatz discussed in the first paragraph of Sec. VI C, Eq. (155), are

$$\hat{U}_1(\theta, i) = e^{-\theta\hat{Z}_{i,i+1}(\hat{a}_i^\dagger\hat{a}_{i+1}-\hat{a}_i\hat{a}_{i+1}^\dagger)}, \tag{156}$$

$$\hat{U}_2(\theta, i) = e^{i\theta\hat{Z}_{i,i+1}(\hat{a}_i^\dagger\hat{a}_{i+1}+\hat{a}_i\hat{a}_{i+1}^\dagger)}, \tag{157}$$

$$\hat{U}_3(\theta, i) = e^{-i\theta\hat{X}_{i,i+1}}, \tag{158}$$

$$\hat{U}_4(\theta, i) = e^{-i\theta\hat{n}_i^2}. \tag{159}$$

While technically not necessary (all other gates together are already able to prepare the ground state by approximating an adiabatic evolution), we find that including $\hat{U}_1$ leads to a lower energy error for the same number of layers compared to when it is not included. All of the above unitaries can be realized with our native gate set using techniques previously discussed; in particular, $\hat{U}_1$ and $\hat{U}_2$ are conditional beam splitters [see Eq. (19)], $\hat{U}_3$ is a single-qubit rotation, and $\hat{U}_4$ can be implemented using a SNAP gate (see Sec. III A 3 or, alternatively, Appendix B for an ancilla-free implementation).

Using these resource unitaries, we construct an $M$-layer ansatz with $N$ field sites and $N+1$ matter sites as

$$|\psi_M(\boldsymbol{\theta})\rangle = \prod_{l=1}^{M}\Bigg[\prod_{i=1}^{N}\hat{U}_3(\gamma_l^i, i)\prod_{i=1}^{N+1}\hat{U}_4(\delta_l^i, i) \times \prod_{i\in\text{even}}^{N}\hat{U}_2(\alpha_l^i, i)\hat{U}_1(\beta_l^i, i) \times \prod_{i\in\text{odd}}^{N}\hat{U}_2(\alpha_l^i, i)\hat{U}_1(\beta_l^i, i)\Bigg]|\phi\rangle, \tag{160}$$

where $|\phi\rangle$ is an easily preparable state, $\boldsymbol{\theta}$ is a vector of the $(4N+1)M$ real-numbered angles in this ansatz, and the products over $i\in$ even (odd) include only the field qubits with an even (odd) index. Thus, every single gate in each layer and at each site has a different angle to be optimized. Figure 22(a) shows one layer of the VQE ansatz.

Since the $\mathbb{Z}_2$ resource unitaries in Eqs. (156)–(159) exactly preserve Gauss's law (in the absence of decoherence), the entire circuit evolution will take place within the Gauss law subspace chosen by the initial state. Restricting the evolution of VQE circuits within a particular subspace is a powerful technique introduced in prior work

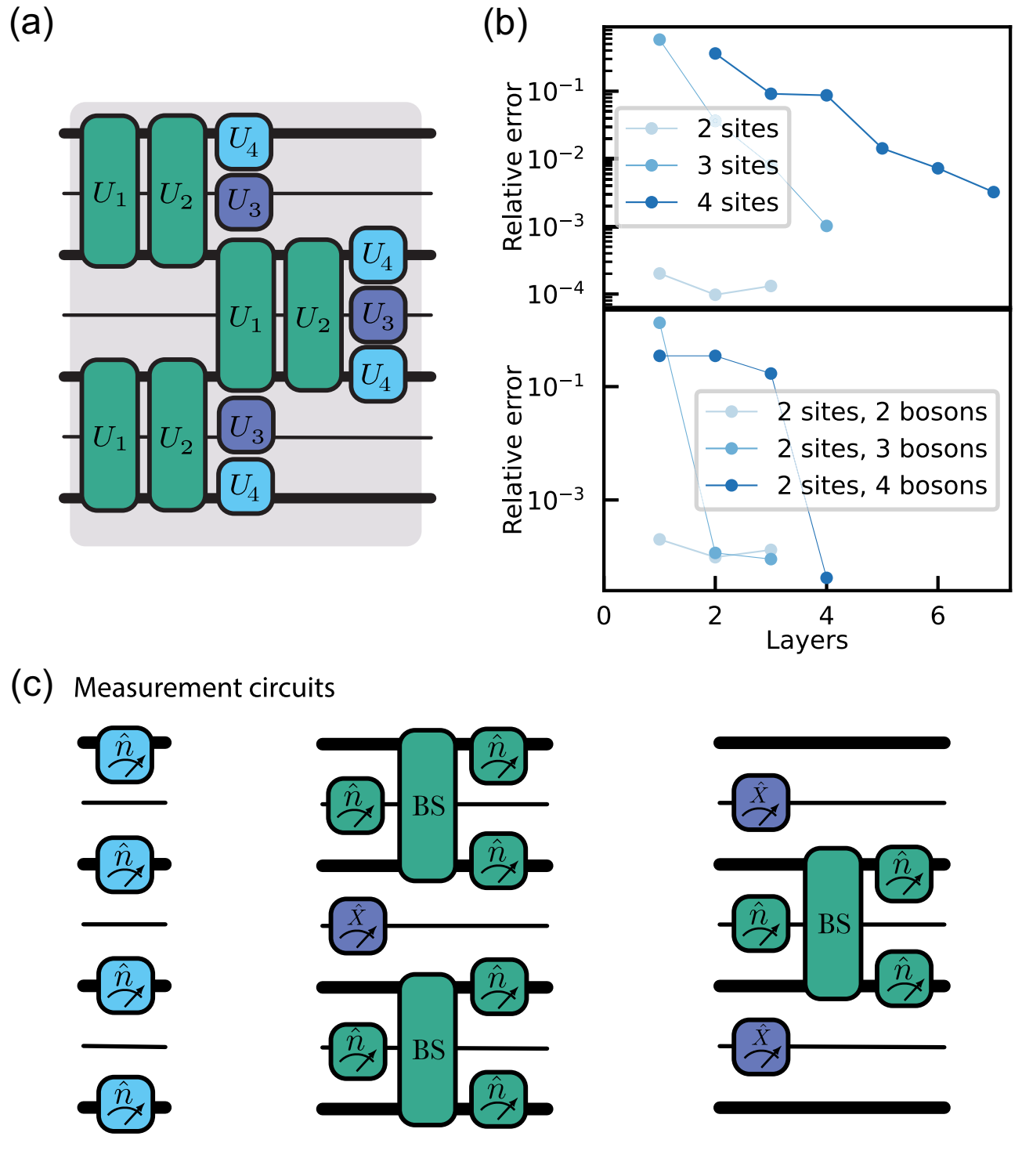

FIG. 22. VQE for the $\mathbb{Z}_2$-Higgs model in $(1+1)$D. (a) Single layer of the VQE ansatz. The different colors refer to the different Hamiltonian terms. (b) Numerical simulation of the VQE without shot noise and decoherence for $U/J = g/J = 1$. We show the relative error $|E_{\text{VQE}} - E_0|/|E_0|$ between the estimated ground-state energy $E_{\text{VQE}}$ obtained via the VQE and the true ground-state energy $E_0$ as a function of the number of layers for a varying number of sites at unit filling (top) and varying numbers of bosons for two sites (bottom). Fluctuations in performance between runs of the VQE (caused by the stochastic optimizer) are small, leading to fluctuations of the relative error of at most a factor of 5 (cf. the deviation of the four-site results from a smooth line; we show the result of a typical single run). (c) Three sets of measurement circuits we use for measuring the energy: gauge-invariant hopping (green), electric-field terms (purple), and the on-site interaction at each site (blue). The colors in panel (c) correspond to the ones in (a). We note that qubit measurement corresponds to a measurement of $\hat{Z}$, while the one for modes corresponds to the measurement of $\hat{n}$.

[276], with many applications to constrained optimization problems such as a maximum independent set [277]. Here, we choose the initial state $|\phi\rangle$ to be a simple-to-prepare state that obeys Gauss's law, specifically, a product state of Fock states with unit occupation.

Because we aim to provide scalable methods adapted for large system sizes, global optimizers such as DIRECT, which were used in previous studies of lattice gauge theories [40], are not well suited here. We compare optimizers (with Python SciPy names COBYLA, Nelder-Mead, CG, trust-constr, BFGS, L-BFGS-B, and SLSQP) and find that, among the optimizers that were tested, SLSQP (or sequential least squares programming) has the fastest convergence rates for the $\mathbb{Z}_2$ model without qubit decoherence or shot noise, measuring the energy by calculating the expectation value of the Hamiltonian matrix directly with the circuit state vector without shot noise.

We find that the VQE circuit (without noise and therefore no sampling error) can estimate ground-state energy with relative error with respect to the true ground-state energy of around $10^{-3}$ for two, three, and four sites using one, four, and seven layers, respectively, as can be seen in Fig. 22(b). The VQE requires, on average, around $10^2$, $10^3$, and $10^4$ circuit evaluations for two, three, and four sites, respectively, an increase that is due to the commensurate increase in the number of Fock states required to represent a larger Hilbert space for unit filling. Increasing the number of layers increases the total number of variational parameters, which can lead to a higher final fidelity; however, this result presents a trade-off—in the presence of noise, reliable results can be extracted only from shallow depth circuits. Thus, the overall fidelity will decrease when a large number of layers is used [278].

Furthermore, we find that the higher the boson filling, the larger the number of layers required, with up to four layers required for four bosons in two sites to reach a relative error below $10^{-3}$. This finding may be related to the increased size of the Hilbert space, i.e., due to the increase in possible combinations for distributing the bosons among a set number of sites.

To measure the expectation value of the Hamiltonian as required for the VQE optimization procedure, we discuss how to measure each term of the Hamiltonian individually [269]. The field term $\langle -g\sum_{i=1}^{L-1}\hat{X}_{i,i+1}\rangle$ is a simple sum of $X$-basis qubit measurements. The on-site interaction term on a single site $i$ can be measured by reading out the photon number of the cavity using the binary search readout [144,211]. This process enables the single-shot readout of $\hat{n}_i$, which can then be used to reconstruct $\langle n_i^2\rangle$. For the gauge-invariant hopping term $-J\sum_{i=1}^{L-1}\langle\hat{Z}_{i,i+1}(\hat{a}_i^\dagger\hat{a}_{i+1}+\mathrm{H.c.})\rangle$, we are faced with the difficulty that we cannot directly read out the beam-splitter term by measuring the cavities in the number basis. Instead, we note the following identity:

$$\begin{aligned}\langle\psi|\hat{a}_i^\dagger\hat{a}_{i+1}+\mathrm{H.c.}|\psi\rangle &= \langle\tilde{\psi}|\mathrm{BS}_{i,i+1}\left(\frac{\pi}{2},\frac{\pi}{4}\right)(\hat{a}_i^\dagger\hat{a}_{i+1}+\mathrm{H.c.})\mathrm{BS}_{i,i+1}^\dagger\left(\frac{\pi}{2},\frac{\pi}{4}\right)|\tilde{\psi}\rangle\\ &= \langle\tilde{\psi}|(\hat{a}_i^\dagger\hat{a}_i-\hat{a}_{i+1}^\dagger\hat{a}_{i+1})|\tilde{\psi}\rangle\\ &= \langle\tilde{\psi}|\hat{n}_i|\tilde{\psi}\rangle-\langle\tilde{\psi}|\hat{n}_{i+1}|\tilde{\psi}\rangle, \qquad (161)\end{aligned}$$

where we have inserted $\mathbb{1}=\mathrm{BS}_{i,i+1}[(\pi/2),-(\pi/4)]\mathrm{BS}_{i,i+1}[(\pi/2),(\pi/4)]$ on either side of the hopping and have cast the expression in terms of the transformed state $|\tilde{\psi}\rangle=\mathrm{BS}_{i,i+1}[(\pi/2),(\pi/4)]|\psi\rangle$. Thus, much like a Hadamard gate facilitates an $X$-basis measurement for qubits, we can measure the expectation value of a bosonic hopping term by "rotating" the modes into the Fock-state basis via a 50:50 beam splitter, followed by a Fock-basis measurement. By correlating this result with that of the gauge-field link $Z$-basis measurement result, one can reconstruct the expectation value of the gauge-invariant hopping.

Only commuting terms can be measured simultaneously; for the $\mathbb{Z}_2$ model, we therefore perform three separate measurement circuits to estimate the energy. In the first circuit in Fig. 22(c), the on-site interaction is measured on all sites simultaneously. In the same circuit, it is possible to measure the field terms, though this step must precede the measurement of the on-site interaction such that the gauge link qubits can then be reset and used for binary readout. Next, we measure half of the gauge-invariant hopping terms, preceded by a measurement of the electric field ($\hat{X}_i$) at every other gauge link. Finally, the other half of the gauge-invariant hopping terms are measured, along with the other half of the field sites.

In summary, in this section we showed that a VQE ansatz for the $\mathbb{Z}_2$-Higgs model that is specifically tailored to the types of operations available in hybrid qubit-boson hardware indeed only requires very few layers to determine the ground state for the small system sizes we can access numerically for proof-of-principle demonstrations. The implementation of this ansatz directly follows from the methods presented in the first part of this paper (Secs. II–IV), which implies that the hybrid qubit-boson hardware will require shallower circuits than qubit hardware, as discussed in Sec. V E. However, more work needs to be done beyond these proof-of-principle results to determine the circuit depths and the number of shots that will be required as a function of system size.

### 2. $U(1)$ quantum link model

In the following, we describe how to carry out a hybrid qubit-boson VQE of the $U(1)$ quantum link model in Eq. (151) in $(1+1)$D. Furthermore, we demonstrate how to detect the first-order phase transition at $\tau=\pi$ in our proposed architecture. While the correspondence between the model in Eq. (151) and quantum electrodynamics in $(1+1)$D is only applicable for an even number of lattice sites due to the staggering of the fermions, we also study an odd number of sites here for benchmarking purposes.

Following the same scheme as for the $\mathbb{Z}_2$-Higgs model, we choose resource unitaries that are generated by the individual terms of the Hamiltonian in Eq. (151). Furthermore, we leverage the Schwinger-boson mapping of the gauge fields as described in Sec. I E, and we map the fermions to qubits using the Jordan-Wigner encoding [which is particularly simple in $(1+1)$D]. With these choices, we construct our VQE ansatz using the following resource unitaries:

$$\hat{U}_{S1}(\theta, i) = e^{i\theta \hat{Z}_i}, \tag{162}$$

$$\hat{U}_{S2}(\theta, i) = e^{-i\theta(\hat{n}^a_{i,i+1} - \hat{n}^b_{i,i+1})}, \tag{163}$$

$$\hat{U}_{S3}(\theta, i) = e^{-i\theta((n^a_{i,i+1})^2 + (n^b_{i,i+1})^2)}, \tag{164}$$

$$\hat{U}_{S4}(\theta, i) = e^{-i\theta(\hat{\sigma}^+_{i+1}\hat{a}^\dagger_{i,i+1}\hat{b}_{i,i+1}\hat{\sigma}^-_i + \text{H.c.})}, \tag{165}$$

$$\hat{U}_{S5}(\theta, i) = e^{-\theta(\hat{\sigma}^+_{i+1}\hat{a}^\dagger_{i,i+1}\hat{b}_{i,i+1}\hat{\sigma}^-_i - \text{H.c.})}. \tag{166}$$

Each of these unitaries can be implemented using techniques described in prior sections. In particular, $\hat{U}_{S1}$ and $\hat{U}_{S2}$ are native gates—see Table II. As explained in Sec. III A 3, $\hat{U}_{S3}$ can be synthesized using a SNAP gate. Finally, the gauge-invariant hopping unitaries $\hat{U}_{S4}$ and $\hat{U}_{S5}$ can be implemented with the strategies described in Sec. IV E 2.

Our ansatz for the wave function is

$$\begin{aligned} |\psi_M(\boldsymbol{\theta})\rangle = \prod_{l=1}^{M}\Bigg[&\prod_{i\in\text{even}}^{N} \hat{U}_{S5}(\alpha_l, i)\hat{U}_{S4}(\beta_l, i) \\ &\times \prod_{i\in\text{odd}}^{N} \hat{U}_{S5}(\alpha_l, i)\hat{U}_{S4}(\beta_l, i) \\ &\times \prod_{i=1}^{N} \hat{U}_{S3}(\gamma_l, i) \prod_{i=1}^{N} \hat{U}_{S2}(\gamma_l, i) \\ &\times \prod_{i=1}^{N+1} \hat{U}_{S1}(\delta_l, i)\Bigg]|\phi\rangle, \end{aligned} \tag{167}$$

for $N+1$ matter sites. Note that, contrary to the ansatz for the $\mathbb{Z}_2$-Higgs model, each site has the same parameter such that there are only $4M$ total parameters.

The expectation value of the Hamiltonian can be measured using correlated measurements. To evaluate the gauge-invariant hopping term, we first use $\hat{\sigma}^+ = \frac{1}{2}(\hat{X} + i\hat{Y})$ to split it into

$$\begin{aligned} &\hat{\sigma}^+_{i+1}\hat{a}^\dagger_{i,i+1}\hat{b}_{i,i+1}\hat{\sigma}^-_i + \text{H.c.} \\ &\quad = \frac{1}{4}(\hat{X}_{i+1}(\hat{a}^\dagger_{i,i+1}\hat{b}_{i,i+1} + \text{H.c.})\hat{X}_i \\ &\qquad + \hat{Y}_{i+1}(\hat{a}^\dagger_{i,i+1}\hat{b}_{i,i+1} + \text{H.c.})\hat{Y}_i \\ &\qquad + \hat{X}_{i+1}(-i\hat{a}^\dagger_{i,i+1}\hat{b}_{i,i+1} + \text{H.c.})\hat{Y}_i \\ &\qquad + \hat{Y}_{i+1}(i\hat{a}^\dagger_{i,i+1}\hat{b}_{i,i+1} + \text{H.c.})\hat{X}_i). \end{aligned} \tag{168}$$

Using our method in Eq. (161), we can map a measurement of the bosonic terms in the above equation to a measurement of the number densities of the two modes. Correlating those number densities with the appropriate qubit measurements, we can therefore measure all four terms in the above expansion. Because we require two circuits to measure all possible lattice sites $i$, we require eight separate measurement circuits for the gauge-invariant hopping. All the other terms are straightforwardly accessible by measuring the values of the boson numbers in the modes [209,211] and the state of the qubits.

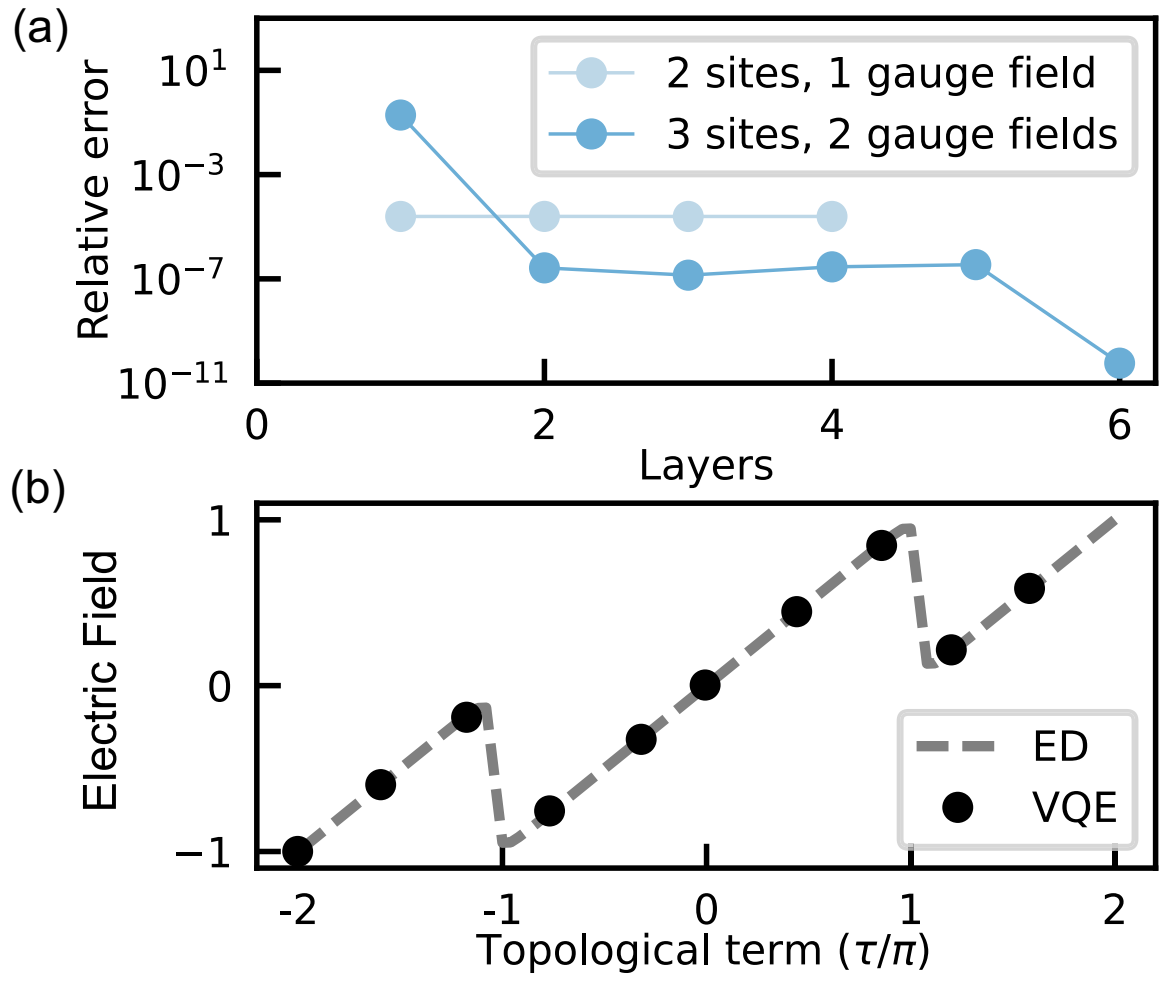

FIG. 23. Numerical VQE results for the $U(1)$ quantum link model in $(1+1)$D. We neglect shot noise and decoherence. (a) Relative error between the estimated ground-state energy obtained via simulating the VQE using Bosonic Qiskit [77] and the true ground-state energy VQE results obtained for two and three sites. (b) VQE compared to exact diagonalization results. The first-order phase transition can be seen at $\tau = \pm\pi$ in the expectation value of the electric field $\frac{1}{2}\sum_{i=1}^{2}\langle\hat{S}^z_{i,i+1}\rangle = \frac{1}{2}\sum_{i=1}^{2}(\langle\hat{n}^a_{i,i+1}\rangle - \langle\hat{n}^b_{i,i+1}\rangle)/2$ in the $(1+1)$D $U(1)$ quantum link model for a three-site system and four layers. The spin length is chosen as $S=1$. $m/J = 1, g/J = 10$.

Using Bosonic Qiskit [77], we implement this ansatz and test the convergence of the VQE with respect to the number of layers for $S=1$ and show the results in Fig. 23. For a single gauge-field site and two matter sites, we initialize the system in the "vacuum state" $|\downarrow 0 \uparrow\rangle$, where the middle label indicates the eigenstate of $S^z$, and we assume perfect energy measurements, i.e., no shot noise and decoherence. We find that one layer is sufficient for obtaining a relative energy error of $10^{-5}$. This finding is related to the fact that the ground state for this system is given by $(1/\sqrt{2})(|\uparrow 1 \downarrow\rangle + |\downarrow 0 \uparrow\rangle)$. For three sites, initialized in $|\downarrow 0 \uparrow 0 \downarrow\rangle$, we find from our numerical evaluations that two layers are required to attain low relative errors of $10^{-7}$, with only a small increase in the total number of iterations from the two-site case.

In the regime of large $g$, the $U(1)$ quantum link model features a first-order phase transition [171,172] at half-integer values of $\tau/2\pi$. This feature can be seen trivially by taking $g \to \infty$, in which case the electric term dominates the Hamiltonian, pinning $\hat{S}^z$ at the integer closest to $\tau/2\pi$. This phase persists for finite $g$ until the mass reaches $M/g \approx 1/3$ [see definitions of $M$ and $g$ in the Hamiltonian in Eq. (151)]. The energy gap between the ground and first excited states is generally exponentially small in system size at first-order phase transitions. Therefore, we expect that this transition will be relatively sharp and visible even for small systems of two or three sites. Hence, it is a good proof-of-principle target for small-scale experiments.

For various values of $\tau$ used to find ground states with the VQE ansatz discussed above [Eq. (162)–(166)], we measure the expectation value of the electric field (in the Schwinger-boson representation) $\langle S^z_{i,i+1}\rangle = \frac{1}{2}\langle \hat{n}^a_{i,i+1} - \hat{n}^b_{i,i+1}\rangle$. In Fig. 23, we indeed find sharp drops corresponding to the first-order phase transition at $\tau = \pm\pi$. The Bosonic Qiskit VQE results match our exact diagonalization results to relative errors of around $10^{-7}$.

Our results show that our VQE approach using hybrid qubit-boson hardware can prepare ground states of the $U(1)$ quantum link model in $(1+1)$D for two- and three-site systems. Although in $(1+1)$D the $U(1)$ quantum link model can be probed with qubit-only hardware by integrating out the gauge fields, it would not generalize to higher dimensions. In that case, each gauge-field link would be encoded in many qubits, and Hamiltonian operations would be highly costly as discussed in Sec. V E. Another possible advantage of our approach is the usage of the Schwinger-boson constraint $\hat{n}_a + \hat{n}_b = 2S$ as a (single) photon loss error detection scheme. This method can be implemented by measuring the joint parity of modes $a$ and $b$; to what extent this improves results from noisy hardware is an interesting direction for future exploration.

### D. VQE with hardware and shot noise

Next, we test the resilience of our methods to decoherence and shot noise in the energy measurements. For this purpose, we simulate the full measurement schemes that would be employed in the experiment and also include qubit decay and dephasing, which are the leading decoherence mechanisms [279] for the small system sizes and unit filling we consider here (i.e., we neglect photon loss). For systems where large photon numbers are possible, mode decay will become appreciable. The model studied in this section is the $\mathbb{Z}_2$-Higgs model, for which we defined the ansatz in Sec. VI C 1. We choose $U/J = g/J = 1$, which we expect to be among the most difficult regimes to simulate due to the balanced competition between all energy terms. We consider three matter sites and two gauge links at unit filling, requiring three modes and two qubits.

We start by explaining how to measure the expectation value of the Hamiltonian in experiment. Next, we benchmark the simultaneous perturbation stochastic approximation (SPSA) optimizer [280,281] (as opposed to SLSQP, used so far) without shot noise, as this is the optimizer of choice for optimizations with shot noise. We investigate how to use averaging over previous iterations to improve convergence of SPSA (still without any noise). Then, we run our improved optimization procedure with shot noise. Upon finding that there is a certain independence to the number of shots in the initial stages of optimization where the gradients are large, we further adapt our procedure to incrementally increase the number of shots during the optimization. Next, we include hardware noise (qubit $T_1$ decay and additional pure $T_\phi$ dephasing) into the numerical simulation. Finally, we error mitigate our results by implementing a procedure to postselect on preservation of Gauss's law.

#### *1. Introducing shot noise in the energy measurement*

We simulate a binary search routine [144,211] for measuring the cavity in the Fock basis. This process naively requires an additional readout qubit per cavity for the measurement (i.e., separate from the transmon qubits that encode the gauge-field links). However, for the $\mathbb{Z}_2$-Higgs model, we can remove this requirement by first measuring the gauge-field links and then reusing the qubits as ancillas for mode readout, cf. Fig. 22(c). For this purpose, we measure and reset the field qubits prior to the cavity measurement. In our VQE simulations, we simulate the full binary search, which enables us to simulate the effective shot noise on the mode measurement that results from the qubit measurements. Because of the shot noise, each experiment requires $\mathcal{O}(1/\epsilon^2)$ circuit executions to estimate $\langle \hat{H}_{\mathbb{Z}(2)}\rangle$ to additive precision $\epsilon$.

#### *2. Using averaging in the stochastic VQE optimization*

In the presence of shot-noise error, VQE optimization is far more challenging [272,278]. Thus, we use the SPSA, which is particularly well suited for the VQE because it only evaluates the Hamiltonian twice per iteration regardless of the dimension of the optimization problem.

In Fig. 24, we show the energy found in each iteration using the SPSA optimizer. The gray dots indicate an optimization without any shot noise, and we see that it does not converge within 7000 iterations.

It has been found that averaging variable parameters can increase the performance of stochastic optimization [282,283]. In our case, we found that the convergence of the SPSA optimizer, which is stochastic, can be improved by periodically choosing parameters for the next iteration that are averages of previous iterations. Therefore, we adapt our implementation of the SPSA optimizer in order to perform a parameter averaging step throughout the optimization: Every $n$ iterations, we use the average of the variable parameters of the last $m$ iterations. It is important not to choose $n$ so small that the averaging is too frequent, resulting in a small sample size, or to choose $m$ too large such that $m \gtrsim n$, which may constrain the optimization too much and lead to trapping in local minima. In Fig. 24, we show data corresponding to $n = 15$ and $m = 7$ in dark blue.

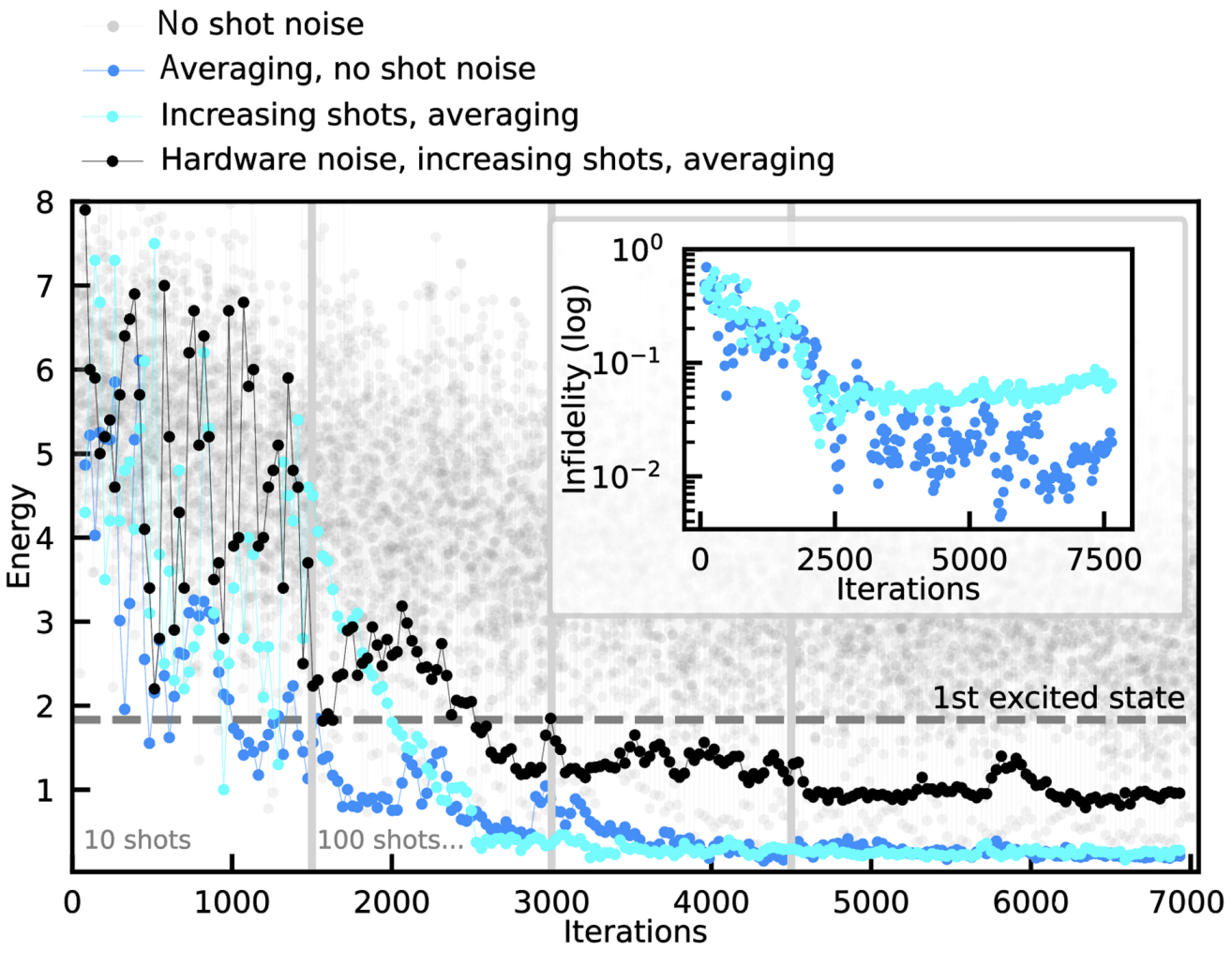

FIG. 24. Numerical VQE results for the $\mathbb{Z}_2$-Higgs model including qubit decay and shot noise. Here, $U/J = g/J = 1$. Exact ground-state energy is $E/J = 0.02$. Energy is plotted in units of $J$, and we show three bosonic sites and two gauge fields, four layers in the ansatz, and optimization using SPSA. For the data without shot noise (gray dots), all iterations are plotted. For the rest of the data, only the iterations with averaged parameters are plotted. When averaging but without including shot noise (dark blue dots), we average the input variable parameters every 15 iterations, averaging over the input variable parameters found in the previous seven iterations. For the light blue data, we include shot noise, increasing the number of shots after every 1500 iterations as indicated by the gray vertical lines. Black dots indicate the same procedure as the light blue data, but with hardware noise, i.e., qubit decay and dephasing. Inset: fidelity with the ground state as a function of the iteration number. Colors match the ones in the main panel.

We find notable improvements using this strategy over the one without averaging, i.e., $n = 1$ and $m = 1$.

#### *3. Adaptive shot budget in the VQE optimization*

We now include shot noise in the estimation of the energy. The number of shots required for good convergence of the VQE is determined by the size of the energy gradient. Generally, gradients are larger in the first iterations and gradually decrease as the minimum is approached, which is evident in the large fluctuation between iterations for the early iterations of less than 2000 in Fig. 24 (black dots). Therefore, it is natural to increase the number of shots used in the energy measurement of later iterations [284–286]. Specifically, we increase the shots per iteration by an order of magnitude every 1500 iterations (starting at 10 shots per iteration). We show the resulting energy trajectory as light blue points in Fig. 24. We find that (for this specific system size) this method yields similar performance to the scheme without shot noise (dark blue). In general, we expect the total number of shots necessary to prepare the ground state with fixed fidelity to increase exponentially with the system size; see, e.g., Ref. [287] for empirical evidence.

#### *4. VQE results including qubit decoherence, averaging, and adaptive shot budget*

Finally, we add the dominant source of experimental noise in the hardware—qubit decay and dephasing—to our simulations in addition to the shot-based measurement of the energy discussed above. This process requires specifying the duration of each of the gates in the VQE ansatz. The ansatz is explained in Sec. VI C 1—see Fig. 22(a) for a circuit diagram. Below Eqs. (156)–(159), we explain how every resource unitary can be realized using native gates. Altogether, the depth of one ansatz layer consists of four conditional beam splitters and a SNAP gate. We assume a value of $\chi = 2\pi \times 0.5$ MHz for the dispersive coupling. In this case, the maximal duration of a conditional beam splitter is approximately 2.25 μs, consisting of two conditional parities that take $T = \pi/\chi = 1$ μs each, and one beam splitter of 250 ns duration (100 ns has been shown to be feasible [91,217]). We do not include duration cancellations, a range of which is possible. For example, two successive conditional beam splitters acting on the same oscillator modes and qubits can cancel each others' conditional-parity operations, leading to a total time of 2.25 μs. We also note that, in principle, the duration depends on the gate parameters, which in experiment could lead to a shorter duration than estimated here. The SNAP gate takes $T = \pi/\chi = 1$ μs [200]. The total duration of one VQE layer is roughly $(4 \times 2.25 + 1) = 10$ μs. Four layers ($T = 40$ μs) are therefore well within the decay time of the qubit ($T_1 = 200$μs) and the mode ($T_1 \approx 1$ ms). Therefore, the probability for the $|1\rangle$ state of a mode to decay is 4%, and for a qubit to decay during the four-layer ansatz circuit, it is 20%. While in most regions of the phase diagram, the occupation of the higher-lying mode states is small—and therefore the effect of mode decay (and other nonidealities such as self-Kerr [192]) is minimal—this is not the case in the superfluid and clump phase when $U/J$ is small. In these cases, mode decay could contribute significantly. We leave the study of this effect for future work and neglect photon loss throughout.

We show the optimization including averaging, adaptive shot budget, and qubit decay in black dots in Fig. 24. Comparing this result to the optimization without qubit decoherence in light blue, we find that qubit decoherence leads to a significant energy error of about $0.86J$. However, we still find that the energy is well below the first excited state. Therefore, we expect a reasonable overlap with the ground state. In order to test this expectation, we plot the fidelity with the ground state in the inset of Fig. 24, using colors matching the main figure. We see that, for the simulations excluding decay and dephasing errors (light and dark blue), the fidelity remains high, following the same trend as the energy.

We also note that the total runtime of the optimization in the presence of noise is small: We find an energy error below the energy of the first excited state after around 3000 iterations with a total shot count of less than $2 \times 10^5$, yielding a total quantum runtime of less than 10 s, if this experiment were to be run on current superconducting hardware that is limited only by gate times.

### 5. Postselection on Gauss's law

The conservation laws obeyed by lattice gauge theories, such as Gauss's law and particle number conservation, can be used to discard unphysical shots in the presence of hardware noise, enabling partial error detection.

In order to use Gauss's law for error mitigation, we must measure the operator $\hat{G}_i$ [see Eq. (2)] for each shot of the experiment. For the $\mathbb{Z}_2$-Higgs model in $(1+1)$D, Gauss's law reads

$$\hat{G}_i = \hat{X}_{i-1,i} e^{i\pi \hat{n}_i} \hat{X}_{i,i+1} = 1, \tag{169}$$

for all $i$. Because Gauss's law commutes with all terms in the Hamiltonian, it is possible to measure Gauss's law for all $i$ as well as the energy of the state in the same shot. This process requires midcircuit measurements and is most often done using the Hadamard test [288–291]. However, the Hadamard test involves ancillary qubits, which we do not have at our disposal. Therefore, we present a method below that avoids ancillary qubits. This method requires additional gates and measurements, compared to the measurement circuits presented in Fig. 22(c). We also discuss that, even for the measurement circuits as described in Fig. 22 (i.e., without additional gates and measurements), Gauss's law can be *partially* checked.

The first scheme is inspired by the schemes presented in Ref. [292] for measuring toric-code stabilizers. The key observation is that the individual parts of the Hamiltonian commute with Gauss's law; therefore, we can check Gauss's law for each measurement individually. For the measurement of the gauge-invariant hopping, replacing the beam splitters in Eq. (161) with conditional beam splitters, a measurement of the number operator of the modes directly yields a measurement of the gauge-invariant hopping. More explicitly,

$$\mathrm{CBS}_{i,i+1}\left(\frac{\pi}{2}, -\frac{\pi}{4}\right)(\hat{n}_i - \hat{n}_{i+1})\mathrm{CBS}_{i,i+1}\left(\frac{\pi}{2}, \frac{\pi}{4}\right) = \hat{Z}_{i,i+1}(\hat{a}_i^\dagger \hat{a}_{i+1} + \mathrm{H.c.}). \tag{170}$$

Hence, conjugating a number measurement of modes $i$ and $i+1$ with conditional beam splitters yields a measurement of the gauge-invariant hopping without having to measure the qubit, as shown by the green part of the circuit in Fig. 25(a). Crucially, one can check that the operators effectively measured by $\hat{n}_i$ and $\hat{n}_{i+1}$ individually also commute with Gauss's law. Because $\hat{X}_{i-1,i}$ commutes with both the gauge-invariant hopping and Gauss's law, it can be measured simultaneously. Gauss's law is then measured by the circuit shown in black and white in Fig. 25(a). It relies on the fact that conjugation of a qubit $\hat{Z}_{i,i+1}$ operator with SQR gates yields $\hat{Z}_{i,i+1} e^{i\pi \hat{n}_i}$, cf. Eq. (29). Similarly, conjugation with Hadamard gates and CNOTs converts $\hat{Z}_{i,i+1}$ into $\hat{X}_{i-1,i}\hat{X}_{i,i+1}$. Taken together, this process converts a measurement of $\hat{Z}_{i,i+1}$ to a measurement of Gauss's law operator $\hat{G}_i$. Hence, these two measurements allow for a full check of Gauss's law while measuring the value of the gauge-invariant hopping in the same shot. This scheme requires additional gates.

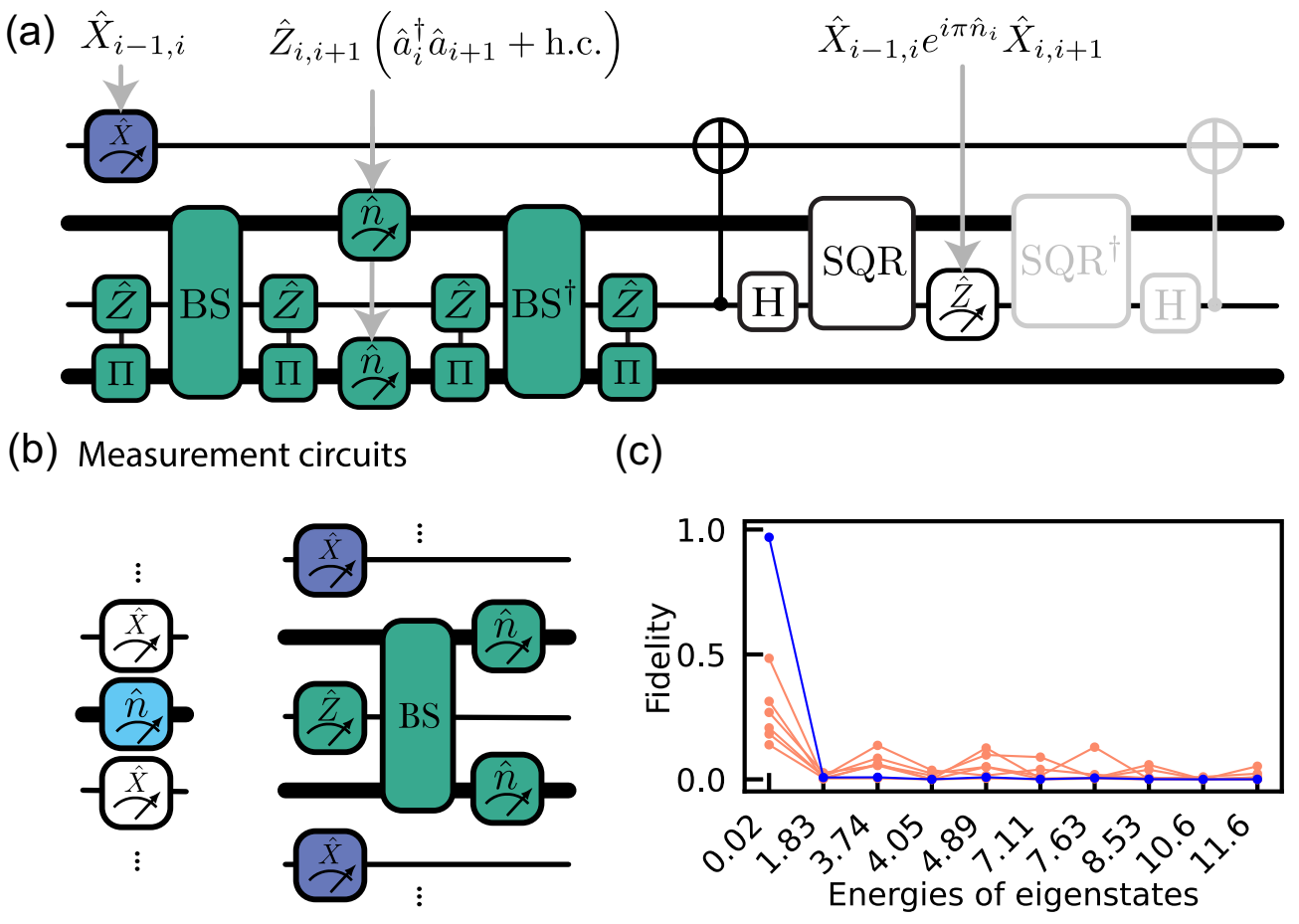

FIG. 25. Ancilla-free postselection methods based on Gauss's law. (a) Measurement circuit of the gauge-invariant hopping and electric-field term [see Fig. 22(c)]. It is adapted here in a way that enables a measurement of Gauss's law at every site, black and white, in the same shot. The gates in gray are only required if more operations are to be carried out on the circuit. Because the measurement of Gauss's law commutes with the one of the gauge-invariant hopping, they can be performed in either order. (b) Measurement circuit of the density in light blue, required for calculating the value of the on-site term. It commutes with $\hat{G}_i$, and we can therefore perform local checks of $\hat{G}_i$ using the measurement results, discarding shots that break Gauss's law. The measurement of the gauge-invariant hopping [see Fig. 22(c)] does not commute with $\hat{G}_i$, and we therefore cannot conduct Gauss's law checks for each site. However, the product $\hat{G}_i\hat{G}_{i+1}$ commutes with these circuits, allowing for postselection upon Gauss's law checks for pairs of sites. (c) Fidelity $|\langle\psi|n\rangle|^2$ of single quantum trajectories of the noisy simulation with the eigenstate $|n\rangle$ in our chosen Gauss's law sector as a function of the eigenenergy. Trajectories that do not have overlap with states outside of the Gauss law sector are shown in blue; others are shown in red. Note that the fidelities of the states shown in red do not sum up to unity because we do not show the overlap with the states outside of our Gauss law sector. Energies are in units of $J$.

In the second scheme, we do not require additional gates and instead propose a form of "partial" error mitigation based on the information we already have access to in the context of the VQE: the measurements of the Hamiltonian terms shown in Fig. 22(c). This scheme does not enable a local check of Gauss's law for every matter site (which is why we call it "partial") and hence only slightly improves the results. These three measurement circuits do not commute; therefore, only one can be used per shot of the experiment. The measurement circuit of the number density [shown in Fig. 22(c)] carries out a photon number measurement on each mode and could thus be combined with the measurement of $\hat{X}$ on each qubit to check Gauss's law at each site [see Fig. 25(a)]. However, the other measurement circuits [see Fig. 22(c)] are not compatible with a measurement of Gauss's law at each site, as some qubits are individually measured along $\hat{Z}$ (which alone does not commute with $\hat{G}_i$ for an overlapping region) and the mode occupations are not resolved for each site. However, in these circuits, we measure $\hat{X}$ for every other gauge link and can furthermore resolve the total photon number in the intermediary pairs of modes. Thus, we can conduct checks of Gauss's law for blocked pairs of sites:

$$\hat{G}_i\hat{G}_{i+1} = \hat{X}_{i-1,i}e^{i\pi(\hat{n}_i+\hat{n}_{i+1})}\hat{X}_{i+1,i+2}. \tag{171}$$

This possibility is illustrated in Fig. 25(b). Thus, we can postselect on valid Gauss's law preserving states for all three measurement circuits needed for estimating the energy. This scheme captures only single Gauss's law violations, not errors on two consecutive sites.

We simulate the full Gauss's law postselection (the first scheme) and focus on its influence on the energy measurement. For this purpose, we take the final VQE parameters found in a noisy simulation (i.e., the final iteration of the black dots in Fig. 24) and run a simulation including hardware noise in the VQE circuit, but we neglect noise in the measurement circuit. We then check whether an error happened in every quantum trajectory of the noisy simulation. The error-free trajectories exhibit high overlap of approximately 96% with the ground state [cf. blue line in Fig. 25(c)]. For trajectories with an error, we project onto the correct Gauss's law sector. The norm of the part of the wave function we project out gives us an estimate for the number of shots that would need to be discarded. We find that 13% of shots need to be discarded. We show in red the overlap with the eigenstates in the correct Gauss's law sector for some such states in Fig. 25(c). Those states still have large overlap with the ground state but also some overlap with excited states. To calculate the average postselected energy, we renormalize the projected states and average over all trajectories. We find a postselected energy of $0.36J$, compared to a result of $0.88J$ without postselection and an exact ground-state energy of $E/J = 0.02$. This finding corresponds to approximately 20% of the difference between the ground-state and first excited-state energy.

In our simulations, we find that 84% of quantum trajectories have no error. This finding is roughly in agreement with the dephasing survival probability of a single qubit, where $T_2 = 200$ μs, only one of the qubits starts in $|+\rangle$, the other starts in $|-\rangle$, and the total circuit duration is $T = 40$ μs, which is $\exp(-T/T_2) \approx 82\%$. The trajectories with no error have energy $0.16J$. The $1 - 84\% = 16\%$ of trajectories in which an error happened therefore have an average energy of $4.66J$. Gauss's law postselection reduces this average energy of the noisy trajectories to $1.41J$, where the remaining energy error results from noise processes within the same Gauss's law sector. We therefore find that removing errors that bring us out of the correct Gauss's law sector drastically reduces the energy error introduced by the noise.

More intricate mitigation methods can be engineered from this Gauss's law check. For example, we note that one can cross-check $\hat{G}_i$ at different sites to not only detect that an error occurred but also resolve its position. This method opens up the possibility to not only postselect on the case of no errors but develop additional novel error-mitigation techniques where the error can be partially corrected. In addition, if the error is spatially disconnected in such a way that one can ascertain that it spread to a region of interest, then it may be desirable to retain this shot for the computation of observables local to this region. In addition to postselection using gauge symmetries [293], VQE implementation for the $\mathbb{Z}_2$-Higgs model proposed here is amenable to other commonly used noise mitigation techniques such as zero-noise extrapolation [294,295] and probabilistic error cancellation [294]. An alternative to postselection is to introduce an additional term in the Hamiltonian that enforces the symmetry constraint by prethermalization [296], or pseudogenerators for gauge protection [297,298].

### E. Measuring observables

Having characterized the hybrid qubit-boson VQE state preparation in terms of the energy error and overlap with the true ground state, we now show how to extract relevant observables in the proposed cQED architecture. While we benchmark the measurement of the observables for the ground state, the measurement schemes can, in principle, be applied to any state.

#### *1. Detecting phases of the $\mathbb{Z}_2$-Higgs model in VQE-prepared ground states*

In Fig. 26, we calculate different gauge-invariant observables in the $\mathbb{Z}_2$-Higgs model for varying on-site interactions, with respect to VQE ground states of a system with four bosonic sites and three gauge-field sites found using noiseless simulations employing the SLSQP optimizer. We compare our findings to exact diagonalization results. We find that all VQE observables agree very well with the exact results, as expected from the low relative energy error found in Fig. 22 for this small system size.

Importantly, despite the small system size, we can see the hallmarks of three qualitatively different states as $U/J$ is scanned [248]. The three states are indicated by the blue (rightmost), red (middle), and yellow (leftmost) lines, representing a Mott-insulating phase, a pair superfluid phase, and a "clump phase"—a phase in which the gauge field perturbatively generates a nearest-neighbor attraction between the bosons, leading to the bosons "clumping" together on the same site [248]. For four sites, we define the "clump order parameter" as a quantity proportional to the probability of all four bosons occupying the same site (defined in the caption of Fig. 26). This observable indeed becomes large for small $U/J$ [248]. A (Mott) insulating state is characterized by a lack of spreading of particles in the system. We observe signatures of this effect in the string-order correlator and pair hopping, which are both small for large $U/J$. Moreover, on-site fluctuations are also small in the insulating state, as expected. Finally, the parity tends towards $-1$, indicating that, on average, there is an odd occupation at each site. By contrast, in a superfluid phase, particles spread far. In the case of the $\mathbb{Z}_2$-Higgs model, single particles stay confined due to the presence of the gauge field, while two particles can spread [248], which we observe in the large pair hopping and fluctuations. Moreover, contributions from states with a single boson are suppressed, which we see from the parity tending towards $+1$.

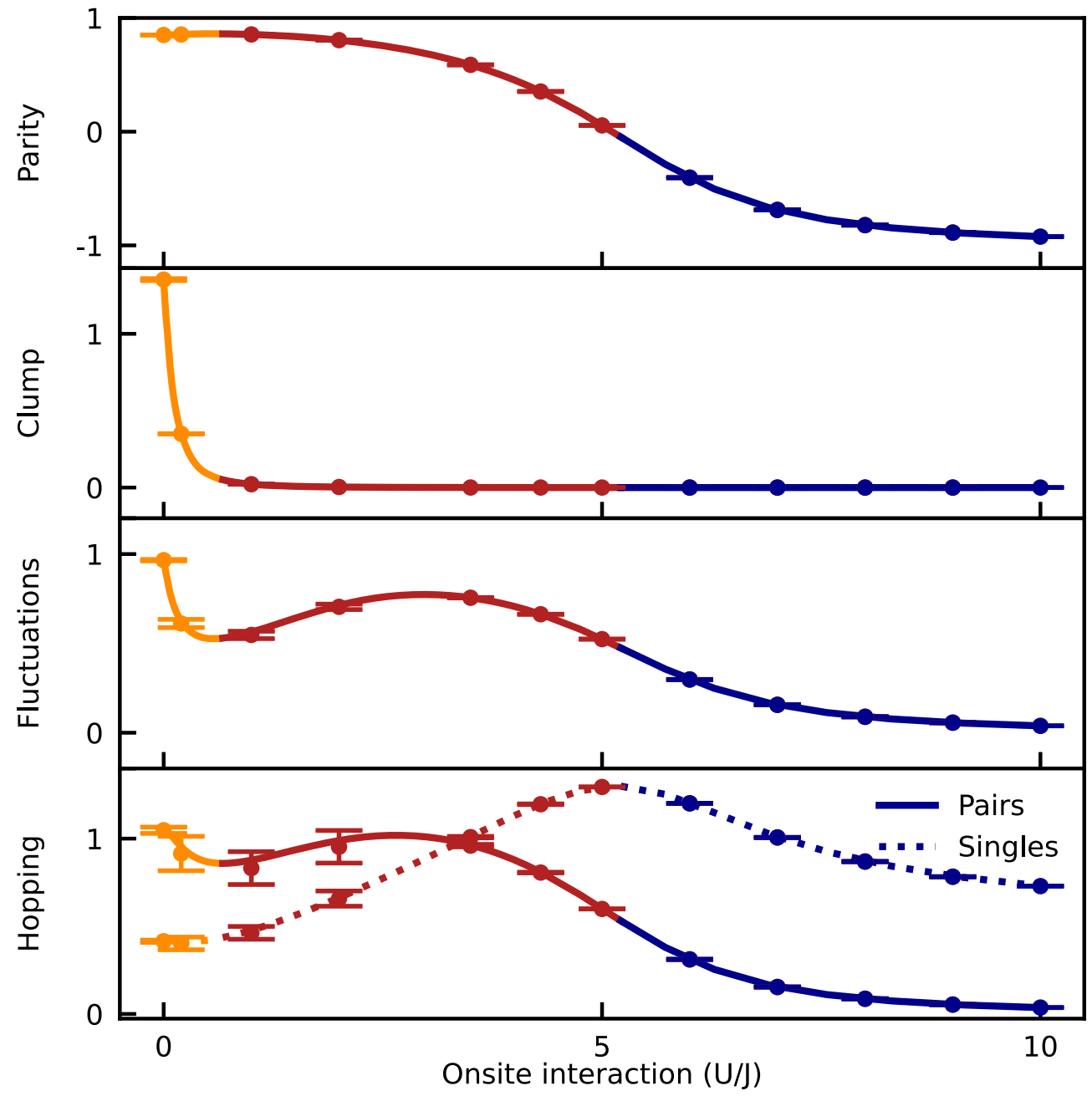

FIG. 26. Correlation functions measured on the VQE ground-state estimate. The observables are defined as follows. Parity: $\sum_i^L \langle e^{i\pi\hat{n}_i}\rangle/L$. Clump: $\sum_i^L \langle \hat{n}_i(\hat{n}_i-1)(\hat{n}_i-2)(\hat{n}_i-3)\rangle/L$. Fluctuations: $\sum_i^L(\langle\hat{n}_i^2\rangle - \langle\hat{n}_i\rangle^2)/L$. Pair hopping: $\sum_i^L\langle\hat{a}_0^{\dagger 2}\hat{a}_i^2 + \text{H.c.}\rangle/L$. String-order correlator (also known as gauge-invariant string-order correlator): $\sum_i^L\langle\hat{a}_0^\dagger\hat{Z}_0\ldots\hat{Z}_i\hat{a}_i + \text{H.c.}\rangle/L$. There are four bosonic sites and three gauge-field sites. Unit filling, i.e., four bosons are in the system, and $g/J = 5$. Error bars indicate standard deviation of five VQE simulations with four layers.

Having shown that signatures of different phases can be found from VQE simulations, we now discuss how the different observables can be measured in our cQED architecture.

### 2. Measuring the string-order correlator

The gauge-invariant string-order correlator is large in a deconfined phase and small in a confined phase. It is defined as $\langle\hat{a}_i^\dagger\hat{Z}_{i,i+1}\ldots\hat{Z}_{j-1,j}\hat{a}_j + \text{H.c.}\rangle$. To measure the value of this correlator on the ground states prepared via the VQE, we need to correlate a measurement of the gauge-field qubits with the measurement of the beam-splitter operator of two bosonic sites. We note that one possibility is to adapt the scheme used in Sec. VI D 1 to measure a long-range hopping through the use of bosonic SWAPs. Ordinarily, such a measurement would require a time linear in the distance $|j-i|$. Alternatively, here we propose a separate strategy to measure this nonlocal observable using homodyne measurements: First, we expand $\hat{a}_i = (\hat{x}_i + i\hat{p}_i)/2$, where $\hat{x}$ and $\hat{p}$ are the bosonic position and momentum operators, respectively, such that

$$\begin{aligned}&\langle\hat{a}_i^\dagger\hat{Z}_{i,i+1}\ldots\hat{Z}_{j-1,j}\hat{a}_j + \text{H.c.}\rangle\\ &= \frac{1}{2}(\langle\hat{x}_i\hat{Z}_{i,i+1}\ldots\hat{Z}_{j-1,j}\hat{x}_j\rangle + \langle\hat{p}_i\hat{Z}_{i,i+1}\ldots\hat{Z}_{j-1,j}\hat{p}_j\rangle). \quad (172)\end{aligned}$$

We must therefore perform correlated measurements of $\hat{x}_i$, $\hat{Z}_{i,i+1}$, and $\hat{x}_j$ (and, separately, $\hat{p}_i$, $\hat{Z}_{i,i+1}$, and $\hat{p}_j$). In order to measure $\hat{x}_i$, we use a Hadamard test implemented with a displacement conditioned on an ancilla, defined as $\text{CD}(\alpha) = e^{(\alpha\hat{a}^\dagger - \alpha^*\hat{a})\hat{Z}^{\text{anc}}}$ (see Table II). Choosing an arbitrary real angle $\theta$ such that $\alpha = i\theta$, we obtain $\text{CD}(i\theta) = e^{i\theta\hat{x}\hat{Z}^{\text{anc}}}$. A measurement of the ancilla after applying the Hadamard test circuit with a small angle $\theta$, shown inside Fig. 27, then yields

$$\langle\hat{Z}^{\text{anc}}\rangle_{\alpha=i\theta} = \langle(\sin(2\theta\hat{x})\rangle \quad (173)$$

$$\approx 2\theta\langle\hat{x}\rangle + O(\theta^3). \quad (174)$$

To measure the first part of the string-order correlator, we perform two Hadamard tests simultaneously in order to correlate sites $i$ and $j$; i.e., after applying the Hadamard circuits on both sites $i$ and $j$, we measure

$$\begin{aligned}&\langle\hat{Z}_i^{\text{anc}}\hat{Z}_{i,i+1}\ldots\hat{Z}_{j-1,j}\hat{Z}_j^{\text{anc}}\rangle_{\alpha_i=\alpha_j=i\theta}\\ &\approx 4\theta^2\langle\hat{x}_i\hat{Z}_{i,i+1}\ldots\hat{Z}_{j-1,j}\hat{x}_j\rangle. \quad (175)\end{aligned}$$

Similarly, we measure the second part of the string-order correlator, i.e., those involving $\hat{p}$ correlations, by using conditional displacements with $\alpha = \theta$.

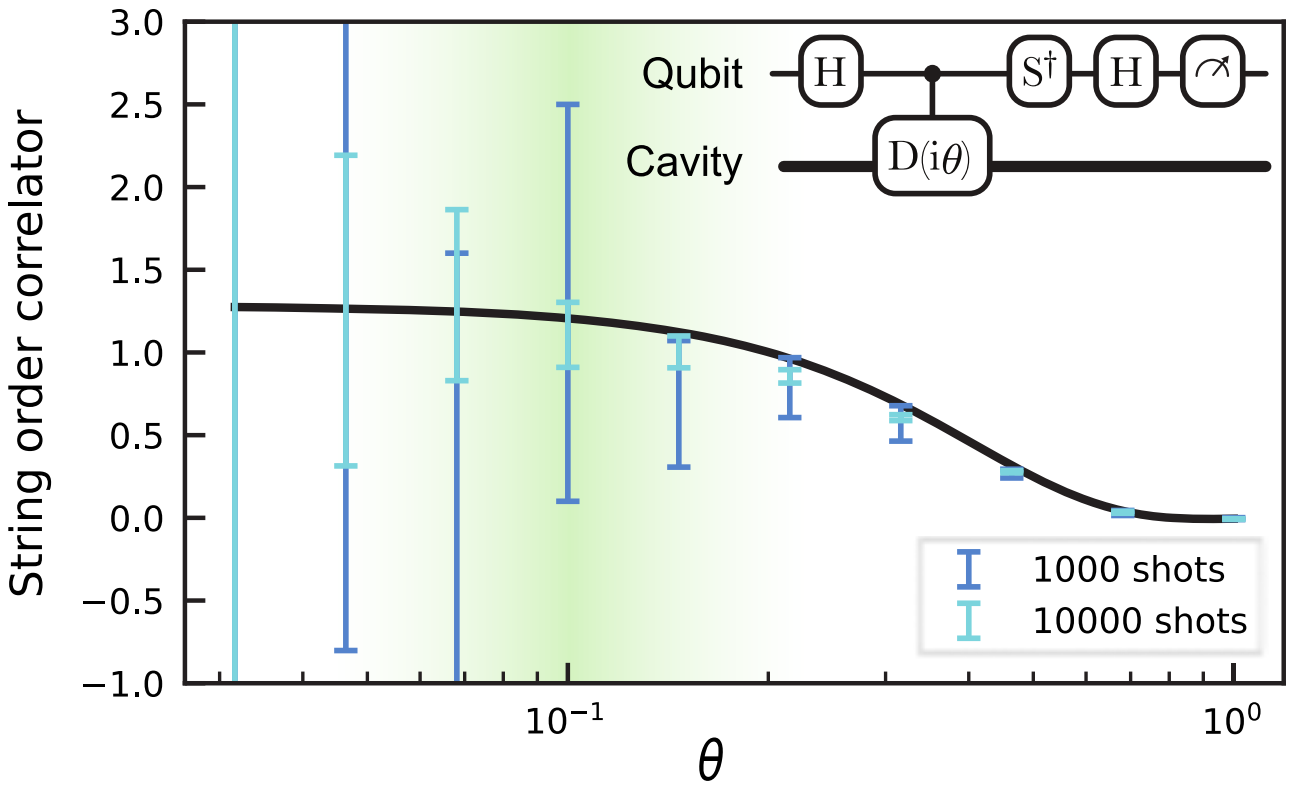

FIG. 27. Measurement of the string-order correlator as a function of the small angle $\theta$ chosen for the conditional displacement. The top-right diagram shows the quantum circuit for measuring $\sin(k\hat{x})$ of an oscillator using the phase kickback of a conditional displacement [211]. In the main plot, we run a full VQE simulation and measure the string-order correlator using the scheme discussed in the main text; i.e., we measure the string-order correlator approximated by the rhs of Eq. (176) by adding two ancilla qubits. The black line is obtained using exact SLSQP optimization, and the error bars include shot noise using SPSA optimization. In both cases, we consider three sites, two gauge-field links, $g/J = 1$, and $U/J = 1$. The green highlighted area indicates the general region that optimizes for accuracy ($\theta$ small) and precision (the smaller $\theta$ is, the more shots are required to precisely measure the result).

In summary, we obtain

$$\begin{aligned}\langle \hat{a}_i^\dagger \hat{Z}_{i,i+1}\ldots\hat{Z}_{j-1,j}\hat{a}_j + \text{H.c.}\rangle &\approx \frac{1}{8\theta^2}\big(\langle \hat{Z}_i^{\text{anc}}\hat{Z}_{i,i+1}\ldots\hat{Z}_{j-1,j}\hat{Z}_j^{\text{anc}}\rangle_{\alpha_i=\alpha_j=i\theta} \\ &\quad + \langle \hat{Z}_i^{\text{anc}}\hat{Z}_{i,i+1}\ldots\hat{Z}_{j-1,j}\hat{Z}_j^{\text{anc}}\rangle_{\alpha_i=\alpha_j=\theta}\big). \end{aligned} \tag{176}$$

To test if we can be in the small $\theta$ regime while still being able to measure the string-order correlator accurately with a reasonable number of shots, we simulate the measurement using Bosonic Qiskit, cf. Fig. 27. We find that a choice of $\theta \approx 0.1$ suffices, requiring approximately $(1/\theta^4) = 10^4$ shots, as expected from the statistical uncertainty of shot noise. Note that, for this measurement, we do not need additional ancillae, as one can recycle gauge link qubits that are not within the support of the string-order correlator for this purpose. Lastly, we remark that one could alternatively use the method of quantum signal processing (see Appendix E) to extract $\langle\hat{x}\rangle$ similarly to Eq. (173) but with a larger value of $\theta$. This result could be achieved by enacting a target function $(2/\pi)\arcsin(\cdot)$ on the signal $\sin(2\theta\hat{x})$; if $2\theta\hat{x} \le 1/2$, this process outputs $(\theta/\pi)\langle\hat{x}\rangle + O(\epsilon)$ while requiring $O(\log(1/\epsilon))$ queries to $\text{CD}(i\theta)$ [299].

### 3. *Measuring the superfluid density*

The superfluid density $\rho_s$ in 1D is given by [300,301]

$$\rho_s = \frac{L}{J}\frac{\partial^2 E_{\text{TPB}}(\phi)}{\partial\phi^2}, \tag{177}$$

where $E_{\text{TPB}}(\phi)$ is the ground-state energy in the presence of "twisted" periodic boundary conditions and $J$ is the hopping matrix element between neighboring sites. The phase $\phi$ is given by the phase difference between two edges; for example, for the $\mathbb{Z}_2$-Higgs model, we have

$$\hat{H}^{\text{TPB}}_{\mathbb{Z}(2)}(\phi) = \hat{H}_{\mathbb{Z}(2)} - J(e^{i\phi}\hat{a}_L^\dagger\hat{Z}_n\hat{a}_1 + e^{-i\phi}\hat{a}_L\hat{Z}_n\hat{a}_1^\dagger), \tag{178}$$

where the entirety of the accumulated phase $\phi$ is incorporated by the hopping term between the two edge sites, where $\hat{H}_{\mathbb{Z}(2)}$ is defined in Eq. (150), $L$ is the number of sites, and $\hat{Z}_n = \prod_{i=1}^{L-1}\hat{Z}_{i,i+1}$ denotes the product of the gauge-field operators on all links of the chain. To measure $\langle\hat{H}^{\text{TPB}}_{\mathbb{Z}(2)}\rangle$, we can follow the general procedure outlined in Sec. VI D 1. The procedure to estimate the stiffness is as follows: We first prepare the ground state for a range of small values of $\phi$. We do so by extending the variational ansatz in Sec. VI C 1 to incorporate edge-to-edge hopping. Once prepared, we estimate Eq. (177) via finite-difference methods or, alternatively, through more sophisticated, noise-resilient techniques such as parameter shift rules [302].

Experimentally, periodic boundary conditions are not required in the hardware to implement edge-to-edge hopping. It is sufficient to synthesize a conditional beam splitter between the two end modes and leverage it for an edge-to-edge hopping term in the Hamiltonian. We can synthesize a beam splitter between the edges of the chain using the bosonic SWAP gate defined in Table III to move the mode information from both edges of the chain to adjacent sites, carry out the required operation between the modes, and SWAP the information back.

## F. Qubit-boson ground-state preparation with quantum signal processing

While above we employed the heuristic VQE approach to approximately prepare the ground state, new developments in an algorithm known as QSP provide an alternative approach to ground-state preparation with provable accuracy and guarantees. Here, we provide a brief comparison of the VQE and QSP for ground-state preparation of the $\mathbb{Z}_2$-Higgs model.

Initially pioneered in Refs. [303,304], QSP provides a systematic method to apply a nearly arbitrary polynomial transformation to a linear operator embedded in a unitary matrix [299], thus furnishing a unifying framework for quantum algorithms [75,299]. In short, QSP works by interleaving the unitary matrix with a sequence of parameterizable SU(2) rotations; see Fig. 28 for an illustration. The output of this sequence is an embedding of a polynomial transformation of the initial linear operator, parametrized by the chosen SU(2) rotations. For a more detailed introduction to QSP, see Appendix E.

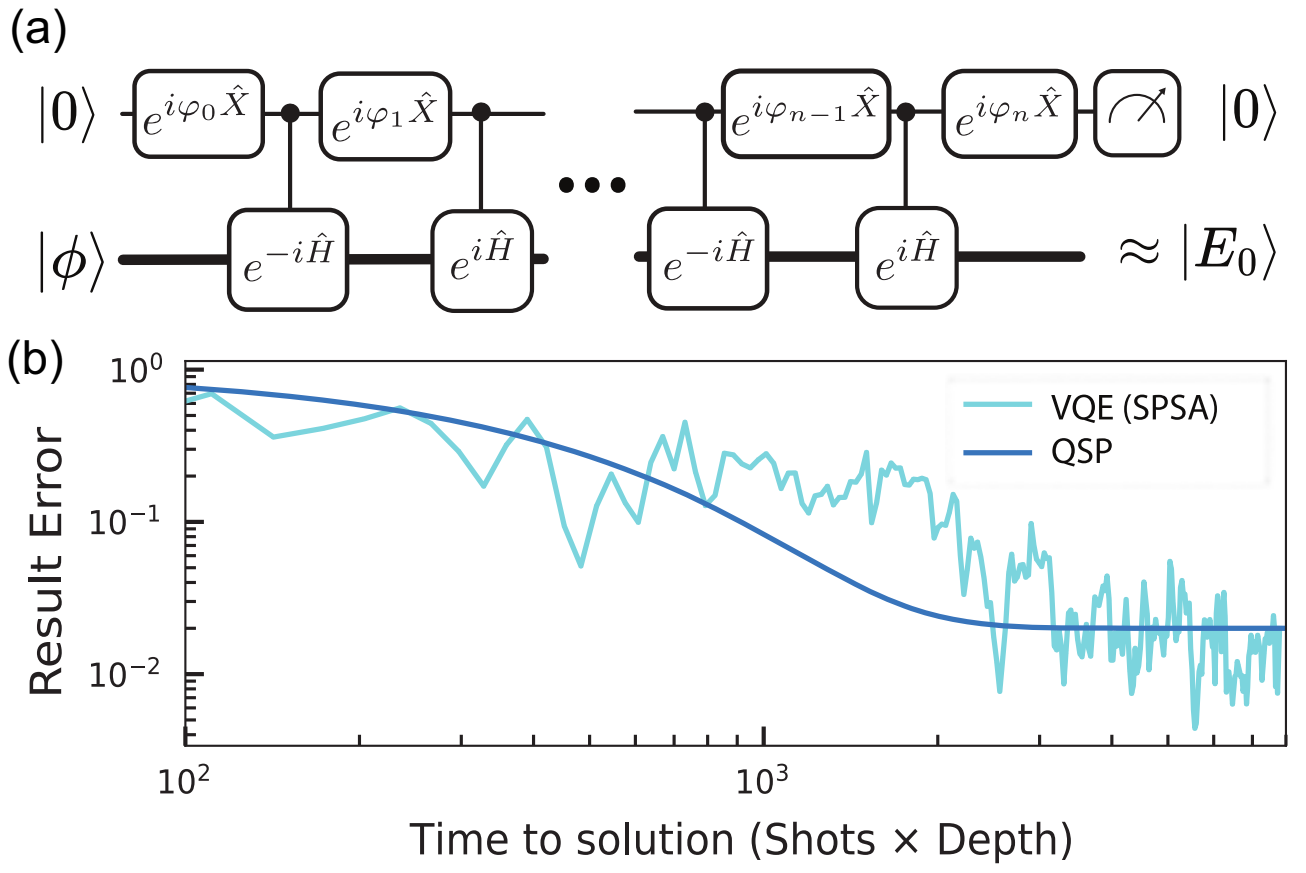

FIG. 28. Considerations for ground-state preparation using quantum signal processing. (a) Circuit for ground-state preparation with quantum signal processing, adapted from Ref. [305]. The parameters $\varphi = (\varphi_0, \varphi_1, \ldots, \varphi_n)$ can be chosen such that, upon postselecting on $|0\rangle$, the resulting state is a good approximation to the ground state $|E_0\rangle$. Note that this circuit interleaves $e^{\pm iH}$, distinct from the usual QSP construction; this modification is included to generate polynomial transformations in $\cos(H/2)$ [i.e., $P(\cos(H/2))$] and better approximate a projector onto the ground-state subspace. (b) Fidelity with the exact ground-state wave function vs time to solution for the VQE and QSP for the $\mathbb{Z}_2$-Higgs model for $U/J = 1 = g/J$, three bosonic sites, and two gauge-field sites. For the VQE, the dominant error is the infidelity with the exact ground state; for QSP, the dominant error is the probability of failure. For details, see Appendix F.

In Appendix F, we illustrate how QSP furnishes an algorithm for ground-state preparation, as initially presented in Ref. [305] for implementation on early fault-tolerant quantum computers. In this scenario, one considers a Hamiltonian $H$ with spectral gap bounded below by $\Delta$ and seeks to prepare its ground state with fidelity greater than or equal to $1 - \epsilon$. As input to this algorithm, one assumes access to an easily preparable initial state that has overlap greater than or equal to $\gamma$ with the ground state. One also assumes access to the time-evolution operators $e^{\pm iH}$, implemented, for instance, with Trotterization. To avoid aliasing, it is assumed that the eigenvalues of $H$ lie in the range $(0, \pi)$; otherwise, one can implement a rescaled time-evolution operator $e^{\pm i(H\Delta t + \phi)}$ to meet this constraint.

With this setup, the algorithm uses QSP to approximate a projector onto the ground state, which is then applied to the initial state to project out its component parallel to the ground state. This algorithm ultimately requires a circuit depth of $O[(1/\Delta)\log(1/\epsilon\gamma)]$, corresponding to that many coherent queries to the controlled time evolution $e^{\pm iH}$. This circuit is repeated $O(1/\gamma^2)$ times to prepare the ground state with high probability. The circuit of this algorithm is illustrated in Fig. 28.

Recent work has shown that QSP is naturally implemented on CV-DV quantum processors [57,306], rendering it applicable to the LGT simulations studied here. To exemplify the requirements of ground-state preparation with QSP, let us consider the $\mathbb{Z}_2$ model with $n = 3$ sites and couplings $J = g = U = 1$, as investigated in Sec. VI D. Let us also use the same initial state as above, which numerically has overlap $\gamma \approx 0.36$ with the ground state. In this scenario, the VQE produces a ground state with infidelity $\epsilon \approx 2 \times 10^{-2}$ (see Fig. 24). Aiming to achieve a similar performance with QSP, our calculations in Appendix F indicate that the QSP approach requires around 50 queries to the controlled time evolution $e^{\pm iH}$. Using a first-order Trotterization as outlined in Sec. VI B, a single such Trotter step would require around three controlled qubit-boson gates per site, for a rough total of greater than or around $10^4$ qubit-boson gates, which is prohibitively more expensive than the VQE.

This prohibitive circuit depth suggests that QSP-based ground-state preparation will only become practical and competitive with the VQE once qubit-boson gate infidelities much less than $10^{-4}$ can be achieved, in contrast to the current qubit-boson gate infidelities of around $10^{-3}$ [201,307]. Moreover, even if such a gate fidelity can be achieved, the number of repetitions of the QSP circuit will generally scale exponentially in the system size without prior knowledge of the ground state (see Appendix F). However, this scaling is to be expected because generic ground-state preparation is QMA hard [308], and of course, the VQE ground-state preparation is also expected to require resources exponential in the system size, without further prior knowledge.

In Fig. 28, we compare the performance of the VQE to QSP, as applied to the above scenario of the $\mathbb{Z}_2$ model. For both methods, we plot the dominant source of error vs time to solution, assuming no gate errors in the underlying circuit. The dominant source of error for the VQE is the infidelity with the ground state, and for QSP, it is the probability of failure (see Appendix F for details); the time to solution is taken to be the circuit depth times the number of shots. For the VQE, increasing the depth (i.e., number of layers) can increase the fidelity with the ground state, but it also leads to a high-dimensional optimization problem that suffers from barren plateaus and requires a high shot count to facilitate optimization [275]. On the other hand, QSP requires a relatively deeper circuit to estimate a projector onto the ground state and a modest number of shots to successfully project out the ground state. Incidentally, the empirical results of Fig. 28 suggest that both dominant sources of error scale similarly as a function of the time to solution, implying that their total resource costs are comparable. This suggestion arises because our VQE circuit has a shallow depth yet requires many shots, whereas QSP has a deeper circuit yet requires fewer shots. We speculate that this seemingly universal behavior arises from the fact that the complexity of ground-state preparation is fundamentally governed by the size of the spectral gap of the Hamiltonian, agnostic to the algorithm used to estimate the ground state. In any event, this analysis implies that the VQE is better suited to settings with low gate fidelities and short coherence times, whereas QSP is more favorable if gates are fast yet circuit repetitions are expensive.

## VII. CONCLUSIONS AND OUTLOOK

To conclude, in this paper, we have developed a resource-efficient framework for implementing fermion-boson quantum simulation of a wide variety of models including lattice gauge theories in $(1+1)$D and $(2+1)$D using qubit-boson quantum computation. For this purpose, we developed a range of compilation tools and subroutines to realize multibody interaction terms using the natively available qubit-boson gate set in circuit QED hardware. In particular, we developed parity-controlled, density-density bosonic gates, and oscillator-mediated multiqubit gates as foundational subroutines, which are useful beyond our simulation motivation. In the context of LGTs, we showed how to implement gauge-invariant hopping terms for transmon and dual-rail qubit fermionic encodings. We also presented a method implementing the magnetic-field term for $U(1)$ gauge fields.

In order to also compare the prefactors of the compilation, we developed an efficient all-qubit algorithm in the Fock-binary encoding for simulating bosonic hopping terms. By performing an end-to-end gate complexity analysis, including gate synthesis errors, we showed that our approach yields an asymptotic improvement from $\mathcal{O}(N_{\max}^{1+\frac{1}{2p}}\log_2(N_{\max})^2)$ to $\mathcal{O}(1)$ scaling in terms of the boson cutoff $S$ for the gauge-invariant hopping term and an $\mathcal{O}(\log(N_{\max}))$ to $\mathcal{O}(1)$ improvement for the magnetic-field term.

We then applied our compilation strategies to some of the most pertinent quantum simulation tasks—ground-state preparation and time evolution, for which we developed a qubit-boson variational quantum eigensolver method. We benchmarked our hybrid qubit-boson approach through numerical simulations of a $\mathbb{Z}_2$-Higgs model and a $U(1)$ quantum link model in $(1+1)$D in Bosonic Qiskit, and we demonstrated that low-depth qubit-boson circuits can prepare states with a high overlap with the ground state for small systems. In particular, in the $U(1)$ quantum link model, we found relative energy errors of less than $10^{-3}$ for just a two-layer ansatz for three matter sites. Separately, we simulated both the time dynamics and the VQE for the $\mathbb{Z}_2$-Higgs model in the presence of the leading source of noise in circuit QED architectures—transmon decay and dephasing. We then developed an approach for postselecting on Gauss's law in VQE circuits, using the fact that each term in the Hamiltonian individually conserves Gauss's law, and we showed that energy errors of 20% of the difference between the ground and first excited states can be reached with this method. We also showed how nontrivial observables can be measured within our approach, including the superfluid stiffness and string-order correlators. Finally, we discussed how quantum signal processing —an alternative approach to eigenstate preparation with performance guarantees—can be implemented in qubit-boson systems, showing that its time to solution for a fixed error is comparable to the VQE.

### A. Prospect for quantum advantage

For simulations in the NISQ era, it is important that errors do not completely destroy the signal to be observed. In cavities, the probability of a Fock state $|N\rangle$ to decay to $|N-1\rangle$ scales linearly with $N$, therefore posing a challenge to the achievable fidelity for large $N$, in particular, in the context of quantum computation (see, e.g., Sec. II E 2 in Ref. [57]). By contrast, when mapping bosons to qubits using the Fock-binary encoding, the probability to be in an encoded Fock state $|N\rangle$ has a decay rate that scales as $\Omega(1)$ and $\mathcal{O}(\log_2(N))$. However, this apparent advantage of the Fock-binary encoding does not directly materialize when considering observables. As we discuss in Appendix G, for both cavities and qubits, even nonlinear observables such as $\langle\hat{n}^2\rangle$ have a decay rate that is $\mathcal{O}(1)$ in $N$. This finding is in analogy to the fact that for many-qubit systems, local observables can have low error even when the global fidelity of the state is essentially zero [309,310]. Therefore, the decay rate of Fock states in cavities, which grows linearly with the Fock-state number, need not lead to particular difficulties for quantum simulation when considering such local observables. Moreover, the effective errors induced by qubit decay in the Fock-binary encoding are highly nonlinear in $N$ and also lead to large jumps in occupations. For example, Fock states with $N=2^n$, $n\in\mathbb{N}$, decay immediately to $N=0$, leading to an effective loss of $2^n$ bosons at once. Hamiltonian evolution with nonlinear terms such as an on-site interaction in the presence of Fock-state-dependent loss can lead to highly correlated errors, which is not the case for bosonic hardware, where the decay rate of $\langle\hat{n}^2\rangle$ for an initial Fock state is a smooth function of the Fock-state number and, in a single decay event, only a single boson is lost. Finally, in cavity QED, boson loss is far slower (decay rate $\kappa\approx 1$ kHz, with some cavities shown to exhibit $\kappa<1$ Hz) than qubit decay (decay rate $\gamma\approx 5$–$10$ kHz). The excellent coherence of cavities is illustrated by the recent experimental preparation of coherent (i.e., definite parity) photon cat states with as many as 1000 photons [187]. Therefore, the differing nature of errors in oscillators, which are closer to physical errors appearing in bosonic systems, can pose advantages for qubit-boson hardware compared to all-qubit simulations of bosonic systems.

As to the prospect of quantum advantage (for a discussion of bosonic quantum computational advantage, see Ref. [311]), we believe that nonequilibrium dynamics in regimes with large boson number fluctuations, starting in Fock states, will be challenging to simulate classically: The widely used time-evolving-block-decimation algorithm for tensor networks scales cubically with the bond dimension and quadratically with the boson number cutoff [99,100]. Monte Carlo methods struggle with nonequilibrium and real-time dynamics, while semiclassical methods such as the truncated Wigner approximation [108] struggle with initial states that are non-Gaussian. Finally, brute-force sparse diagonalization methods struggle with large systems—we expect dynamics of system sizes of $L > 20$ bosonic sites with an average of two bosons per site to be challenging to simulate when considering initial states leading to large boson fluctuations such as $|0\cdots 0LL0\cdots 0\rangle$, in which case the cutoff needs to be chosen to be essentially given by the number of bosons in the system, $2L$, leading to a Hilbert space size of $\binom{3L-1}{2L}$. The largest simulations to date have reached Hilbert space sizes of less than $2^{40}$ [98]. Hence, we expect to surpass classical simulability using sparse matrix methods for $L \approx 16$. This expectation is in contrast to qubit-only simulations, where systems of $L \gtrsim 50$ are needed to reach beyond-classical regimes [312].

### B. Future directions

Our study opens a promising pathway towards using native qubit-boson hardware to simulate boson-fermion-gauge-field systems. Compared to qubit-based hardware, this approach will be most advantageous when simulating systems with strong boson number fluctuations, for example, in chaotic field theory evolution [313], as such a case requires large cutoffs and, consequently, the synthesis of square-root factors (when acting with creation and annihilation operators) that are expensive to synthesize in all-qubit hardware. A key advantage of our digital approach compared to classical or analog methods is its versatility—our approach can simulate topologies that do not match the hardware topology. For example, a 3D model can, in principle, be mapped to the 2D array at the expense of needing to implement long-range entangling gates with SWAPs. Alternatively, holographic approaches could be employed [314,315].

Extending our approach to non-Abelian gauge theories, most importantly, $SU(N)$ to simulate the weak and strong forces, is most pertinent. To this end, our Schwinger-boson approach can be relatively straightforwardly generalized [176,316], and we expect similar scaling advantages to materialize as we found for $U(1)$ fields.

Many of our compilations were derived by hand. Finding more systematic qubit-boson compilation procedures, building on initial work [317–319], could further lead to new efficient compilations.

Many applications of qubit-boson quantum simulation, such as vibronic excitations of molecules or phonon-electron interactions in condensed-matter physics, show similar decoherence processes to the one present in qubit-boson hardware. This similarity opens up the direction of hybrid qubit-boson digital simulation of open-system dynamics, which would be an interesting topic to explore next. In noisy qubit hardware, it would not be immediately possible as the physical loss processes do not resemble boson loss in Fock-binary encodings as discussed above.

Looking into the future beyond noisy simulations, our work motivates the study of bosonic error-correcting codes, i.e., codes that directly encode a bosonic logical operator [110], which may be concatenated with qubit codes [320] or combined with analog error mitigation techniques [321]. Furthermore, it would be interesting to explore whether our proposed hybrid qubit-boson approach also yields advantages for equilibrium state preparation away from the ground state. To that end, quantum signal processing [75], methods based on filtering [322–325], or thermalization [326,327] could be employed. Another interesting direction is to develop explicit benchmarks of dynamical quantum simulations of, e.g., string breaking and particle scattering problems [26,51] to better compare hybrid qubit-boson and qubit-based approaches.

While challenging in cavity QED hardware, an implementation in photonic platforms [328] could be attempted, or a purely native approach for fermion-boson problems could be explored by combining our approach with digital fermionic quantum processing in neutral atom arrays [53,74,329], where bosons can be encoded in (the mechanical) harmonic oscillator degrees of atoms in tweezers [57,73].

## ACKNOWLEDGMENTS

This research was funded by the U.S. Department of Energy, Office of Science, National Quantum Information Science Research Centers, Co-design Center for Quantum Advantage (C2QA) under Contract No. DE-SC0012704. C2QA led this research. We also acknowledge support from the U.S. Department of Energy, Office of Science, National Quantum Information Science Research Centers, Quantum Systems Accelerator. S. K. is supported with funds from the Ministry of Science, Research and Culture of the State of Brandenburg within the Centre for Quantum Technologies and Applications (CQTA). J. M. M. acknowledges support from the National Science Foundation Graduate Research Fellowship under Grant No. 2141064. L. F. is supported by the Deutsche Forschungsgemeinschaft (DFG, German Research Foundation) as part of the CRC 1639 NuMeriQS–Project No. 511713970 and under Germany's Excellence Strategy–Cluster of Excellence "Color meets Flavor" (CmF) EXC

3107–533766364 and Cluster of Excellence "Matter and Light for Quantum Computing" (ML4Q) EXC 2004/2–390534769. We acknowledge helpful discussions with Marko Cetina, Zohreh Davoudi, Alexey Gorshkov, Fabian Grusdt, Mohammad Hafezi, Or Katz, and Torsten Zache.

S. M. G. is a consultant for, and equity holder in, Quantum Circuits, Inc. (part of D-Wave Quantum, Inc.).

## APPENDIX A: IMPLEMENTING FERMIONIC HAMILTONIAN TERMS

### 1. Hopping

In this appendix, we remind the reader of an existing approach for digitally simulating fermionic Hamiltonian terms, which we show is feasible in hybrid qubit-boson hardware and simulation.

To take into account the negative sign in the many-body wave function appearing when two fermions exchange position, we employ the JW [330] mapping (although many mappings would be possible, such as Bravyi-Kitaev [331], parity basis [332], or log-depth versions of the JW encoding [333,334]). The JW mapping fixes a reference site, defining a one-dimensional chain embedded in a $D$-dimensional volume and trailing a set of $\hat{Z}$ operators with each application of a fermionic creation or destruction operator along the chain:

$$\hat{c}_i^\dagger = \left(\bigotimes_{k=1}^{i-1} \hat{Z}_k\right) \otimes \hat{\sigma}_i^+, \qquad \hat{c}_i = \left(\bigotimes_{k=1}^{i-1} \hat{Z}_k\right) \otimes \hat{\sigma}_i^-, \tag{A1}$$

where $\hat{\sigma}^- = [(\hat{X} + i\hat{Y})/2], \hat{\sigma}^+ = [(\hat{X} - i\hat{Y})/2]$.

For a one-dimensional system and only nearest-neighbor hopping, the strings of $\hat{Z}$ operations cancel out, and $\hat{c}_i^\dagger \hat{c}_{i+1} + \text{H.c.}$ reduces to $\hat{\sigma}_i^+ \hat{\sigma}_{i+1}^- + \text{H.c.}$ However, for $D > 1$, the strings of $\hat{Z}$ only cancel if the hopping is between neighboring sites along the JW chain, but they do not cancel if the two sites are not nearest neighbors along the JW chain.

A highly parallelized way of implementing hopping terms $e^{i2\theta(\hat{c}_i^\dagger \hat{c}_j + \text{H.c.})}$ between sites $i$ and $j$ that are not adjacent along the JW chain is a fermionic SWAP network [226] (see, e.g., Ref. [229] for a large-scale implementation). This idea uses FSWAP gates,

$$\text{FSWAP}_{i,j} = \text{SWAP}_{i,j} \cdot \text{CZ}_{i,j}, \tag{A2}$$

to bring every pair of fermions to Jordan-Wigner adjacent sites in exactly depth $N$, where $N$ is the number of fermions. When fermions are on adjacent sites, hopping is implemented using iSWAP($\theta$) gates. In this way, all hopping gates can be implemented in depth $N$, using $\mathcal{O}(N^2)$ FSWAP gates and $N$ iSWAP($\theta$) gates. This value can be reduced to depth $\sqrt{N}$ for nearest-neighbor hopping on the 2D lattice.

We now comment on the transpilation of the required gates in terms of $ZZ(\theta)$ and single-qubit gates.

$\text{SWAP}_{i,j}$ fully exchanges the contents of sites $i$ and $j$ on which it is applied, and CZ is the conditional-Z gate. CZ applies $\hat{Z}$ on the second qubit only if the first is in $|1\rangle$ (which happens to be symmetric). This case corresponds to the unitary $CZ_{1,2} = \frac{1}{2}((\mathbb{1}_1 + \hat{Z}_1)\mathbb{1}_2 + (\mathbb{1}_1 - \hat{Z}_1)\hat{Z}_2)$. We have that $CZ_{1,2} = e^{i(\pi/4)} e^{-i(\pi/4)\hat{Z}_1} e^{-i(\pi/4)\hat{Z}_2} e^{i(\pi/4)\hat{Z}_1\hat{Z}_2}$.

The Jordan-Wigner-adjacent hopping must be implemented, which is given by

$$\begin{aligned}\text{iSWAP}_{i,j}(\theta) &= e^{i\theta(\hat{\sigma}_i^+ \hat{\sigma}_{i+1}^- + \text{H.c})} \\ &= e^{-i\theta(\hat{X}_i \hat{X}_{i+1})} e^{-i\theta(\hat{Y}_i \hat{Y}_{i+1})}.\end{aligned} \tag{A3}$$

Both $e^{-i\theta(\hat{X}_i \hat{X}_{i+1})}$ and $e^{-i\theta(\hat{Y}_i \hat{Y}_{i+1})}$ can be implemented by combining the $ZZ_{i,j}(\theta)$ with single-qubit rotations using that $\hat{X} = \hat{H}\hat{Z}\hat{H}$ and $\hat{Y} = \hat{S}\hat{H}\hat{Z}\hat{H}\hat{S}^\dagger$, where $S = e^{i\pi/4} R^Z(\pi/2)$. Furthermore, we note that, when following $e^{-i\theta(\hat{X}_i \hat{X}_{i+1})}$ directly with $e^{-i\theta(\hat{Y}_i \hat{Y}_{i+1})}$, both dual-rail qubits have $\hat{H}\hat{S}\hat{H}$ applied to them, which reduces to $\frac{1}{2}[(1+i)\mathbb{1} + (1-i)\hat{X}] = R^0(\pi)$, a rotation around the $\hat{X}$ axis. Therefore, the iSWAP operation will require single qubit rotations and $ZZ(\theta)$ gates. These are the only two operations required to implement fermionic hopping using fermionic SWAP networks.

### 2. On-site potential

On-site potential terms for fermions take the form $\sum_i \mu_i \hat{n}_i$, where $\hat{n}_i = \hat{c}_i^\dagger \hat{c}_i$. The time evolution of this term, $e^{-i\mu_i t \hat{n}_i}$, can be mapped to single-qubit phase gates acting on each site individually because $\hat{n} = (\mathbb{1} - \hat{Z})/2$. These terms are not affected by the Jordan-Wigner mapping.

### 3. On-site and intersite interactions

Intersite interactions take the form $\hat{H} = \sum_{\langle i,j\rangle} V_{i,j} \hat{n}_i \hat{n}_j$. The time evolution under this Hamiltonian, $e^{iJt\sum_{\langle i,j\rangle} \hat{n}_i \hat{n}_j}$, can be mapped to $ZZ_{i,j}(V_{i,j}t/2)$ gates combined with single-qubit phase gates in the JW mapping using $\hat{n}_{\text{qubit}} = (\mathbb{1} - \hat{Z})/2$. For multicomponent models such as the Fermi-Hubbard model in which each site can, for example, have a spin-up fermion and/or a spin-down fermion, the different components are mapped onto different qubit sites. Therefore, on-site interactions $\hat{H} = U\sum_{\langle i\rangle} \hat{n}_{i,\alpha} \hat{n}_{i,\beta}$ between different components $\alpha$ and $\beta$ can also be implemented using the $ZZ_{i,j}(Ut/2)$ gate.

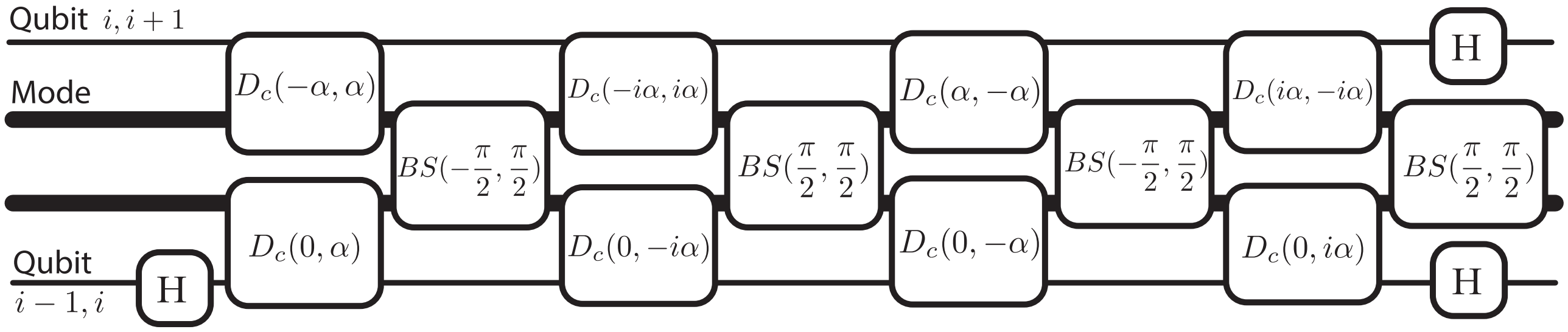

FIG. 29. Circuit for implementing the unitary $V$ defined in Eq. (B4). This circuit enables an ancilla-free implementation of on-site interactions in the $(1+1)$D $\mathbb{Z}_2$-Higgs model for the choice $\alpha = \frac{1}{2}\sqrt{(\pi/2)}$. In the above, $D_c(\beta,\gamma) = |0\rangle\langle 0| \otimes D(\beta) + |1\rangle\langle 1| \otimes D(\gamma)$ is an asymmetric conditional displacement (see Ref. [57] for details), where $D(\cdot)$ is an unconditional displacement as defined in Table II.

## APPENDIX B: ANCILLA-FREE IMPLEMENTATION OF BOSONIC ON-SITE INTERACTIONS IN THE (1+1)D $\mathbb{Z}_2$-HIGGS MODEL

In this appendix, we discuss the ancilla-free implementation of time evolution for the bosonic on-site interaction term in the $\mathbb{Z}_2$-Higgs model. Our motivation for developing this scheme is that, in the $(1+1)$D case in Sec. VI, each transmon dispersively coupled to a cavity encodes a matter degree of freedom and is therefore not "available" as an ancilla for the realization of on-site interactions using the method of Sec. III A 3. While experiments have successfully incorporated multiple transmons per cavity [202], such a strategy is challenging and outside of the proposed architecture in Fig. 4.

In this section, we discuss a workaround for realizing this term that relies on Gauss's law. For simplicity, we describe this strategy in the context of the $(1+1)$D $\mathbb{Z}_2$-Higgs model, but note that an extension to $(2+1)$D is possible. In particular, we develop an exact, ancilla-free implementation of arbitrary on-site interactions using one qubit per cavity while also preserving the quantum information in each qubit.

This approach leverages a combination of SNAP gates, Mølmer-Sørensen-like multiqubit gates discussed in Sec. II F, and Gauss's law constraints. The key idea is to introduce an identity in the form of Gauss's law, for which we use the convention

$$\hat{G}_i = \hat{X}_{i-1,i}\, e^{i\pi\hat{n}_i}\, \hat{X}_{i,i+1} = +\mathbb{1} \tag{B1}$$

to rewrite

$$e^{-iU\hat{n}_i^2 t} = e^{-i(\hat{X}_{i-1,i}e^{i\pi\hat{n}_i}\hat{X}_{i,i+1})U\hat{n}_i^2 t}, \tag{B2}$$

where the symbols are as defined in Sec. III A 3.

To synthesize the right-hand side using native gates, we recast it as

$$e^{-i\hat{X}_{i-1,i}\hat{X}_{i,i+1}\sum_n \theta_n |n\rangle\langle n|_i} = \hat{V}^\dagger e^{-i\hat{Z}_{i,i+1}\sum_n \theta_n |n\rangle\langle n|_i}\hat{V} \tag{B3}$$

in terms of the Fock states $|n\rangle$ with eigenvalue $n$, and the angles are defined as $\theta_n = e^{i\pi n}Un^2 t$. The right-hand side of the equality in Eq. (B3) corresponds to a SNAP gate defined in Table II, conjugated by the two-qubit unitary

$$\hat{V} = \hat{H}_{i,i+1} e^{i(\pi/4)(\mathbb{1}_{i-1,i} - \hat{X}_{i-1,i})\hat{Z}_{i,i+1}}, \tag{B4}$$

where $\hat{H}_{i,i+1}$ is a Hadamard gate on the qubit linking sites $i$ and $i+1$, and $\mathbb{1}_{i,i+1}$ the identity. Therefore, synthesis of this term reduces to the task of implementing $\hat{V}$, which is possible in the proposed architecture through the use of the Mølmer-Sørensen-like scheme introduced in Ref. [57]. The exact circuit that realizes $\hat{V}$ is shown in Fig. 29.

While this approach enables us to implement arbitrary on-site interactions without additional ancillas, it is important to note that its reliance on Gauss's law can result in the propagation of photon-loss and qubit dephasing errors as effective gate errors. However, such effects can be at least partially mitigated by postselecting results on states that satisfy or partially satisfy Gauss's law (which we discuss in Sec. VI D 5). However, note that this will come at the cost of an increased number of shots that scales exponentially with system size.

## APPENDIX C: REALIZING THREE-WAVE MIXING HAMILTONIANS WITH DRIVEN SIDEBAND TRANSITIONS

In this appendix, we outline one possible method for realizing the three-wave mixing terms needed for gauge-invariant fermonic hopping as described in Sec. IV E 2. The goal here is to realize the two Hamiltonians

$$E = (ab^\dagger\sigma_b^+ + a^\dagger b\sigma_b^-)/2, \tag{C1}$$

$$F = i(ab^\dagger\sigma_b^+ - a^\dagger b\sigma_b^-)/2. \tag{C2}$$

We focus on two nodes of the proposed experimental architecture as shown in Fig. 30. Here, an ancilla qubit $q$ (transmon) at frequency $\omega_q$ is coupled to an oscillator $a$ with vacuum Rabi rate $2g_{qa}$. Oscillator $a$ is coupled to oscillator $b$ with a tunable beam-splitter interaction $g_{\rm bs}(t)$ activated by a pumped three-wave mixing element with pump frequency $\omega_p$. The second transmon, coupled to cavity $b$, can be dynamically decoupled during the interaction.

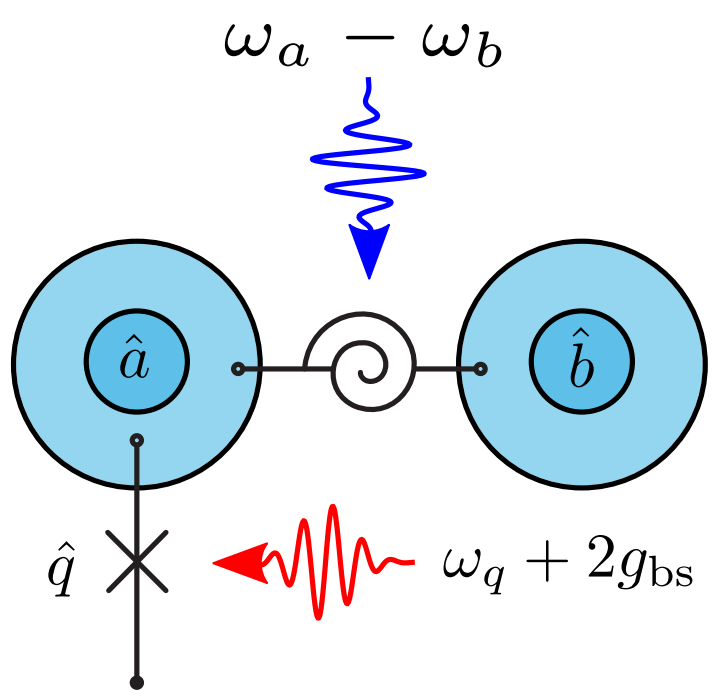

FIG. 30. Realization of an effective three-wave mixing term. We show two nodes of the proposed superconducting architecture consisting of a transmon, two cavities, and a tunable beam-splitter interaction. The second transmon coupled to cavity $b$ is not shown. The SNAIL mixing element is driven at $\omega_a - \omega_b$ to hybridize modes $a$ and $b$. In addition, a sideband drive at frequency $\omega_q + 2g_{\rm bs}$ is applied to the transmon, activating the desired three-wave mixing interaction.

By activating the beam splitter between modes $a$ and $b$ while simultaneously driving a sideband on qubit $q$, we can achieve the desired three-wave coupling. Here, the sideband frequency should be $\omega_q + 2g$, blue-detuned by the mode splitting between $a$ and $b$, when the beam splitter is activated.

In the lab frame, expanding the junction's cosine potential to fourth order, the relevant effective Hamiltonian for the three modes can be approximated as

$$\begin{aligned} H_0/\hbar = {} & \omega_a a^\dagger a + \omega_b b^\dagger b + \omega_q q^\dagger q \\ & + g_{\rm bs}(a^\dagger b e^{-i\omega_p t} + a b^\dagger e^{i\omega_p t}) \\ & + g_{qa}(a^\dagger q + a q^\dagger) - \frac{E_J}{4!}(\varphi_q(q + q^\dagger))^4, \end{aligned} \tag{C3}$$

where $\varphi_q = (2E_C/E_J)^{\frac{1}{4}}$ and $\omega_p$ is the beam-splitter pump frequency.

To analyze this Hamiltonian, we first apply a dispersive transformation on modes $q$ and $a$. Assuming $g \ll \Delta_{qa}$, we expand to first order in $(g_{qa}/\Delta)$ to find

$$\begin{aligned} H_1/\hbar = {} & \omega_a a^\dagger a + \omega_b b^\dagger b + \omega_q q^\dagger q \\ & + g_{\rm bs}(a^\dagger b e^{-i\omega_p t} + a b^\dagger e^{i\omega_p t}) \\ & - \frac{E_J}{4!}\left(\varphi_q(q + q^\dagger) + \varphi_q \frac{g_{qa}}{\Delta_{qa}}(a + a^\dagger)\right)^4, \end{aligned} \tag{C4}$$

where we have redefined $\omega_a$, $\omega_b$, and $\omega_q$ to be the normalized mode frequencies.

Driving the beam splitter on resonance corresponds to a SNAIL pump frequency of $\omega_p = \omega_b - \omega_a$. Moving into the rotating frame of modes $a$, $b$, and $q$, we find

$$\begin{aligned} \tilde{H}_1/\hbar = {} & g_{\rm bs}(a^\dagger b + a b^\dagger) - \frac{E_J}{4!}(\varphi_q(q e^{-i\omega_q t} + q^\dagger e^{i\omega_q t}) \\ & + \varphi_q \frac{g_{qa}}{\Delta_{qa}}(a e^{-i\omega_a t} + a^\dagger e^{i\omega_a t}))^4. \end{aligned} \tag{C5}$$

From here, we make a Bogoliubov transformation $\tilde{H}_2 = U^\dagger \tilde{H}_1 U$, where $U = \exp((\pi/4)(a^\dagger b - a b^\dagger))$, leading to

$$\begin{aligned} \tilde{H}_2 = {} & g_{\rm bs}(b^\dagger b - a^\dagger a) - \frac{E_J}{4!}(\varphi_q(q e^{-i\omega_q t} + q^\dagger e^{i\omega_q t}) \\ & + \varphi_q \frac{g_{qa}}{2\Delta_{qa}}((a + b)e^{-i\omega_a t} + (a^\dagger + b^\dagger)e^{i\omega_a t}))^4. \end{aligned} \tag{C6}$$

Finally, we can account for the mode splitting of $2g_{bs}$ by moving into the rotating frame of $a$ and $b$ at frequencies $\mp g$ to find

$$\begin{aligned} \tilde{H}_3 = {} & -\frac{E_J}{4!}\Big[\varphi_q(q e^{-i\omega_q t} + q^\dagger e^{i\omega_q t}) \\ & + \varphi_q \frac{g_{qa}}{2\Delta_{qa}}(a e^{-i(\omega_a - g)t} + b e^{-i(\omega_a + g)t} \\ & + a^\dagger e^{i(\omega_a - g)t} + b^\dagger e^{i(\omega_a + g)t})\Big]^4. \end{aligned} \tag{C7}$$

Here, we see that by activating a resonant beam splitter between modes $a$ and $b$, the transmon mode $q$ has an effective four-wave coupling to both mode $a$ and $b$, seen by the qubit at effective frequencies $\omega_a \pm g$. Thus, we can drive the qubit with a sideband drive at frequency $\omega_d = \omega_q + 2g$ to activate the desired three-wave mixing term. To see this case, we add a drive term with strength $\xi$,

$$\begin{aligned} \tilde{H}_3 = {} & -\frac{E_J}{4!}\Big[\varphi_q(q e^{-i\omega_q t} + q^\dagger e^{i\omega_q t}) \\ & + \varphi_q \frac{g_{qa}}{2\Delta_{qa}}(a e^{-i(\omega_a - g)t} + b e^{-i(\omega_a + g)t} + a^\dagger e^{i(\omega_a - g)t} \\ & + b^\dagger e^{i(\omega_a + g)t} + \xi e^{-i\omega_d t} + \xi^* e^{i\omega_d t})\Big]^4 \end{aligned} \tag{C8}$$

and expand the four-wave mixing term. We find the desired resonant process

$$\tilde{H}_{\rm sideband}/\hbar = -E_J \varphi_q^3 \left(\frac{g_{qa}}{2\Delta_{qa}}\right)^2 (\xi a b^\dagger q^\dagger + \xi^* a^\dagger b q). \tag{C9}$$

To avoid driving other undesired terms, the strength of the process must be small with respect to the mode splitting, $2g_{\rm bs}$. Further analysis of this process in an experimental setting will be needed to solidify its use as a viable path to realizing effective three-wave mixing. Other methods could also be used, such as a dedicated four-wave mixing element coupling modes $q$, $b$, and $a$.

## APPENDIX D: QUBIT GATE COUNTS FOR SIMULATING A BEAM-SPLITTER HAMILTONIAN

In this appendix, we explain how to obtain the number of two-qubit gates necessary for compiling a beam splitter between two modes $e^{-i\theta\hat{H}}$ with

$$\hat{H} = \hat{a}^\dagger b + \hat{a}\hat{b}^\dagger \tag{D1}$$

onto a qubit-only quantum computer within the Fock-binary encoding. We do this by generalizing the calculation presented in Appendix F of Ref. [57], which shows the compilation of a displacement on a single mode $e^{-i(\alpha\hat{a}^\dagger-\alpha^*\hat{a})}$, where $\alpha$ is a complex number.

The beam splitter can be divided into four parts according to $H = (H_{\text{even,even}} + H_{\text{even,odd}} + H_{\text{odd,even}} + H_{\text{odd,odd}})$. The four parts contain matrix elements that are square roots of products of even and/or odd numbers. For example, $H_{\text{even,odd}}$ only contains matrix elements with value $\sqrt{x\times y}$, where $x$ is even and $y$ is odd. This case corresponds to those parts of the Hamiltonian where $\hat{a}^\dagger\hat{b}$ acts on Fock states $|\text{odd}\rangle|\text{odd}\rangle$ transforming them to $\sqrt{\text{even}\times\text{odd}}|\text{even}\rangle|\text{even}\rangle$, and $\hat{a}\hat{b}^\dagger$ acts on Fock states $|\text{even}\rangle|\text{even}\rangle$ transforming them to $\sqrt{\text{even}\times\text{odd}}|\text{odd}\rangle|\text{odd}\rangle$:

$$\hat{a}^\dagger\hat{b}|2j+1\rangle|2l+1\rangle = \sqrt{2(j+1)\times(2l+1)}|2(j+1)\rangle|2l\rangle, \tag{D2}$$

$$\hat{a}\hat{b}^\dagger|2(j+1)\rangle|2l\rangle = \sqrt{2(j+1)\times(2l+1)}|2j+1\rangle|2l+1\rangle. \tag{D3}$$

The action of $H_{\text{even,odd}}$ rotates between the states $|2j+1\rangle|2l+1\rangle$ and $|2(j+1)\rangle|2l\rangle$.

Using a first-order Trotter formula, we can therefore decompose the time evolution under a beam splitter into four parts,

$$e^{-i\theta\hat{H}} = e^{-i\theta\hat{H}_{\text{even,even}}}e^{-i\theta\hat{H}_{\text{even,odd}}}e^{-i\theta\hat{H}_{\text{odd,even}}}e^{-i\theta\hat{H}_{\text{odd,odd}}} + \mathcal{O}(\theta^2). \tag{D4}$$

This result reduces the problem to compiling time evolution under the four parts individually. However, they do not commute, and therefore, there is an error. To make this error arbitrarily small, we use a $(2p)$ order Trotter formula $\mathcal{S}(\theta)$:

$$\mathcal{S}(\theta) = \prod_{j=1}^{N_{\text{exp}}/4} e^{-i\theta\hat{H}_{\text{even,even}}\tau_{4j-3}}e^{-i\theta\hat{H}_{\text{even,odd}}\tau_{4j-2}} \times e^{-i\theta\hat{H}_{\text{odd,even}}\tau_{4j-1}}e^{-i\theta\hat{H}_{\text{odd,odd}}\tau_{4j}}, \tag{D5}$$

where the formula is generic but is specified in Eq. (11) of Ref. [239], where $\tau_{4j-m}$ are also defined. This formula yields error $||(\mathcal{S}(\theta)) - e^{-i\theta\hat{H}}|| \le \epsilon$ with a number of elementary exponentials $N_{\text{exp}}$ (i.e., one of the four constituent exponentials above), which, using Eq. (65) in Ref. [244], yields

$$N_{\text{exp}} \le 2^3 5^{p-1}\left\lceil pN_{\max}\theta\left(\frac{5}{3}\right)^p\left(\frac{N_{\max}\theta}{\epsilon}\right)^{\frac{1}{2p}}\right\rceil, \tag{D6}$$

for a $(2p)$ order Trotter formula, where $N_{\max}$ is the cutoff. To obtain the total number of gates, we now need to determine the number of gates for each of the exponentials.

Crucially, these four parts are one-sparse matrices within the Fock-binary encoding and can be decomposed into direct sums of terms that are a constant times a $\hat{\sigma}^x$ Pauli matrix. This approach will enable us to map the evolution under this Hamiltonian to a rotation on a single ancilla qubit. For example, in the case of $H_{\text{even,odd}}$, the amplitudes of $|2j+1\rangle|2l+1\rangle$ and $|2(j+1)\rangle|2l\rangle$ will be mapped to the amplitudes of $|0\rangle$ and $|1\rangle$ of an ancilla qubit, and a single qubit rotation will effectively carry out the action of $H_{\text{even,odd}}$. We explain this process in more detail in the following.

We divide the state of the two modes into even and odd components, where we assume that each Fock state is encoded via the Fock-binary encoding into qubits:

$$|\Psi\rangle_B = \sum_{nm} a_{n,m}|n\rangle|m\rangle \tag{D7}$$

$$= \sum_{j=0}^{\frac{N_{\max}-1}{2}}\sum_{l=0}^{\frac{N_{\max}-1}{2}}(a_{2j,2l}|2j\rangle|2l\rangle + a_{2j+1,2l+1}|2j+1\rangle|2l+1\rangle + a_{2j+1,2l}|2j+1\rangle|2l\rangle + a_{2j,2l+1}|2j\rangle|2l+1\rangle). \tag{D8}$$

The preparation of this qubit state is, in general, not a trivial task; however, we do not consider the complexity of this task here.

We now discuss the case where $H_{\text{even,odd}}$ is applied. We add an ancilla qubit in $|0\rangle$. Grouping terms that will not be affected by the operation $a_{2j+1,2l}|2j+1\rangle|2l\rangle + a_{2j,2l+1}|2j\rangle|2l+1\rangle$ into $|\psi_{\text{rest}}\rangle$, we can write the state as

$$|\Psi\rangle_B = \sum_{j=0}^{\frac{N_{\max}-2}{2}}\sum_{l=0}^{\frac{N_{\max}-1}{2}}(a_{2(j+1),2l}|2(j+1)\rangle|2l\rangle + a_{2j+1,2l+1}|2j+1\rangle|2l+1\rangle + |\psi_{\text{rest}}\rangle)|0\rangle, \tag{D9}$$

where we have neglected $|0\rangle|2l\rangle$ in the first ($|\text{even}\rangle|\text{even}\rangle$) term of the sum, which is allowed because $H_{\text{even,odd}}$ transforms $|\text{even}\rangle|\text{even}\rangle$ states with $\hat{a}\hat{b}^\dagger$, canceling $|0\rangle|2l\rangle$; we have also neglected $|N_{\max}\rangle|2l+1\rangle$ in the second ($|\text{odd}\rangle|\text{odd}\rangle$) term of the sum, which is allowed because $H_{\text{even,odd}}$ transforms $|\text{odd}\rangle|\text{odd}\rangle$ states with $\hat{a}^\dagger\hat{b}$, taking this state to $\sqrt{(N_{\max}+1)\times(2l+1)}|N_{\max}+1\rangle|2l\rangle$, which is unphysical for a cutoff of $N_{\max}$.

The two amplitudes in the sum not contained in $|\psi_{\text{rest}}\rangle$ are those that are connected by $H_{\text{even,odd}}$, and we wish to transfer them to the ancilla qubit. For this purpose, we flip the ancilla qubit controlled on the joint parity of the mode registers. The least significant qubit of a register encoding a Fock state in the binary is in $|0\rangle$ if the parity is even and $|1\rangle$ if it is odd. Therefore, we apply a Toffoli gate to the ancilla qubit, controlled on the two least-significant bits of the mode registers. In this way, we obtain

$$|\psi\rangle_B = \sum_{j=0}^{\frac{N_{\max}-2}{2}} \sum_{l=0}^{\frac{N_{\max}-1}{2}} (a_{2(j+1),2l}|2(j+1)\rangle|2l\rangle|0\rangle + a_{2j+1,2l+1}|2j+1\rangle|2l+1\rangle|1\rangle) + |\psi'_{\text{rest}}\rangle. \quad \text{(D10)}$$

Now, we complete the transfer of the coefficients by removing the dependency on the mode Fock state registers by transforming them to the same value, which can be completed by modifying them, dependent on the value of the ancilla qubit. As there is freedom to choose which value to set them to, we choose to set the register to the value that will be used later inside the square-root, which we know from Eqs. (D2) and (D3) to be $2(j+1)\times(2l+1)$.

For this purpose, we perform a 1-controlled incrementer on the first mode binary register controlled on the ancilla, which brings $a_{2j+1,2l+1}|2j+1\rangle|2l+1\rangle|1\rangle$ to $a_{2j+1,2l+1}|2(j+1)\rangle|2l+1\rangle|1\rangle$, and a 0-controlled CNOT on the least-significant bit of the second mode controlled on the ancilla, which brings $a_{2(j+1),2l}|2(j+1)\rangle|2l\rangle|0\rangle$ to $a_{2(j+1),2l}|2(j+1)\rangle|2l+1\rangle|0\rangle$, leading the state

$$\sum_{j,l}|2(j+1)\rangle|2l+1\rangle(a_{2(j+1),2l}|0\rangle + a_{2j+1,2l+1}|1\rangle) + |\psi''_{\text{rest}}\rangle. \quad \text{(D11)}$$

A controlled incrementer with the construction presented in Appendix F of Ref. [57] requires $23n-12$ CNOT gates, where $n$ is the size of the register (representing the first mode) to be incremented. Therefore, so far, the number of CNOT gates (with the additional 0-controlled CNOT gate, and six CNOT gates to implement the Toffoli gate) to transfer the amplitudes to the ancilla qubit is

$$N_{\text{tr}} = 23n - 5. \quad \text{(D12)}$$

We then apply the operation

$$\sum_{j,l}|2(j+1)\rangle|2l+1\rangle\langle 2l+1|\langle 2(j+1)| \times R_0(-it\sqrt{2(j+1)\times(2l+1)}). \quad \text{(D13)}$$

Crucially, one can check that $|\psi''_{\text{rest}}\rangle$ does not evolve under this operation.

### 1. Calculating square roots on the fly

It is cheaper to first multiply the values contained in the $a$ and $b$ registers and then take their square root. The cost for a multiplication of the values contained inside two registers both of size $n$ is

$$N_\times = 9n^2 + 6n \quad \text{(D14)}$$

(see Ref. [121]). Let us encode the result of the multiplication in a register of size $2n$. The cost of the calculation of a square root of the value contained inside a register of size $2n$ is

$$N_{\text{SQRT}} = (270m+126)(2n)^2 + (228m+96)(2n)) - 12m \quad \text{(D15)}$$

(see Ref. [121]), where $m$ is the number of iterations needed to compute the reciprocal square root. Results with $10^{-8}$ error, for example, have been obtained using $m=3$ Newton iterations [121].

The rotation $R_0$ must be implemented using the value of this square root as the phase. For this purpose, for each qubit of the binary representation of the square root, we apply the following. If $(\sqrt{p})_j = \hat{n}_j = (\mathbb{1}_j - \hat{Z}_j)/2$ represents the value of the $j$th bit in the fixed point representation of the square root in increasing order of significance, and $k$ is the index of the ancilla qubit, then

$$e^{-i\sqrt{p}\hat{X}_k t}. \quad \text{(D16)}$$

We implement this in parallel, on all the parts of the superposition, using the following operation:

$$\prod_{j=0}^{(2n)-1} e^{-i2^j\hat{n}_j\hat{X}_k t}. \quad \text{(D17)}$$

Thus, we can implement $e^{-itH_{\text{even,odd}}}$ by applying this circuit to the ancilla qubit. A controlled single-qubit rotation can be implemented using two controlled NOT gates and single-qubit rotations. Thus, the cost of the simulation of this term would involve

$$N_{\mathrm{cR}} = 2 \times (2n). \tag{D18}$$

We then disentangle the ancilla and restore the mode registers to prepare the application of the other three parts of the Hamiltonian, which follows the same steps as the amplitude transfers but in reverse and which therefore costs $N_{\mathrm{tr}}$.

Therefore, the total cost of a single exponential, e.g., $H_{\mathrm{odd,odd}}$, is

$$N_{\mathrm{CNOT/Exp}} = (2 \times N_{\mathrm{tr}}) + N_{\times} + N_{\mathrm{SQRT}} + N_{\mathrm{CR}}, \tag{D19}$$

where $N_{\mathrm{CNOT/Exp}}$ scales as $\mathcal{O}(\log_2(N_{\max})^2)$.

The total number of $N_{\mathrm{CNOT/Exp}}$ per beam splitter depends on the target error. For this purpose, we use Eq. (65) of Ref. [244], similarly to our discussion in Sec. VIII B in Ref. [57]. There are various contributions to the final error of implementing $e^{-i\theta\hat{H}}$. First, there is the Trotter error from the division of the Hamiltonian into noncommuting even and odd terms shown in Eq. (D5). We set this error contribution to $\epsilon_{\mathrm{Trot}} = \epsilon/2$. Second, there is the error coming from the square-root calculation, which is multiplied by the Hamiltonian prefactor $\theta$. To set this error also to $\epsilon/2$ total error, we choose $m$ such that the error per square root is $\epsilon/(2\theta N_{\mathrm{exp}})$. Therefore, to calculate the total number of gates for a given $N_{\max}$, $\theta$, $\epsilon$, $p$, we first calculate $N_{\mathrm{exp}}$ from Eq. (D6). From this result, we calculate the target error per square root $\epsilon/(2\theta N_{\mathrm{exp}})$, which in turn determines the order $m$ of the Newton iterations. Next, we evaluate the number of gates per square root, $N_{\mathrm{SQRT}}$, and the number of gates per exponential, $N_{\mathrm{gates,exp}}$. Finally, the number of gates for a single beam splitter is given by $N_{\mathrm{exp}} \times N_{\mathrm{CNOT/Exp}}$. Asymptotically, this method scales as $\mathcal{O}(N_{\max}^{1+(1/2p)} \log_2(N_{\max})^2)$, assuming that, for large $N_{\max}$, the scaling of $m$ is weak with $N_{\max}$. This scaling is almost linear in $N_{\max}$.

### 2. Precalculated square roots

Next, we discuss classically precalculating the square roots, storing them in a qubit register and then performing controlled rotations. While this at first seems attractive, this method requires as many controlled rotations as there are possible Fock states, and each controlled rotation would need to be controlled on the mode registers being in a specific Fock state, i.e., multiqubit controlled operations.

The operation will need to be controlled on the value of registers of modes $a$ and $b$. Assuming both registers have the same length $n = \log(N_{\max})$, where $N_{\max}$ is the maximum Fock state, the total length of the combined register on which the operation needs to be controlled is $2n$. One n-Toffoli gate, i.e., a qubit rotation controlled on a qubit register of size $n$ being in state $|\psi\rangle$, has a theoretical lower bound of $2n$ CNOT gates [335]. Therefore, the number of CNOT gates for a (2n)-Toffoli gate is $2(2n)$. For each value of the $a$ and $b$ registers, we need to first apply the (2n)-Toffoli gate, then apply the single qubit rotation $R_0(-it\sqrt{2(j+1)\times(2l+1)})$ to the ancilla [coming from Eq. (D13)], and then uncompute by applying the exact same (2n)-Toffoli gate again. Therefore, the total CNOT cost of each check-and-apply operation (or Toffoli rotation) is

$$N_{\mathrm{CNOT/tR}} = 2(2(2n)). \tag{D20}$$

The total number of times this search-and-apply operation needs to be implemented corresponds to the total number of possible states of the registers $a$ and $b$. Each register can have $2^n$ possible states. Therefore, the joint register of size $2n$ can have $2^{2n}$ possible states, which must all be checked at the cost of $N_{\mathrm{CNOT/tR}}$ for each check-and-apply operation. The total CNOT cost of covering all possible combinations with the Toffoli rotation is

$$N_{\mathrm{CNOT/c,tR}} = (2^{2n}) \times 2(2(2n)). \tag{D21}$$

Therefore, the total cost of implementing a single exponential when precalculating the square roots and controlling the rotations on the state of the registers is

$$\begin{aligned} N_{\mathrm{CNOT/Exp}} &= (2 \times N_{\mathrm{CNOT/transfer}}) + N_{\mathrm{CNOT/c,tR}} \\ &= 6n(2^{2n}) + 46n - 10. \end{aligned} \tag{D22}$$

The total number of CNOTs per beam splitter is $N_{\mathrm{exp}} \times N_{\mathrm{CNOT/Exp}}$. The leading-order term of $N_{\mathrm{CNOT/Exp}}$ is $6n(2^{2n})$ with $n = \log_2(N_{\max})$; therefore, $N_{\mathrm{CNOT/Exp}}$ scales as $\mathcal{O}(N_{\max}^2 \log_2(N_{\max}))$.

Asymptotically, this method scales as $\mathcal{O}(N_{\max}^{3+(1/2p)} \times \log_2(N_{\max}))$, which is worse than the almost-linear scaling in the *in situ* square-root setup. For small cutoffs, it is advantageous, however, to precalculate the square roots and check which one to apply because the prefactors for calculating them *in situ* are so large. The two methods are compared in Fig. 17.

### 3. Controlled parity, controlled beam splitter

A controlled beam splitter CBS could be implemented very simply by controlling every single CNOT on the control qubit, which can be done by transforming the CNOTs into Toffoli gates at the cost of six CNOT per Toffoli gate. Therefore, this process increases the gate count by a factor of 6, which is a lower bound as it does not count the additional CNOTs that come from controlling the single qubit gates.

However, a cheaper way is to use the controlled parity trick, i.e., sandwiching the beam splitter between two $\exp(i(\pi/2)\hat{\sigma}^z\hat{n})$ gates. Decomposing $\hat{n}$ in binary, $\hat{n} = \sum_{k=0}^{\lceil \log_2(N_{\max}) \rceil} 2^k(1-\hat{\sigma}_j^z)/2$. Each $e^{-i\theta\sigma_1^z\sigma_2^z}$ gate takes two CNOT gates to implement ($\mathrm{CNOT}_{1,2} R_Z(2\theta)_2 \mathrm{CNOT}_{1,2}$). Hence, a conditional parity requires

$$N_{\mathrm{CNOT/CP}} = 2\left\lceil \log_2(N_{\max}) \right\rceil \tag{D23}$$

CNOT gates, and only an additional $4\lceil \log_2(N_{\max}) \rceil$ CNOT gates are required to turn a BS into a CBS gate.

## APPENDIX E: QUANTUM SIGNAL PROCESSING

In this appendix, we first review the technique of QSP [303,304,336]. In short, QSP is a parametric algorithm for generating a polynomial transformation of a subsystem. Subsequently, we will discuss how QSP may be generalized to apply a polynomial transformation to a linear operator, a procedure known as a quantum eigenvalue transform (QET) [75,303,304]. This ability to polynomially transform an arbitrary linear operator is incredibly powerful and has been shown to provide a universal framework from which nearly all known quantum algorithms may be derived [75,299].

### 1. Quantum signal processing

QSP [303,304,336] works by interleaving a signal operator $W$ and a signal processing operator $S$. These operators are taken to be SU(2) rotations about different axes, where the signal operator is a rotation through a fixed angle $\theta$ and the signal processing operator is a rotation through a variable angle $\phi$. Conventionally, $W$ is taken to be an $x$ rotation and $S$ a $z$ rotation.

Following standard conventions, we define the signal operator as

$$W(x) = \begin{bmatrix} x & i\sqrt{1-x^2} \\ i\sqrt{1-x^2} & x \end{bmatrix}, \tag{E1}$$

which is an $x$ rotation through an angle $\theta = -2\cos^{-1} x$ ($x \in [-1, 1]$). Similarly, we define the signal processing operator as

$$S(\phi) = e^{i\phi Z}, \tag{E2}$$

which is a $z$ rotation through an angle $-2\phi$. Then, introducing a set of QSP phases $\vec{\phi} = (\phi_0, \phi_1, \ldots, \phi_d) \in \mathbb{R}^{d+1}$, one can construct the QSP sequence $U^{\vec{\phi}}$, which is defined as the following interspersed sequence of $W$ and $S$:

$$U^{\vec{\phi}} = S(\phi_0) \prod_{k=1}^{d} W(x) S(\phi_k) = e^{i\phi_0 Z} \prod_{k=1}^{d} W(x) e^{i\phi_k Z}. \tag{E3}$$

Incidentally, the matrix elements of this sequence are polynomials in $x$, parametrized by the QSP phases $\vec{\phi}$. In particular, Ref. [336] proved that

$$\begin{aligned} U^{\vec{\phi}} &= e^{i\phi_0 Z} \prod_{k=1}^{d} W(x) e^{i\phi_k Z} \\ &= \begin{bmatrix} P(x) & iQ(x)\sqrt{1-x^2} \\ iQ^*(x)\sqrt{1-x^2} & P^*(x) \end{bmatrix}, \end{aligned} \tag{E4}$$

where $P(x)$ and $Q(x)$ are polynomials that obey

1. $\deg(P) \le d, \quad \deg(Q) \le d-1$
2. $P(x)$ has parity $d$ mod 2, $Q(x)$ has parity $(d-1)$ mod 2
3. $|P(x)|^2 + (1-x^2)|Q(x)|^2 = 1, \ \forall\ x \in [-1, 1]$. (E5)

The reverse is also true: For polynomials $P(x)$ and $Q(x)$ that obey these conditions, the corresponding phases $\vec{\phi}$ can be determined efficiently with a classical algorithm [337].

This result indicates that we can prepare polynomials of $x$ by projecting into a matrix element of $U^{\vec{\phi}}$, such as $P(x) = \langle 0|U^{\vec{\phi}}|0\rangle$. This polynomial may be explicitly obtained by measuring the transition probability of a state evolving under $U^{\vec{\phi}}$, that is, the probability $p = |\langle 0|U^{\vec{\phi}}|0\rangle|^2 = P(x)^2$. Moreover, for a target polynomial of degree $d$, the corresponding QSP circuit is constructed from $O(d)$ quantum gates [i.e., depth $O(d)$].

### 2. Quantum eigenvalue transformation

The construction of QSP may be generalized to apply a polynomial transformation to an operator, that is, to its eigenvalues, which is fittingly known as the quantum eigenvalue transformation. To understand this heuristically, note the following interpretation of QSP: QSP begins with a matrix $W(x)$ that block encodes $x$ as $x = \langle 0|W(x)|0\rangle$, and by interspersing this matrix with rotation operators parametrized by $\vec{\phi}$, QSP generates an operator that block encodes a polynomial transformation of $x$ as $P(x) = \langle 0|U^{\vec{\phi}}|0\rangle$. Thus, in terms of block encodings, this procedure performs the polynomial transformation $x \mapsto P(x)$.

Paralleling this scenario, one may instead begin with a unitary $U$ that block encodes a matrix $A = \sum_\lambda \lambda|\lambda\rangle\langle\lambda|$:

$$U = \begin{bmatrix} A & \cdot \\ \cdot & \cdot \end{bmatrix} \tag{E6}$$

(this is only possible if $\|A\| \le 1$; if this condition is not met, one can instead block encode a rescaled matrix $(1/\|A\|)A$). More generally, we assume that $A$ can be accessed with a projector $\Pi$ as $A = \Pi U \Pi$. Equation (E6) uses the simple projector $\Pi = |0\rangle\langle 0| \otimes I$.

In the QET, one intersperses $U$ with a rotation operator that acts as a $z$ rotation within each eigenspace of $A$. This operator may be expressed as $\Pi_\phi := e^{i\phi(2\Pi - I)}$ for a rotation angle $\phi$ and can be straightforwardly realized with access to a $\Pi$-conditional NOT operation [299]. Paralleling the QSP sequence of Eq. (E4), one defines the following QET sequence, which effectively performs QSP within each eigenspace and thus outputs a polynomial transformation of $A$:

$$U^{\vec{\phi}} = \Pi_{\phi_0} \prod_{k=1}^{d} U\Pi_{\phi_k}$$
$$\Rightarrow \Pi U^{\vec{\phi}} \Pi = \sum_{\lambda} P(\lambda)|\lambda\rangle\langle\lambda| = P(A), \qquad \text{(E7)}$$

where, as before, $P(A)$ is a degree $d$ polynomial that obeys the conditions of Eq. (E5). In terms of block encodings, QET maps the matrix $A$ to a polynomial transformation $P(A)$. As above, for a target polynomial of degree $d$, the circuit depth is $O(d)$.

In practice, the QET sequence can be applied to an initial state $|\psi\rangle$ to prepare a state $(1/\sqrt{\langle\psi|P(A)^2|\psi\rangle})P(A)|\psi\rangle$, with successful preparation occurring with probability $p = \langle\psi|P(A)^2|\psi\rangle$. In the next section, we use this strategy to prepare the ground state of a Hamiltonian.

## APPENDIX F: GROUND-STATE PREPARATION WITH QUANTUM SIGNAL PROCESSING

In the main text, we used VQE to estimate the ground state of the $\mathbb{Z}_2$ gauge theory and the Schwinger model. However, recent work suggests that QSP provides an alternative approach to ground-state preparation with guaranteed fidelity and success probability [305]. It has also been shown that QSP can naturally be performed on hybrid CV-DV quantum processors [306], rendering it applicable to lattice gauge theory simulations on these platforms. Note that we use "QSP" and "QET" interchangeably here, as their distinction is only in the argument they apply to.

In the setting of ground-state preparation, the goal is to prepare an approximation to the ground state $|\psi_0\rangle$ of a Hamiltonian $H$ with fidelity $1-\epsilon$ and probability of success $1-\vartheta$. We assume access to an easily preparable state $|\phi\rangle$ that has overlap $|\langle\psi_0|\phi\rangle| \geq \gamma$ with the ground state, and we also knowledge that the spectral gap of $H$ is at least $\Delta$.

This problem was studied in the context of near-term quantum devices in Ref. [305]. There, the authors present a ground-state preparation algorithm that applies QSP to the time-evolution operator $U = e^{-iH}$. This algorithm is a more feasible input model than applying QSP directly to the Hamiltonian because a block encoding of the time-evolution operator may be realized as a controlled-$U$ and can be implemented with Trotterization, for instance, whereas achieving a block encoding of the Hamiltonian is significantly more challenging. With this input model, QSP produces polynomials in the argument $\cos(H/2)$, i.e., polynomials $P(\cos(H/2))$.

As an overview, the algorithm presented in Ref. [305] prepares the ground state by first constructing a QSP polynomial that approximates a projector onto the ground-state subspace. Specifically, this polynomial approximates a step function that filters out energies greater than the ground-state energy. The precise construction of this polynomial necessitates that it has degree $\tilde{O}((1/\Delta)\log(1/\epsilon))$. This QSP sequence is then applied to the initial state $|\phi\rangle$, which filters out its contributions from high-energy states and thus projects out an approximation to the ground state. This protocol succeeds if we measure the correct block of the QSP sequence, the probability of which we boost by repeating this protocol numerous times to achieve a sufficiently high success probability.

Let us study this case more closely to determine the number of queries required for this algorithm. First, we note that Ref. [305] assumes access to the time-evolution operator $e^{-iH}$, where the eigenvalues of $H$ lie in the range $[\eta, \pi-\eta]$ for some $\eta > 0$; if this is not satisfied, one can instead simulate $e^{-i(Ht+\phi)}$ for a time $t$ and phase $\phi$ such that the effective Hamiltonian $\tilde{H} = Ht + \phi$ obeys this constraint. For instance, we consider the operator $e^{-i[(\pi/2)+((\pi/2)-\eta)(H/\|H\|)]}$, such that the effective Hamiltonian $\tilde{H} = (\pi/2) + [(\pi/2)-\eta](H/\|H\|)$ is guaranteed to have eigenvalues in the range $[\eta, \pi-\eta]$.

Moreover, we also assume knowledge of a parameter $\mu$ that is at least $\Delta/2$ away from the ground-state energy $E_0$ and the first-excited-state energy $E_1$: $E_0 \leq \mu - \Delta/2 < \mu + \Delta/2 \leq E_1$ (where $\Delta$ is the aforementioned lower bound on the spectral gap). If no $\mu$ is known *a priori*, Ref. [305] provides an additional algorithm to determine such a value. The goal then is to construct a QSP polynomial that provides an approximation to a step function whose discontinuity occurs at a value $x' \in (\mu - \Delta/2, \mu + \Delta/2)$, such that this step function filters out eigenvalues greater than the ground-state energy. If we want the polynomial to suffer an error of at most $\varepsilon$, then it must obey

$$\begin{aligned} |P(x) - 1| &\leq \varepsilon, \qquad x \in [\sigma_+, \sigma_{\max}], \\ |P(x)| &\leq \varepsilon, \qquad x \in [\sigma_{\min}, \sigma_-], \\ |P(x)| &\leq 1, \qquad x \in [-1, 1], \end{aligned} \qquad \text{(F1)}$$

where

$$\sigma_\pm = \cos\left(\frac{\mu \mp \Delta/2}{2}\right),$$
$$\sigma_{\min} = \cos\left(\frac{\pi - \eta}{2}\right), \qquad \sigma_{\max} = \cos\left(\frac{\eta}{2}\right). \qquad \text{(F2)}$$

[Recall that the argument of our polynomial is $\cos(H/2)$.]

Because the step function exhibits a discontinuity, it can only be approximated by a polynomial away from the discontinuity at $x'$, i.e., outside of the range $(x' - \delta/2, x' +$ $\delta/2)$ for some width $\delta$. It is well established that this function can be approximated to within error $\varepsilon$ by a polynomial of degree $O[(1/\delta)\log(1/\varepsilon)]$ [246]. To satisfy the above constraints, we desire the width of the region around the discontinuity to be $\delta = \sigma_+ - \sigma_- = O(\Delta)$. Hence, it will suffice to choose a QSP polynomial of degree $O[(1/\Delta)\log(1/\varepsilon)]$. Equivalently, this is the depth of the QSP sequence, i.e., the number of coherent queries to the controlled-$U$ operator. For completeness, we note that, very recently, Ref. [338] demonstrated how to reduce this query complexity by a constant factor by rescaling the spectral gap.

The action of this QSP polynomial on the initial state $|\phi\rangle$ is to retain the amplitude of the ground state up to a factor of greater than or equal to $1-\varepsilon$ while rescaling the magnitude of the component orthogonal to the ground state by a factor of less than or equal to $\varepsilon$. Schematically, we may write the initial state as $|\phi\rangle = \gamma|\psi_0\rangle + \sqrt{1-\gamma^2}|\psi_\perp\rangle$, where $|\psi_\perp\rangle$ accounts for components orthogonal to the ground state. Then, upon application of the QSP sequence, $|\phi\rangle$ is mapped to a state like $|\phi'\rangle = (1-\varepsilon)\gamma|\psi_0\rangle + \varepsilon\sqrt{1-\gamma^2}|\psi_\perp\rangle$ up to normalization (here, we have assumed the worst-case error). We desire that the corresponding normalized state has fidelity $\langle\psi_0|\phi'\rangle/\langle\phi'|\phi'\rangle \geq 1-\epsilon$ with the ground state, for which it suffices to select $\varepsilon = \sqrt{\epsilon}(\gamma/\sqrt{1-\gamma^2}) = O(\sqrt{\epsilon}\gamma)$.

The above QSP polynomial is only applied to the initial state if the correct block of the QSP sequence is accessed, which occurs with probability of at least $\langle\phi'|\phi'\rangle = (1-\varepsilon)^2\gamma^2 + (1-\gamma^2)\varepsilon^2 = O(\gamma^2)$. We can use repetition to boost this success probability to a value close to 1. To ensure that the probability of failure is $\leq \vartheta$ after $N$ repetitions, we require $(1-O(\gamma^2))^N \leq \vartheta$, for which $N = O[(1/\gamma^2)\log(1/\vartheta)]$ suffices.

In summary, this algorithm for QSP-based ground-state preparation prepares an approximation to the ground state with fidelity greater than or equal to $1-\epsilon$ by repeatedly applying QSP to an initial state. Each QSP instantiation uses one copy of the initial state $|\phi\rangle$ and requires $O[(1/\Delta)\log(1/\epsilon\gamma)]$ coherent queries to the controlled time-evolution operator $U = e^{-iH}$. In order to ensure success probability greater than or equal to $1-\vartheta$, we repeat this procedure $O[(1/\gamma^2)\log(1/\vartheta)]$ times.

### 1. Application to the $\mathbb{Z}_2$-Higgs model

Let us now look at the requirements of QSP-based ground-state preparation in a practical application. We desire a rough order-of-magnitude estimate on the gate depth of the QSP circuit. Consider the $\mathbb{Z}_2$ Hamiltonian with $n=3$ sites and couplings $J=1=g=U$, as studied in Sec. VI C. Numerics indicate that the normalized effective Hamiltonian $\tilde{H} = (\pi/2) + [(\pi/2)-\eta](H/\|H\|)$ (with $\eta = 0.05$) has a spectral gap $\Delta \approx 0.24$, which necessitates a width $\delta = \sigma_+ - \sigma_- \approx 0.089$. In this setting, the VQE in Sec. VI C begins with an initial state $|\phi\rangle$ that numerically has overlap with the ground state $\gamma \approx 0.36$ and ultimately produces an approximation of the ground state with fidelity $\epsilon \approx 2\times10^{-2}$. To achieve a similar performance with QSP-based ground-state preparation, we must construct a polynomial approximation to the desired step function with error $\varepsilon = \sqrt{\epsilon}(\gamma/\sqrt{1-\gamma^2}) \approx 0.055$. Empirical calculations of the polynomial constructions outlined in Refs. [246,339] indicate that such a polynomial may be constructed with degree $d \approx 50$.

As we discussed in the main text, the QSP circuit for this polynomial is rather deep and expensive, significantly more than our VQE approach. For instance, employing even the first-order Trotterization as outlined in Sec. VI B, a controlled time evolution would require around three controlled qubit-boson gates per site, for a total of greater than or around $10^4$ qubit-boson gates in series. Thus, this approach will only be practical and competitive with VQE once qubit-boson gate infidelities much less than $10^{-4}$ can be achieved, in contrast to the current qubit-boson fidelities of less than or around 0.9995. Moreover, in this Trotterization approach, the circuit depth will grow linearly with the number of sites, $n$. In addition, the number of repetitions of the QSP circuit will generally scale exponentially as $O(1/\gamma^2) = 2^{O(n)}$ without *a priori* knowledge of the ground state, which is to be expected because generic ground-state preparation is QMA hard [308].

These points indicate that, at least for NISQ-era machines, the VQE provides a more viable approach to ground-state preparation than QSP-based algorithms, despite the elegance and provably accurate performance of the latter. For more details on using QSP to prepare ground states of lattice gauge theories, see Ref. [338], which provides a comprehensive exposition of this subject.

## APPENDIX G: COMPARISON OF THE IMPACT OF NOISE BETWEEN QUBIT-BOSON AND QUBIT HARDWARE

In this appendix, we discuss the influence of boson mode decay on observables and compare it to the influence of qubit decay in a Fock-binary encoding of a boson mode. The primary problem we address here is the following: It can be shown that under mode decay, when starting in a Fock state $|N\rangle$, the probability $p_N$ to remain in this state decays as

$$p_N(t) = \exp(-\gamma N t) \tag{G1}$$

with time $t$. This result seems to pose challenges to our approach as it seems to imply that high Fock states cannot be useful for quantum simulation. However, in this appendix, we show that the decay rate of some relevant *observables* does not increase with $N$. Moreover, an alternative approach using qubits in a Fock-binary encoding does not yield qualitative advantages and even leads to highly nonlinear effective bosonic noise processes.

Consider a Markovian noise process with no Hamiltonian evolution and a set of Lindblad operators $\hat{L}_j$ with rate $\gamma$. In this case, the time evolution of an observable $\hat{O}$ is given by

$$\frac{\mathrm{d}}{\mathrm{d}t}\hat{O} = \gamma\left(\sum_j \hat{L}_j^\dagger \hat{O}\hat{L}_j - \frac{1}{2}\{\hat{L}_j^\dagger \hat{L}_j, \hat{O}\}\right). \tag{G2}$$

In the following, we consider the time evolution of the occupation number operator $\hat{n}$ and its square $\hat{n}^2$ starting from a Fock state $|N\rangle$.

### 1. Oscillator

For a bosonic mode encoded in an oscillator, the leading loss process is mode decay $\hat{L} = \hat{a}$ with rate $\kappa$. For the time evolution of the mode occupation, $\hat{O} = \hat{n}$, we then find from Eq. (G2) that

$$\langle \hat{n}(t)\rangle = \langle \hat{n}(0)\rangle e^{-\kappa t}. \tag{G3}$$

Moreover, inserting $\hat{O} = \hat{n}^2$, we find

$$\frac{\mathrm{d}}{\mathrm{d}t}\langle \hat{n}^2(t)\rangle = -2\kappa\langle \hat{n}^2(t)\rangle + \kappa\langle \hat{n}(t)\rangle. \tag{G4}$$

Inserting Eq. (G3) and solving the resulting inhomogeneous differential equation, we find

$$\langle \hat{n}^2(t)\rangle = (\langle \hat{n}^2(0)\rangle - \langle \hat{n}(0)\rangle)e^{-2\kappa t} + \langle \hat{n}(0)\rangle e^{-\kappa t}. \tag{G5}$$

Hence, $\langle \hat{n}^2(t)\rangle$ decays with a rate less than $2\kappa$. This upper bound is independent of which Fock states the initial state occupies, showing that even when large Fock states are occupied, meaningful observables can be extracted in the presence of mode decay.

### 2. Qubits in Fock-binary encoding

In the Fock-binary encoding, the bosonic mode is represented by a set of $N_q$ qubits, introducing a cutoff $N_{\max} = 2^{N_q} - 1$ in the boson number. The boson number operator is given by

$$\hat{n} = \sum_{j=0}^{N_q-1} 2^j \hat{n}_j^q, \tag{G6}$$

with $\hat{n}_j^q = |1\rangle\langle 1|$ the projector on the $|1\rangle$ state of the qubit. To compare to the oscillator decay, we consider qubit decay $\hat{L}_j = \hat{\sigma}_j^-$ with rate $\gamma$. First, we again consider the decay of the expectation value of the number operator, finding

$$\langle \hat{n}(t)\rangle = \langle \hat{n}(0)\rangle e^{-\gamma t}. \tag{G7}$$

Surprisingly, we find the same expression as for the mode, with $\kappa$ replaced by $\gamma$. Mathematically, this result is due to the fact that $\hat{n}$ is linear in $\hat{n}_j^q$. More interestingly, the evolution of the square number operator

$$\hat{n}^2 = \sum_{i,j=0}^{N_q-1} 2^{i+j}\hat{n}_i^q \hat{n}_j^q \tag{G8}$$

reduces to the evolution of the correlations $\hat{n}_i^q \hat{n}_j^q$. Inserting $\hat{O} = \hat{n}_i^q \hat{n}_j^q$ into Eq. (G2), we find

$$\langle \hat{n}_i^q \hat{n}_j^q(t)\rangle = \langle \hat{n}_i^q \hat{n}_j^q(0)\rangle e^{-2\gamma t} \quad \text{for } i \neq j. \tag{G9}$$

The case $i = j$ reduces to the evolution of $\hat{n}_i^q$ because $(\hat{n}_i^q)^2 = \hat{n}_i^q$. Splitting Eq. (G8) into these two cases, we find

$$\langle \hat{n}^2(t)\rangle = e^{-2\gamma t}\sum_{i\neq j=0}^{N_q-1} 2^{i+j}\langle \hat{n}_i^q \hat{n}_j^q(0)\rangle + e^{-\gamma t}\sum_{i=0}^{N_q-1} 4^i \langle \hat{n}_i^q(0)\rangle. \tag{G10}$$

From this expression, we see that the decay of the squared number operator depends strongly on the initial state. For instance, when starting in a Fock state with $N = 2^n$ and $n$ integer, we find $\langle \hat{n}^2(t)\rangle = N^2 e^{-\gamma t}$, i.e., a decay with rate $\gamma$. By contrast, a Fock state $N = 2^n - 1$ evolves according to

$$\langle \hat{n}^2(t)\rangle = N^2 e^{-2\gamma t} + (e^{-\gamma t} - e^{-2\gamma t})\frac{1}{3}(N^2 + 2N), \tag{G11}$$

which for large $N$ and short time leads to a decay rate $\frac{5}{3}\gamma$. This result shows that, in the Fock-binary encoding, the influence of qubit decay is highly nonlinear in the Fock-state number. In more complicated dynamics of superpositions of Fock states, this will lead to results that are hard to interpret as the constituent Fock-state coefficients will decay with different rates that are not simple functions of $N$. To illustrate this case further, we show the $1/e$ decay time of $\langle \hat{n}^2\rangle$ in Fig. 31. While the oscillator mode decay leads to a smooth effective decay rate of $\langle \hat{n}^2\rangle$ as a function of $N$, the qubit decay leads to highly erratic behavior.

Notably, bosonic modes stored in high-Q cavities exhibit decay rates $\kappa$ that are roughly 10 times lower than the decay rates $\gamma$ observed in typical qubits. Consequently, despite the fact that for $\kappa = \gamma$ the Fock-binary encoding results in the effective decay time of $\langle \hat{n}^2\rangle$ being up to twice as long, in reality, bosonic modes in cavities exhibit a significantly longer effective decay time for this observable.

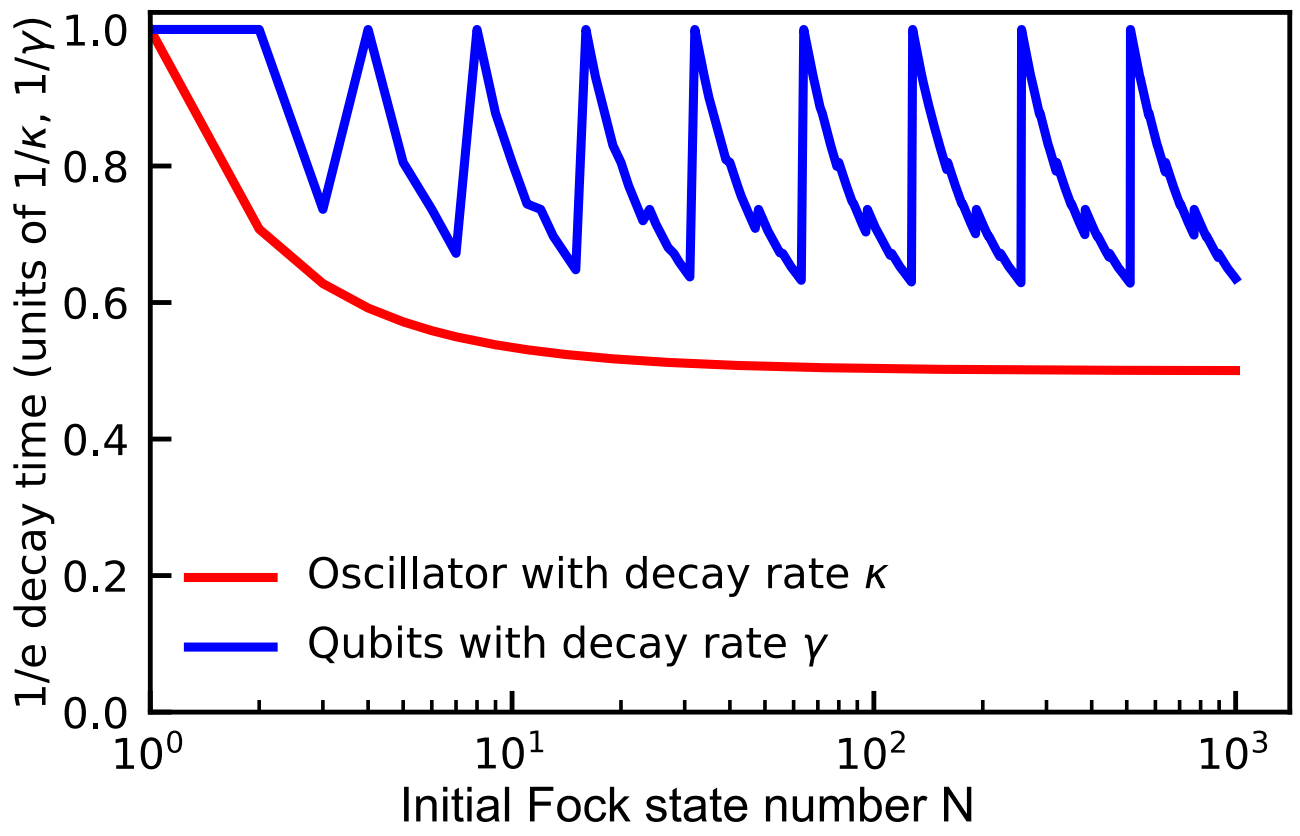

FIG. 31. Effective decay time of $\langle\hat{n}^2\rangle$. We start from an initial Fock state with occupation $N$ and evolve under mode decay or qubit decay. Note that $\langle\hat{n}\rangle$ exhibits a decay time of $1/\kappa$ and $1/\gamma$, respectively, independent of $N$.

## APPENDIX H: ESTIMATE FOR CUTOFF-DEPENDENT GATE FIDELITIES

In this appendix, we calculate the expected fidelity for two of the most crucial qubit-boson gates (beam splitter and controlled parity) taking into account photon loss (dephasing has been shown to be negligible [187]) for the cavity modes, and decay and dephasing for the transmon qubit. This initial noise model does not include other nonidealities such as Kerr and cross-Kerr terms. In particular, we are interested in determining the dependence of the fidelity on the largest Fock state $K$ involved in the dynamics. Note that, while strictly speaking, this is not a cutoff as there is no physical hard barrier preventing the system from occupying higher states, in many settings it is possible to exclude occupation of higher states, e.g., when simulating dynamics with a conserved global occupation number. Below, we find that this noise model does not, in principle, prohibit the relatively high-fidelity realization of these gates.

We use a Lindblad master equation to model the dynamics,

$$\dot{\rho}(t) = -i[\hat{H},\hat{\rho}(t)] + \sum_i \gamma_i\left(\hat{L}_i\rho L_i^\dagger - \frac{1}{2}\{\hat{L}_i^\dagger\hat{L}_i,\hat{\rho}\}\right), \quad \text{(H1)}$$

where $\gamma_i$ are the rates of the jump operators $\hat{L}_i$. We take into account amplitude damping $\hat{L}_\mathrm{a} = \hat{\sigma}^-$ and phase damping $\hat{L}_\mathrm{p} = \hat{\sigma}^z$ if a qubit is involved, and mode decay $\hat{L}_\mathrm{d} = \hat{a}$. The corresponding times are noted $T_a$, $T_p$, and $T_d$. As shown in many works, e.g., Ref. [187], mode dephasing $\hat{L} = \hat{n}$ can be neglected to a good approximation. For transmons in most cases, $T_2 \approx T_1$, implying a dephasing rate of $\gamma_p = \gamma_a/2$. Finally, these incoherent errors usually dominate gate infidelities in superconducting qubit architectures [217].

Concretely, to obtain an estimate for immediately accessible fidelities with existing hardware, we use qubit lifetimes of $T_a = 200$ μs [195] and cavity lifetimes of $T_d = 2$ ms [91]. To obtain an estimate for realistic fidelities, which will be reached using state-of-the-art developments, we use the numbers for tantalum transmon qubits of $T_1 = 2$ ms [237], and cavity lifetimes demonstrated in 2023 of $T_d = 34$ ms [187]. We also note that, in 2020, there were demonstrations of very long-lived photon states in cavities, which would extend these predictions even further, $T_d = 2$ s [188], although quantum control of such long-lived cavities using ancilla qubits has not been demonstrated (but it has been demonstrated in cavities with lifetimes above 20 ms [340]).

We define the fidelity as the average return fidelity:

$$\mathcal{F} = \frac{1}{\mathcal{D}}\sum_{j=1}^{\mathcal{D}}\langle j|e^{i\hat{H}t}(\mathcal{L}(|j\rangle\langle j|))e^{-i\hat{H}t}|j\rangle, \quad \text{(H2)}$$

where $\mathcal{L}(\cdot)$ denotes evolution until time $t$ under the Lindbladian and $\mathcal{D}$ is the total Hilbert space dimension. We set $\mathcal{D}$ by choosing a maximum Fock state at which to cut off the local Hilbert space of each oscillator. We then obtain the cutoff-dependent fidelity.

### 1. Conditional parity

We consider a conditional-parity gate implemented with time evolution for a period $t = \pi/\chi$ under the Hamiltonian

$$\hat{H}_\mathrm{BS} = \frac{\chi}{2}\hat{n}\hat{Z}. \quad \text{(H3)}$$

We show the dependence of the fidelity as a function of the maximum Fock state for various combinations of $\chi$ (see, e.g., Refs. [195,341]) in Fig. 32(a), the mode decay rate and qubit decay rate, using the numbers quoted above. We find that, when high qubit-boson couplings can be combined with the longest-lived cavities and transmons, fidelities greater than or equal to 99.995% may be achieved, even when the maximum Fock-state number is up to 20.

### 2. Beam splitter

We also consider a beam-splitter gate with Hamiltonian

$$\hat{H}_\mathrm{BS} = g_\mathrm{BS}(\hat{a}^\dagger b + \mathrm{H.c.}), \quad \text{(H4)}$$

where couplings up to $g_\mathrm{BS} = 2\pi 25$ MHz have been demonstrated with linear inductive couplers [90]. To be specific, we consider time evolution until $t = \pi/4g_\mathrm{BS}$, corresponding to a $\sqrt{\mathrm{iSWAP}}$ gate in the 0–1 subspace.

As for the conditional-parity gate, we show the fidelities for beam splitters in Fig. 32(b). In particular, our numerical result for maximum Fock state 1 and the green configurations is similar to the one experimentally achieved in

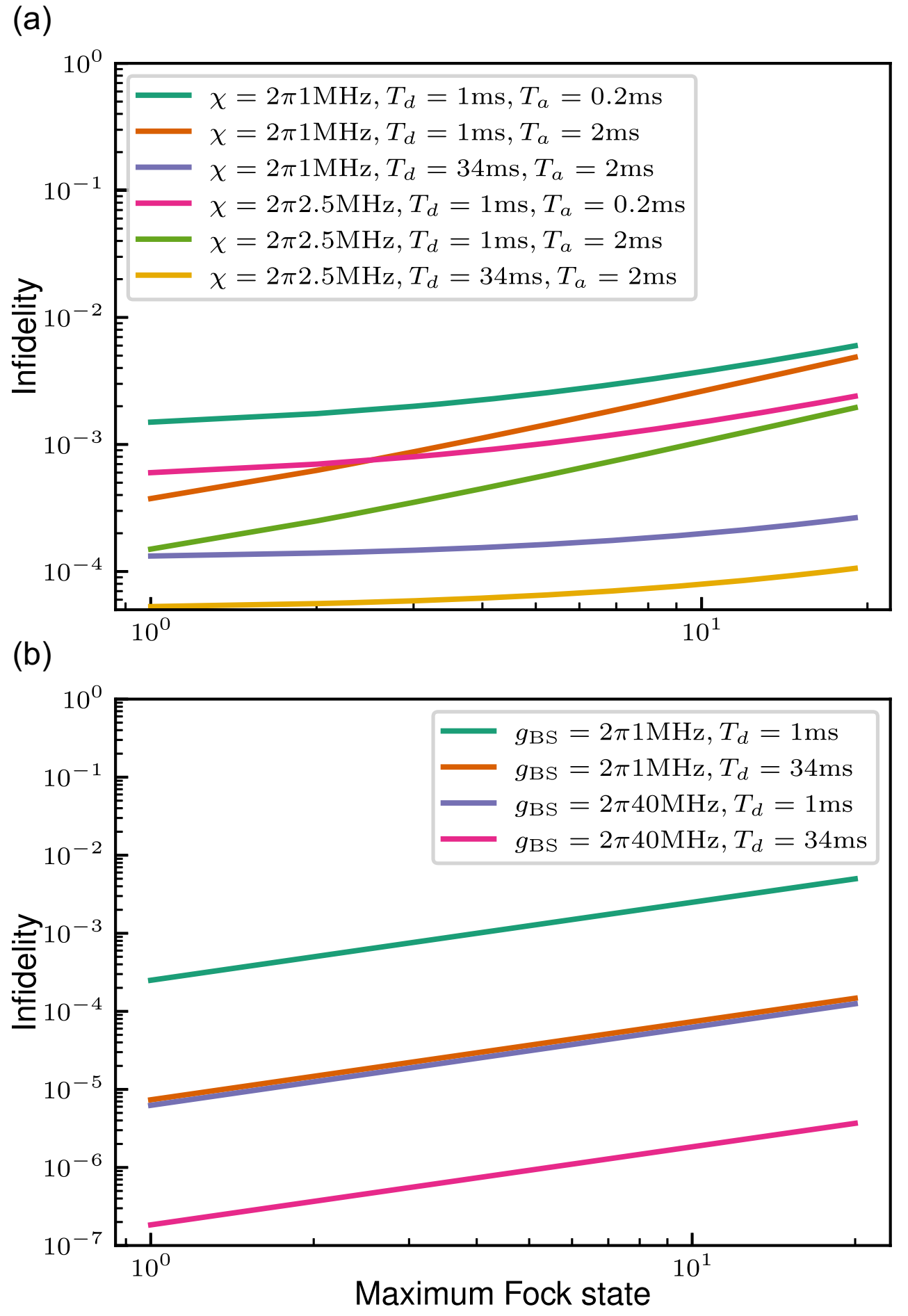

FIG. 32. Fidelities of bosonic gates as a function of maximum Fock state. (a) Conditional parity. The infidelities for maximum Fock state 10 are $8 \times 10^{-3}$, $5 \times 10^{-3}$, $5 \times 10^{-4}$, $1.4 \times 10^{-3}$, $8 \times 10^{-4}$, and $8 \times 10^{-5}$ from top to bottom. (b) Beam splitter. The infidelities for maximum Fock state 10 are $1.4 \times 10^{-3}$, $1 \times 10^{-4}$, $4 \times 10^{-5}$, and $3 \times 10^{-6}$, from top to bottom.

Ref. [217], showing that our model is realistic (our infidelity is slight larger, which comes from the fact that we consider the full Hilbert space and not just the $|01\rangle$ and $|10\rangle$ subspaces). For the most optimistic configurations, we find fidelities exceeding 99.999% up to Fock state 10.

### 3. $Z_2$ model circuits

We show in Fig. 33 the circuit fidelity for simulating the $Z_2$ lattice gauge theory up to maximum Fock state 20 in the bunch phase, using several different combinations of beam splitter and controlled parity parameters. We find that even currently achievable decay times achieve high fidelities.

In these estimates, we make the common assumption that maximum-mode Fock states are set to $|20\rangle$ and transmons are limited by decay (i.e., dephasing is set to half of the lifetime). We plot a range of results for weak to strong dispersive couplings from $2\pi 1$ MHz [209] to $2\pi 2.5$ MHz [90], and beam-splitter strengths for $g_{\text{BS}}$ from $2\pi 1$ MHz

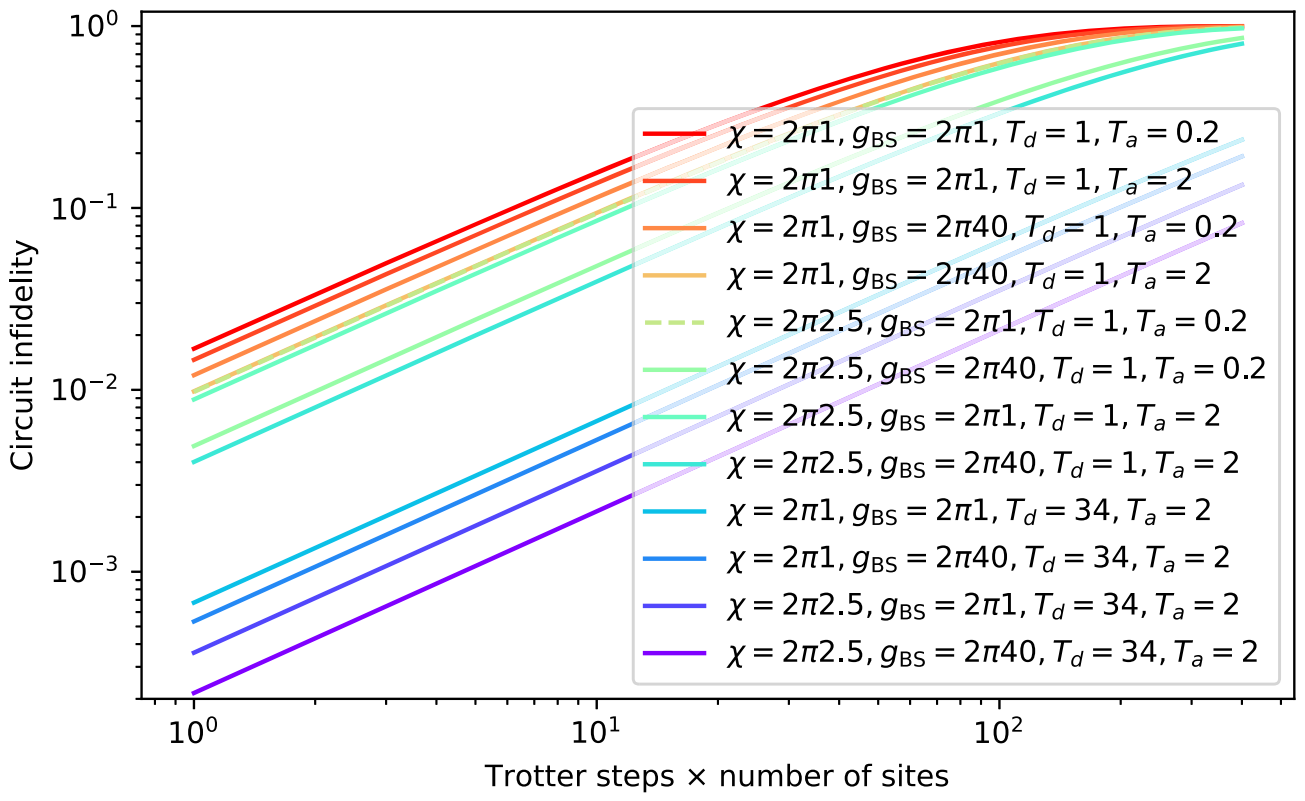

FIG. 33. Proposal for quantum advantage over classical and qubit methods. We show circuit fidelity for time evolution under the $Z_2$ lattice gauge theory, considering a maximum 20 bosons per site, as well as dispersive coupling $\chi$ and beam-splitter strength $g_{\text{BS}}$ in MHz. Cavity and transmon lifetimes, $T_d$ and $T_a$, respectively, are given in ms. Note that this infidelity is due to experimental noise, so it does not include Trotter error.

[91] to $2\pi 40$ MHz [90]. We also include existing, easily accessible experimental lifetimes for the cavity of $T_d = 1$ ms [91] and the transmon of $T_a = 200$ μs [195], as well as individually demonstrated cavity lifetimes of $T_d = 34$ ms [187] and recently demonstrated tantalum transmon qubit lifetimes of $T_1 = 2$ ms [237].

We find that there is a large jump in fidelities when the cavity-mode lifetime increases from 1 ms to 34 ms, as can be seen by the lower four blue and purple lines in Fig. 33 compared to the upper green, orange, and red lines. The two optimal fidelities are found for a dispersive interaction of $\chi = 2\pi \times 2.5$ MHz, and state-of-the-art cavity and transmon lifetimes. Throughout the results, we find very little dependence on the beam-splitter fidelity, which is due to the conditional parity having a consistently lower fidelity (because it includes the transmon, which usually has a lower lifetime than the cavity mode) and also to the circuit for the $Z_2$ model gauge-invariant hopping requiring twice as many conditional parities as beam splitters. Lastly, we observe from Fig. 33 that, among the lower circuit fidelities, the lowest half have the weaker dispersive interaction $\chi = 2\pi \times 1$ MHz, as opposed to $\chi = 2\pi \times 2.5$ MHz, which makes sense as the gate operations will be slower at smaller $\chi$.

For the pessimistic lower-bound transmon and cavity lifetimes and weak dispersive and beam-splitter couplings that we chose for the benchmarks and proof-of-principle experimental proposals, the circuit fidelity for three Trotter steps and 20 sites is 62% if we restrict the maximum Fock state to $|10\rangle$. If the maximum Fock state in every operation is $|20\rangle$, we obtain a final circuit fidelity of 38%. For state-of-the-art experimentally demonstrated tantalum transmons [237] of lifetimes over $T_a = 2$ ms, and larger $\chi = 2\pi \times 2.5$ MHz [90], we have 60% final circuit fidelity for three

Trotter steps at maximum Fock state 20. Including the same LINC technology for the beam splitters, we achieve $g_{BS} = 40$ MHz [90], which leads to 48% circuit fidelity for ten Trotter steps and maximum Fock state 20. Alternatively, using weak dispersive and beam-splitter strengths but with state-of-the-art transmon ($T_a = 2$ ms [237]) and cavity lifetimes of $T_d = 34$ [187] (note that the highest state-of-the-art cavity lifetime demonstrations have $T_1 \geq 2$ s [188], and we do not use these here as other sources of noise may appear at this stage), for maximum Fock state $|20\rangle$, it is possible to obtain a staggering 88% fidelity for ten Trotter steps and 77% fidelity for 20 Trotter steps, which we predict is well beyond classical methods. With tantalum transmons, long-lived cavity modes, and the LINC technology demonstrated in Ref. [90], we predict final circuit fidelities for 20 sites, each with maximum occupation $|20\rangle$ as may arise in the $Z_2$ lattice gauge theory in the bunching regime, of an impressive 92% for 20 Trotter steps and 96% for ten Trotter steps.

Despite all these lifetimes and coupling strengths having already been individually demonstrated experimentally, these fidelity results are clearly beyond classical capabilities, which should encourage future developments in this hardware such as incorporating the already demonstrated very-long-lived photon states in cavities, $T_d = 2$ s [188], which would extend these predictions even further.